\documentclass[english]{spimufcphdthesis}

\usepackage[utf8]{inputenc}
\usepackage{minitoc}

\usepackage{amsmath}

\declarethesis[]{Tunable Bloch surface waves devices}{12 December 2017}{XXX}

\addauthor[tatiana.kovalevich@femto-st.fr]{Tatiana}{Kovalevich}

\addjury{Emiliano}{Descrovi}{Rapporteur}{Professeur
 à L'École Polytechnique de Turin}
\addjury{Rachel}{Grange}{Rapporteur}{Professeur à L'École Polytechnique Fédérale de Zurich}
\addjury{Hans Peter}{Herzig}{Examinateur}{Professeur à L’École Polytechnique Fédérale de Lausanne}
\addjury{Matthieu}{Roussey}{Examinateur}{Professeur à L'université de l'Est de la Finlande}
\addjury{Hervé}{Maillotte}{Examinateur}{Directeur de Recherche CNRS, UBFC, FEMTO-ST}
\addjury{Thierry}{Grosjean}{Co-Directeur de thèse}{Chargé de Recherche CNRS, UBFC, FEMTO-ST}
\addjury{Maria-Pilar}{Bernal}{Directeur de thèse}{Directeur de Recherche CNRS, UBFC, FEMTO-ST}

\thesisabstract[english]{This thesis is devoted to develop tunable devices on the base of one-dimensional photonic crystals (1DPhC) which can sustain Bloch surface waves (BSWs). First, we explore the possibilities to control the BSW propagation direction with polarization of incident light. In this case we manufacture additional passive structures such as gratings on the top of the 1DPhC, which are working both as a BSW launcher and polarization–controlled “wave-splitters”. We test this type of launcher in air and in water as an external medium. Then, we demonstrate the tunability of the BSW by adding an active layers into the multilayer stack. Here a crystalline X-cut thin film lithium niobate (TFLN) is used to introduce anisotropic properties to the whole 1DPhC. Different ways to manufacture 1D PhCs with LiNbO$_3$ on the top would be described. Finally, we explore the concept of the electro-optically tuned BSW.}

\thesiskeywords[english]{Bloch surface waves, gratings, polarization-selective devices, thin films, lithium niobate, photonic crystals}

\thesisabstract[french]{Cette thèse est consacrée au développement de dispositifs accordables sur la base de cristaux photoniques unidimensionnels qui peuvent supporter des ondes de surface de Bloch (BSW. Tout d'abord, nous explorons les possibilités de contrôler la direction de propagation des BSW par le biais de la polarisation de la lumière incidente. Dans ce cas, nous gravons sur le dessus du cristal photonique 1D des structures passives de type réseau, qui permettent à la fois de coupler la lumière incidente aux BSWs et de se comporter comme une lame séparatrice ultracompacte contrôlée par la polarisation lumineuse. Nous avons testé ce type de coupleur sur des cristaux photoniques 1D fonctionnant dans l’air et dans l’eau. Ensuite, nous démontrons l'accordabilité des BSWs en ajoutant une fine couche active dans la structure photonique multicouche. Il s’agit d’un film mince de niobate de lithium monocristallin qui permet d’introduire des propriétés anisotropes dans le cristal photonique 1D. Différentes façons de fabriquer des cristaux photoniques 1D  contenant du niobate de lithium monocristallin ont été développées dans le cadre de ce travail. Ces travaux nous ont permis d’explorer le concept de contrôle électro-optique des BSWs.}

\thesiskeywords[french]{Les ondes de surface de Bloch, le(s) réseau(x), les dispositifs sélectifs en polarisation, les films minces, le niobate de lithium, les cristaux photoniques}

\Set{speciality}{Optics}

\declareupmtheorem{mytheorem}{My Theorem}{List of my Theorems}

\begin{document}

%%--------------------
%% The following line does nothing until
%% the class option 'nofrontmatter' is given.
%\frontmatter

%%--------------------
%% The following line permits to add a chapter for "acknowledgements"
%% at the beginning of the document. This chapter has not a chapter
%% number (using the "star-ed" version of \chapter) to prevent it to
%% be in the table of contents

%%%%%%%%%%%%%%%%%%%%%%%%%%%%%%%%%%%%%%%%%%%%%%%%%%%%%%%%%%%%%%%%%%%%%%%%%%%%%%%%%%%%%%%%%%%%%%%%%%%%%%%%%%%%%%%%%%%%%%%%%%%%%

\chapter*{Acknowledgements.}

I am using this opportunity with pleasure to acknowledge everyone who supported me throughout the PhD studies. I am thankful for aspiring guidance and friendly advices during the work.  First and foremost, I would like to express my sincere gratitude to my supervisors for the opportunity to get enrolled into the PhD study on the very exiting and interesting topic, for their continuous support and help during my PHD study with great patience, motivation and knowledge. I appreciate all their contributions of time, ideas and fundings to make my PhD finish successfully. I'm genially inspired by my first supervisor Maria-Pilar Bernal, as a great example of a woman in science, who is managing to be not only a bright scientist, but also a great director, manager, and also a mother of a family. I'm tremendously grateful to my co-supervisor Mr. Thierry Grosjean for his unbelievable and bright ideas and guidance though all this tree years. 
He is always available for the scientific discussions. He is approachable with just a knock on his door. I also want to acknowledge my co-supervisor Mr. Hans Peter Herzig from EPFL university. Without his support it would be hardly possible to realize manufacturing and characterization of many photonic crystals required for this PhD.

I am grateful to the jury members, Prof. Emiliano Descrovi, Prof. Rachel Grane, Prof. Matthieu Roussey, Dr. Hervé Maillotte, for spending time to correct my thesis and providing their critical reviews on it.

I would additionally like to thank Prof. Matthieu Roussey for manufacturing 1D PhCs, valuable comments in paper writing, all the support and optics-related discussions.

Many thanks to Dr. Valery Konopsky from Institute of spectroscopy and Russian Academy of science for his clear and valuable explanations of the Bloch surface waves basis and for his guidance in understanding of fundamentals of one-dimensional photonic crystals.

I would also like to thank the engineers in FEMTO-ST, MIMENTO for their help in the experiments. Many thanks to Miguel Suarez for help in the set up maintenance, to Roland Salut and Gwenn Ulliac for FIB manufacturing, to Loran Robeer, Djaffar Belharet, Ludovic, Marina, Tristan, David for advises and help with clean room processes.

Many thanks to Dr. Myun-Sik Kim and Dr. Richa Dybey from EPFL for valuable discussions and help with experiments. Their support made the collaboration between FEMTO-ST and EPFL really strong.

Thanks to Prof. Huihui Lu for his guidance and assistance in FDTD simulations.

Many thanks to Mme. Sophie Marguier for her support in administrative affairs for the period of my stay and for her valuable advices on all other kind of matters.

I am also thankful to Mme. Val{\'e}rie Fauves and Dr. Christophe Gorecki for their help in administrative work.

Also many thanks to the friends I met in FEMTO-ST for their help and accompany during the last three years. Thanks to Remi, Masha, Anurupa, Etienne, Bogdan, Mengjia, Luis, Florant, Nyha, Gautie, Wendy, Vahan. Thanks to Anton for being a great friend in Switzerland during all my work and tourism visits. We had a really good time together and it will always be a precious memory for me. 

Special thanks to Olga, who came to France together with me and to Ayham for their especially precious support. We shared a lot of happy moments.

Finally, I want to express my greatest gratitude to my parents and my sister for their continuous love and support throughout the difficult phases. I always miss you, my dearest people, and you are always there for me despite any distance.

At the end, I would like to acknowledge COLEGIYM SMYLE for financial support of the work and NANO-LN company for providing such a necessary thin film lithium niobate wafers.

%%%%%%%%%%%%%%%%%%%%%%%%%%%%%%%%%%%%%%%%%%%%%%%%%%%%%%%%%%%%%%%%%%%%%%%%%%%%%%%%%%%%%%%%%%%%%%%%%%%%%%%%%%%%%%%%%%%%%%%%%%%%%%%%%%%%%%%%%%%%%%%%%%%%%%%

%%--------------------
%% Include a general table of contents
\dominitoc
\tableofcontents
\mtcsetrules{*}{off}

%%--------------------
%% The content of the PhD thesis
\mainmatter

\chapter*{General introduction.}

The experimental property of Bloch surface waves (BSW) in periodic layered media has been studied first in 1978 by Yeh, Yarif and Cho \cite{Yeh:1977}. It has been shown that a truncated dielectric multilayer can sustain surface waves under particular illumination conditions. BSWs are non-radiative electromagnetic waves confined at the interface between a truncated periodic dielectric multilayer and a surrounding media \cite{yu:13}.

The main idea underlying this work is to use the advantages of Bloch Surface Waves (BSW), such as low losses \cite{Angelini:14} and high field confinement \cite{wu:14} at the one dimensional photonic crystal (1DPhC) surface for various optical devices. Depending on the materials used for the multilayer, the BSW based platform can be applied to integrated optics, biosensing or particle trapping. The main goal is to develop new designs of 1DPhC with tunable properties for these applications. 

We investigate both passive and active tunable devices. 
	
In the first case we change the properties of the incident light, such as polarization, in order to control the propagation direction of the BSW. Additionally instead of a standard coupling through the prism in Kretschmann configuration we use a grating as a BSW launcher with incident light at normal incidence. We reach a higher degree of control of Bloch Surface Waves by controlling the direction of the flow of light at the sub-wavelength from the macro-scale. We introduce new functions with simplified technologies. Finally, we reach a high degree of integration. Avoiding the use of prism allows us to create a compact, plug-and-play overall system with no need of searching for a good coupling angle.
	
In the second case we create 1DPhCs loaded by an active top layer material. We fabricate different nanostructures on the top of the multilayers, such as gratings, waveguides and 2D photonic crystals and study the propagation properties of the BSW, excited on the top of the 1DPhC. 
	
As an active material of multilayer we choose lithium niobate. This is a material, which is sometimes called 'the silicon of photonics' \cite{Manzo:13}. It is an artificial ferroelectric crystal belonging to the 3 m crystallographic group \cite{Kosters:09}. It is characterized by large pyroelectric, piezoelectric, acousto-optic, nonlinear and electro-optic coefficients and is one of the key materials for the fabrication of integrated optical devices \cite{weis:85}. LiNbO$_3$ is widely used in photonics, with a broad range of applications ranging from acoustic-wave transducers and filters in mobile telephones, to optical modulators and wavelength converters in fiber telecommunication systems, to name just a few. Recent advances in the micro and nano-structuration of the LiNbO$_3$, involving domain engineering by electric field poling techniques as well as ion-exchange processes and etching techniques, for instance FIB milling, enable the fabrication of both linear and nonlinear functionalities. Therefore we select this material for our studies.

This manuscript is divided in 7 chapters.

In a first chapter, a brief overview of BSW history and applications as well as a comparison of BSWs to some similar techniques will be provided. I will introduce the state of the art and motivation of this work. It concerns mainly control of BSW propagation on the surface of 1D photonic crystal and different ways to achieve the tunability. 

Analytical and numerical methods used to predict a BSW behavior will be described in the 2nd chapter. We will also introduce the main equipment and the technological process used for fabrication of the devices. 

The following two chapters would be dedicated to the polarization control and grating coupling of BSW.
 
I will describe a grating coupler which can also work as a “wave-splitter”, depending on the polarization of the incident light. Here, the polarization dependent tunability and directional excitation of the BSWs is demonstrated. The switch/coupler is designed to work as a highly miniaturized plug-and-play element for BSW based all-optical integrated platform.

In the 4th chapter I will show results obtained during preliminary experiments on particle manipulation by grating coupled BSWs in water. These tests are the first steps on the way to create a BSW based device for long distance directional guidance of particles on a chip for biological applications.  

The second part of this work is dedicated to LiNbO$_3$ BSW devices. We investigate different ways to introduce all the range of interesting optical properties of this material into the BSW sustaining platform.

In the chapter 5 we demonstrate anisotropic properties of the first 1DPhC with thin film of LiNbO$_3$ as a part of the multilayer. We have manufactured and characterized for the first time the 1DPhC with a crystalline X-cut thin film LiNbO$_3$ (TFLN). The 1DPhC is able to sustain Bloch surface waves (BSW) at the TFLN and SiO$_2$ interface. Additional tunability of excitation angle was observed due to anisotropic properties of TFLN.

In the chapter 6 the manufacturing and characterization process of the first 1DPhCs with TFLN with the BSW at the TFLN and air interface will be described. The devices were manufactured in 2 different configurations: on the membrane and on the glass support.  The characterization was performed in the visible part of spectrum. Such a designs of the 1DPhC open up the possibility of creating BSW based active tunable devices and allows to explore the nonlinear properties of LiNbO$_3$.  Additional clean room processes which were developed in our group in order to improve the top surface quality of 1DPhC in order to work at higher wavelengths (1550nm) will be also described here.

Chapter 7 will contain the information about first passive structures (waveguides) on TFLN based 1DPhC, which were manufactured by Focused Ion Beam (FIB) milling. Clear coupling and guiding of the BSW was observed. In addition, we have developed a specific electrode deposition technique, which can be used in the future for electro-optical modulation of BSWs. Here we will demonstrate the feasibility of creating electrically tuned BSW based devices.

The manuscript will end with a general conclusion, where also the perspectives for future work would be described.

\chapter{State of the art for tunable BSW devices.}

Bloch surface waves (BSWs) \cite{Yeh:1977} are electromagnetic surface waves excited at the interface between a truncated periodic dielectric multilayer and a surrounding medium. Recent developments in the thin film deposition make it possible to grow multilayer media with well-controlled periodicities and layer thicknesses. This opens a way to an actual creation of photonic crystals suitable for BSW propagation. Repeatable high precision clean room processes are now successfully used to generate BSW-based 'flatland' and optical devices. In this chapter the overview of BSW's history and comparison of BSW with surface plasmon polaritons (SPP) and with on-silicon photonics will be provided. We will explain how already existing systems may be improved by additional nano-structuring or by implementing new materials into the multilayer. 

\vspace*{0.2cm}
\minitoc

\section{Bloch surface waves. Origin and applications.}

\begin{figure}[!t]
\begin{center}
\includegraphics[width=4in]{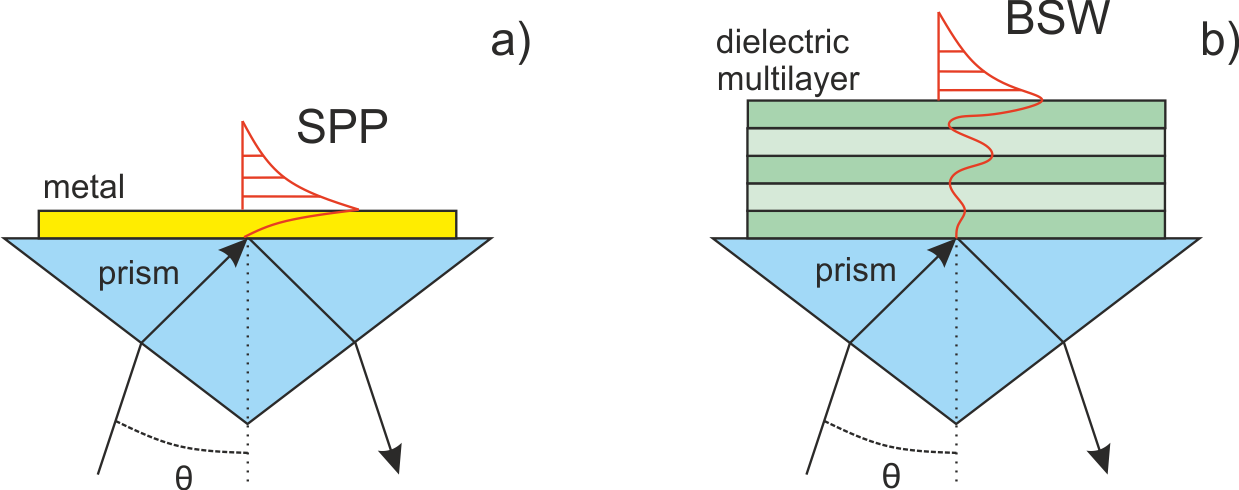}
\caption{Scheme of surface plasmon polariton (SPP) (a) and Bloch surface wave (b) both excited in Kretschmann configuration.}
\label{fig:Fig1_1_SPP_BSW}
\end{center}
\end{figure}

Periodic optical media and specifically stratified periodic structures play an important role in a number of applications, including high reflective mirrors \cite{Turner:1966}, phase matching in nonlinear optical applications \cite{Helmy:2006}, optical birefringence \cite{Tyan:1997} and many others. Such periodic structures may be also applied for the excitation of optical surface states - Bloch surface waves. These waves are electromagnetic surface waves excited at the interface between a truncated periodic dielectric multilayer and a surrounding medium and share some common characteristics with SPPs, which are nowadays pivotal in a large variety of applications ranging from optical circuitry to biological sensors \cite{Tanaka:2003,Homola:2008}. Their dispersion is located within forbidden bands of the 1DPhC, beyond the light line of the external homogeneous medium. As a consequence, the field envelope develops inside the 1DPhC and decays exponentially in the homogeneous medium. The sketch of SPP and BSW excited in Kretschmann configuration is shown in Fig. \ref{fig:Fig1_1_SPP_BSW}.

\begin{figure}[!b]
\begin{center}
\includegraphics[width=3.5in]{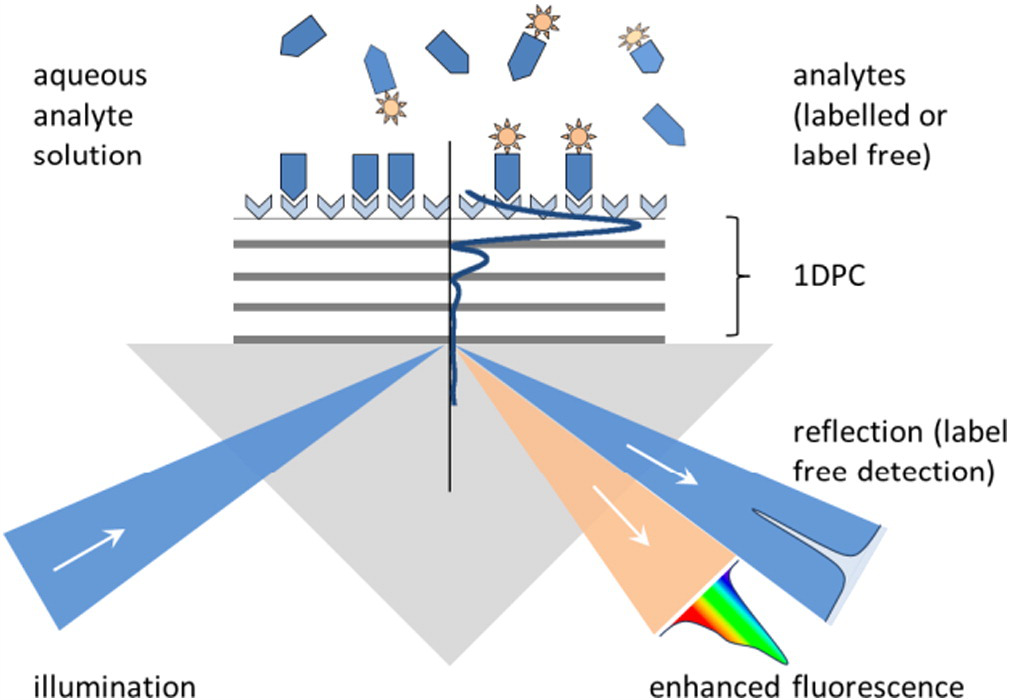}
\caption{Scheme of a Bloch surface wave optical biosensor for label-free and fluorescence detection \cite{Munzert:2017}.}
\label{fig:Fig1_2_BSW_for_biosensor}
\end{center}
\end{figure}

However, SPPs undergo  important limitations due to the ohmic losses of metal such as modest propagation length and heat dissipation \cite{kuttge:2008}. Being evanescent surface waves which are produced at the free top surface of a non-absorbing dielectric multilayer stack \cite{sinibaldi:12,konopsky:07,giorgis:10,guo:10,rivolo:12}, BSWs show dramatically enhanced propagation lengths up to the millimeter range \cite{Dubey:j16} (only few tens of micrometers for SPPs \cite{weeber:01}) and provide new optical opportunities such as the possibility to obtain TE or TM-polarized surface waves (SPPs are limited to TM polarization)\cite{konopsky:2010}. Owing to these new properties, BSWs have found numerous applications in vapor sensing \cite{michelotti:10}, fluorescence detection \cite{Fluo1,Fluo2,Fluo3} and imaging \cite{frascella:13,angelini:13,sinibaldi:14}, and integrated optics \cite{R1,R2,E1,E2,yu:14,descrovi:10}. Being inspired from already existing SPP applications, BSWs adopt and improve biosensing techniques \cite{E3,sinibaldi:15,konopsky:13,farmer:12}. A scheme of a Bloch surface wave optical biosensor for label-free and fluorescence detection is shown in Fig. \ref{fig:Fig1_2_BSW_for_biosensor}. 

\begin{figure}[!t]
\begin{center}
\includegraphics[width=3in]{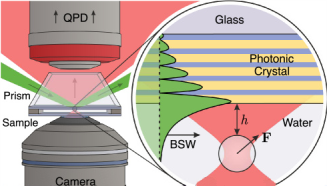}
\caption{Scheme of optical trapping. A particle is trapped in optical tweezers (red) in the vicinity of a 1DPhC, and its displacement is measured when BSW radiation (green) is turned on. In the inset, a magnified view is presented, and the BSW electric field distribution is indicated in green. \cite{Shilkin:2015}.}
\label{fig:Fig1_optforses}
\end{center}
\end{figure}

A wide choice of dielectric materials also gives a big freedom for BSWs to be used in different scientific domains: for example, in studies of unique material's properties, such as absorption studies of graphen monolayers \cite{Sreekanth:2012}, in surface imaging and in nonlinear optics, for phase-matched third-harmonic generation via doubly resonant optical surface modes \cite{Konopsky:2016} and second harmonic generation in polymeric nanowires \cite{Wang:2017}. BSW can also be used in optical trapping and manipulation \cite{Shilkin:2015} for direct measurements of forces induced by Bloch surface waves (see Fig. \ref{fig:Fig1_optforses}).

\begin{figure}[!b]
	\begin{center}
		\includegraphics[width=4in]{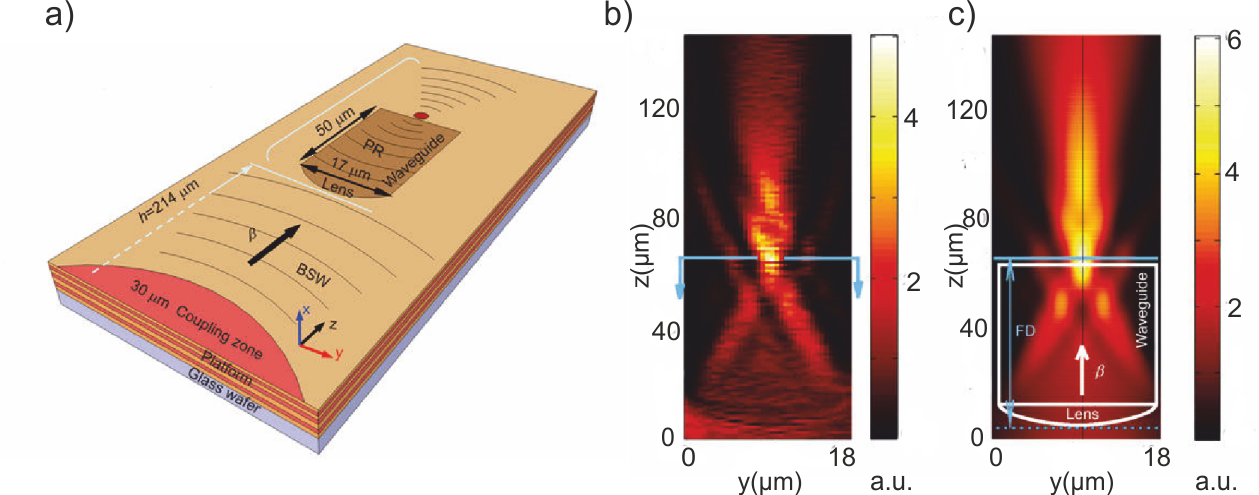}
		\caption{(a) Schematic of the 2D lens: the BSW propagating on the surface of the platform is focused by an ultra-thin photoresist layer patterned as a planoconvex lens appended to a waveguide. (b) Measured and (c) simulated near-field intensity, showing light propagation through the 2D planoconvex lens. A focused spot is observed 60 $\mu$m from the first interface of the lens, and the positions of the 2D planoconvex lens and the attached waveguide are marked by closed solid lines. $\beta$ is the propagation constant of the BSW \cite{yu:14}.}
		\label{fig:Fig1_3_flatland1}
	\end{center}
\end{figure}

\begin{figure}[!t]
	\begin{center}
		\includegraphics[width=\linewidth]{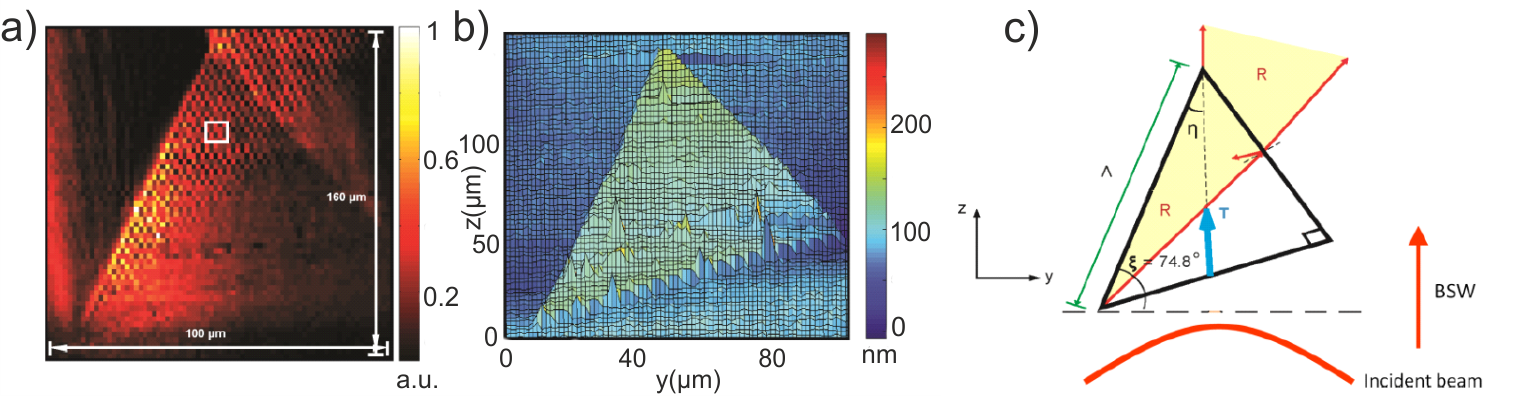}
		\caption{(a) Measured near-field intensity, showing BSW propagation through the 2D right-angled isosceles prism. (b) Prism topography. (c) Schematics of BSW propagation with respect to the prism \cite{yu:13}.}
		\label{fig:Fig1_3_flatland2}
	\end{center}
\end{figure}

\begin{figure}[!b]
	\begin{center}
		\includegraphics[width=\linewidth]{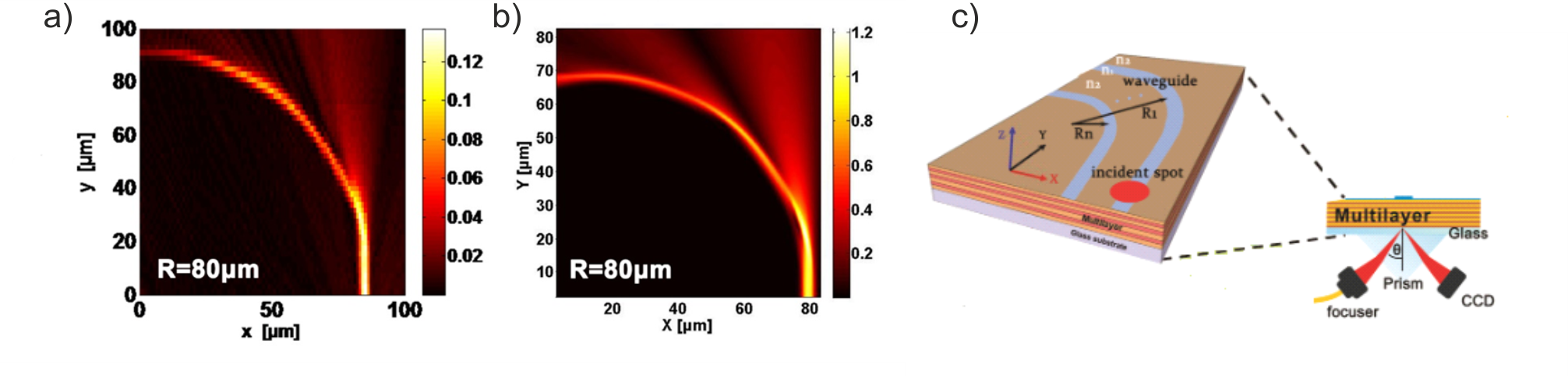}
		\caption{Experimental (a) and theoretical (b) intensity distribution of bending waveguides having different radii values. R=80$\mu$m. (c) Schematics of BSW coupling into the curved waveguides\cite{wu:14}.}
		\label{fig:Fig1_3_flatland3}
	\end{center}
\end{figure}

Long propagation distance of BSW naturally leads to the idea of creating on-chip all optical integrated components, already known as a concept of BSW based 'flatland' optics. At the moment it is possible to focus [Fig. \ref{fig:Fig1_3_flatland1}], to deviate [Fig. \ref{fig:Fig1_3_flatland2}] and to guide [Fig. \ref{fig:Fig1_3_flatland3}] light in free 2D space.

In 2014 L. Yu et. al. \cite{yu:13, yu:14} proposed and demonstrated the concept where the dielectric multilayer platform, wherein the manipulation of BSWs can be achieved and various 2D photonic components can be reproduced. These flat photonic components can have arbitrary shapes, which are generally more difficult to fabricate and control in 3D. They demonstrated that 2D photonic components can be implemented by coating an in-plane shaped ultra-thin polymer layer on a multilayer. The presence of the polymer modifies the local effective refractive index, enabling a direct manipulation of the BSW. By locally shaping the geometries of the 2D photonic components, the BSW can be diffracted, focused, coupled and resonated with 2D freedom. A multiheterodyne SNOM was used to monitor the near-field behavior on the examined platform and a lens-shaped or prism-shaped polymer layer. They demonstrated that the shaped polymer layer can be considered as a complete 2D component, which allows arbitrary reshaping of BSW propagation on the top of multilayer platform.

Also in 2014 X. Wu et. al. studied BSWs propagation in curved waveguides.
They demonstrated the propagation of light in ultra-thin curved polymer waveguides having different radii fabricated on a BSWs sustaining multilayer. A phase-sensitive multi-parameter near-field optical measurement system (MH-SNOM), which combines heterodyne interferometry and SNOM, is used for the experimental characterization. They experimentally shown that when light goes through the curved part of the waveguide, energy can be converted into different modes. The superposition and interference of different modes lead to a periodically alternating bright and dark beat phenomenon along the propagation direction.

These all-dielectric chips with multilayers may be compared with already widely used ones in on silicon photonics integrated optics. In comparison to BSW based chips this field has a much longer history and as a consequence more developed fabrication technology. At this moment BSWs cannot compete with on-chip guided optics, in terms of wafer scaled manufacturing and light coupling, even though at the sub-wavelength scale they provide a high degree of freedom and accuracy in light manipulation. Bloch surface waves based components may bring higher degree of integration and may reach in a near future the centimeter scale.

\section{Tunable Bloch surface waves. Passive and active devices.}

In this work we use the advantages of BSWs such as high field confinement at the one dimensional photonic crystal (1DPhC) surface and long propagation length in order to create different optical devices. We design various multilayers, which are additionally nano-structured or include active materials. The main goal is to develop new tunable designs of 1DPhC for integrated optics, particle manipulation and, as a perspective, for biosensing and nonlinear applications. 

It can be seen that a new generation of optoelectronic and biomedical technologies is expected to emerge from the development of on-chip BSW-based integrated systems \cite{R3,R4,R5,R6,R7}. A large panel of dielectric micro-elements, such as 2D-lens \cite{yu:14}, 2D prism and grating \cite{yu:13}, ridge waveguides \cite{wu:14,descrovi:10}, ring resonators \cite{R8,R9,menotti:15,dubey:16}, is nowadays available for controlling the propagation of BSWs, which opens the perspective of the engineering of optical circuits and functions in ultra-compact and ultrafast architectures. These tiny 2D optical components, deposited at the top surface of the 1DPhC, modify the BSW propagation direction by locally patterning the effective refractive index \cite{R10,R11,R12}. Reaching dynamical tunability with this technological approach would require either real-time modification of the component shapes or their refractive indices. 

BSW coupling is also a crucial step in the development of dielectric surface optical functions. Directional coupling of an incident beam to a BSW can be achieved using a bulky prism coupler in the Kretschmann configuration \cite{kovalevich:16} or more rarely by using Otto architecture \cite{michelotti:10}. Such techniques lead to BSW launching in a straightforward way but they suffer from low compactness and tuning the BSW propagation direction requires the rotation of both the prism and the input laser beam, which remains hard to implement. Tuning the propagation direction of optical surface waves in a ultra-compact and versatile architecture would greatly extend the level of manipulation of BSW but it remains a real challenge.

In 2014 A. Angelini et.al. \cite{Angelini:14} studied focusing and extraction of light mediated by Bloch surface waves. They presented the use of surface modes on patterned dielectric multilayers to deliver electromagnetic power from free-space to localized volumes and vice versa. Due to low-loss energy transfer proper periodic ring structures were shown to provide a subwavelength focusing of an external radiation onto the multilayer surface. Also the radiated power from emitters within the ring center was shown to be efficiently beamed in the free-space, with a well-controlled angular divergence.

In principle, when an external radiation of a specific wavelength $\lambda$ is impinging on the circular grating from air, it undergoes diffraction according to the Bragg's law. The first-order (+1) diffracted radiation has a wavevector component $k_{T+1}$ parallel to the multilayer surface as given by $k_{T+1} = k_{T_0} + K$, where $k_{T_0}$ is the wavevector component of the incident radiation parallel to the multilayer surface. When $k_{T+1}$ matches the BSW wavevector, energy coupling between the incident radiation and the surface mode can occur. 

\begin{figure}[!t]
	\begin{center}
		\includegraphics[width=5.4in]{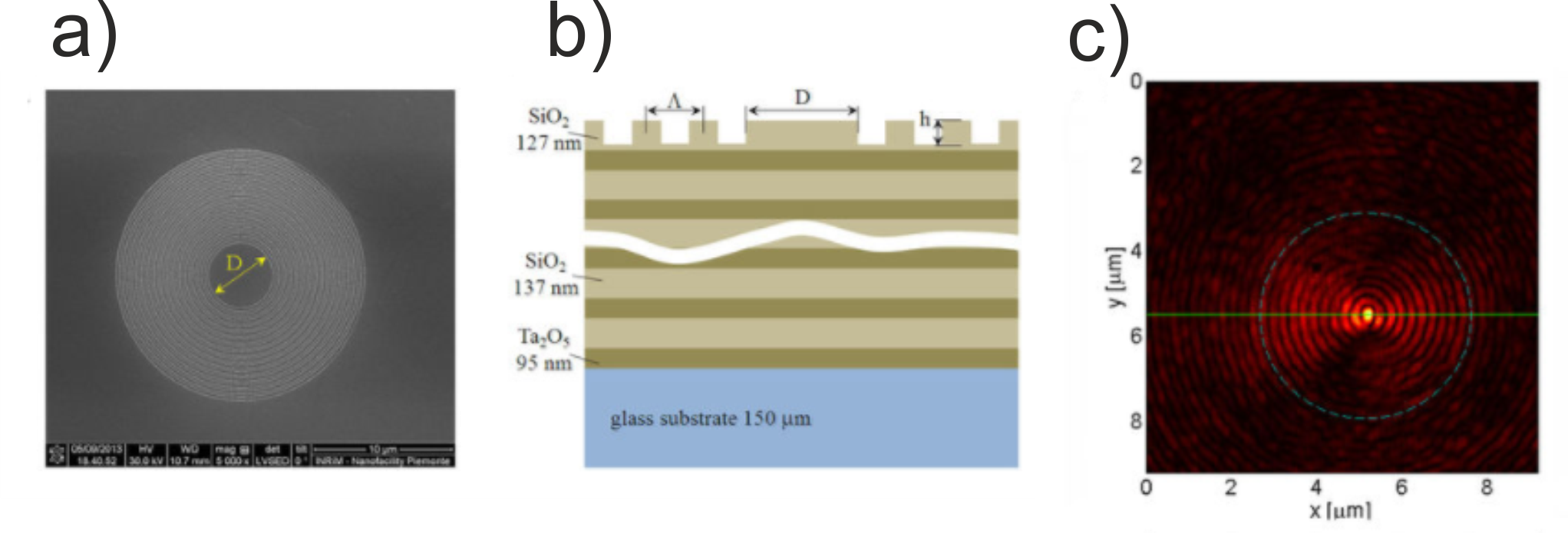}
		\caption{(a) Scanning Electron Microscope image of the circular grating fabricated on the multilayer structure; (b) Sketch of the cross-sectional view of the patterned multilayer. The stack sequence is  glass-[Ta$_2$O$_5$-SiO$_2$]$\times$6-Ta$_2$O$_5$-SiO$_2$-air.  The grating period is $\Lambda = 520~nm$ and the inner spacer diameter can be either $D = 5~\mu m$ or $D = 8~\mu m$. The height of the grating is $h = 100~nm$; (c) Amplitude distribution of the leakage radiation associated to a converging laser BSW \cite{Angelini:14}. }
		\label{Grat_Emiliano}
	\end{center}
\end{figure}

Following this considerations it has been demonstrated that a linear grating can couple BSW on a planar multilayer [Fig. \ref{Grat_Emiliano}].

Grating-based couplers can offer the opportunity to launch optical guided modes and tune their propagation direction with incident polarization and in micrometer scale architectures. Such kind of couplers are routinely produced for the excitation of SPP \cite{koev:12} and waveguide mode \cite{taillaert:2003} but they have been hardly investigated for launching BSWs \cite{scaravilli:16,Kang:16}. 

Therefore, in the first part of this work, we will study a double cross grating as a new way to couple BSW with a capability of switching the propagation direction of the excited surface waves with incident polarization. We will use this new way of coupling for integrated optics application. We will also consider this way of coupling for optical trapping and manipulation applications.

At the moment, different designs of 1DPhCs sustaining BSWs have already been proposed.
All designs are based on multilayers formed by alternating low and a high refractive index media. A wide range of materials was used in order to create multilayers which are able to sustain BSWs. For instance SiO$_2$/TiO$_2$ \cite{li:13}, Ta$_2$O$_5$/SiO$_2$ \cite{konopsky:07,konopsky:13} as well as SiN$_x$/SiO$_2$ \cite{E5,E6,Ballarini:11,Liscidini:apl09} and other materials were used for multilayer fabrication, offering a large panel of configurations for different applications at different wavelengths.

In this work we will use Si$_3$N$_4$, SiO$_2$ and Al$_2$O$_3$, TiO$_2$ and LiNbO$_3$ for multilayer fabrication.

By manipulating the index contrast between successive layers, the BSW propagation can be indeed controlled. So far, only amorphous isotropic materials \cite{konopsky:07,li:13,Ballarini:11,Liscidini:apl09} have been used in the fabrication of the multilayers and only passive BSW based devices have been investigated. In order to move towards anisotropic and active tunable devices, new materials have to be involved in the multilayer structure fabrication. 

Lithium niobate (LiNbO$_3$) is a well-known high refractive index birefringent material with tunable optical properties. In addition to ferro-electrical, piezo-electrical, and thermo-electrical properties, LiNbO$_3$ is transparent over a wide wavelength range (350 nm $-$ 5200 nm), and presents nonlinear optical polarizability and Pockels effect, which yields a unique opportunity for fabrication of new BSW sustaining tunable devices. 

In this work, we propose a 1DPHC with LiNbO$_3$ as a part of the multilayer structure. Being embedded in the multilayer LiNbO$_3$ brings anisotropic properties to the whole 1DPhC. The use of LiNbO$_3$ as a top part of the multilayer allows nonlinear, electro- and thermo- optical properties of the structure. Therefore in this work we demonstrate fabrication steps on the way to various LiNbO$_3$ based photonic crystals.

\chapter{Multilayer modeling methods and fabrication tools.}

In order to predict light behavior inside the photonic crystal various modeling techniques can be used. In this work we rely on numerical and semi-analytical calculations. Thus, different methods of light propagation analysis used in this work will be described in this chapter. Also we describe some of the major fabrication tools used for sample preparation. 

\vspace*{0.2cm}
\minitoc

\section{Impedance approach.}

A few analytical methods enables the direct and fast calculation of the light distribution in multilayer stacks, such as the impedance approach \cite{konopsky:2010, Brekhovskikh:12,Delano:69}. This method requires a very short computation time and it is very specific to the simulation of the light propagation within multilayers made of flat interfaces. 

Here PhCs are considered as materials that possess a periodic modulation of their refraction indices on the scale of the wavelength of light \cite{Yablonovitch:93}. Those materials perform photonic band gaps that are similar to the electronic band gaps for electron waves traveling in the periodic potential of the crystal. In both cases, there are frequency intervals in which wave propagation is forbidden. This analogy may be extended to include surface levels, which can exist in band gaps of electronic crystals. In photonic crystals, they correspond to optical surface waves with dispersion curves located inside the photonic band gap.
   
According to \cite{konopsky:2010} the 'characteristic impedance' of an optical medium is the ratio of the electric field amplitude to the magnetic field amplitude in this medium, i.e. $Z_{char}=E/H=1/n$. 

The concept of impedance in the optics of homogeneous layers is based on a mathematical analogy with impedance in electrical circuits. In the case of electrical circuits the wave propagates through transmission line sections with different electrical properties. In the case of optics the wave propagates through the layers with different optical properties. For reflection from a plane interface, we will use the 'normal impedance' \cite{konopsky2009registration} $Z$, which is the ratio of the tangential components of the electric field to the magnetic field:

%
%The concept of impedance in the optics of homogeneous layers is based on a mathematical analogy between a cascade of transmission line sections and a multilayer optical coating. For reflection from a plane interface, we will use the 'normal impedance' \cite{konopsky2009registration} $Z$, which is the ratio of the tangential components of the electric field to the magnetic field:

\begin{equation}
Z=\frac{E_{tan}}{H_{tan}},
\label{eq:refname1}
\end{equation}
     
The impedances for the TE-polarized wave (in which the electric field vector is orthogonal to the incident plane) and for the TM-polarized wave (in which the electric field vector is parallel to the incident plane ) are correspondingly: 

\begin{equation}
Z_{TE}=\frac{E_x}{H_y}=\frac{1}{n\cos(\theta)}, ~~~\mbox{for the TE wave}
\label{eq:refname2}
\end{equation}

\begin{equation}
Z_{TM}=\frac{E_y}{H_x}=\frac{\cos(\theta)}{n}, ~~~\mbox{for the TM wave}
\label{eq:refname3}
\end{equation}

\subsection{Dispersion curve.}

\begin{figure}[!t]
\begin{center}
\includegraphics[width=4in]{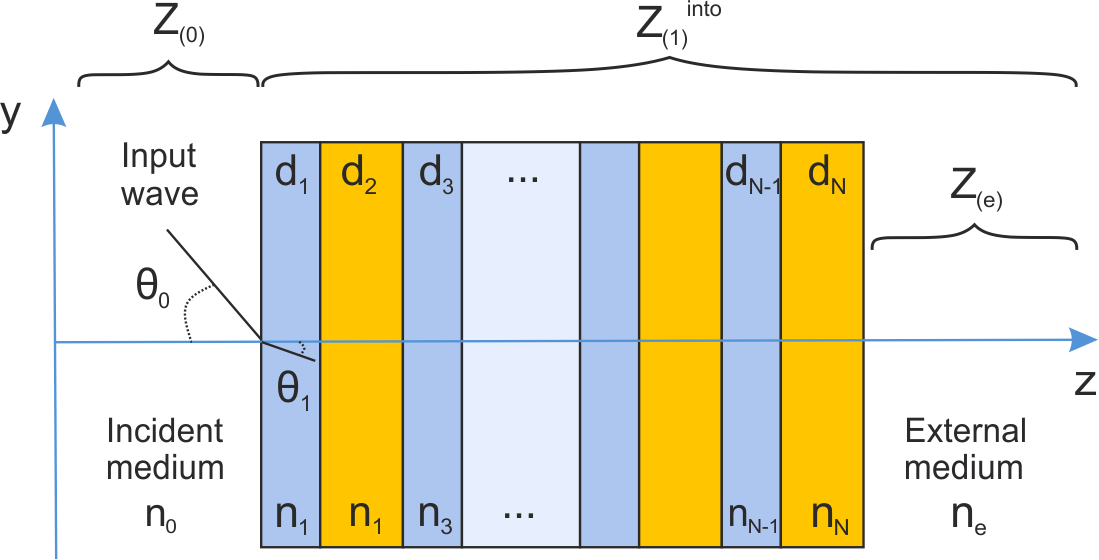}
\caption{Reflection and transmission for a multilayer with N layers.}
\label{fig:Fig2_1_1DPhC_impedance}
\end{center}
\end{figure}

Let us consider the structure as shown in Fig. \ref{fig:Fig2_1_1DPhC_impedance}.
 
If we have $N$ number of layers, it will be convenient to use an effective refractive index of the surface wave $n_{\textnormal{eff}} = n_0 \sin (\theta_0) = k_y (\lambda / 2 \pi )$ as an angle variable (as the mode exists). It is a unified angle variable for all layers, since according to Snell's law $n_0\sin(\theta_0)=n_j\sin(\theta_j)$, for any $j$. Here, $n_0$ is refractive index of the incident media, $\theta_0$ is the incident angle, $k_y$ - wave vector component along $Y$ axis. The impedance of each layer has the following form: 

\begin{equation}
Z_{TE(j)}=\frac{1}{n_j \cos(\theta _j)}=\frac{1}{n_j \sqrt{1-(n_{\textnormal{eff}}/n_j})^2}, ~~~\mbox{for the TE wave}
\label{eq:refname4}
\end{equation}

\begin{equation}
Z_{TM(j)}=\frac{\cos(\theta _j)}{n_j}=\frac{\sqrt{1-(n_{\textnormal{eff}}/n_j})^2}{n_j}, ~~~\mbox{for the TM wave}
\label{eq:refname5}
\end{equation}

The longitudinal component (along $Z$) of the wave vector in each layer reads:

\begin{equation}
k_{z(j)}=\frac{2 \pi}{\lambda} n_j \cos(\theta_j) =\frac{2 \pi}{\lambda} n_j \sqrt{1-(n_{\textnormal{eff}}/n_j})^2, ~~~\mbox{for TE and for TM polarizations}
\label{eq:refname6}
\end{equation}
 
If the multilayer is made up of $N$ plane-parallel, homogeneous, isotropic dielectric layers (with refractive indices $n_j$ and geometrical thicknesses $d_j$, where $j = 1, 2, ..., N$) between semi-infinite incident $(0)$ and external $(e)$ media (see Fig. \ref{fig:Fig2_1_1DPhC_impedance}). The input impedance of a semi-infinite external medium $(e)$ and layers from $N$ to $j$ may be calculated by the following recursion relation   \cite{Brekhovskikh:12}:

\begin{equation}
Z_{(j)}^{into}=Z_j \frac{Z_{(j+1)}^{into} - i Z_{(j)} \tan(\alpha_j)}{Z{(j)} - i Z_{(j+1)}^{into} \tan(\alpha_j)}
\label{eq:refname7}
\end{equation}

where $\alpha_j = k_{z(j)} d_j = (2\pi/\lambda) n_j \cos(\theta_j) d_j$; $j = N, N–1, ..., 2, 1$ and $Z_{(N + 1)}^{into} = Z_{(N + 1)} = Z_{(e)}$, $n_{N + 1} = n_e$ while $d_{N + 1} = d_e = 0$ ($Z_{(N + 1)}^{into}$ is the input impedance of a semi-infinite external medium).

Here, $Z_{(j)}$ is the normal impedance of medium $j$, given by \ref{eq:refname4} or \ref{eq:refname5}. Hereafter, the use of $R$ and $Z$ (without subscripts s or p) means that the equation holds for both polarizations when the corresponding impedances $Z_{TE}$ or $Z_{TM}$ are inserted.
  
In this case the equations for the reflection coefficients of the TE- and TM-polarized waves from any complex multilayer have the following form: 
 
\begin{equation}
R=\frac{Z_{(1)}^{into}-Z_{(0)}}{Z_{(1)}^{into}+Z_{(0)}}   
\label{eq:refname8}
\end{equation}

Fresnel's formulas for multilayer transmission coefficients are as follows: 

\begin{equation}
T_{TE}=\prod_{j=0}^{j=N} T_{{TE}(_{~j}^{j+1})},~~ \mbox{with}~~ T_{{TE}(_{~j}^{j+1})}=-\frac{Z_{{TE}(j+1)}^{into}+Z_{{TE}(j+1)}}{Z_{{TE}(j+1)}^{into}+Z_{{TE}(j)}}e^{i \alpha_{j+1}}   
\label{eq:refname9}
\end{equation}

and

\begin{equation}
T_{TM}=\prod_{j=0}^{j=N} T_{{TM}(_{~j}^{j+1})},~~ \mbox{with}~~ T_{{TM}(_{~j}^{j+1})}=-\frac{n_j Z_{{TM}(j)}}{n_{j+1} Z_{{TM}(j+1)}} \frac{Z_{{TM}(j+1)}^{into}+Z_{{TM}(j+1)}}{Z_{{TM}(j+1)}^{into}+Z_{{TM}(j)}} e^{i \alpha_{j+1}}   
\label{eq:refname10}
\end{equation}

where $T_{(_{~j}^{j+1})}$ are the transmission coefficients at an interface between the $j^{th}$ layer and the $(j + 1)^{th}$ layer.

The dispersion of a 1DPhC can be presented as optical field enhancement $\log(I_e/I_0)$ (i.e. as $\log {\lvert T_{(_0^e)} \rvert}^2$ in the external medium near the structure), as a function of the vacuum wavelength and the incidence angle. 

\subsection{Optimization of layer's thicknesses.}

Before plotting the dispersion curve in the case of alternating low and high refractive index materials, thicknesses of layers should be optimized. We consider a stack of only two layers with refractive indices $n_1$ and $n_2$ and thicknesses $d_1$ and $d_2$. The layers are surrounded by incident ($n_0$) and external ($n_e$) media.

According to \cite{konopsky:2010} a quarter-wavelength thickness of layers in the 1DPhC does not provide the maximum extinction ratio (E.R.) per layer. This is important in order to minimize overall thickness of the whole stack. Optimized values can be obtained by finding the maximum of the following function:

\begin{equation}
    E.R. = f(d_1 , d_2) = \left| \frac{\ln \left|T_{(_{~1DPhC(j)}^{1DPhC(j+3)})}\right|^2}{d_1 + d_2} \right|
\label{eq:refname11}
\end{equation}

where $T_{(_{~1DPhC(j)}^{1DPhC(j+3)})}$ is a multilayer transmission coefficient for one period of the 1DPhC (two layers) and it can be found from the following expression:

\begin{equation}
T_{(_{~1DPhC(j)}^{1DPhC(j+3)})}=e^{i(\alpha_1+\alpha_2)} \frac{(Z_2+Z_{1DPhC}^{into})(Z_1+Z_{1DPhC+Z_1}^{into})}{(Z_1+Z_{1DPhC}^{into})(Z_2+Z_{1DPhC+Z_1}^{into})}  
\label{eq:refname12}
\end{equation}

Here, the optimization should be done for a chosen polarization (TE or TM), at the fixed wavelength $\lambda$, and at the fixed expected $n_{\textnormal{eff}}$ for fixed internal and external medium ($n_0$ and $n_e$) and for fixed materials ($n_1$ and $n_2$). $Z_1$ and $Z_2$ are defined by \ref{eq:refname4} or \ref{eq:refname5} for the first and second materials respectively. 

$Z_{1DPhC}^{into}$ and $Z_{1DPhC+Z_1}^{into}$ are defined as following:

\begin{equation}
Z_{1DPhC}^{into}=-\frac{i}{2} \frac{(Z_2^2-Z_1^2)\tan(\alpha_1)\tan(\alpha_2)\pm \sqrt{s}}{Z_2\tan(\alpha_1)+Z_1\tan(\alpha_2)},
\label{eq:refname13}
\end{equation}

where

\begin{equation}
s=-4Z_1 Z_2 (Z_2 \tan(\alpha_1)+Z_1\tan(\alpha_2))( Z_1 \tan(\alpha_1)+Z_2\tan(\alpha_2) )+[(Z_2^2-Z_1^2)\tan(\alpha_1)\tan(\alpha_2)]^2
\label{eq:refname14}
\end{equation}
 
and 
 
\begin{equation}
Z_{1DPhC+Z_1}^{into}=Z_1 \frac{Z_{1DPhC}^{into}-iZ_1 \tan(\alpha_1)}{Z_1-iZ_{1DPhC}^{into}\tan(\alpha_1)} 
\label{eq:refname15}
\end{equation}

\subsection{Optimization of number of layers.}

The optimal number of pairs in multilayer for each wavelength depends on total extinction in the layers. It means that the reflectance ($\left|R\right|^2=0$). 

For the multilayer which is made only out of dielectric materials it is important to include the imaginary part of refractive indices for all the materials. For example $n_{Al_2O_3}=1.67+0.001i$ at $\lambda=632.8$nm. The real part of $n$ can be obtained from ellipsometry measurements, refractive index database or elsewhere \cite{refrind, Palik:98} for various dielectric materials. Meanwhile the imaginary part of $n$ equals to 0 for the most of dielectrics, such as SiO$_2$, LiNbO$_3$, TiO$_2$, etc. However, in the multilayer stack the losses on the boundaries between layers occurs. Therefore the values of imaginary part of refractive indices should be chosen arbitrary and verified lately experimentally. 

In the theoretical calculations for all-dielectric 1DPhC it is necessary to take this losses into account in order to observe a reflectance dip at the BSW excitation angle.

\section{Other methods. RCWA. FDTD.}

The impedance approach is a very quick modeling tool and it's perfectly adapted for analysis of BSWs excited in Kretschmann configuration and for multilayer without any defects. In the case when there is and additional nano-structuring of the surface on the multilayer other modeling techniques should be used.

\subsection{Rigorous coupled-wave analysis.}

Rigorous coupled-wave analysis (RCWA) is a semi-analytical method in computational electro-magnetics that is most typically applied to solve scattering from periodic planar structures (such as standard gratings). It is a Fourier-space method so devices and fields are represented as a discrete sum of spatial harmonics \cite{Moharam:81,Neviere:02,Moharam:82}.

In this work RCWA is used in order to derive optimal grating parameters on the top of 1DPhC. Here the space was divided in three areas: two semi-infinite homogeneous and isotropic regions (below and above the grating) and one modulated region (grating part). In homogeneous regions the field was calculated as a sum of plane-waves (diffracted orders). In the grating area the field was calculated as a sum of Bloch modes.

More detailed explanation about RCWA analysis and light propagation in periodic media can be found elsewhere \cite{Moharam:81,Neviere:02}. The homemade code of FEMTO-ST Institute written by Dr. Philippe Boyer was used to analyze the grating coupling of BSW \cite{Kovalevich:17}. RCWA calculation method is a fast tool designed to work with periodic structures (particularly gratings). It is faster than FDTD calculations, though, in the case of our code can be applied only for infinite grating coupler. We start with this method in order to minimize calculation time for grating coupler optimization. The concept of BSW grating coupler would be described in the following chapter.  

\subsection{Finite-difference time-domain method.}

Finite-difference time-domain (FDTD) method is a numerical analysis technique used for modeling computational electrodynamics (finding approximate solutions to the associated system of differential equations). 

The FDTD method belongs in the general class of grid-based differential numerical modeling methods (finite difference methods). The time-dependent Maxwell's equations (in partial differential form) are discretized using central-difference approximations to the space and time partial derivatives. The resulting finite-difference equations are solved in either software or hardware in a leapfrog manner: the electric field vector components in a volume of space are solved at a given instant in time; then the magnetic field vector components in the same spatial volume are solved at the next instant in time; and the process is repeated until the desired transient or steady-state electromagnetic field behavior is fully evolved \cite{kunz1993finite}.   

This method is now widely employed to simulate light-matter interaction in nano-optics \cite{Taflove:95,RSOFT,lumerical} and it is used to calculate the total electromagnetic field in any computational volume that contains the structure under study.

%\textit{An electromagnetic wave is represented by a 3D array: $E_x$, $E_y$, $E_z$, $H_x$, $H_y$ and $H_z$. This unit is called a Yee-cell and a picture of the standard FDTD cartesian Yee cell is shown in Fig. \ref{fig:Fig_2_Yee_cell.png}. Electric and magnetic field components are interleaved both in space and time, thus those can be solved sequentially in a leapfrog manner. Calculation is repeated until the desired number of time steps is reached. Maxwell’s time dependant curl equations are solved at each point of the space and time.}
	
%	\begin{figure}[!t]
%		\begin{center}
%			\includegraphics[width=4.5in]{Fig_2_Yee_cell.png}
%			\caption{3D Yee cell showing the electric (E) and magnetic (H) field components in space.}
%			\label{fig:Fig_2_Yee_cell.png}
%		\end{center}
%	\end{figure}

The FDTD method can be used to solve very complicated 3D problems, but it needs a large amount of memory and the computation time is increasing substantially. Depending on the designed structure, simulations may easily require tens of thousands timesteps for a proper simulation result. Luckily, in most of the cases a 2D option, using the effective index approximation, offers a good tool for a raw and a fast optimization with shorter computation time. Faster computation time can be reached in 2D cases, because one direction in the design is assumed to be infinite. This assumption removes all the derivatives in this assumed direction from Maxwell’s equations.

In this work we use commercially available software R-SOFT Full Wave package for studies of field distribution inside the 1DPhC with finite structures on top, such as finite grating and decoupling grooves. These FDTD simulations are performed to verify final field distribution in the 1DPhC with a finite grating on top, after the optimized grating parameters were found with RCWA method. This simulations are also used for BSW coupling efficiency through the grating.

\vspace{1.5cm}

In summary, in this work three major methods are used. Every time for the multilayer design and optimization the calculations with impedance approach are performed. This is a specifically designed fast method for quick mapping of dispersion curves. We estimate layers' thicknesses required for 1DPhC at particular wavelength, predict the values of incident angles at which BSW excitation occurs and calculate effective refractive index of BSWs with this approach. 
For optimization of grating parameters, such as grooves depth and width, RCWA method is used, as a specifically optimized tool for this task. 
Whenever we need to obtain a precise field distribution inside the multilayer (already designed and optimized with quick methods) we use FDTD simulations, which take a long time but provide a precise information about field parameters at every point of the structure under study.

\section{Sample fabrication tools.}

The whole idea of the BSW became possible only when technology developed well enough to deposit thin (order of several hundreds nanometers) homogeneous layers of dielectric materials. Therefore, in this section we briefly describe multilayer deposition techniques, which were used for sample fabrication. These techniques are - plasma-enhanced chemical vapor deposition (PECVD) and atomic layer deposition (ALD).

Another crucial equipment which was used for samples fabrication is focused ion beam (FIB). It was used for multilayer characterization of all the samples as well as for nano-structuring of the top layer surface.

Deep reactive ion etching (DRIE) also plays a key role for LiNbO$_3$ related part of studies. Thin films of lithium niobate always come on the Si wafer support. In order to integrate this material in 1DPhC this support layer have to be removed. 

We would focus our attention in a bit more details on major fabrication tools. Though, it is important to mention that many various additional processes were used for sample fabrication, such as lithography, precise dicing, sputtering,  wet etching, scanning electron microscopy, etc.

\subsection{Plasma-enhanced chemical vapor deposition.}

	\begin{figure}[!b]
		\begin{center}
			\includegraphics[width=5.1in]{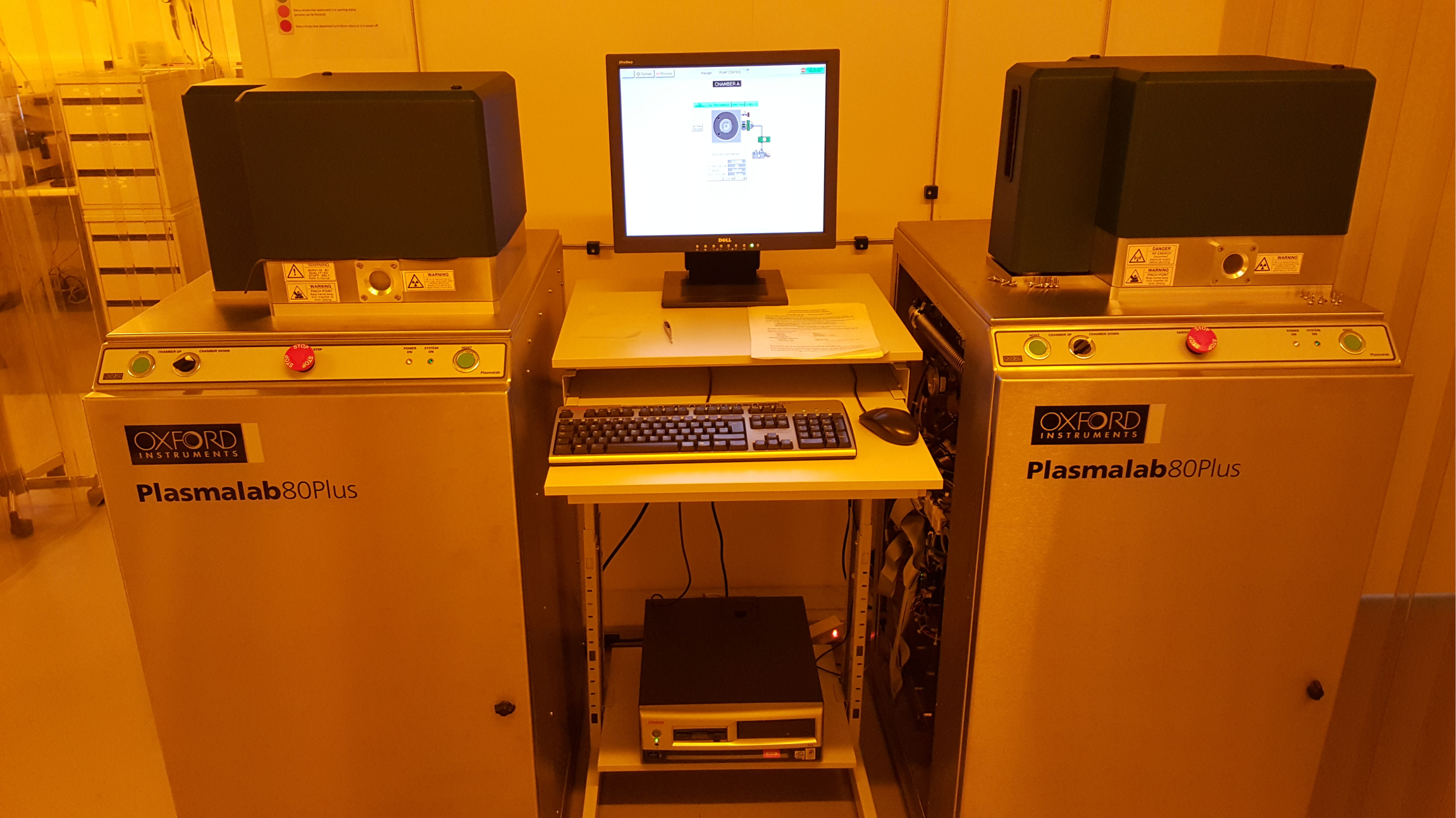}
			\caption{PECVD equipment (PlasmaLab80+).}
			\label{PlasmaLab80+ PECVD.png}
		\end{center}
	\end{figure}

Plasma-enhanced chemical vapor deposition (PECVD) is a chemical vapor deposition process used to deposit thin films from a gas state (vapor) to a solid state on a substrate. PECVD is a process by which thin films of various materials can be deposited at lower temperature than that of standard Chemical Vapor Deposition (CVD) \cite{Batey:89}.

In PECVD processes, deposition is achieved by introducing reactant gases between parallel electrodes—a grounded electrode and an RF-energized electrode. The capacitive coupling between the electrodes excites the reactant gases into a plasma, which induces a chemical reaction and results in the reaction product being deposited on the substrate. The substrate, which is placed on the grounded electrode, is typically heated to 250$^\circ$C to 350$^\circ$C, depending on the specific film requirements. In comparison, CVD requires 600$^\circ$C to 800$^\circ$C. The lower deposition temperatures are critical in many applications where CVD temperatures could damage the devices being fabricated.

The films typically deposited using PECVD are silicon nitride (Si$_x$N$_y$), silicon dioxide (SiO$_2$), silicon oxy-nitride (SiO$_x$N$_y$), silicon carbide (SiC), and amorphous silicon ($\alpha$-Si). Silane (SiH$_4$), the silicon source gas, is combined with an oxygen source gas to form silicon dioxide or a nitrogen gas source to produce silicon nitride.

In this work Oxford Plasmalab80+ equipment [Fig. \ref{PlasmaLab80+ PECVD.png}] is used to deposit silicon nitride and silicon dioxide of the multilayer platform \cite{yu:13,wu:14}. Film thickness possible to deposit are in the range from a few nm to 2$\mu$m. Film stress can be controlled by high/low frequency mixing techniques. Deposition can be performed at temperatures from 80$^\circ$C to 340$^\circ$C.

For our study the deposition was performed by Dr. Myun-Sik Kim from optics and photonics technology laboratory, Ecole Polytechnique F\'{e}d\'{e}rale de Lausanne (EPFL).

\subsection{Atomic layer deposition.}\label{ALD}

Atomic layer deposition (ALD) is a cyclic coating method to fabricate thin films. This method is a modification of a chemical vapor deposition method and the basic idea of ALD process is to pulse two precursor vapors in a reaction chamber periodically \cite{Leskela:02}. 

In an ideal case, an atomic layer of material is grown in one cycle. This is possible due to a self-limiting process, which prevents more atoms from adsorbing on the surface. Due to the saturation of each reaction steps, many benefits can be achieved when compared to other thin film deposition technologies: the film thickness can be controlled very accurately (less than 1 nm scale) and the optical quality of films is very high, which enables to coat, e.g., high quality thin film stacks very accurately. Furthermore, ALD provides the possibility to grow conformal coatings around different kind of samples, like gratings and fibers. Also, large chambers can be used in mass production. 

Moreover, low fabrication temperature enables to coat samples which have preprocessed components, like replicated polymer devices, which do not withstand high temperatures. Disadvantages of ALD are a slow growth rate, volatile precursors, and despite an already large amount of dielectric materials available, the deposition of metals remains a challenging issue \cite{Miikkulainen:13}.

Generally, thermal ALD processes are done in lower temperatures than CVD processes. Fabrication temperature of some oxides and nitrides can be further decreased even near to room temperature by using a plasma-assisted processes.

\begin{figure}[!t]
	\begin{center}
		\includegraphics[width=3.5in]{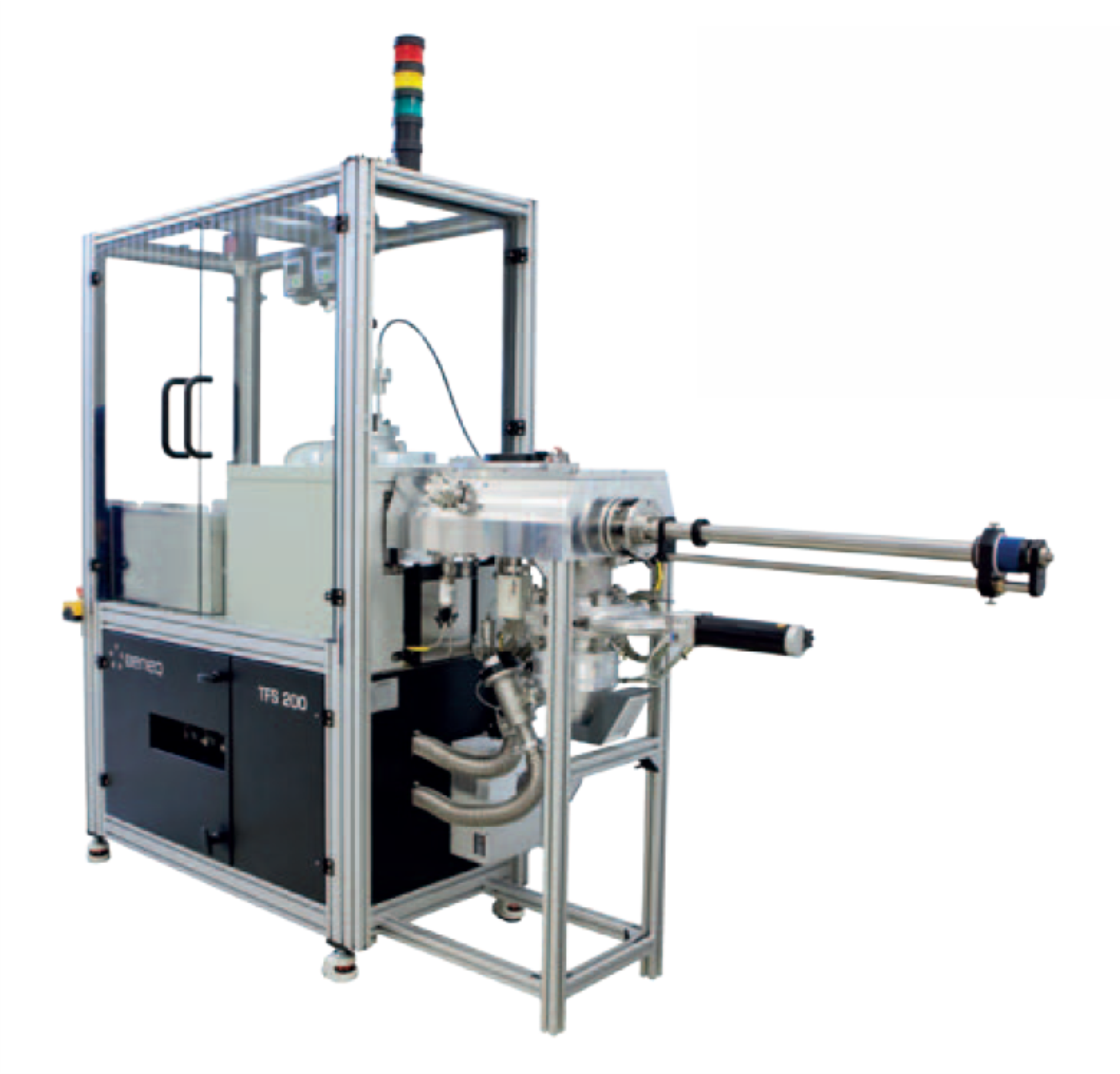}
		\caption{ALD tool (Beneq TFS 200).}
		\label{TFS_200_ALD_tool_beneq_tfs_200_ald}
	\end{center}
\end{figure}

In this work two different ALD grown materials are used: titanium dioxide (TiO$_2$) and aluminum oxide (Al$_2$O$_3$). Fabrication processes are thermal for both materials at a temperature of 120$^\circ$C, which results in an amorphous material. In a case of TiO$_2$, titanium tetrachloride (TiCl$_4$) and water (H$_2$O) and for Al$_2$O$_3$ trimethyl aluminum (TMA) and water (H$_2$O) are used as the precursors. The growth rate for TiO$_2$ is 0.07 nm/cycle and for Al$_2$O$_3$ it is 0.12 nm/cycle. The used ALD machine was ALD TFS 200 by Beneq [Fig. \ref{TFS_200_ALD_tool_beneq_tfs_200_ald}].

For our study the deposition was performed by Dr. Markus H\"{a}yrinen from Institute of Photonics, University of Eastern Finland.

\subsection{Focused ion beam.}

Focused ion beam (FIB) is a technique used particularly in the semiconductor industry, materials science and increasingly in the biological field for site-specific analysis, deposition, and ablation of materials. A FIB setup is a scientific instrument that resembles a scanning electron microscope (SEM). However, while the SEM uses a focused beam of electrons to image the sample in the chamber, a FIB setup uses a focused beam of ions instead.

FIB systems use a finely focused beam of ions (usually gallium) that can be operated at low beam currents for imaging or at high beam currents for site specific sputtering or milling. The gallium (Ga+) primary ion beam hits the sample surface and sputters a small amount of material, which leaves the surface as either secondary ions (i+ or i-) or neutral atoms (n$_0$). The primary beam also produces secondary electrons (e-). As the primary beam rasters on the sample surface, the signal from the sputtered ions or secondary electrons is collected to form an image.

\begin{figure}[!b]
	\begin{center}
		\includegraphics[width=2.8in]{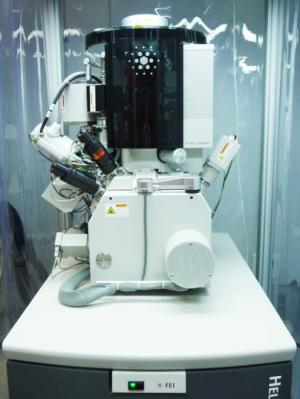}
		\caption{Dual Beam SEM$/$FIB equipment (FEI Helios 600i).}
		\label{Dual_Beam_SEM_FIB_FEI_Helios_600i}
	\end{center}
\end{figure}

At low primary beam currents, very little material is sputtered and modern FIB systems can easily achieve 5 nm imaging resolution. At higher primary currents, a great deal of material can be removed by sputtering, allowing precision milling of the specimen down to a sub- micrometer or even a nano-scale.

If the sample is non-conductive, a low energy electron flood gun can be used to provide charge neutralization. Also samples may be imaged and milled with a conducting surface coating.

Because of the sputtering capability, the FIB is used as a micro- and nano-machining tool, to modify or machine materials at the micro- and nano-scale. Commonly the smallest beam size for imaging is 2.5–6 nm. The smallest milled features are somewhat larger (10–15 nm) as this is dependent on the total beam size and interactions with the sample being milled.

FIB tools are designed to etch or machine surfaces, an ideal FIB might machine away one atom layer without any disruption of the atoms in the next layer, or any residual disruptions above the surface. Yet currently because of the sputter the machining typically roughens surfaces at the sub-micrometer length scales. A FIB can also be used to deposit material via ion beam induced deposition. FIB-assisted chemical vapor deposition occurs when a gas, such as tungsten hexacarbonyl (W(CO)6) is introduced to the vacuum chamber and allowed to chemisorb onto the sample. By scanning an area with the beam, the precursor gas will be decomposed into volatile and non-volatile components; the non-volatile component, such as tungsten, remains on the surface as a deposition. This is useful, as the deposited metal can be used as a sacrificial layer, to protect the underlying sample from the destructive sputtering of the beam. From nanometers to hundred of micrometers in length, tungsten metal deposition allows metal lines to be put right where needed. Other materials such as platinum, cobalt, carbon, gold, etc., can also be locally deposited \cite{Orloff:03,Giannuzzi:04}.

The FIB and the SEM are available separately as stand-alone instruments. Both beams can be used for imaging, while the FIB is also able to modify samples at micro- and nano-scales. The ability of the FIB to modify samples is a disadvantage when using the FIB for imaging, as each scan of the FIB damages the sample. Hence, integrating both columns has a tremendous advantage, as the FIB modifications can be continuously monitored and examined with the SEM.

In this work FIB-SEM tool was used for multilayer characterization. In order to check number of layers, homogeneity and thicknesses of deposited materials we milled a small opening on the sample surface by FIB and then the image of the cross-section was acquired  by SEM. It was also used for all the nano-structuring of the top surface of multilayer. All the gratings, waveguides, grooves and 2DPhCs were fabricated by FIB milling \cite{Lacour:05}. In order to have a good image resolution and to avoid charging effects \cite{Yogev:08,Samantaray:08} during the FIB-SEM process a thin layer of metal (Cr or Al) was deposited on the top of the samples. The metal was removed dy wet etching after FIB milling or characterization was finished. Also for some samples in-situ Pt deposition was used. For example for characterization of dimensions of manufactured gratings or for the alignment marks deposition. The alignment marks were necessary for milling of complicated structures, such as 2DPhCs or big gratings. This marks also serves as a reference surface during FIB milling of dielectrics in order to avoid shifts during long fabrication process.

In this work the nano-structuring and sample characterization was made by Dr. Roland Salut and Dr. Gwenn Ulliac from FEMTO-ST institute on Dual Beam SEM / FIB FEI Helios 600i equipment [Fig. \ref{Dual_Beam_SEM_FIB_FEI_Helios_600i}].

\subsection{Deep reactive ion etching.}

For 1DPhC with thin film LiNbO$_3$ as a top layer deep reactive ion etching (DRIE) was used. The main goal of this process is to remove Si support layers from TFLN sample.

	\begin{figure}[!b]
	\begin{center}
		\includegraphics[width=2.8in]{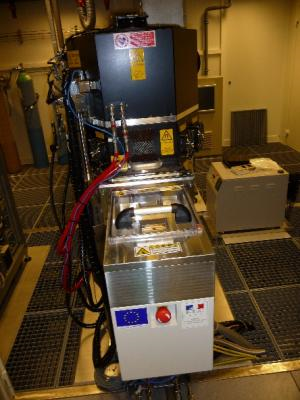}
		\caption{Equipement RAPIER SPTS.}
		\label{Equipement_RAPIER_SPTS}
	\end{center}
\end{figure}

Silicon can be etched in an anisotropic (or isotropic) and dry way with the technology called “deep RIE”. This method combines deposition and etching plasma assisted processes. While the silicon is etched, the process enables to deposit a fluorinated component on the pattern walls in order to passivate them. Etching-deposition cycles with given gases and time enable to etch deeply and anisotropically the silicon independently of its orientation.

The Si etching in this work was performed on RAPIER SPTS equipment [Fig. \ref{Equipement_RAPIER_SPTS}] at FEMTO-ST institute. The machine is dedicated to the silicon and silica ($\le 1 \mu m$) etching with available gases: SF$_6$, C$_4$F$_8$, Ar, He, N$_2$ and O$_2$. The machine is equipped by real-time endpoint detection system (Claritas).

\section{Conclusion.}

Thus we have introduced all the modeling methods which have been used in this work.
For all the 1DPhCs which will be described later, the optimization is firstly made with impedance approach and then the performance of the finalized crystals with additional nano-structuring is confirmed by RCWA and FDTD simulations. Also we have described major deposition and nano-structuring tools, used in this work.

\chapter{Polarization tunability.}

In this part of our work we propose a highly miniaturized grating based BSW coupler which is gathering launching and directional switching functionalities in a single element. This device allows to control with polarization the propagation direction of Bloch surface waves at subwavelength scale, thus impacting a large panel of domains such as optical circuitry, function design, quantum optics, etc. Here instead of standard coupling through the prism we use a crossed grating configuration. Thus we reach a new coupling feature, such as a highly miniaturized launching. In this case light can be launched orthogonally with respect to the multilayer and therefore we do not need to adjust the system to according to the BSW coupling angle as in Kretschmann configuration. Also selectivity of grating towards polarization together with sensitivity of BSW towards polarization gives us an opportunity to have directional propagation of light on the chip without any changes of experimental conditions, except the polarization of the incident light. This comes as a very handy tool for integrated optics components.

\vspace*{0.2cm}
\minitoc

    \section{Concept of grating coupling.}

  In order to develop various dielectric surface optical functions we study BSW coupling through the grating. The standard coupling of an incident beam to a BSW can be achieved using a prism coupler in the Kretschmann \cite{kovalevich:16} or sometimes in Otto \cite{michelotti:10} configurations. Such techniques allow an easy BSW launching but they are quite bulky. In these cases tuning the BSW propagation direction requires the rotation of both the prism and the input laser beam. Grating-based couplers can be used to launch optical guided modes and tune their propagation direction with incident polarization in micrometer scale architectures. Such kind of couplers are routinely produced for the excitation of SPP \cite{koev:12,Kim:05,Romanoto:09} and waveguide modes \cite{taillaert:2003} but they have been hardly investigated for launching BSWs \cite{Angelini:14,E7}. 
  
     \begin{figure}[!t]
\begin{center}
\includegraphics[width=4.5in]{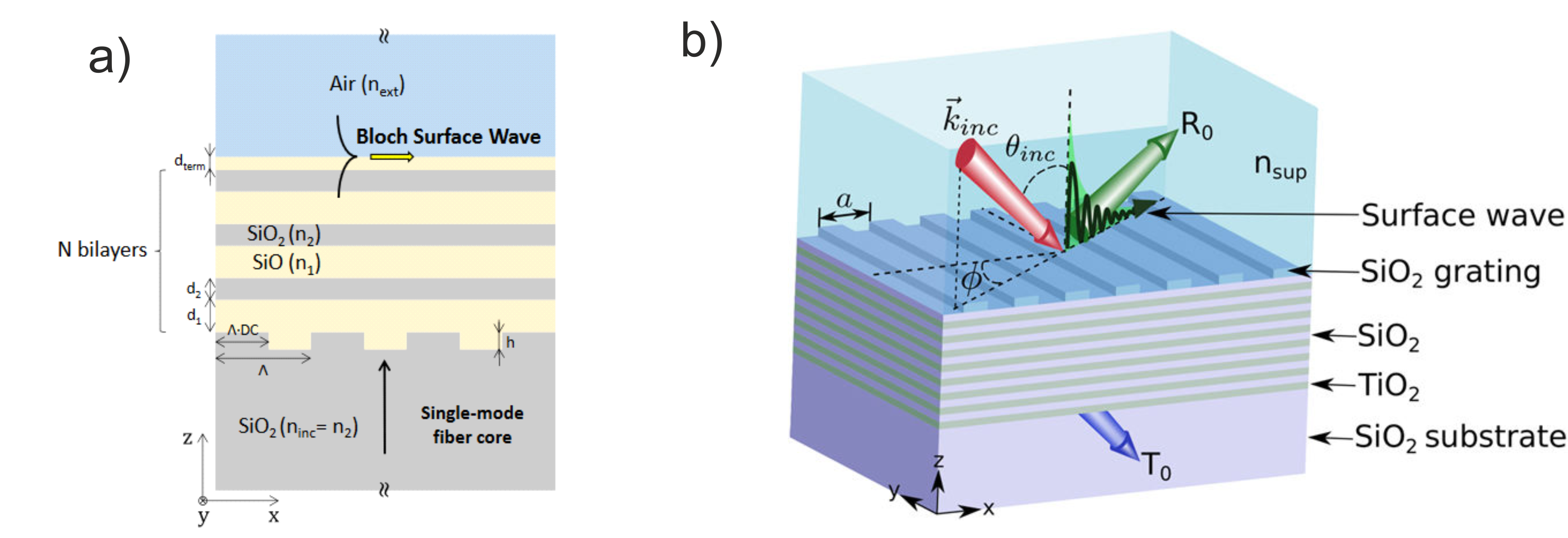}
\caption{(a) 2D schematic of the on-fiber-tip diffraction grating idealized configuration \cite{scaravilli:16}; (b) Schematic of a grating coupling technique to excite leaky Bloch surface waves on the surface of a dielectric multilayer\cite{Koju:17}}
\label{fig:Fig3_1copling_grat_prizm}
\end{center}
\end{figure}
  
  The idea of grating coupling for BSWs was firstly proposed in 2016 by  Scaravilli et. al. \cite{scaravilli:16} and by Kang et. al \cite{Kang:16} for biosensing applications. The grating coupler was manufactured on the fiber tip and then layers of alternating high and low refractive index materials were deposited on the top of the grating [Fig. \ref{fig:Fig3_1copling_grat_prizm}(a)]. 
  
Later in 2017 Koju et. al. \cite{Koju:17} have numerically studied leaky Block-like surface waves azimuthally generated by a grating coupler. Through computational simulations they have shown that moderate-Q leaky BSWs on a dielectric multilayer surface with periodic corrugation can be used to significantly enhance the sensitivity of biosensors. The polar incident angle was fixed to a specific value and then swept over the azimuthal angle to fulfill the phase matching requirement in order to excite leaky BSWs was made [Fig. \ref{fig:Fig3_1copling_grat_prizm}(b)].
  
\begin{figure}[!b]
\begin{center}
\includegraphics[width=4
in]{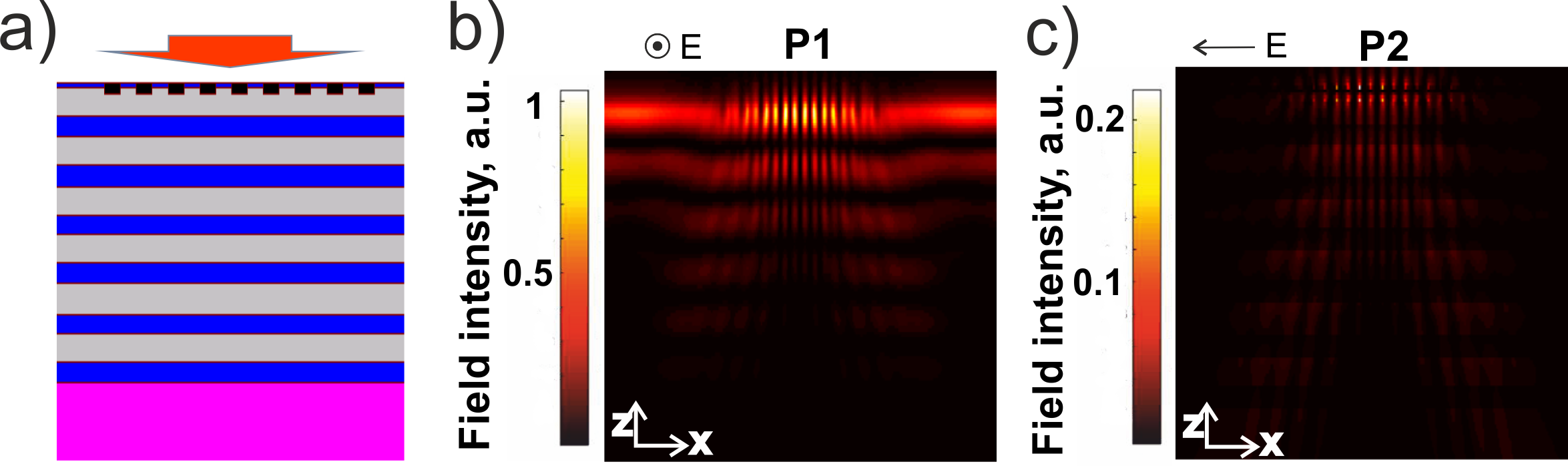}
\caption{(a) Schematic of a grating coupling with normal incident light. (b) FDTD simulations for polarization P1. (c) FDTD simulations for P2.}
\label{fig:Fig3_2_copling_grat_field}
\end{center}
\end{figure}
  
In our work, we consider a grating coupler mostly for the applications in integrated optics. It will be milled on the top of 1DPhC and the light will be coupled at the normal incident angle. We demonstrate a double cross grating as a BSW coupler capable of switching the propagation direction of the excited dielectric surface wave under control with incident polarization. Highly miniaturized grating based BSW coupler can provide launching and directional switching functionalities in a single element. This device allows to control with polarization the propagation direction of Bloch surface waves at sub-wavelength scale, thus impacting a large panel of domains such as optical circuitry, function design, quantum optics, etc \cite{Kovalevich:17}. 
  
Depending on the multilayer design, BSWs can be excited by TE or by TM polarized incident beam. Let us consider a multilayer, which is designed to support a TE polarized surface wave at the given wavelength with accordingly designed grating coupler on the top. Next, let's consider the normal incidence angle for the input beam with general state of polarization. Thus in the plane of the top surface of 1DPhC there would be both TE and TM components (with respect to the grating): polarization P1 - when the electric field vector is parallel to the grating and polarization P2 - when the electric field vector is orthogonal to the grating (as sown in Fig. \ref{fig:Fig3_3_copling_crossgrat}). Under these experimental conditions it can be clearly seen that the light with polarization P1 couples in BSW and propagates along the surface of the 1DPhC, which is not the case for the polarization P2 [Fig. \ref{fig:Fig3_2_copling_grat_field}]. 

Figure \ref{fig:Fig3_2_copling_grat_field}(b) and (c) shows a field profile in 1DPhC for the grating illuminated by TE and TM polarized light respectively. FDTD simulations were performed with Fullwave (R-soft) commercial software for grating and 1DPhC designed to work at $\lambda=1550$nm for TE polarization. (Optimization of structure's parameters will be shown later.)  

 \begin{figure}[!t]
\begin{center}
\includegraphics[width=3in]{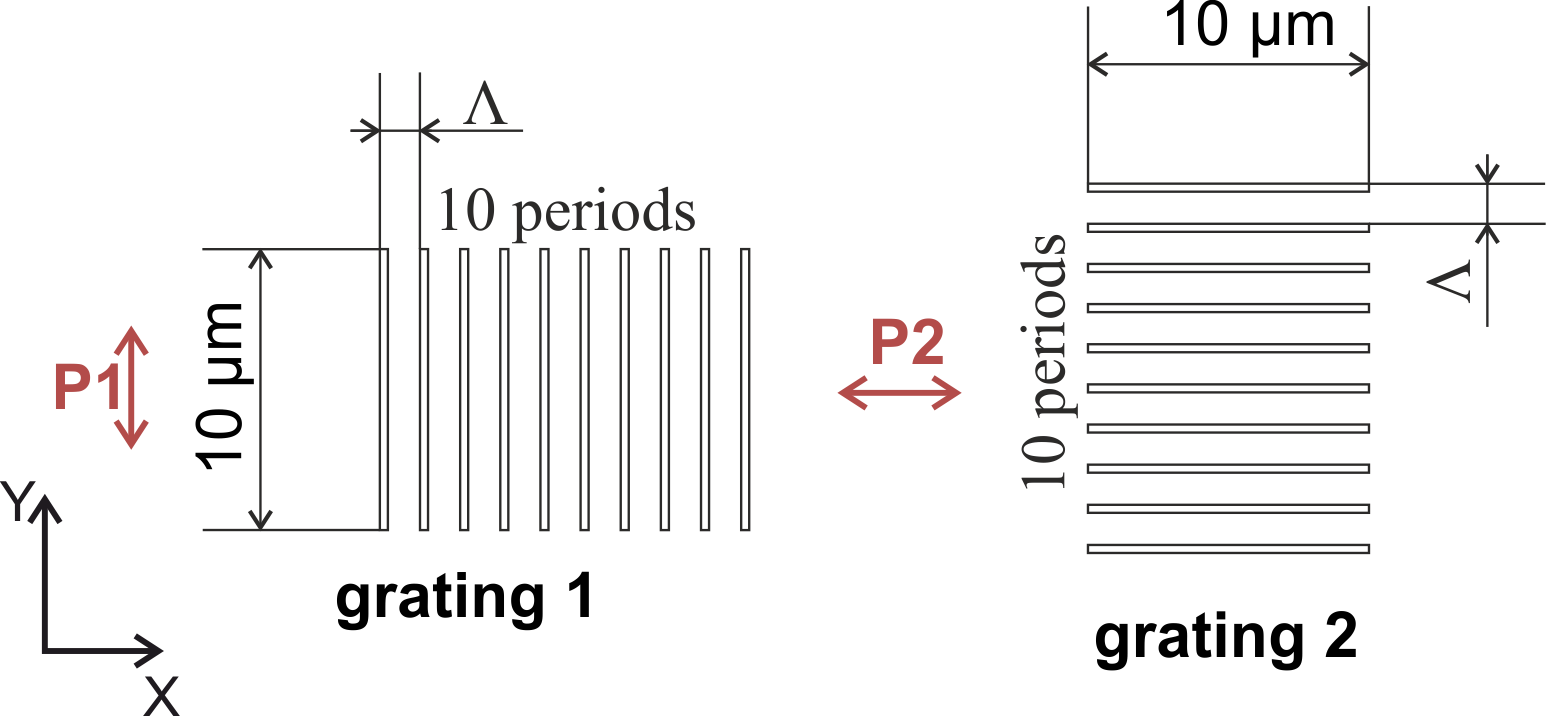}
\caption{Polarization with respect to gratings}
\label{fig:Fig3_10_gratscetch}
\end{center}
\end{figure} 
        
\begin{figure}[!b]
\begin{center}
\includegraphics[width=4.3in]{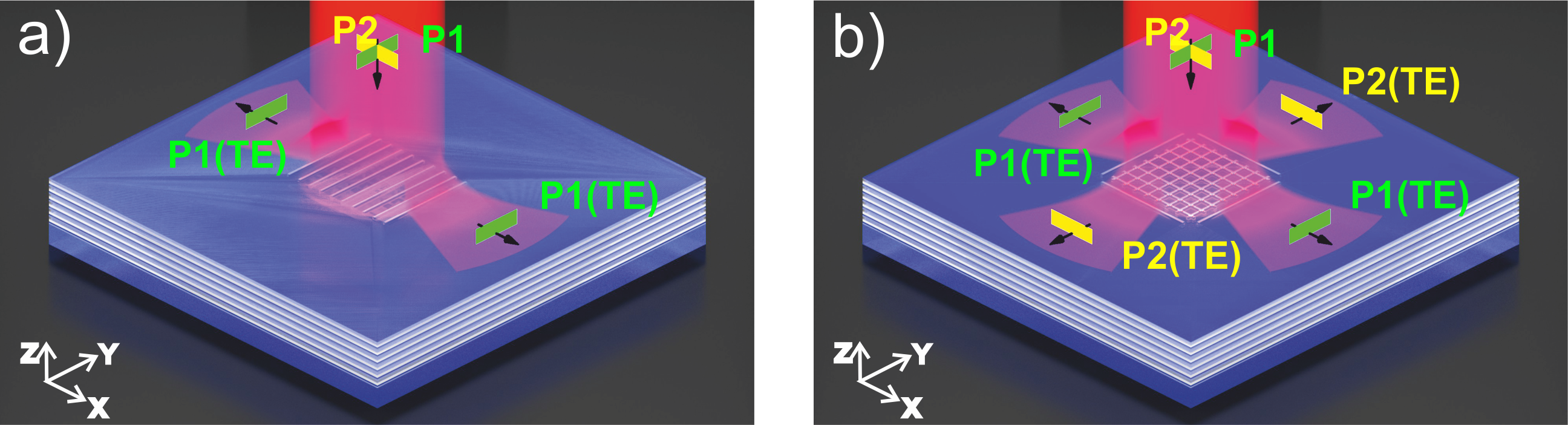}
\caption{(a) Illustration of the BSW coupling at the single grating; (b) Illustration of the BSW coupling at the inter-crossed grating.}
\label{fig:Fig3_3_copling_crossgrat}
\end{center}
\end{figure}    

It can be seen that for the single grating 1 [Fig. \ref{fig:Fig3_10_gratscetch}] the BSW coupling occurs only for the polarization P1. However, a cross-grating can couple both incident polarizations to TE polarized BSWs of their own direction \cite{bogaerts:07,barwicz2007polarization}. This way the cross-grating coupler acts as a polarization dependent beam-splitter and, in a general case with random incident light polarization, it allows the propagation of BSWs in orthogonal directions [Fig. \ref{fig:Fig3_3_copling_crossgrat}].

Here, grating 1 and 2 [Fig. \ref{fig:Fig3_10_gratscetch}] are the gratings with periodicity along X and Y direction respectively. Thus, polarization P1 is the polarization with the electric field vector parallel to the grating 1 (TE polarization with respect to the grating 1); and P2 is the polarization with the electric field vector parallel to the grating 2 (TE polarization with respect to the grating 2).

    \section{Multilayer configuration.}

As it was already mentioned, the 1DPhC is designed to support a TE polarized BSW at the wavelength of 1550 nm (TM or TM-like modes do not exist). The platform is composed of dielectric layers with alternating refractive indices. Six pairs of silicon dioxide and silicon nitride, with refractive indices of 1.45+0.001i and 1.79+0.001i at $\lambda=1550$ nm, respectively, were deposited by PECVD on a glass wafer ($n_g=1.501$). The thicknesses of the layers are 492 nm and 263 nm, respectively. An 80-nm-thick layer of silicon nitride is deposited on top of the 1DPhC. PECVD multilayer deposition was made in EPFL optics and photonics technology laboratory.

  \begin{figure}[!b]
	\begin{center}
		\includegraphics[width=4.2in]{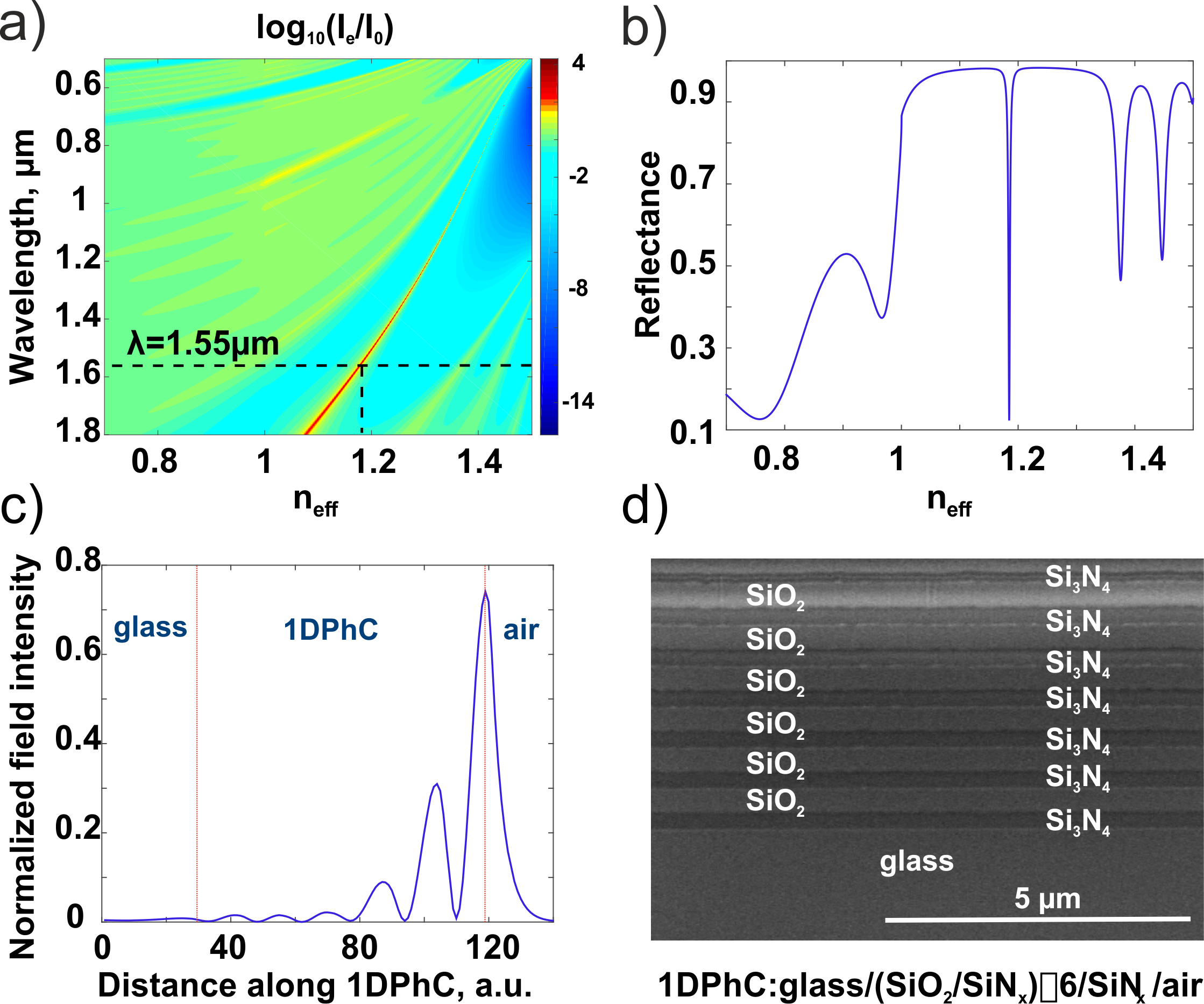}
		\caption{(a) Band gap diagram of the 1DPhC; (b) Calculated reflectance at 1550 nm wavelength; (c) Field profile in the 1DPhC for the BSW at 1550 nm wavelength (FDTD calculations); (d) FIB-SEM image of 1DPhC.}
		\label{fig:Fig3_4_ML}
	\end{center}
\end{figure}

We modeled the structure as for the prism coupler using the impedance approach \cite{konopsky:2010} described in chapter 2. The dispersion curve for this 1DPhC is shown in the Fig. \ref{fig:Fig3_4_ML}(a), where $I_0$ is a field intensity of the incident light and $I_e$ is a field intensity at 1DPhC top layer and air interface. Indeed, for the wavelength of 1550 nm the BSW occurs inside the band gap when the effective refractive index of BSW ($n_{BSW}$) equals to 1.186. The reflectance minimum shown in the Fig. \ref{fig:Fig3_4_ML}(b) indicates that the designed multilayer is optimized and works at 1550nm wavelength. Typical field profile within the 1DPhC for the BSW mode is shown in Fig. \ref{fig:Fig3_4_ML}(c), where the evanescent decay at the 1DPhC/air interface can be clearly observed.

The platform is fabricated using plasma-enhanced chemical vapor deposition. In order to verify the quality and the thickness of the deposited layers and to minimize the damage of the sample, Focused ion beam - Scanning Electron Microscopy (FIB-SEM) measurements were done [Fig. \ref{fig:Fig3_4_ML}(d)]. FIB was used in order to mill a small area on the sample surface, thus giving the access for the SEM measurements of layers thicknesses. All the layers of silicon oxide and silicon nitride can be clearly observed. Measured layer thicknesses are within 50 nm deviation from requested.

    \section{Grating optimization.}

    	\subsection{Optimization according to the grating formula.}
        
      For the grating couplers, the coupling occurs when the wave-vector matching condition is fulfilled as below \cite{Loewen:97}:

\begin{equation}
k_{\parallel}= k_{BSW}\pm mK,
\label{eq:refname1}
\end{equation}

\noindent where $k_{\parallel} = k\cdot sin\theta$ with $k$ being the incident wave-vector and $\theta$ the incident angle, $k_{BSW}$ is the BSW wave-vector, $K$ is the grating vector (grating wavenumber) defined as $K=2\pi / \Lambda$. $\Lambda$ is the period of the grating. $m$ is the diffraction order number. We consider a grating, whose first diffraction order couples into the surface wave at a normal incidence, i.e., $\theta=0$.
Therefore, the condition $k_{BSW}=K$ leads to the grating period $\Lambda$ equal to $\lambda_{BSW}$, obtained by $\lambda/n_{BSW}=1.306~\mu m$. 

\begin{figure}[!b]
	\begin{center}
		\includegraphics[width=3in]{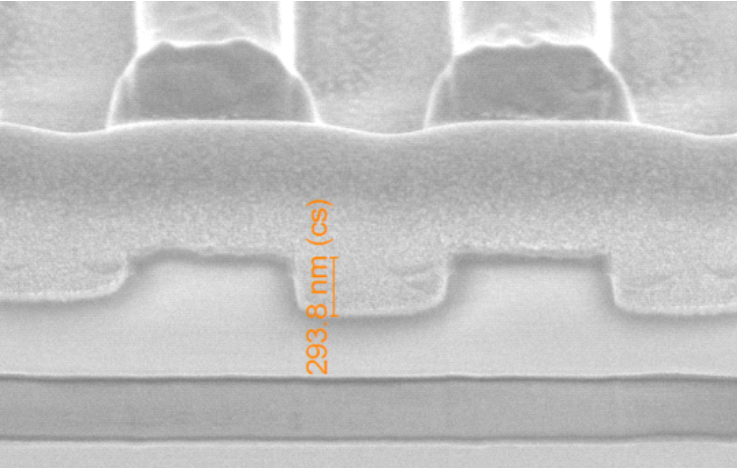}
		\caption{FIB-SEM image of the analytically optimized grating cross-section for the Sample 1}
		\label{fig:Fig3_7_grat1}
	\end{center}
\end{figure}

The advantage or this approach resides in its simplicity and it gives us such an important parameter of the grating as a period ($\Lambda$). Though it does not provide the information about the width and the depth of the grooves. Therefore we have arbitrarily chosen the width $a=0.653~\mu m$ as a half of the period and the depth $h=0.3~\mu m$ (parameters for the Sample 1). 

As an additional test we perform FDTD simulations with Fullwave (R-soft) commercial software for the designed multilayer and grating parameters optimized with grating formula [Fig. \ref{fig:Fig3_2_copling_grat_field}(b,c)].

For the designed grating, we performed FDTD simulations in order to see detailed field profile inside the sample. Indeed, coupling into the BSW is observed [Fig. \ref{fig:Fig3_2_copling_grat_field}(b)] when the grating is illuminated by TE polarized light (with respect to the grating) at the normal incidence. The BSW can be seen as a field enhancement along the sample top surface. The field intensity decreases at each pair of multilayer in such a way, that a standard for BSW field profile in the 1DPhC (see Fig. \ref{fig:Fig3_4_ML}(c)) can be observed. Also we may see the transmitted light, propagating along $Z$ axis. 

If we keep the same simulation parameters and only change the polarization from TE to TM, the BSW is no longer coupled [Fig. \ref{fig:Fig3_2_copling_grat_field}(c)]. We may only see 
the transmitted light, propagating along $Z$ axis. This proves the selectivity of designed grating coupler towards polarization.

The designed grating was manufactured by FIB milling on the top of the 1DPhC [Fig. \ref{fig:Fig3_7_grat1}]. In order to avoid charging effect 100 nm of Cr were deposited on the multilayer by sputtering (by PLASSYS MP 500 sputtering system). 

The bombardment of the charged species to the surface of an insulator can cause sample charging. The bombardment of the insulator with Ga$^+$ will cause the specimen to accumulate excess positive charge. Thus, any emitted secondary electrons will be attracted back to the surface, and will not be detected. If the sample charges significantly, charge reduction methods may be required. \cite{Giannuzzi:06} In this work we eliminate the charging dy coating samples with metal layer (Cr, Pt or Al).

 \subsection{RCWA grating studies.}
 
\begin{figure}[!b]
	\begin{center}
		\includegraphics[width=4.2in]{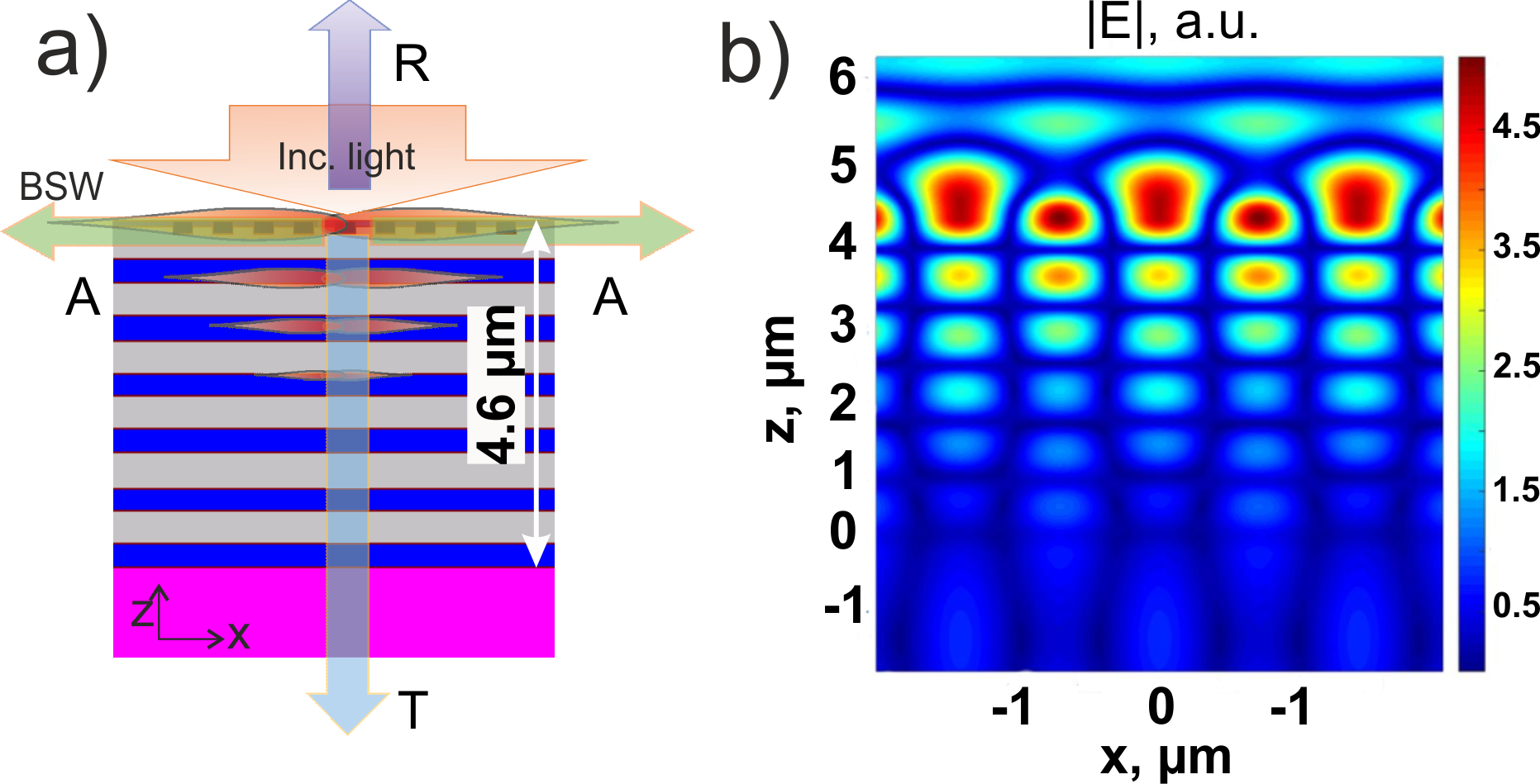}
		\caption{Schema of light propagation in 1DPhC for RCWA, (T) - light transmitted through the sample, (R) - light reflected from the sample surface, (A) - absorbed light, which represents losses and light coupled to BSW; (b) Field profile in 1DPhC for optimized grating parameters (RCWA simulations).}
		\label{fig:Fig3_5_RCWA_light}
	\end{center}
\end{figure}  
        
In order to take into account the influence of the groove size in the grating on BSW coupling, additional calculations with RCWA method were performed. As we have mentioned before RCWA method is a fast tool adapted specifically to work with periodic structures. Therefore we find optimal grating parameters with RCWA and lately verify the field distribution in the obtained structure with FDTD. We consider experimental conditions, where the light comes orthogonally to the sample surface. Some part of light is transmitted (T) through the sample, some is reflected (R) and some is absorbed (A) [Fig. \ref{fig:Fig3_5_RCWA_light}(a)].

In RCWA simulations reflected (R) and transmitted (T) light is calculated in the far field. Absorption (A) is calculated as $1-R-T$, taking into account that the total energy in the system is equal to 1. Therefore (A) represents all the losses in the system. The BSW is an evanescent wave and can not propagate to the far field, where R and T are detected. Hence, the light coupled in BSW also considered as losses or absorption. Here we can conclude that the absorption part is responsible for the BSW coupling.  Therefore such parameters of the grating as a period ($\Lambda$), depth ($h$) and width($a$) of the grooves were optimized imposing maximum absorption. 
               
All the nano-structuring on the top of 1DPhC in this work is done by FIB milling. It means that depth of the grooves should be smaller or the same as the width. This criteria allows to avoid high conicity of the walls (what is a standard issue of FIB milling process \cite{lacour:05}). Therefore as a starting point for RCWA analysis we set $h=a$. 

\begin{figure}[!t]
	\begin{center}
		\includegraphics[width=\linewidth]{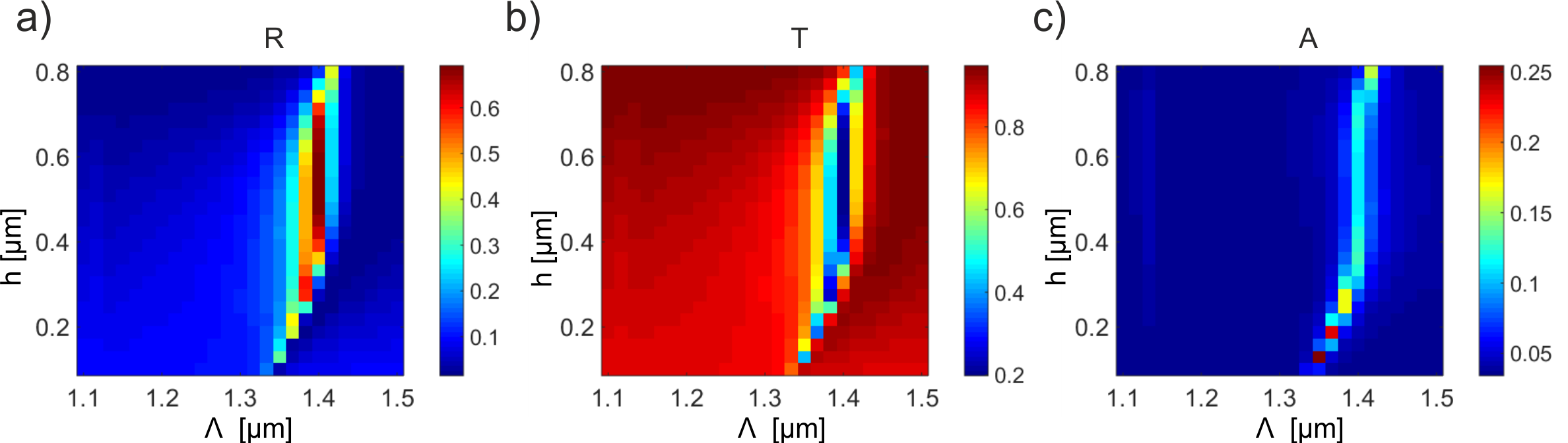}
		\caption{(a) Reflection $R(\Lambda,h)$; (b) Transmission $T(\Lambda,h)$; (c) Absorption $A(\Lambda,h)$ as a function of period and width of the grooves. Depth is equal to width.}
		\label{fig:Fig3_6_RCWA_RTA_h_equal_a}
	\end{center}
\end{figure}

\begin{figure}[!b]
	\begin{center}
		\includegraphics[width=3.7in]{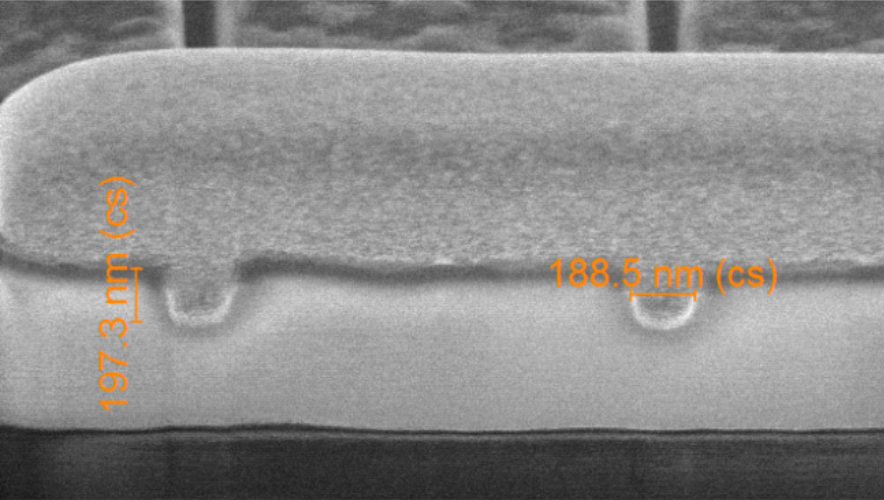}
		\caption{FIB-SEM image of the analytically optimized grating cross-section for the Sample 2.}
		\label{fig:Fig3_8_grat2}
	\end{center}
\end{figure} 

For the simulations we use the above defined multilayer. The grating is illuminated by TE polarized light with the wavelength $\Lambda = 1550~nm$ at the normal incidence. The grating is considered to be infinite along $X$ direction [Fig. \ref{fig:Fig3_5_RCWA_light}(a)]. We calculate reflection (R), transmission (T) and absorption (A) as a function of period ($\Lambda$ ranging from 1.1 to 1.5 $\mu m$) and depth of the grooves ($h$ ranging from 0.1 to 0.8 $\mu m$), with the width $a$ being fixed to 0.18 [Fig. \ref{fig:Fig3_6_RCWA_RTA_h_equal_a}].

From Fig. \ref{fig:Fig3_6_RCWA_RTA_h_equal_a} can be seen that reflection reaches its maximum and transmission reaches its minimum at the same value of period. Meanwhile the absorption reaches it's maximum at different values of the period. It means that for the grating configuration, where T=T$_{max}$ the maximum amount of light at 1550 nm at the normal incidence would be transmitted to the far field through the multilayer. At A=A$_{max}$ we have the maximum amount of losses, which also include the surface wave. 

To prove that the optimization should be done according to the maximum of absorption we plot the intensity profile within the 1DPhC [Fig. \ref{fig:Fig3_5_RCWA_light}b] with a grating parameters corresponding to the maximum of reflection ($R_{max}$) and to the maximum of absorption ($A_{max}$). The modulus of electric field at the interface of 1DPhC and external medium (air) is $\lvert E \rvert _{BSW}=3300~a.u.$ and $\lvert E \rvert _{BSW}=4707~a.u.$ for ($R_{max}$) and ($A_{max}$) correspondingly.

\begin{figure}[!t]
	\begin{center}
		\includegraphics[width=3.6in]{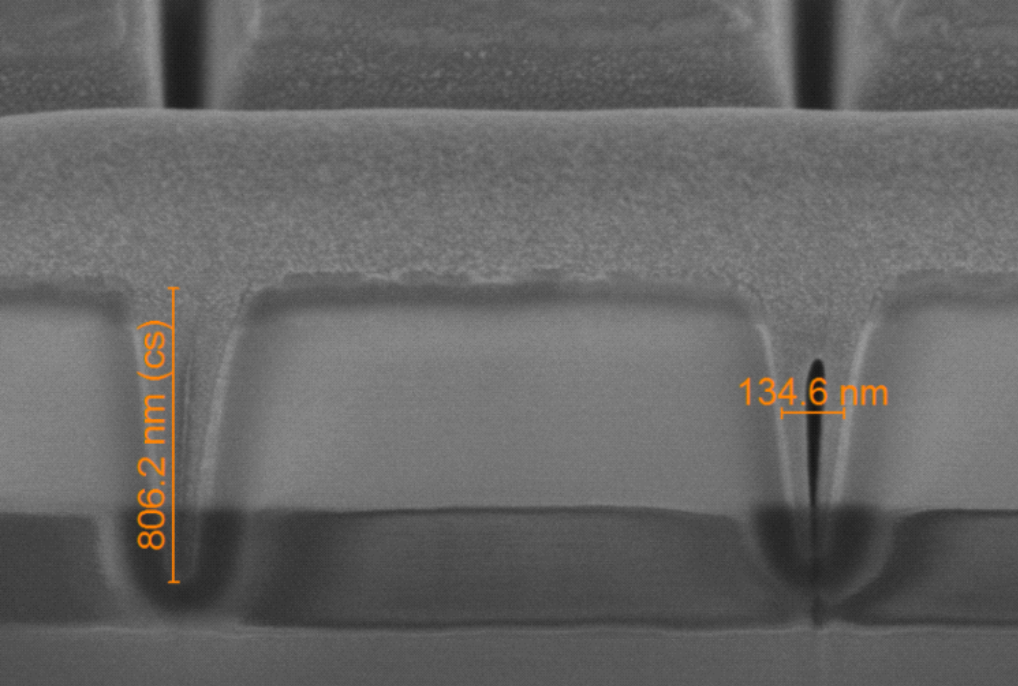}
		\caption{FIB-SEM image of the analytically optimized grating cross-section for the Sample 3.}
		\label{fig:Fig3_9_grat3}
	\end{center}
\end{figure} 

\begin{table}[!b] 
	\begin{center}
		\begin{tabular}{l|ccc} 
			\hline\noalign{\smallskip} 
			& $\Lambda,~\mu m$ & $h,~\mu m$ & $a,~\mu m$\\\noalign{\smallskip}\hline\noalign{\smallskip} 
			$Sample~1$ & 1.306 & 0.3 & 0.653\\ 
			$Sample~2$ & 1.352 & 0.186 & 0.186 \\ 
			$Sample~3$ & 1.367 & 0.7125 & 0.136 \\ 
			\noalign{\smallskip}\hline 
		\end{tabular} 
	\end{center}
	\caption{Grating parameters.}
	\label{tab:gratings_comp} 
\end{table}   

For the Sample 2 we choose the parameters of grating with $h=a$ where we reach the maximum of absorption $A_{max}=0.2587$: period $\Lambda=1.3523~\mu m$, depth and width $h=a=0.186~\mu m$.
        
For the Sample 3 we consider the general case, when we do not limit ourselves to the criteria $h=a$ and make a sweep over all possible values. We achieve maximum absorption $A_{max}=0.3387$ for period $\Lambda=1.367~\mu m$, depth $h=0.7125~\mu m$ and width $a=0.136~\mu m$.

The manufacturing of the grating and image of the grating profile for the Sample 2 and Sample 3 was done in the same way as for the Sample 1 by means of FIB [Fig. \ref{fig:Fig3_8_grat2}] and [Fig. \ref{fig:Fig3_9_grat3}]. Thus we obtained 3 samples with 3 different sets of grating parameters (see Table \ref{tab:gratings_comp}). Here we can see that the numerically achieved value for the grating period is slightly different from the analytically predicted one. The difference is due to overall refractive index change at the top layer of 1DPhC, which appears after nano-structuring. When we take into account the size of the grooves, we also take into account presence of air inside of the structures.  Still all the optimization methods give us the period value around $1.3~\mu m$. Also from [Fig. \ref{fig:Fig3_9_grat3}], we see the expected conicity from FIB milling. 

The conicity of the grating walls after the FIB milling can lead to several consequences.
Firstly, it leads to the prism effect, when the walls reflect the light into 1DPhC, or back to the incident medium, what can decrease the coupling efficiency \cite{Burr:08}.  Meanwhile, the conical shape of walls means that the grating generates not only the 0 diffracted order, but also some others. In terms of the wavelength at which the grating is designed to work (1550 nm in our case) it again means decrease of coupling efficiency of incident light into BSW. However, the grating also provides a spectrum of incident angles and wavelengths at which the phase matching conditions still can be achieved. In the real experimental conditions we have Gaussian distribution of the input beam, what means, that we have a spectrum of incident angles. Thus coupling losses for 1550 nm light at the normal incidence can be compensated and the overall system becomes more robust towards the wavelength or incident angle shift.  
 
   \begin{figure}[!b]
 	\begin{center}
 		\includegraphics[width=5.8in]{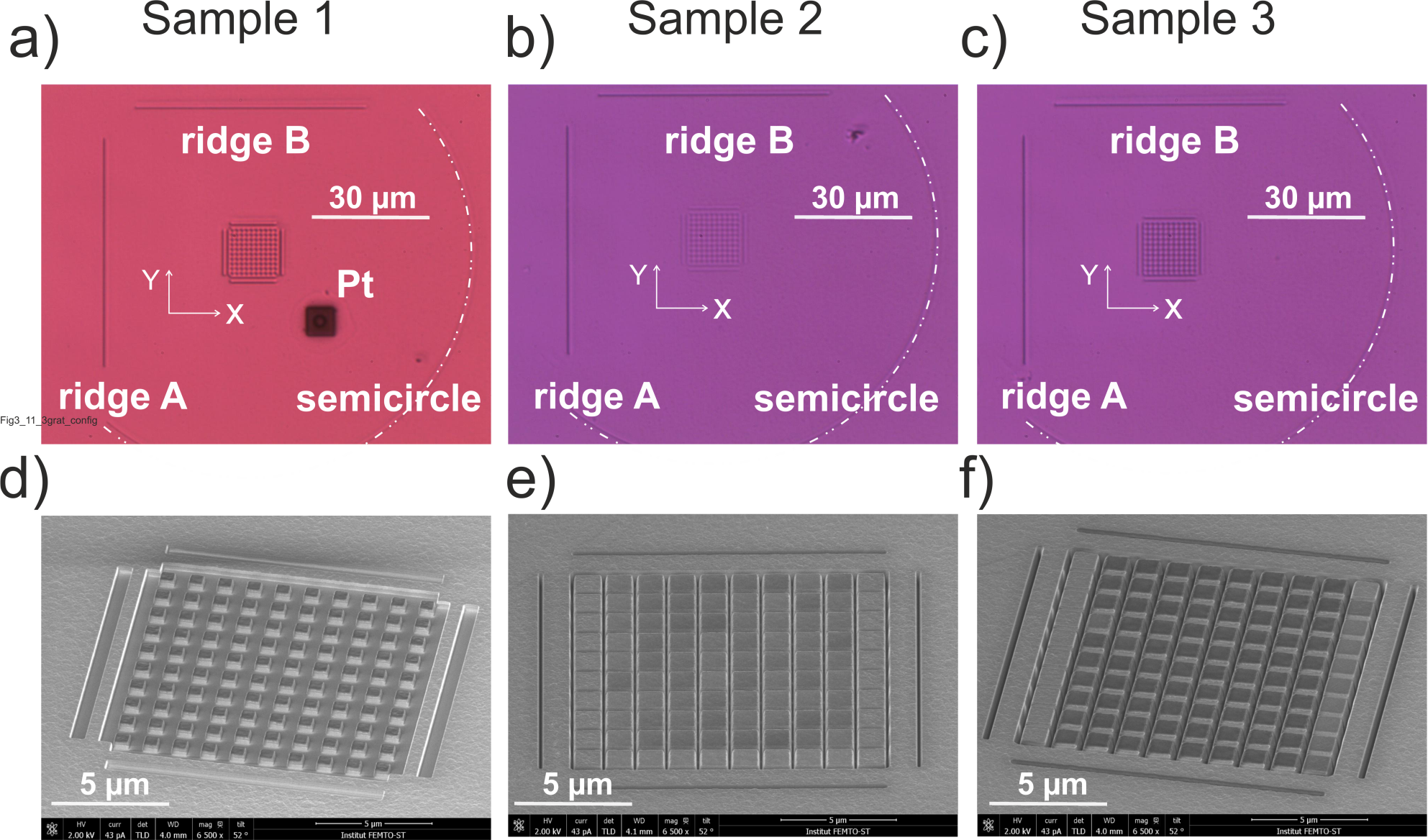}
 		\caption{(a),(b),(c) Microscope images of gratings and decouplers for Sample 1,2 and 3 respectively; (d), (e), (f) SEM images of cross-gratings for Sample 1,2 and 3 respectively.}
 		\label{fig:Fig3_11_3grat_config}
 	\end{center}
 \end{figure}

\subsection{Grating manufacturing.}
         
We have prepared 3 samples with gratings with different parameters according to Table \ref{tab:gratings_comp}. Ten grooves of 10 $\mu$m length in (X) and (Y) directions were made [Fig. \ref{fig:Fig3_11_3grat_config}]. SEM images of gratings in this work are tilted for better visualization.

In order to detect the presence of the BSW additional grooves (200 nm width, 400 nm depth) were milled by FIB at the distance of $30~\mu m$ from the grating together with a semicircle which is at the distance $50~\mu m$ [Fig. \ref{fig:Fig3_11_3grat_config} (a-c)]. These grooves and the semi-circle work as decouplers. At the Fig. \ref{fig:Fig3_11_3grat_config}(a) an additional area with a thin layer of Pt can be seen. Pt mark is made to be used as a reference during the milling process. The milling process is a cycle composed of milling steps followed by drift corrections. The drift correction consists in scanning the mark and to compare this image to the reference image made before starting the process. The correction is made automatically by the software and is based on Fast Fourier Transformation. The reference mark is a circle milled on a platinum rectangle in order to ensure a good stability of the image parameters (sharpness, brightness and contrast) as it will be scanned several tens times during the cycle. For Samples 2 and 3 Pt reference layer is deposited outside of the semicircle and it is not shown at the microscope images.

 \section{Experimental.}
         
 \subsection{Setup.}
    
\begin{figure}[!t]
	\begin{center}
		\includegraphics[width=\linewidth]{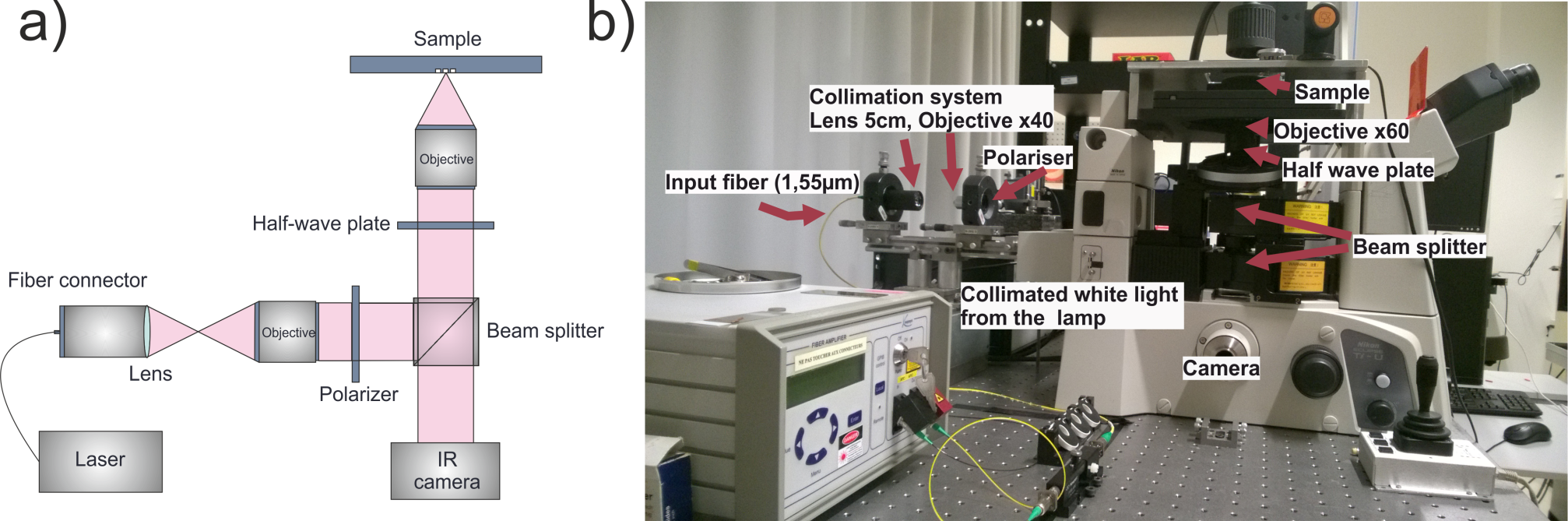}
		\caption{(a) Schema of experimental setup for sample characterization; (b) Optical table with the setup.}
		\label{fig:Fig3_12_setup}
	\end{center}
\end{figure}

The schema of experimental setup for sample characterization is shown in the Fig. \ref{fig:Fig3_12_setup}. The light was focused through the objective (NA=0.65) at the sample surface. We adjust the incident beam diameter with respect to the top objective's front lens in such a way, that the diameter of the incident beam is smaller than the diameter of the front lens of the focusing objective. Thus, we obtain a light spot that can cover the grating. Polarization was rotated by a half-wave plate. The signal was detected in reflection mode by the infrared camera (Xenics XEVA-2232).      

The set up is built on the base of inverted microscope NIKON Eclipse Ti-U. The HP 8164A Tunable Laser is used as a light source (1460 nm - 1580 nm). The wavelength was set to $1550~nm$ and output power to 0.5 mW. Sample was held by micro-manipulating platform for alignment purposes.

\begin{figure}[!t]
\begin{center}
\includegraphics[width=\linewidth]{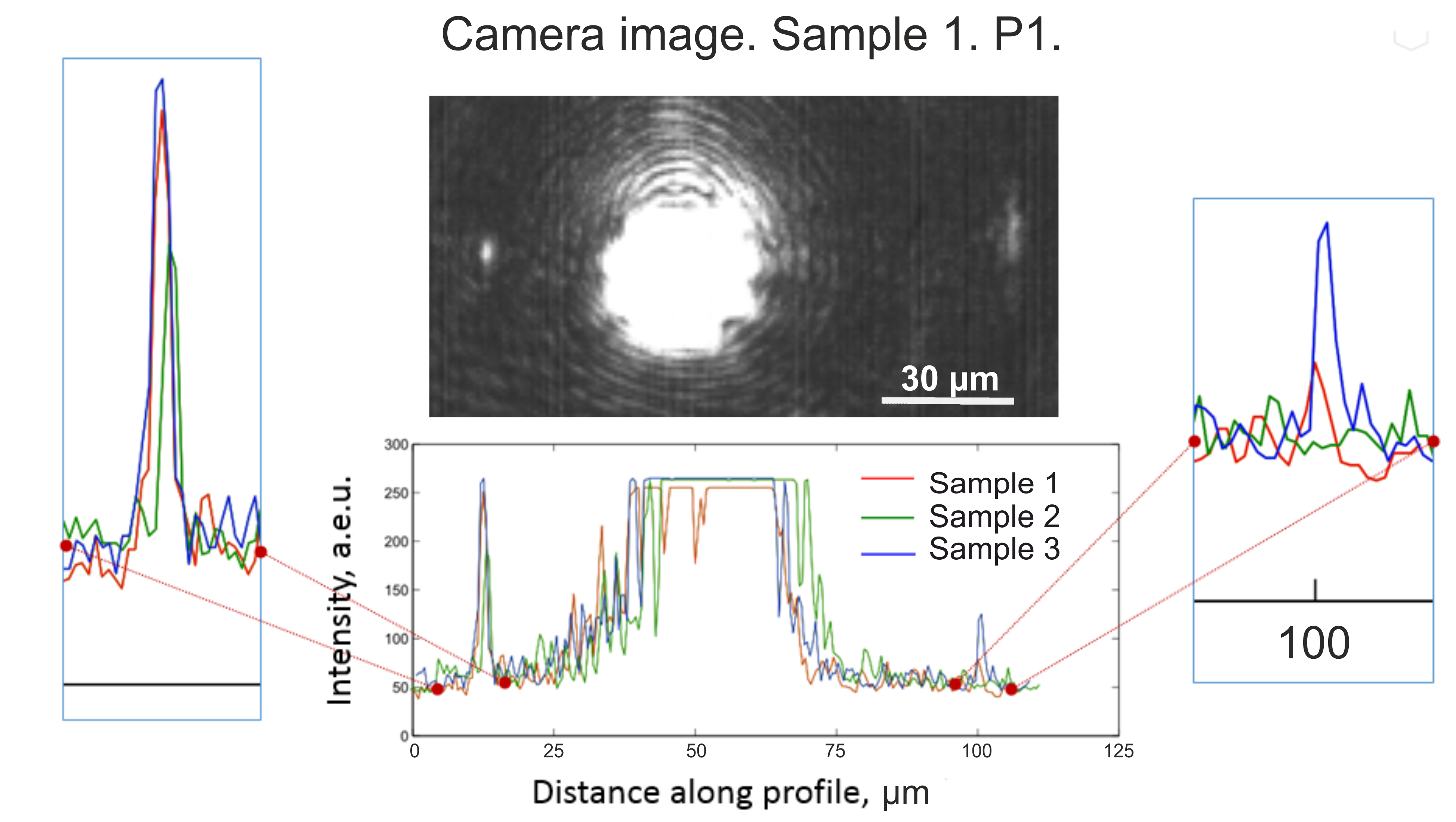}
\caption{Camera image of the illuminated grating and decouplers (left and right) and image profile for all three samples.}
\label{fig:Fig3_14_compare_exp_grat}
\end{center}
\end{figure} 
       
\subsection{Directional BSW guidance.}
 
As a first step of the optical characterization we have tested all 3 samples at one chosen polarization P1. In Fig. \ref{fig:Fig3_14_compare_exp_grat} we may see that light couples into the BSW at the grating area, propagates along the surface of 1DPhC and decouples at ridge A and at the semicircle. Decoupling at the ridge is stronger due to the losses along the propagation of the BSW. In order to achieve clear image of the decoupled light the coupling area was saturated. The image profile for all 3 samples for polarization P1 is shown at the Fig. \ref{fig:Fig3_14_compare_exp_grat}. Here it can be seen that the BSW coupling occurs for all three samples, but experimentally Sample 3 gives us the best coupling conditions. Therefore we choose sample 3 for all future analysis. Also the decoupling at $50~\mu m$ distance is quite week ($\times 3$ intensity decrease for Sample 3). Therefore for directionality studies we focus at the data collected from the grooves at the distance $30~\mu m$. 
        
\begin{figure}[!t]
\begin{center}
\includegraphics[width=5in]{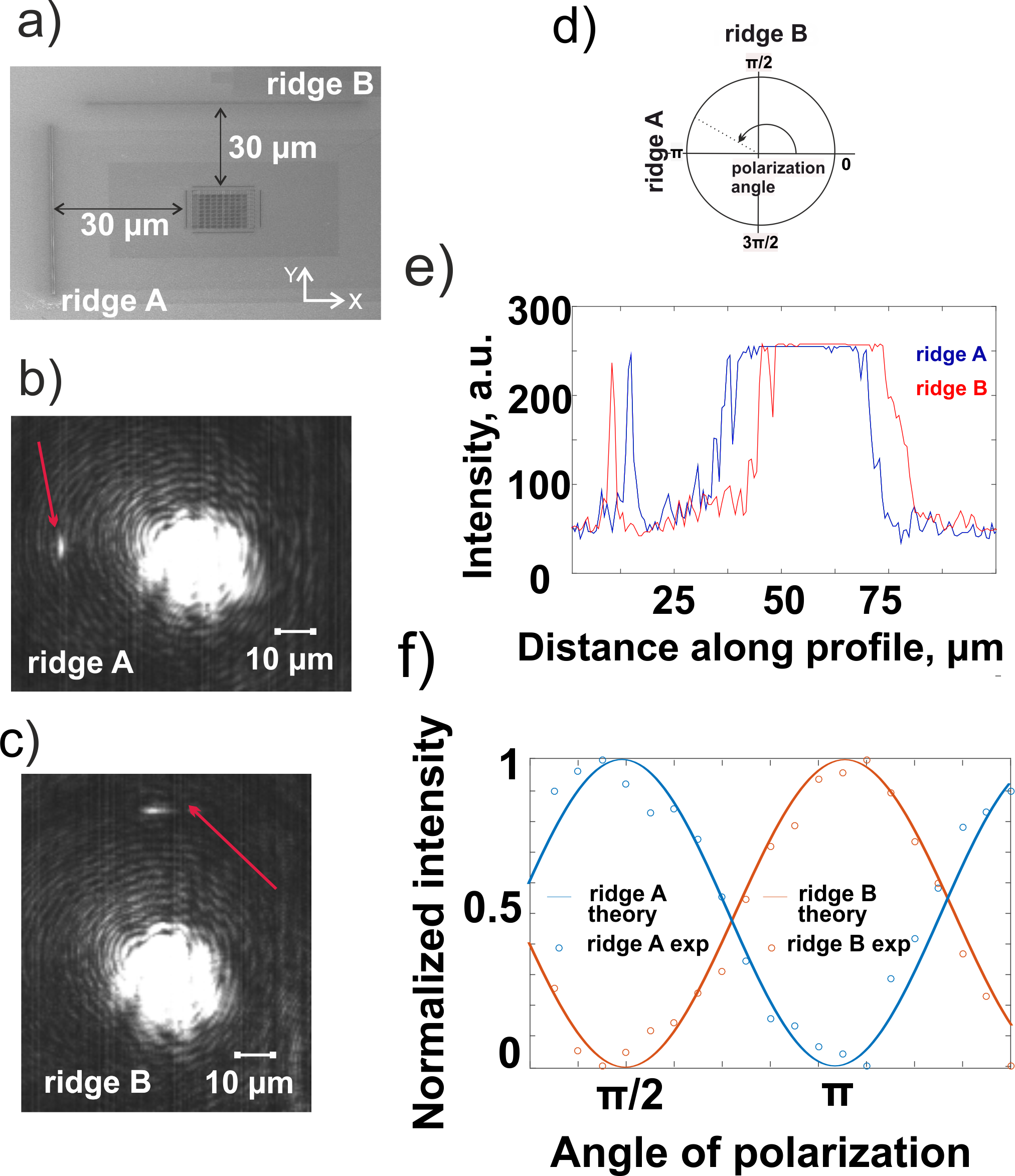}
\caption{(a) SEM image of the Sample 3; (b) Camera image of the BSW coupled along $X$ direction (by P1); (c) Camera image of the BSW coupled along $Y$ (by P2); (d) Intensity profile of camera image cross-sections for both BSW coupling directions; (e) Intensity of the decoupled light at the ridge A and B depending on polarization of the incident beam; (f) Sketch for polarization angle change.}
\label{fig:Fig3_15_TE_TM_for_grat_3}
\end{center}
\end{figure}

By illuminating the cross-grating by light with polarization P1 [Fig. \ref{fig:Fig3_3_copling_crossgrat}(b)] the grating 1 with periodicity along the X-axis [Fig. \ref{fig:Fig3_15_TE_TM_for_grat_3}(a)] works as a BSW coupler, BSW propagates along the surface and partially decouples at the ridge A [Fig. \ref{fig:Fig3_15_TE_TM_for_grat_3}(b)]. By illuminating the cross-grating by light with polarization P2 [Fig. \ref{fig:Fig3_3_copling_crossgrat}(b)] the grating 2 with periodicity along the Y-axis [Fig. \ref{fig:Fig3_15_TE_TM_for_grat_3}(a)] works as a BSW coupler, BSW partially decouples at the ridge B [Fig. \ref{fig:Fig3_15_TE_TM_for_grat_3}(c)]. The cross-sections of the images for both gratings are shown at Fig. \ref{fig:Fig3_15_TE_TM_for_grat_3}(e), where the intensity of decoupled light can be seen.  

The diameter of the incident light is bigger than a grating in order to provide better BSW coupling. At the Fig. \ref{fig:Fig3_15_TE_TM_for_grat_3}(e) a subtle difference in the distance from the pick to the incident beam spot can be observed. This occurs due to some decentering between the grating and input beam during manipulations for polarization rotation. 
      
Figures \ref{fig:Fig3_15_TE_TM_for_grat_3}(b) and \ref{fig:Fig3_15_TE_TM_for_grat_3}(c) show the limit positions when we switch the polarization of the incident beam from the state where only P1 or P2 is present. The incident beam covers the grating, light couples into the BSW, propagates along the sample surface and partially decouples on the groove, what can be seen as a bright line at the side from the incident beam spot. If both polarizations (P1 and P2) are present in the incident beam it is possible to control the amount of energy propagating in each direction [Fig. \ref{fig:Fig3_15_TE_TM_for_grat_3}(e)]. 
 
 The angle of polarization was changed as it is shown at Fig. \ref{fig:Fig3_15_TE_TM_for_grat_3}(d). We have only P2 polarization component at 90$^\circ$ angle and only P1 polarization component at 180$^\circ$.  
 
 We varied the angle of polarization over 300$^\circ$ range with a 15$^\circ$ step. Experimental data for normalized intensity detected at the ridge A and ridge B is shown as blue and yellow circles at Fig. \ref{fig:Fig3_15_TE_TM_for_grat_3}(e). Cosine function fits for data collected from both grooves are shown as solid lines. The 180$^\circ$ shift is observed between functions picks. Thus we detect maximum intensity at the ridge B (only P2, 90$^\circ$ angle) when there is a minimum intensity of light at the ridge A. Similarly we obtain maximum intensity at the ridge A (only P1, 180$^\circ$ angle) when there is a minimum intensity at the ridge B. Once we rotate the polarization in such a way that both P1 and P2 are present we can see that BSW splits in two directions. The amount of light delivered to the grooves is defined as a cosine function of the polarization angle with a shift over $\pi$ between two functions. The BSW splits equally in orthogonal directions at 135$^\circ$ and 225$^\circ$.

\section{Coupling efficiency estimations.}
     
\subsection{Numerical calculations.}
   
     \begin{figure}[!t]
   	\begin{center}
   		\includegraphics[width=3.8in]{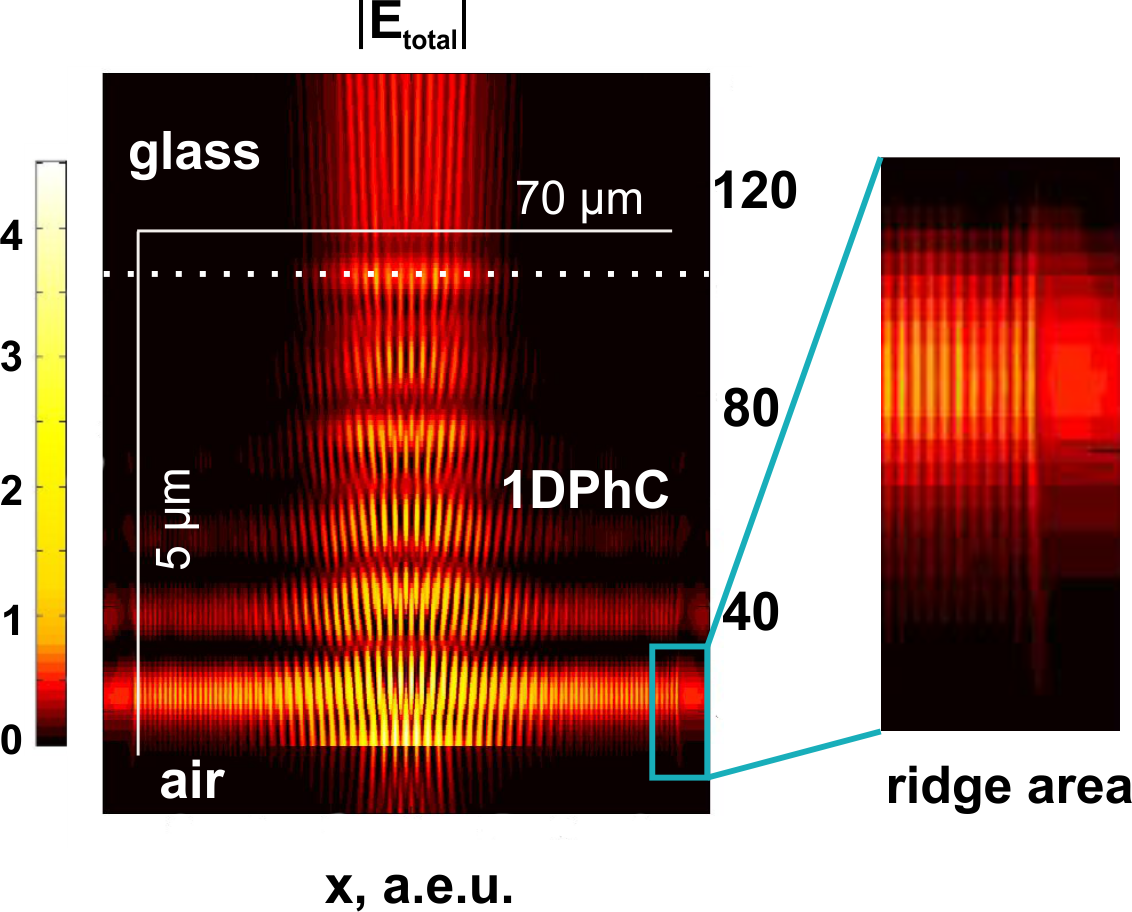}
   		\caption{Field profile of 1DPhC for the Sample 3 with optimized grating parameters and diffusers made by FDTD simulations.}
   		\label{fig:Fig3_17_field_FDTD}
   	\end{center}
   \end{figure} 
   
   \begin{figure}[!t]
   	\begin{center}
   		\includegraphics[width=\linewidth]{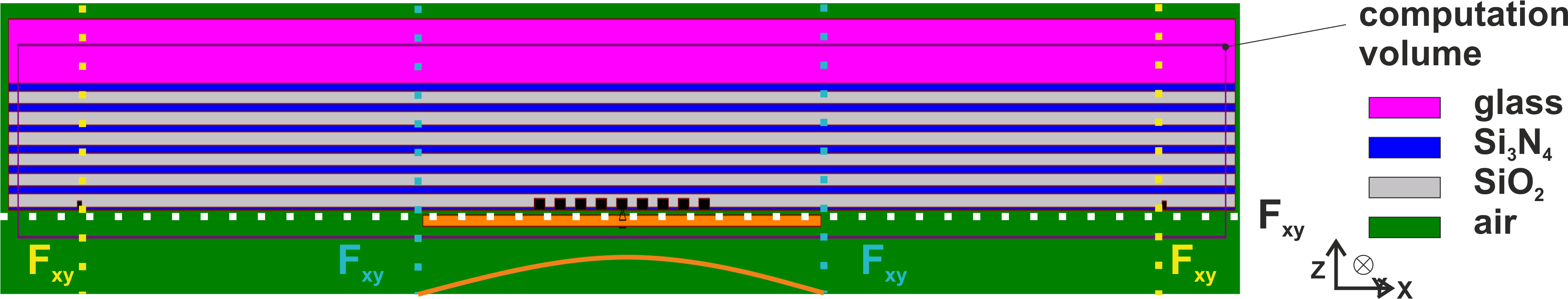}
   		\caption{Multilayer, grating and decouplers for the Sample 3 in R-SOFT simulations.}
   		\label{fig:Fig3_18_flux_explanation}
   	\end{center}
   \end{figure}

The BSW is not totally decoupled at the ridge and keeps propagating for the distance more than 30 $\mu$m, what can be seen from the field map in 1DPHC. The map should be calculated in the sample with a finite grating. Impedance approach or RCWA method can't provide this calculations. Therefore FDTD method was used for simulations [Fig. \ref{fig:Fig3_17_field_FDTD}]. In order to estimate the coupling efficiency of the device numerical simulations were performed. The coupling was calculated as a ratio of the flux of the Poynting vector over two surfaces: the BSW coupling area (blue dashed lines in the Fig. \ref{fig:Fig3_18_flux_explanation}) and the incident light area  (white dashed line in the Fig. \ref{fig:Fig3_18_flux_explanation}). According to these calculations about 18\% of energy is coupled into the BSW with this grating configuration. This value is quite competitive with standard ways of BSW excitation (only 6\% for BSW prism coupler in Kretschmann configuration \cite{yu:14}).

\begin{figure}[!b]
	\begin{center}
		\includegraphics[width=\linewidth]{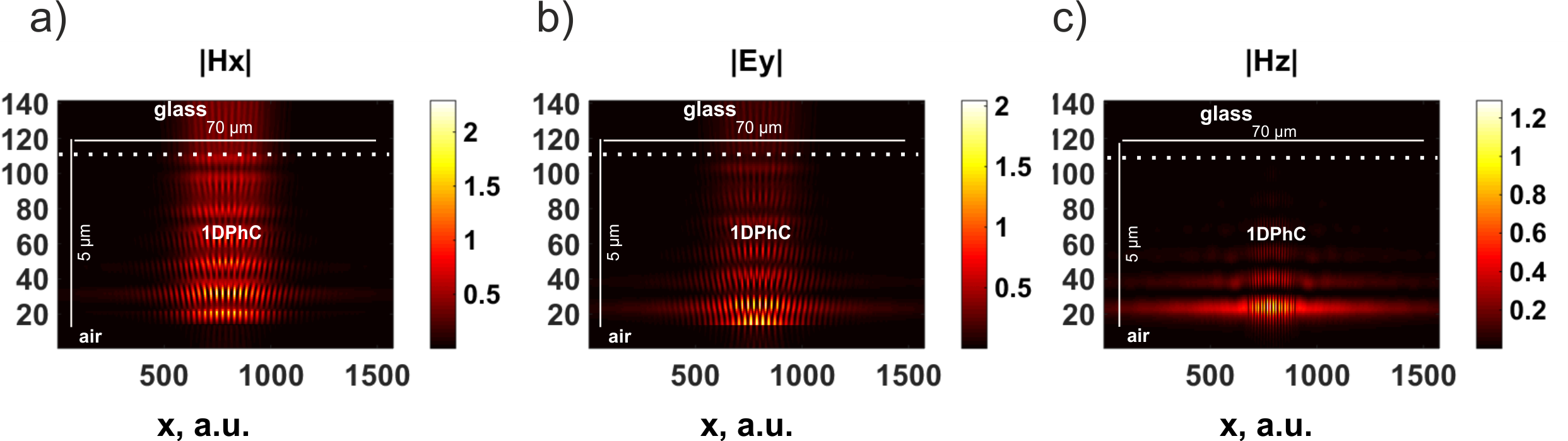}
		\caption{Field profile inside the 1DPhC without decoupling ridges for (a) $H_x$; (b) $E_y$; (c) $H_z$ field components.}
		\label{FDTD_no_ridge}
	\end{center}
\end{figure}

Indeed, let us consider the grating and the decouplers of the Sample 3 and perform the FDTD simulations for Gaussian incident wave with TE polarization (approach to the experimental conditions). From these simulations we can get the field map over the calculation volume for electric and magnetic components of the field $E_y$, $H_x$ and $H_z$ (see Fig. \ref{FDTD_no_ridge}). 

Then we extract the data about the values of $E_y$, $H_x$ and $H_z$ along the coordinate, where we want to calculate the flux of the Poynting vector (blue, yellow and white dashed lines at the Fig. \ref{fig:Fig3_18_flux_explanation}). In the Fig. \ref{fig:Fig3_18_flux_explanation} green area is an external air media, purple area is a glass substrate on which the multilayer was deposited (blue layers - Si$_3$N$_4$, gray layers - SiO$_2$), orange line is the launch area, violet line indicates computational volume. PML conditions were applied from all the sides.

\begin{figure}[!t]
	\begin{center}
		\includegraphics[width=\linewidth]{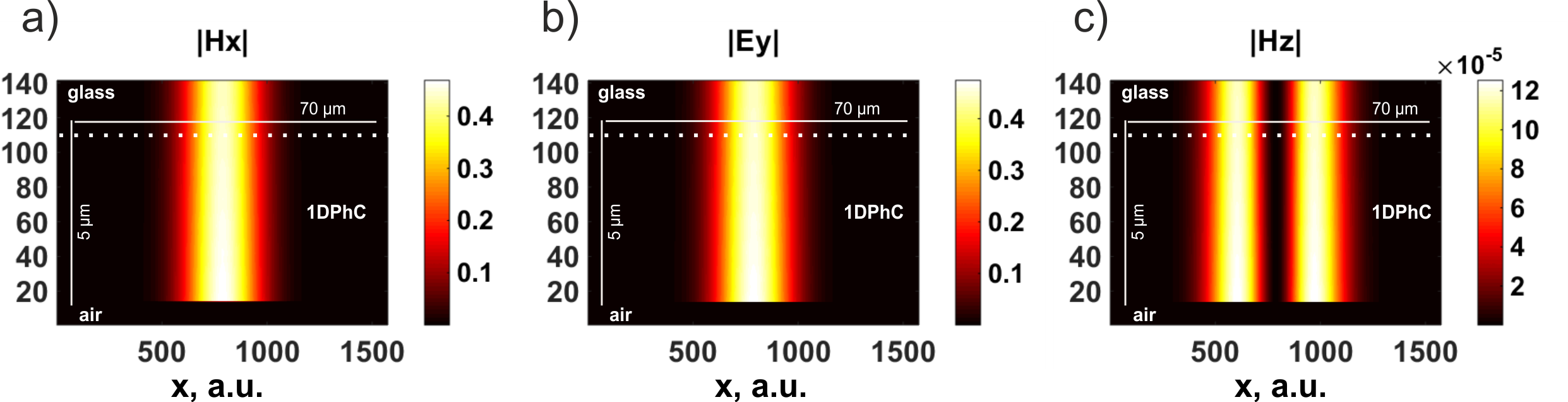}
		\caption{Field profile inside the computation volume without 1DPhC for (a) $H_x$; (b) $E_y$; (c) $H_z$ field components.}
		\label{FDTD_nothing}
	\end{center}
\end{figure}

Poynting vector represents the directional energy flux (the energy transfer per unit area per unit time) of an electromagnetic field. Flux of the Poynting vector through $YZ$ plane (blue line) gives us an information about the energy carried by BSW. Flux of the Poynting vector through $XY$ plane (white line) gives us the information about the energy carried by incident light. $E_y$, $H_x$ and $H_z$ along the white line were retrieved from the simulations without any structures inside the computation volume (see Fig. \ref{FDTD_nothing}). This way we avoid the energy transfer through plane of the white line introduced by light reflected from the 1DPhC. 

Flux of the Poynting vector for our case can be calculated as following:

\begin{equation}
F_{yz}=\int\limits_{S} \overline{\pi} d\overline{S} = L_y \int \overline{\pi_x} dz = - \frac{1}{2} L_y \int Re \left\{E_y H_z^*  \right\} dz  
\label{eq:refname3_1}
\end{equation}

\begin{equation}
F_{xy}=\int\limits_{S} \overline{\pi} d\overline{S} = L_y \int \overline{\pi_z} dx = - \frac{1}{2} L_y \int Re \left\{E_y H_x^*  \right\} dx  
\label{eq:refname3_1}
\end{equation}

Where $\overline{\pi}$ is a Poynting vector and $L_y$ is the length of the calculation volume along $Y$ axis.  

For our case we define coupling efficiency as a ratio between the energy in the input beam and the energy carried by BSW. In terms of flux of the Poynting vector coupling efficiency would be:

\begin{equation}
Eff=\frac{F_{yz}}{F_{xy}}   
\label{eq:refname3_1}
\end{equation}

If we calculate the efficiency just next to the coupling area we would have about 18\% in the particular case of the Sample 3 \textit{($F_{xy}$ along the blue line in Fig. \ref{fig:Fig3_18_flux_explanation})}. Though there are losses during the propagation of the BSW. Therefore less light (energy) is delivered to the decoupling groove (12\% to the both sides - $F_{xy}$ along the yellow line in Fig. \ref{fig:Fig3_18_flux_explanation}).
       
\subsection{Experimental data analysis.}
    
\begin{figure}[!ht]
\begin{center}
\includegraphics[width=3.5in]{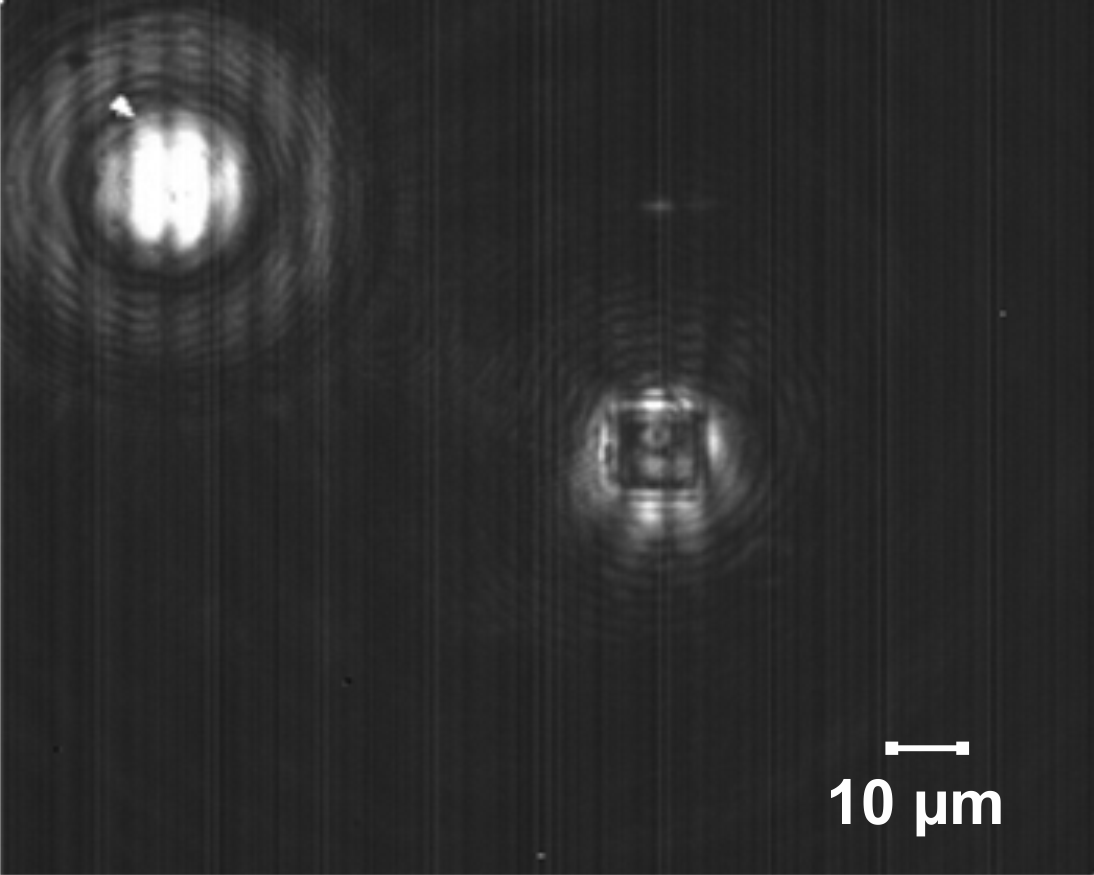}
\caption{Non-saturated camera image image of reflected and decoupled light.}
\label{fig:Fig3_19_1_im_to_process}
\end{center}
\end{figure}

In current experimental configuration, it is hard to measure the coupling efficiency directly. Therefore, in order to check the correlation between numerical results and experimental data, we compare the amount of decoupled light from experiment and simulations by normalizing it with respect to the reflected light. For proper quantitative analysis, we use non-saturated image of reflected and decoupled light [Fig. \ref{fig:Fig3_19_1_im_to_process}]. Figure \ref{fig:Fig3_19_1_im_to_process} shows the grating coupling area (in the center). Incident light reflects from the grating. In this case we use light polarization P2. BSW decouples at the ridge B (bright spot on the top of the grating). In the left top corner the reflection from the half wave plate is shown. The half wave plate is slightly tilted. Thus the reflection from it and from the grating can be resolved at the image. 

\begin{figure}[!b]
	\begin{center}
		\includegraphics[width=4.5in]{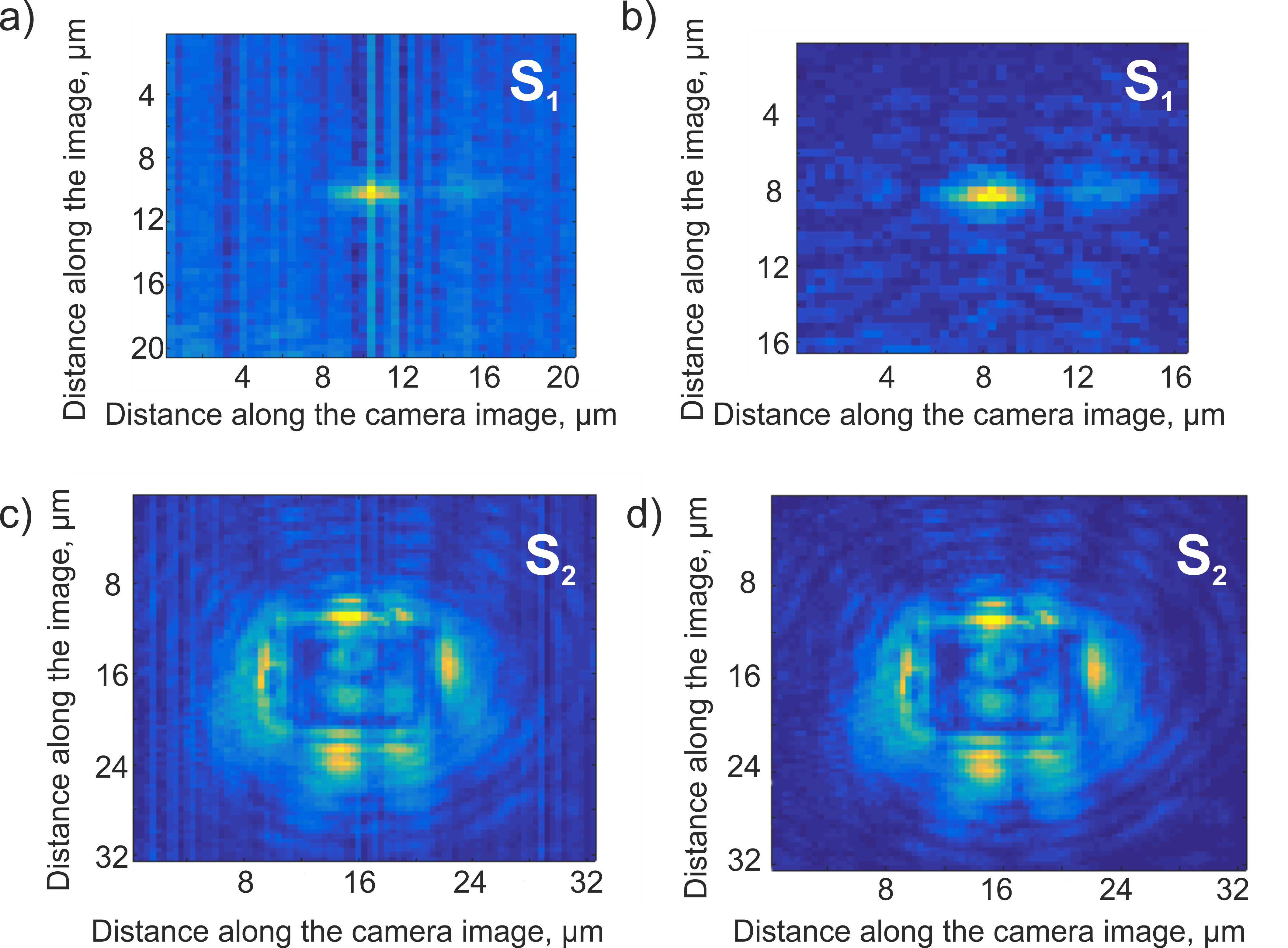}
		\caption{(a)Data from original image around the grove area; (b) Data from original image around the grating area; (c) Processed image without the noise for the grove area; (d) Processed image without the noise for the grating area.}
		\label{fig:Fig3_19_procesed_pics2}
	\end{center}
\end{figure} 

Quantitative data about the reflectance and about the amount of light decoupled at the grove was retrieved by processing the data from the Fig.  \ref{fig:Fig3_19_1_im_to_process}. Using Matlab we select the area of grating and the area around the ridge B [Fig. \ref{fig:Fig3_19_procesed_pics2}(a),(c)]. These images contain vertical stripes, which are introduced by camera and can be considered as a noise. To eliminate this problem we subtract the noise from the image. For this purpose we took a separate image without structures and illumination. As a result we obtained two clear images [Fig. \ref{fig:Fig3_19_procesed_pics2}(b),(d)]. In the first estimation we may say that the intensity of the light decoupled from the ridge B is equal to the intensity sum over the surface $S1$ [Fig. \ref{fig:Fig3_19_procesed_pics2}(b)] and the reflectance is equal to the intensity sum over the surface $S2$ [Fig. \ref{fig:Fig3_19_procesed_pics2}(d)]. Thus the ratio of intensities over the surface of the decoupling area and reflectance from the incident beam area would be $R_I=S_1/S_2$, which is equal to $0.65\%$.

\begin{figure}[!t]
\begin{center}
\includegraphics[width=\linewidth]{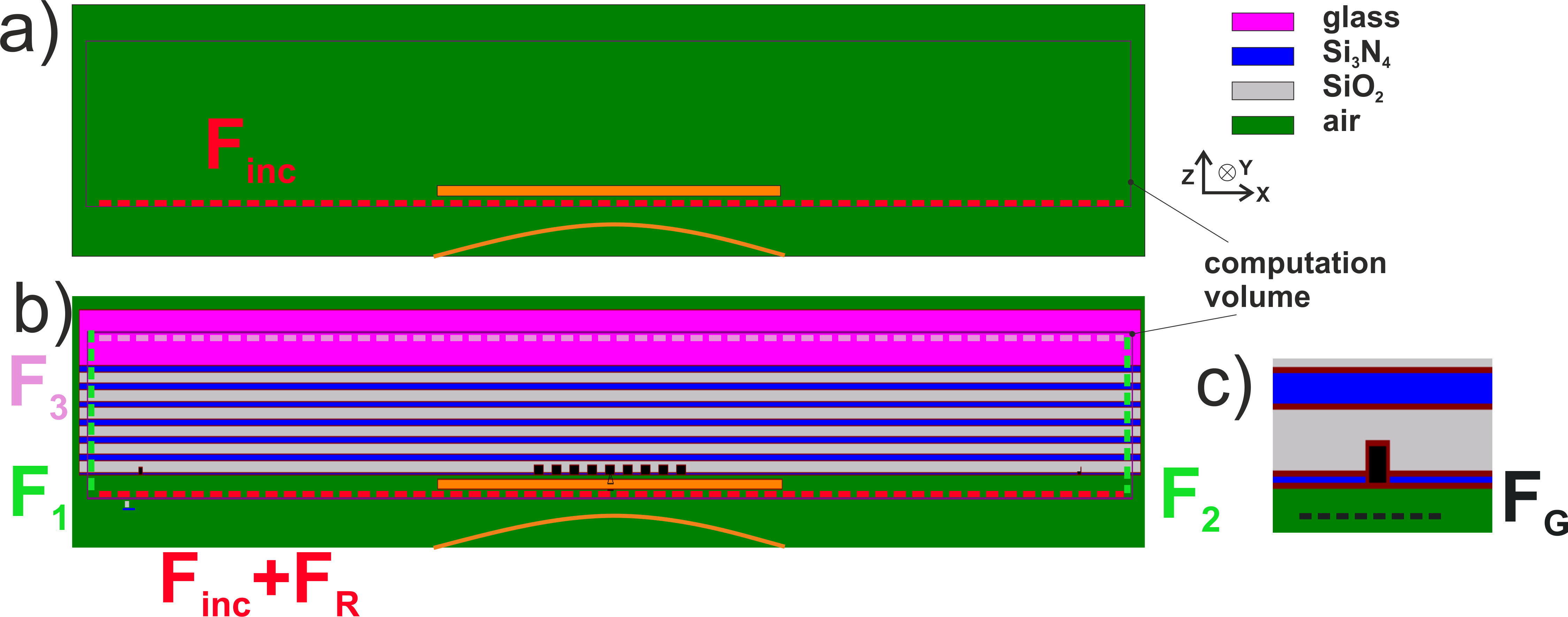}
\caption{(a) System and flux line for input energy calculations; (b) System and flux line for energy coupled to BSW and transmitted (multilayer without groves); (c) System and flux line for decoupled energy.}
\label{fig:Fig3_18_decupl_explanation}
\end{center}
\end{figure} 

\begin{figure}[!b]
	\begin{center}
		\includegraphics[width=\linewidth]{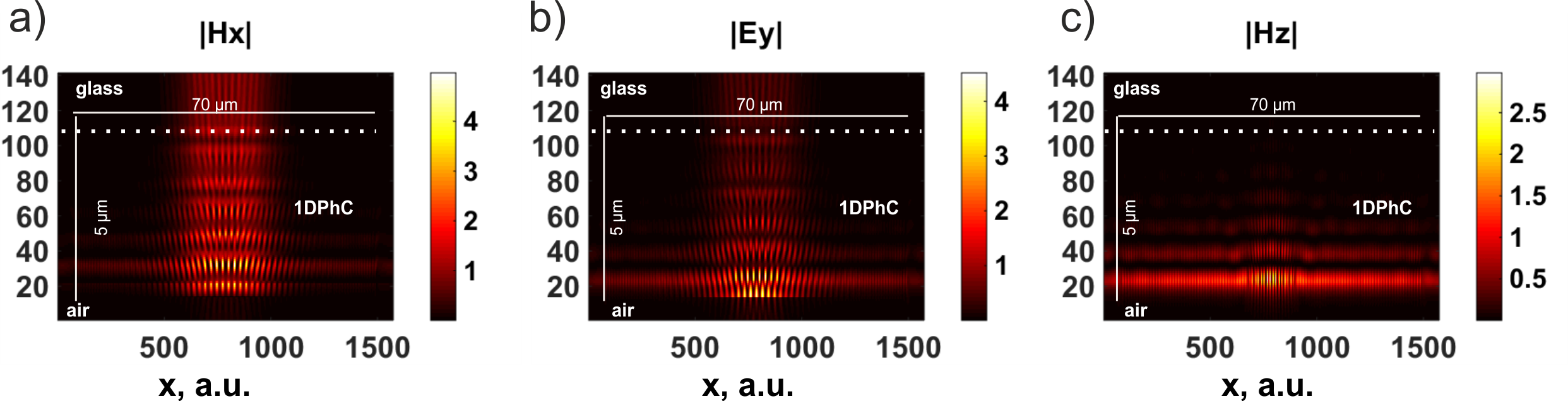}
		\caption{Field profile inside the 1DPhC with decoupling ridges for (a) $H_x$; (b) $E_y$; (c) $H_z$ field components.}
		\label{FDTD_with_ridge}
	\end{center}
\end{figure} 

To estimate $R_I$ numerically we use the energy conservation law: energy, brought by the incident light would be equal to the energy, carried by reflected, transmitted and coupled in BSW light.  In terms of the flux of the Poynting vector it would mean, that $F_{inc}=F_1+F_2+F_3+F_R$. $F_{inc}$ - represents the input energy (can be calculated as a flux of the Poynting vector along the red line at Fig. \ref{fig:Fig3_19_procesed_pics2}(a), where we don't have any structures); $F_1$ and $F_2$ ($F_1=F_2$) - represent the energy cared by BSWs (can be calculated as a flux of the Poynting vector along the green lines at Fig. \ref{fig:Fig3_19_procesed_pics2}(b)); $F_3$ - represents the energy in transmitted light (can be calculated as a flux of the Poynting vector along the pink line at Fig. \ref{fig:Fig3_19_procesed_pics2}(b)); $F_R$ is reflection related part of Poynting vector flux through the plane of the red line at the Fig. \ref{fig:Fig3_19_procesed_pics2}(b). The Poynting vector flux through the closed system gives an information about the total energy in the system. Therefore we use the closed system as it shown at Fig. \ref{fig:Fig3_19_procesed_pics2}(b).
Taking into account all the above listed considerations $F_R$ may be calculated as following: $F_R=F_{inc}-2F_1-F_3$. 

Finally $R_I=F_g/F_R$, where $F_g$ is a flux of the Poynting vector through the surface in front of the grove (see Fig. \ref{fig:Fig3_19_procesed_pics2}(c)). $F_g$ gives us estimation about decoupled light. The ratio between decoupled light at the ridge and reflected light ($R_I$) from numerical simulations is 0.80\%. Field maps from FDTD analysis for $H_x$,  $E_y$ and  $H_z$ components used for $R_I$ calculations within the structure under study are shown at Fig. \ref{FDTD_with_ridge}

Result from numerical calculations and result from experimental data are in a good agreement, therefore we can assume that coupling efficiency of the device, which used for experiments, is close to the numerically obtained ones. Here we can conclude that our estimations of coupling efficiency are correct.

\section{Conclusion.}
		
In summary, in this chapter we have demonstrated, designed and fabricated a grating coupler on a BSW sustaining platform. Grating parameters were analytically and numerically optimized by different methods in order to obtain the best coupling conditions for a given 1DPhC. 
The light is launched orthogonally to the multilayer. Due to a special grating configuration we demonstrate directionality of the BSW propagation depending on polarization of the incident light. The structure was experimentally realized on the surface of the 1DPhC crystal by FIB milling. Estimated coupling efficiency of the design was calculated, which in theory reaches 18\% for our particular case. Experimental results are in a good agreement with a theory. The investigated configuration can be successfully used as a BSW launcher in on-chip all-optical integrated systems and work as a surface wave switch or modulator.   

With a grating launcher and the BSW propagation direction control with polarization of the incident light we reach a higher degree of control of Bloch surface waves. We obtain control of the propagation direction from the macro-scale down to the nano-scale. Also with the inter-crossed grating coupler we have new functionalities with simplified technologies and high degree of integration.

\chapter{BSW for particles manipulation.}

In this chapter we will apply the BSW in the field of optical trapping and particles manipulation. Additionally we use the concept of the grating coupler for BSW propagation control in water. We design the 1DPhC which is able to sustain the surface wave with water as an external media and the corresponding cross-grating launcher and investigate the interaction between the BSW and latex beads in the given configuration.
This design can be also used to push particles in different directions depending on the polarization of the incident light as a perspective. This particle manipulation concept can find some applications in biomedical field. 
%Control of light propagation with polarization always remains an interesting topic for studies in optics. It means that it is possible to achieve a tunability without changing anything in the setup or in the sample (device) under consideration. Input beam polarization is the only parameter to control, which is relatively easy task to do. In this chapter we will describe some concepts where polarization control of BSWs may be used and preliminary steps made on the way to realization of these concepts.

\vspace*{0.2cm}
\minitoc

\section{Tunability in water. Concept.}
    
\begin{figure}[!ht]
\begin{center}
\includegraphics[width=4in]{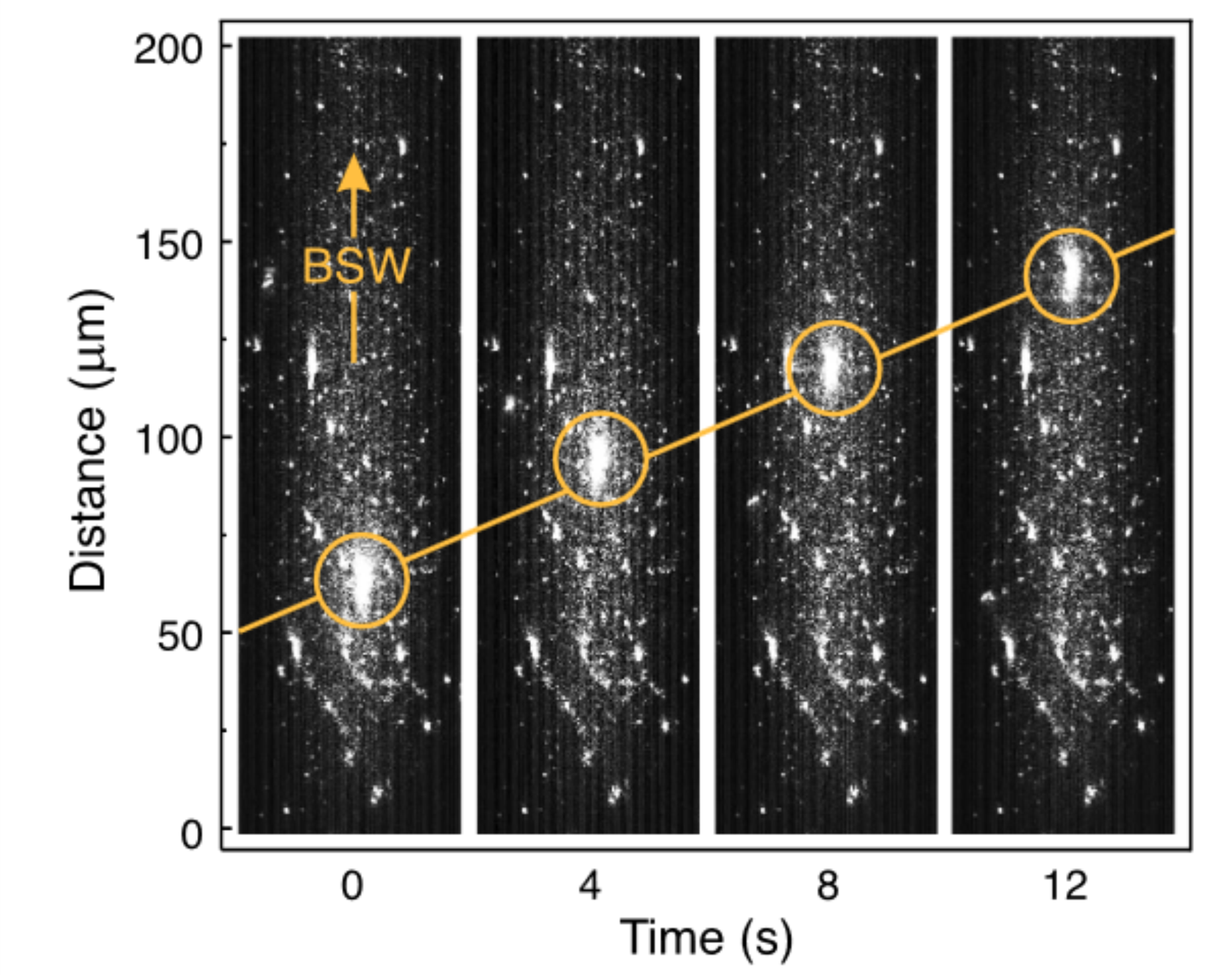}
\caption{Time-sequenced set of microimages of a particle at the PC
surface. The bright area on the images corresponds to the BSW propagation region, and the bright spots inside the circles correspond to the micro-particle propelled by the BSW evanescent field \cite{Shilkin:2015}.}
\label{fig:Fig_4_1_pushing_bids_prizm}
\end{center}
\end{figure}

In 2015 Shilkin et. al. experimentally demonstrated the interaction between a single dielectric micro-particle and the evanescent field of the Bloch surface wave in a one-dimensional photonic crystal \cite{Shilkin:2015}. The Bloch surface wave-induced forces on a 1 $\mu m$ polystyrene sphere were measured by photonic force microscopy. They proved that particles can be pushed by BSW over long distances (100 $\mu m$) as it is shown in Fig. \ref{fig:Fig_4_1_pushing_bids_prizm}. The results demonstrated the potential of 1D photonic crystals for the optical manipulation of micro-particles and suggested a novel approach for utilizing light in lab-on-a-chip devices.    
    
The manipulation of particles was induced by the BSW excited in Kretschmann configuration. Light at the wavelength of 532 nm was coupled through the prism and we may see the bright area on the images (see Fig. \ref{fig:Fig_4_1_pushing_bids_prizm}), which corresponds to the BSW propagation region.

In this part of the work we propose a concept of the device which would combine the capability of a cross-grating coupler to guide BSW in an arbitrary direction depending on polarization of the incident light and the possibility to manipulate particles. 

Indeed the results achieved in our previous work \cite{Kovalevich:17} together with achievements of Shilkin et. al. \cite{Shilkin:2015} proves the possibility of creation of such a device.

\subsection{1DPHC and grating design in water.}
       
We design and fabricate the 1DPhC which sustains the  BSW at the multilayer/water interface. The crystal consists of three pairs of silicon nitride and silicon oxide layers with additional thin layer of titanium oxide on the top. The whole stack was made on the glass support. For Si$_3$N$_4$ and SiO$_2$ layers fabrication the plasma-enhanced chemical vapor deposition (PECVD) is used. The top layer of TiO$_2$ is deposited by atomic layer deposition (ALD). PECVD multilayer deposition was made in EPFL optics and photonics technology laboratory. TiO$_2$ deposition was performed in the Institute of Photonics (University of Eastern Finland). The thicknesses of all the dielectrics are optimized according to impedance approach as described in Chapter 2.1.2. in such a way that the whole stack sustains the BSW at 808 nm wavelength. This wavelength was selected due to the small absorption in water. Designed layer's thicknesses are 160 nm for Si$_3$N$_4$, 240 nm for SiO$_2$ and 20 nm for TiO$_2$. Refractive indices used for simulations are $n_{glass} = 1.5106$, $n_{Si_3N_4} = 1.8065 + 0.001i$, $n_{SiO_2} =   1.4532 + 0.001i$, $n_{TiO_2}=2.5178+0.001i$ and $n_{water} = 1.3141$ at $\lambda=808$ nm.   
       
\begin{figure}[!t]
\begin{center}
\includegraphics[width=5.3in]{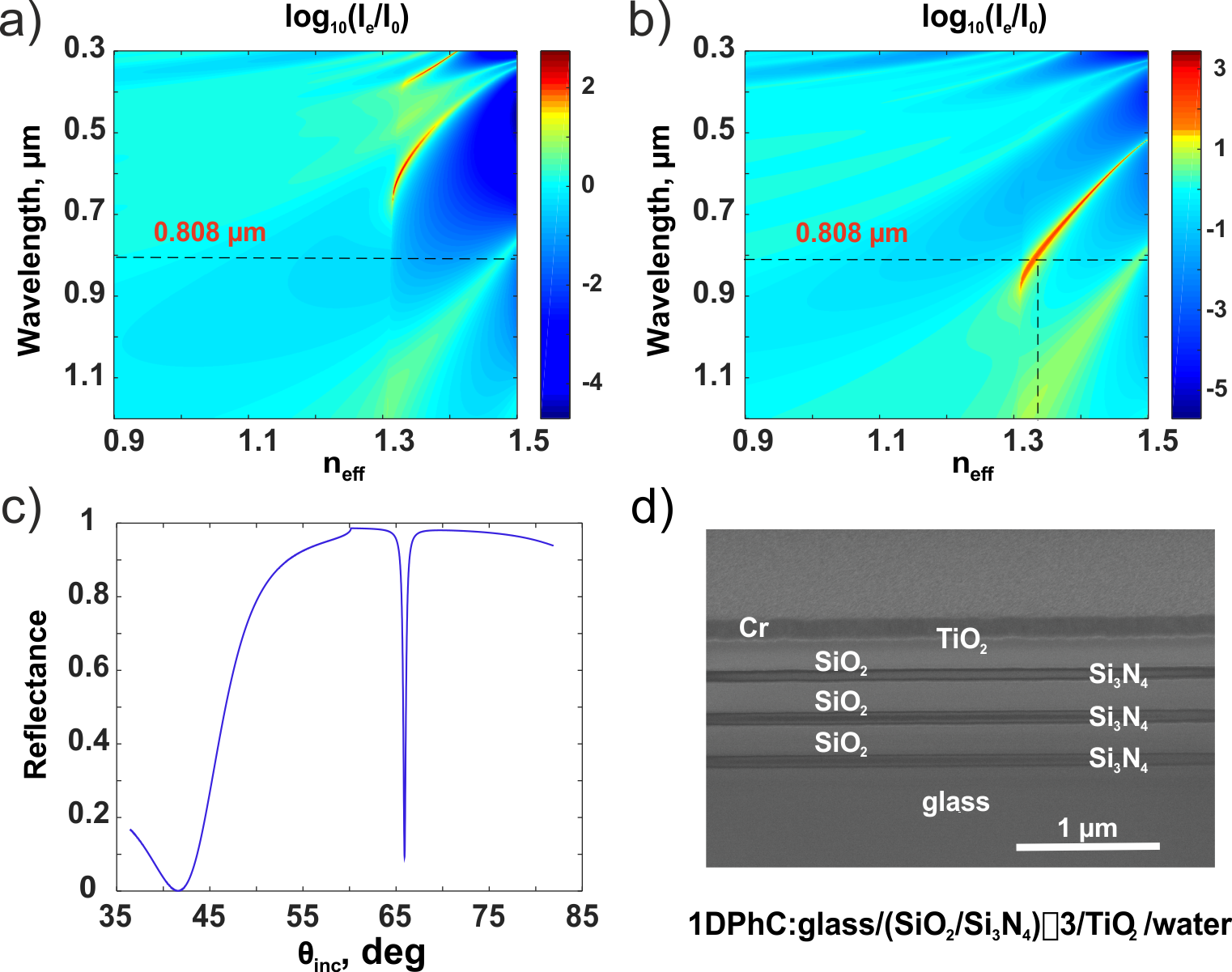}
\caption{(a) Band gap diagram of the 1DPhC without TiO$_2$; (b) Band gap diagram of the 1DPhC with TiO$_2$; (c) Calculated reflectance at 808 nm wavelength; (d) FIB-SEM image of 1DPhC.}
\label{fig:Fig_4_2_ML}
\end{center}
\end{figure}

Dispersion curves for the 1DPhC without and with TiO$_2$ top layer are shown at the Fig. \ref{fig:Fig_4_2_ML}(a) and (b) respectively. The multilayer is designed in such a way that BSW does not exist for $\lambda=808~nm$ if there is no titanium oxide [Fig. \ref{fig:Fig_4_2_ML}(a)] but it is supported by the 1DPhC with additional 20 nm of TiO$_2$ [Fig. \ref{fig:Fig_4_2_ML}(b)]. This feature of the design can be used in future for additional nano-structuring of the top surface. For example, it would be possible to achieve strong BSW confinement in a TiO$_2$ 2D waveguide.

The profile of the fabricated multilayer is shown in the Fig. \ref{fig:Fig_4_2_ML}(d). The SEM image was taken during the FIB process.  The FIB was used in order to make a small opening on the sample surface, in order to reveal the multilayer for SEM. All the layers of silicon oxide and silicon nitride can be clearly observed. Measured layer thicknesses are within 40 nm deviation. Additional 100 nm of Cr were deposited on the top of 1DPhC by sputtering in order to avoid charging effects. After characterization, Cr was removed by wet etching. 

For multilayer optimization in Kretschmann configuration the BSW related reflectance dip occurs at 65$^\circ$ incident angle ($\rm{n_{BSW}}=1.375$). Reflectance dumps down to 0.09 for 3 pairs of Si$_3$N$_4$ and SiO$_2$ according to calculations made by impedance approach [Fig. \ref{fig:Fig_4_2_ML}(c)]. 

The period for the grating coupler was defined in the same way as described in Chapter 3.3.1: $\Lambda=\lambda/n_{BSW}=808/1.375=587~nm$. The depth of the grooves ($h$) was chosen as 250 nm, width of the grooves ($a$) was chosen as 300 nm. 

\begin{figure}[!t]
\begin{center}
\includegraphics[width=5.5in]{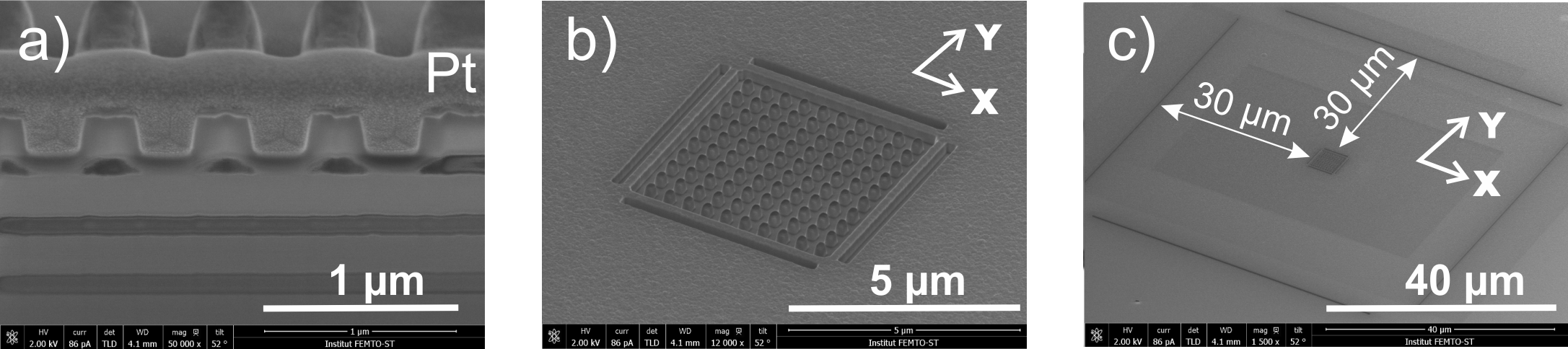}
\caption{(a) FIB-SEM image of the analytically defined grating cross-section for the BSW in water; (b) SEM image of the manufactured grating; (c) SEM image of the final sample for BSW in water.}
\label{fig:Fig_4_3_grat_in_water}
\end{center}
\end{figure}

The  grating was manufactured by FIB milling. SEM image of the grating cross-section is shown at the Fig. \ref{fig:Fig_4_3_grat_in_water}(a). For this image the thin layer of Pt was deposited in-situ on the top of the 1DPhC for high resolution of the cross-section. Ten grooves of 6 $\mu$m length in (X) and (Y) directions were made  [Fig. \ref{fig:Fig_4_3_grat_in_water}(b)]. 

In order to detect the presence of the BSW additional grooves (200 nm width, 400 nm depth) were milled by FIB at the distance of 30 $\mu$m from the grating [Fig. \ref{fig:Fig_4_3_grat_in_water}(c)]. SEM images of the grating are tilted for better visualization.

\subsection{Chamber preparation}
       
\begin{figure}[!b]
 	\begin{center}
 		\includegraphics[width=\linewidth]{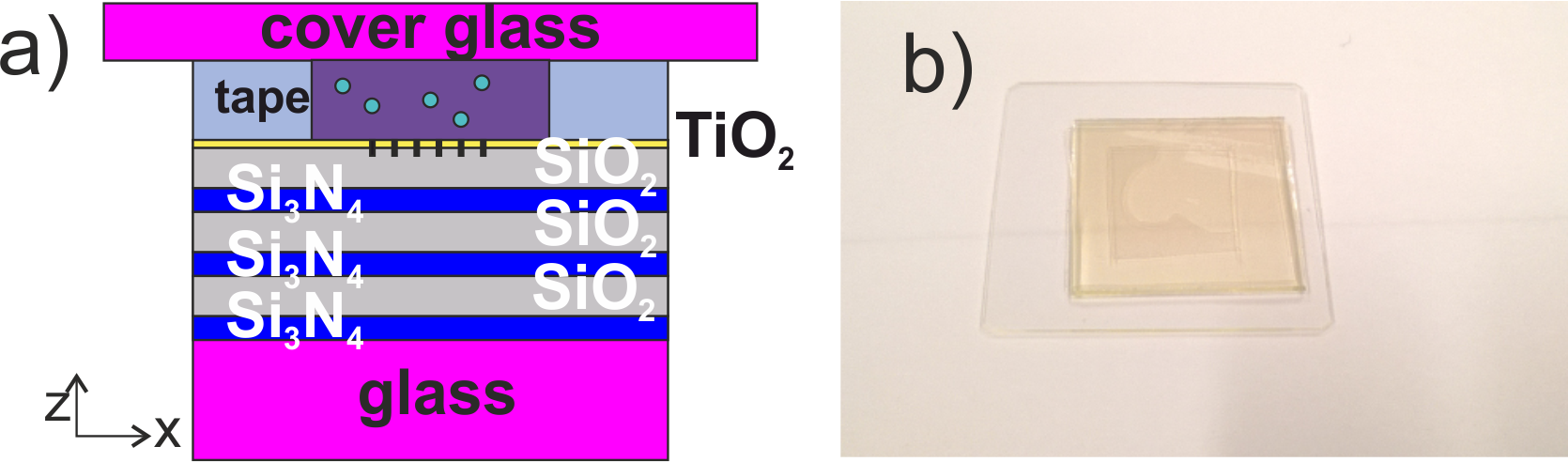}
 		\caption{(a) Schematics of the sample with a chamber; (b) Sample with a chamber and solution with beads in the chamber.}
 		\label{fig:Fig_4_4_sample_on_prism}
 	\end{center}
 \end{figure}

In order to realize particle manipulation by BSW we create a chamber on the top of 1DPhC. For this purpose we use double face tape ARseal 90880. The square 1x1 cm$^2$ is cut out of tape. The prepared tape with a hole, which would work as a chamber, is glued to 1.5x1.5 cm$^2$ multilayer chip with a grating coupler in the center. The thickness of the tape is 147 $\mu m$. The chamber is filled by the solution with 1 $\mu m$ diameter latex beads with a micro-pipette. The chamber is isolated from the top by 122 $\mu m$ thick microscope cover glass. Schematics of the sample is shown in Fig. \ref{fig:Fig_4_4_sample_on_prism}(a). The photograph of the sample is shown in Fig. \ref{fig:Fig_4_4_sample_on_prism}(b).    
 
The 1DPhC was designed to work in the environment with refractive index equal to $n_{water} = 1.3141$. In order to avoid sticking of the latex beads to each other special CTAB (cetyltrimethyl ammonium bromide, prepared at Institute UTINAM-UMR 6213, UBFC) solution was used. Refractive index of the solution together with particles was measured by RESO Reflectometer (METLER TOLEDO). At $\lambda = 589.3 nm$ refractive index of water is 1.3329. The measured value of refractive index of the solution is 1.3330, and the refractive index of the solution with particles at the same wavelength is 1.3341. From these measurements we may conclude that the refractive index of experimental media (CTAB with latex beads) is close to values used in analytical calculations (1.3141).  
 
Before testing the grating coupling the BSW was excited on the 1DPhC/solution interface in Kretschmann configuration.

\subsection{Experimental setup for BSW grating coupling in water.}

\begin{figure}[!t]
\begin{center}
\includegraphics[width=5.2in]{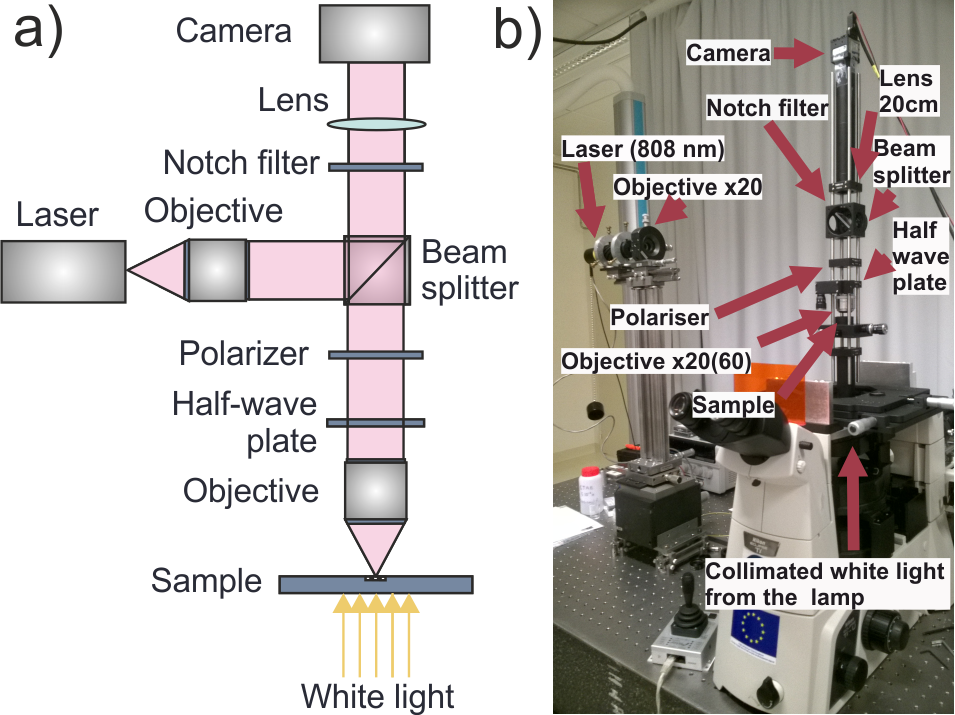}
\caption{(a) Schema of experimental setup for sample characterization; (b) Optical table with the setup.}
\label{fig:Fig_4_5_setup}
\end{center}
\end{figure}      
       
The schema of experimental setup of sample characterization for BSW grating coupling in water is shown at the Fig. \ref{fig:Fig_4_5_setup}. The light from a diode laser (808 nm, maximum power 500 mW) is collected through the long distance objective (x20, NA=0.40).Then it passes through the polarizer and half wave plate (which can be rotated for polarization control). It if finally focused at the sample surface through the long distance objective (with magnification 20, NA=0.40, or with magnification 60, NA=0.70). Light goes through the microscope cover glass on the top of the sample, through the solution with particles and comes to the grating at the normal incidence. Light reflected from the sample is collected by the camera (uEye, UI-2240SE-C-HQ). Particles are imaged by using the collimated white light from the inverted microscope NIKON Eclipse Ti-U lamp in transmission mode. To cut away the 808 nm wavelength from the reflected light for better particles visualization a Notch filter was used (NF808-34, Thorlabs). Both the input beam and white light have to pass through the chamber therefor the tape and the microscope cover glass were chosen with a smallest thickness available for sample preparation in order to reduce optical aberrations.

 \subsection{BSW manipulation of latex beads.}      
       
\begin{figure}[!t]
\begin{center}
\includegraphics[width=\linewidth]{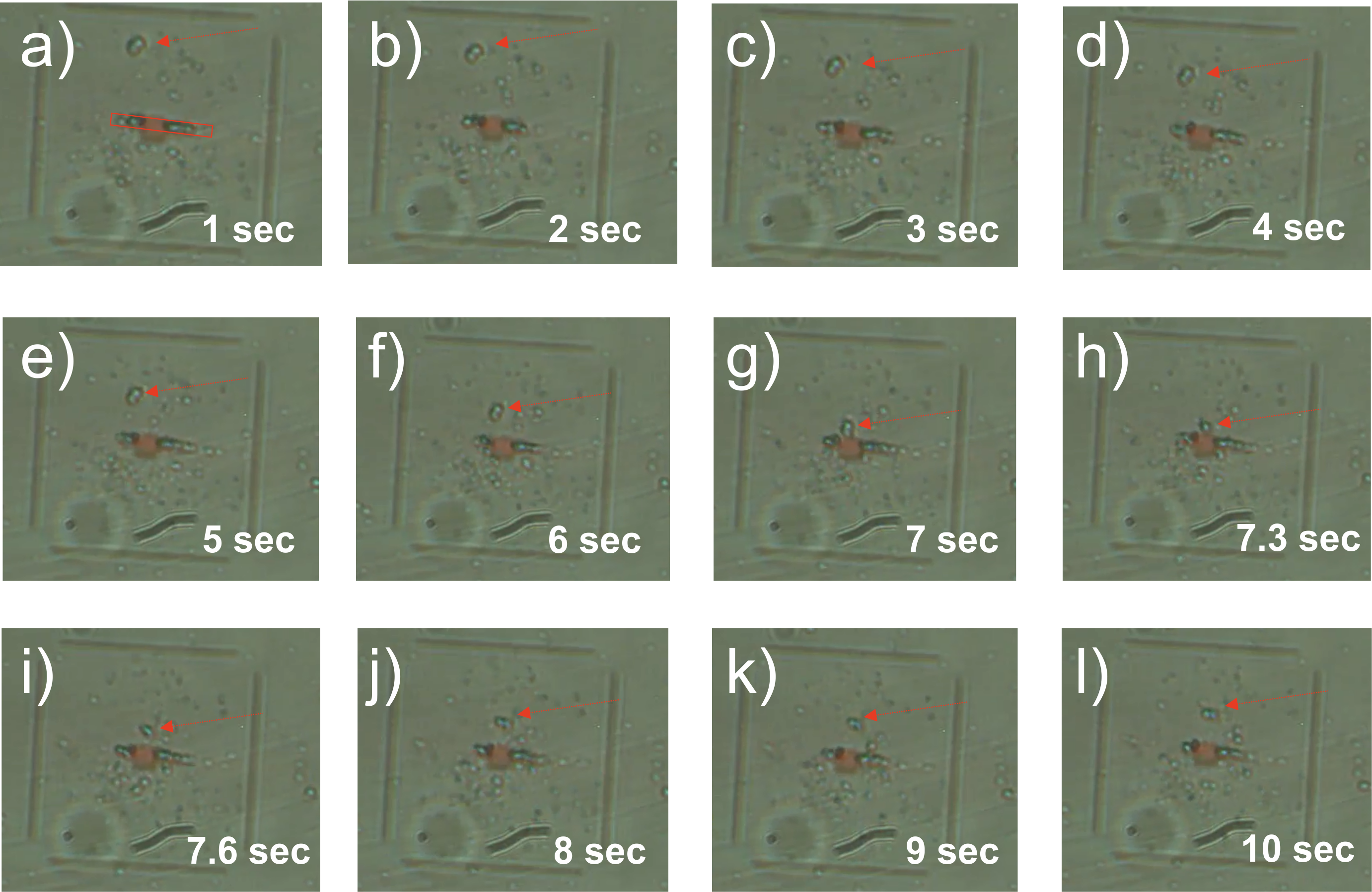}
\caption{(a) Sample visualization with x20 objective; Illustration of the grating, latex beads and rectangular shape of the incident beam; Demonstration of beads attraction by focused light (a-g); Illustration of beads, pushed away from the grating by BSW (g-i).}
\label{fig:Fig_4_6_pusing_bids}
\end{center}
\end{figure}

The aperture of the laser has a rectangular form (parallel beam divergence - 12$^\circ$, perpendicular beam divergence - 30$^\circ$), therefore during our first experiments we have rectangular shape of the incident beam [Fig. \ref{fig:Fig_4_6_pusing_bids}(a)]. Focused light creates a field gradient, which attracts particles [Fig. \ref{fig:Fig_4_6_pusing_bids}(a-g)]. They stay aligned along the incident beam spot. The 808 nm wavelength was filtered for better visualization of latex beads. For the images on Fig. \ref{fig:Fig_4_6_pusing_bids} the objective with magnification 20 was used.

When the BSW couples on the grating it generates a field gradient by evanescent tail and pushes the attracted  particles away [Fig. \ref{fig:Fig_4_6_pusing_bids}(g-i)]. And then the process repeats. Under described experimental conditions and for given 1DPhC with a grating coupler we have indicated 11 $\mu$m distance on which latex beads were moved by BSW. Once the particles are pushed at the distance far enough to escape the influence of the BSW, the gradient force of the input beam bring them back to the grating and the pushing - attracting process repeats all over again. 

This is the first demonstration of particles manipulation by the BSW excited on the grating under the normal light incidence, but for better guidance performance as well as for the directional propagation control with polarization the experimental conditions should be improved. The incident beam shape should be circular in order to provide required symmetry of energy distribution and the spot should cover all the grating area.

\begin{figure}[!t]
\begin{center}
\includegraphics[width=4.4in]{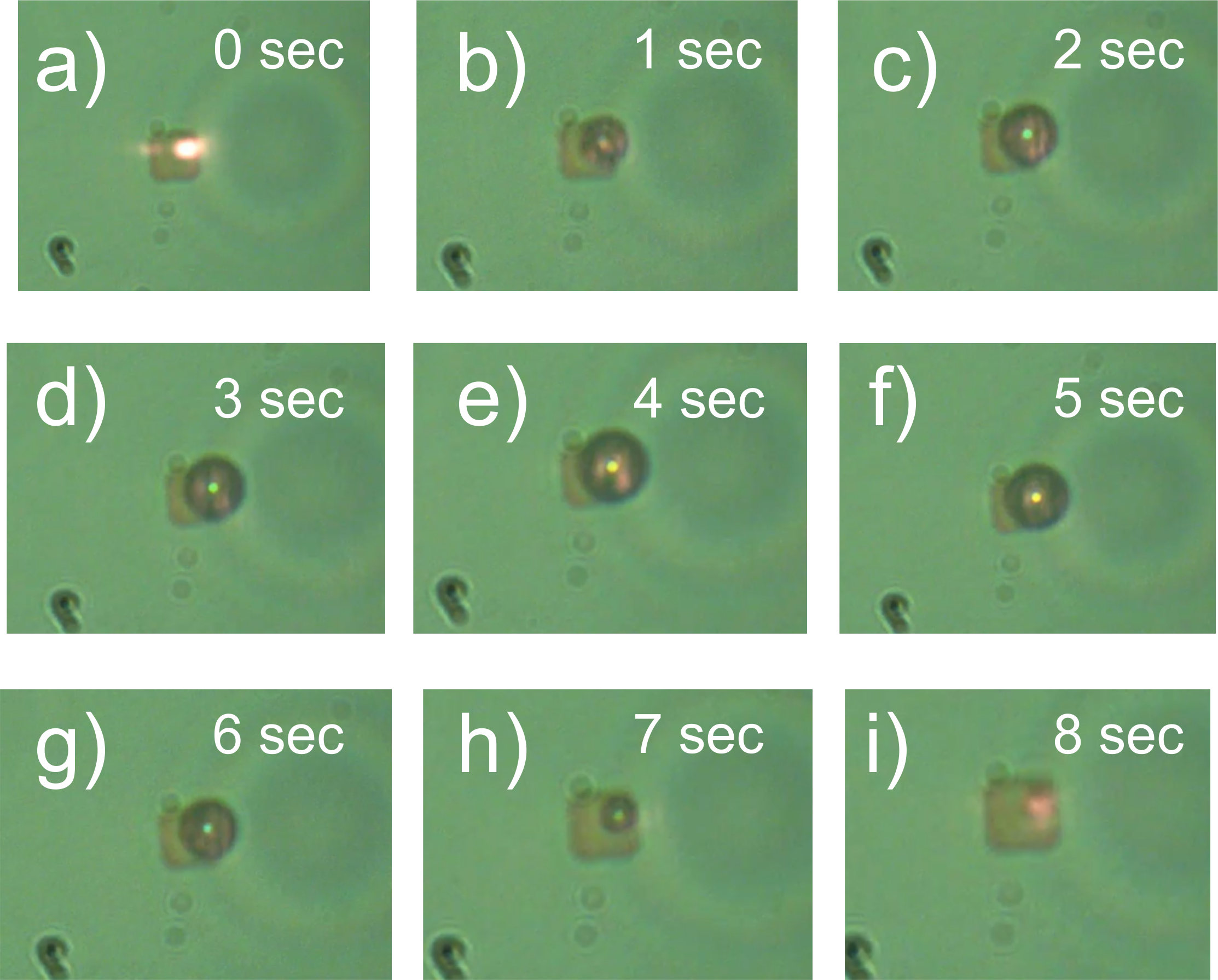}
\caption{(a) Sample visualization with x60 objective; Demonstration of expansion (b-f) and return to the original state (g-i) of the polymer.}
\label{fig:Fig_4_7_polymer_expattion}
\end{center}
\end{figure}

In this work we succeeded in reshaping the beam spot at the focal plane of the grating by varying the distance between the collecting objective and laser as well as by changing the magnification of the focusing objective (from x20 to x60). We obtained almost circular input beam shape as it can be seen at Fig. \ref{fig:Fig_4_7_polymer_expattion}(a). The beam size still requires corrections.
         
\begin{figure}[!b]
\begin{center}
\includegraphics[width=3.7in]{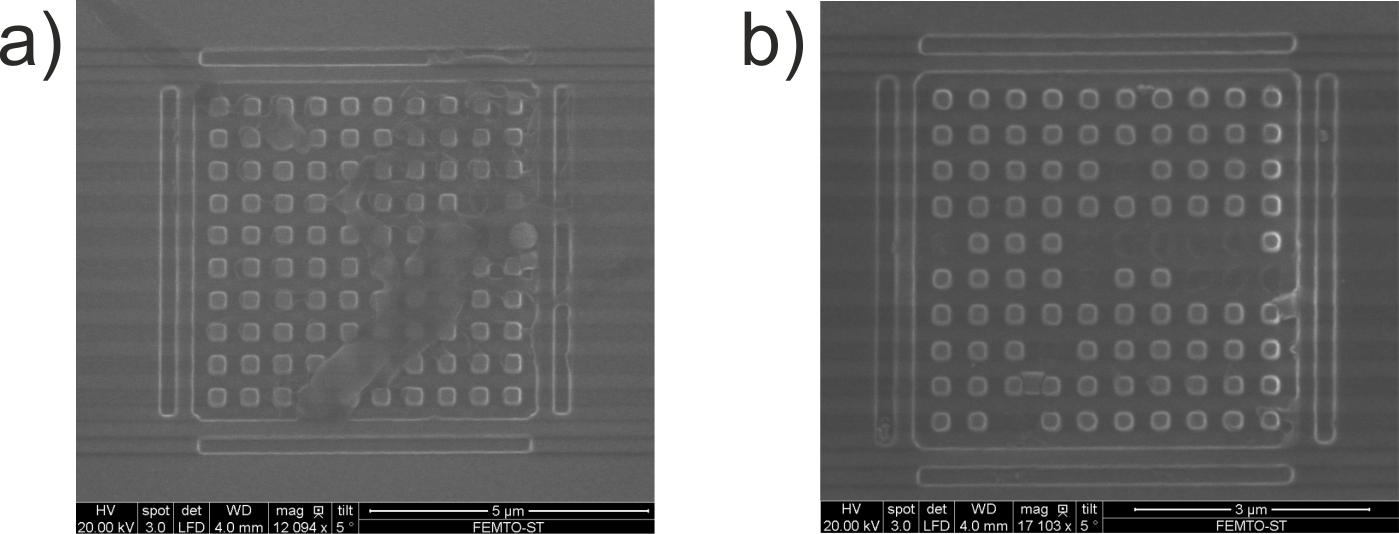}
\caption{(a) Grating contamination by glue from the tape after the forth cleaning time; (b) Burned or broken grating after the sixth cleaning time.}
\label{fig:Fig_4_8_after_cleaning}
\end{center}
\end{figure} 

Also it should be mentioned that the chamber fabrication with a double face tape has some disadvantages. After 3-4 days the liquid in the chamber escapes through the micro-channels which appear at the interface between the multilayer and tape or cover glass and tape. In this case the sample may be dissembled and reused. 

To remove the tape minimum 3 hours in acetone are required (additional ultrasound bath at 45kHz frequency is helpful). Once the tape is softened the microscope cover glass can be removed. The remained residuals of glue from the tape should be gently wiped by a cotton stick without touching the central area where the grating is. After one more hour in acetone together with ultrasound bass and rinsing in water the standard cleaning in piranha solution for 2 minutes should be done in order to remove organic contamination. Then sample should be rinsed in water and ethanol and dried by nitrogen gun. After all this cleaning proses the sample can be reused. 

After several times reusing the sample, grating still can get contaminated up to the point that it is impossible to clean [Fig. \ref{fig:Fig_4_8_after_cleaning}(a)] or can even get broken  [Fig. \ref{fig:Fig_4_8_after_cleaning}(b)].
     
Experimentally, the presence of contamination can be easily detected. When there is a thin layer of some polymers on the grating the BSW cannot be coupled. Moreover the polymer absorbs a lot of energy from the input beam and expands [Fig. \ref{fig:Fig_4_7_polymer_expattion}(b-f)]. After reducing the power of the laser, the polymer returns to it's original state [Fig. \ref{fig:Fig_4_7_polymer_expattion}(g-i)].

\section{Conclusion.}

In this chapter we have underlined an attractive direction for the further investigation of BSW dependency on polarization. In order to realize polarization guidance of particles we demonstrate the latex beads interaction with BSW in water for the case of grating coupling. We have shown the feasibility to use a cross-grating coupler for lab-on-a-chip trapping devices.

\chapter{LiNbO$_3$ and BSW.}

Previously discussed methods of polarization tunability of BSWs describe passive devices, where, in order to reach the change in surface wave propagation we have to modify incident beam parameters. This modulation is easy to achieve, and with additional nano-structuring of the 1DPhC, we can obtain a drastic change in the BSW propagation direction. However, active tunable devices require the use of active tunable materials and lithium niobate is a perfect candidate for this role. 

LiNbO$_3$ is transparent over a wide wavelength range (350 nm - 5200 nm) and has high refractive index. It is a nonlinear birefringent material with tunable optical properties due to the Pockels effect together with ferro-electrical, piezo-electrical, and thermo-electrical effects \cite{Wong:02}. This material was widely investigated at FEMTO-ST for creation of ultra-compact sensing devices \cite{Lu:13,Lu:12,Roussey:06}. Various techniques for LiNbO$_3$ processing such as precise dicing\cite{Courjal:15} and polishing \cite{bassignot2012acoustic} were developed. Therefore, we chose this material for our studies.

In this chapter, we present an original type of one dimensional photonic crystal that includes one anisotropic layer made of a lithium niobate thin film. We demonstrate the versatility of such a device sustaining different Bloch Surface Waves, depending on the orientation of the incident wave. By varying the orientation of the illumination of the multilayer, we measured an angle variation of 7$^\circ$ between BSWs corresponding to the extraordinary and ordinary index of the lithium niobate thin film. The potential of such a platform opens the way to novel tunable and active planar optics based on the electro- and thermo-optical properties of lithium niobate. 

\vspace*{0.2cm}
\minitoc

\section{Thin film lithium niobate as a part of 1DPhC.}

\begin{figure}[!t]
	\centering
	\includegraphics[width=4.4in]{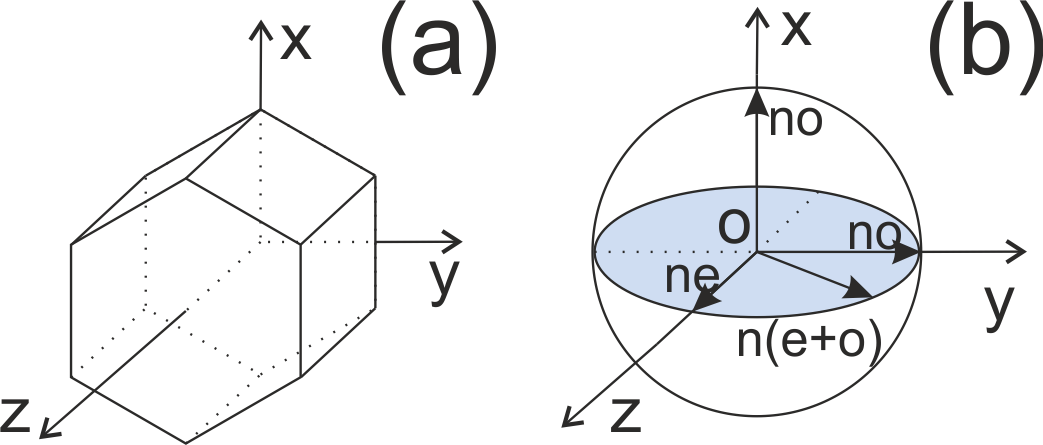}
	\caption{(a) Sketch of the crystalline directions of LN; (b) Sketch of index ellipsoid of LN.}
	\label{fig:Fig_5_1_LN_crystallineXcut}
\end{figure}

\begin{figure}[!b]
	\centering
	\includegraphics[width=4in]{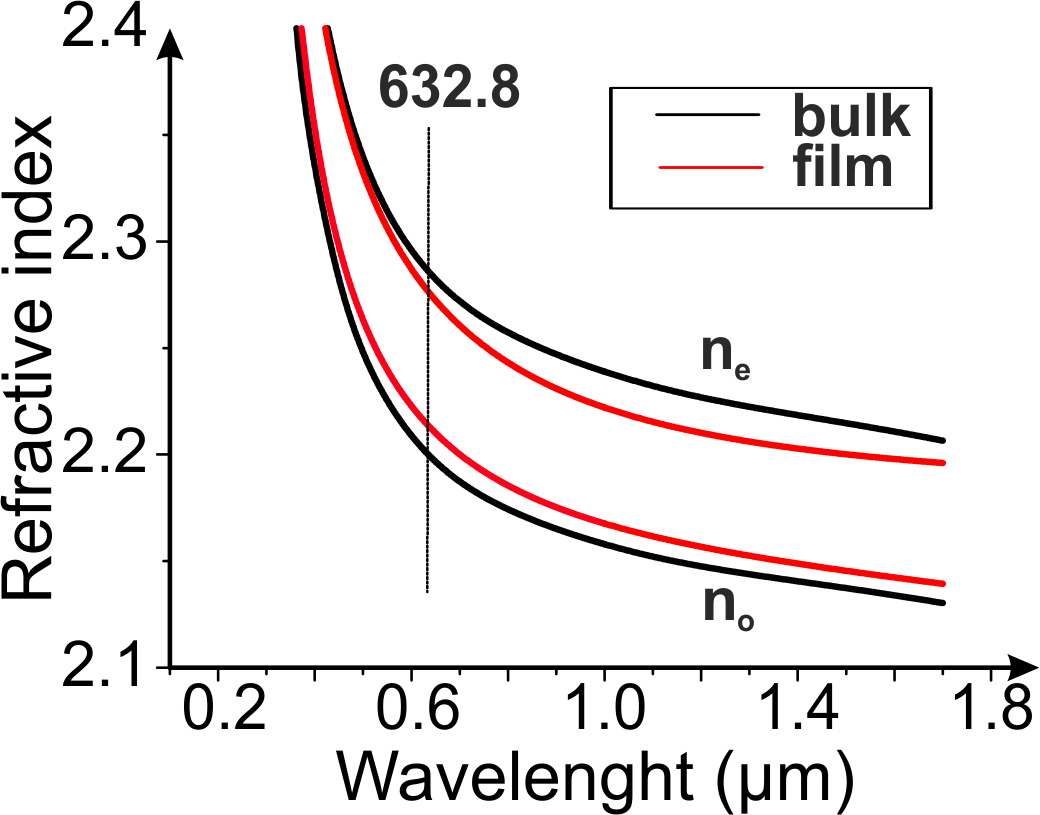}
	\caption{Measured refractive indices for LN bulk and TFLN.}
	\label{fig:Fig_5_2_elipsoLN}
\end{figure}

The tunability of BSWs may be controlled by the effective index change of the multilayers, which can be obtained by the refractive index variation of a single layer, here, the LiNbO${_3}$ layer. Optical properties of LiNbO$_3$ can be modified by an external physical signal, e.g., by the application of an electric field \cite{Roussey:06} or by variation of temperature \cite{Li:03}. In this work, we study a multilayer that contains a thin film of lithium niobate (TFLN). Here, the LiNbO$_3$ layer is an X-cut single-crystal thin film of 700 nm thickness. In this case the change of refractive index can be achieved, not only using the electro or thermo-optical properties of the film, but also by rotating the sample in such a way that light propagates along the ordinary or the extraordinary axis. In that case, the refractive index change will vary from $n_o$ to $n_e$, with $n_o$ being the ordinary refractive index and $n_e$ the extraordinary refractive index of the TFLN. Figures \ref{fig:Fig_5_1_LN_crystallineXcut}(a) and \ref{fig:Fig_5_1_LN_crystallineXcut}(b) show the crystalline directions and the sketch of the index ellipsoid of an X-cut LiNbO$_3$. Thus, the use of this material as a part of the multilayer stack brings anisotropic properties to the whole 1DPHC in the ZOY plane. 

Lithium niobate possess anisotropic properties due to it's trigonal crystal system, which lacks inversion symmetry. This material has negative uniaxial birefringence which depends slightly on the stoichiometry of the crystal and on temperature. It's chemical formula belongs to a part of the crystallographic group $R_{3C}$. According to crystalline axis X, Y and Z, the refractive index of the crystal is $n_o$, $n_o$ and $n_e$ respectively. 

1DPhCs that sustain surface waves require sub-wavelength thickness layers. In the case of lithium niobate this requirement is challenging, especially if the material has to be single-crystal. Nowadays technologies such as sputtering, evaporation or epitaxial growth of LiNbO$_3$ only allow amorphous or polycrystalline films \cite{Joshkin:03}. In order to achieve the desired LN properties and to avoid light scattering on the crystal-grain  boundary a single-crystal X-cut TFLN bonded to an SiO$_2$ layer on LN substrate \cite{Han:15} has been used. The TFLN samples were provided by NANOLN company.

The TFLN was prepared using He$^+$ ion implantation technique which induces changes in the refractive index of LiNbO$_3$. The lattice damage caused by the ion implantation could be reduced by annealing under oxygen atmosphere \cite{Han:15}. However, for a precise determination of the TFLN refractive indices, a detailed spectroscopic ellipsometry (SE) study was performed. The ellipsometry measurements were done in the Center for Physical Sciences and Technology (Lithuania) by Ieva Baleviciute. Generalized SE measurements were done using dual rotating-compensator ellipsometer RC-2 (J. A. Woollam Co. Inc.) at incidence angles from 45$^\circ$ to 85$^\circ$ in steps of 5$^\circ$ in the 300 nm – 1700 nm spectral range. An optical response from the TFLN sample was analyzed using a multilayer optical model, i.e., a semi-infinite LN substrate, SiO$_2$ and TFLN. The SiO$_2$ refractive index was taken from CompleteEase database (J. A. Woollam Co. Inc.). To model the LN thin film refractive indices, the Sellmeier dispersion was used:

\begin{equation}
n_{o,e}=\sqrt{\left(\epsilon(\infty)+\frac{A_{o,e}\lambda^2}{\lambda^2-B_{o,e}^2}-E_{o,e}\lambda^2\right)},
\label{eq:refname1}
\end{equation}

\begin{figure}[!t]
	\centering
	\includegraphics[width=4.5in]{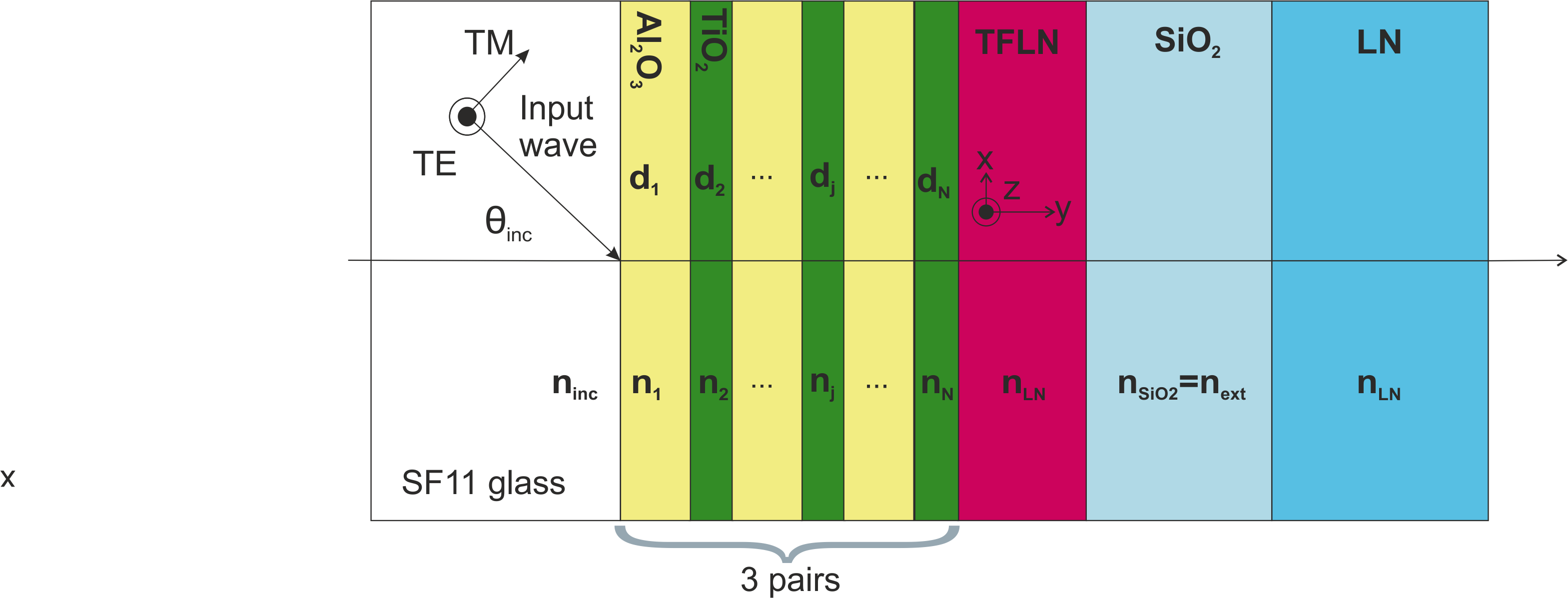}
	\caption{1DPhC structure.}
	\label{fig:Fig_5_4_1DPhC}
\end{figure}

where $\varepsilon(\infty)$ is the high-frequency offset, $A$ is the amplitude, $B$ is the oscillator energy, and $E$ is the position of a pole in the infrared region. The determined Sellmeier parameters are $A_o = 2.209, A_e = 2.016, B_o = 0.22799, B_e = 0.22422, E_o = 0.03245, E_e = 0.02915, \varepsilon_0(\infty) = 2.650$ and $\varepsilon_e(\infty) = 2.607$. Bulk lithium niobate has a trigonal symmetry and displays negative uniaxial birefringence ($n_e - n_o=-0.0861$ at $\lambda=632.8$~nm). Measured refractive indices of TFLN slightly differ from their bulk values, and birefringence decreases to - 0.0629 value. A decrease in a birefringence indicates that the LN thin film suffers from He$^+$ ion implantation \cite{Han:15}. For further calculations we will use the values of ordinary and extraordinary refractive indices of TFLN at $\lambda= 632.8$~nm, which are $n_o=2.277\pm0.005$ and $n_e=2.2141\pm0.005$, respectively [Fig. \ref{fig:Fig_5_2_elipsoLN}].

\section{Design of 1DPhC with TFLN.}

\begin{figure}[!t]
	\centering
	\includegraphics[width=3in]{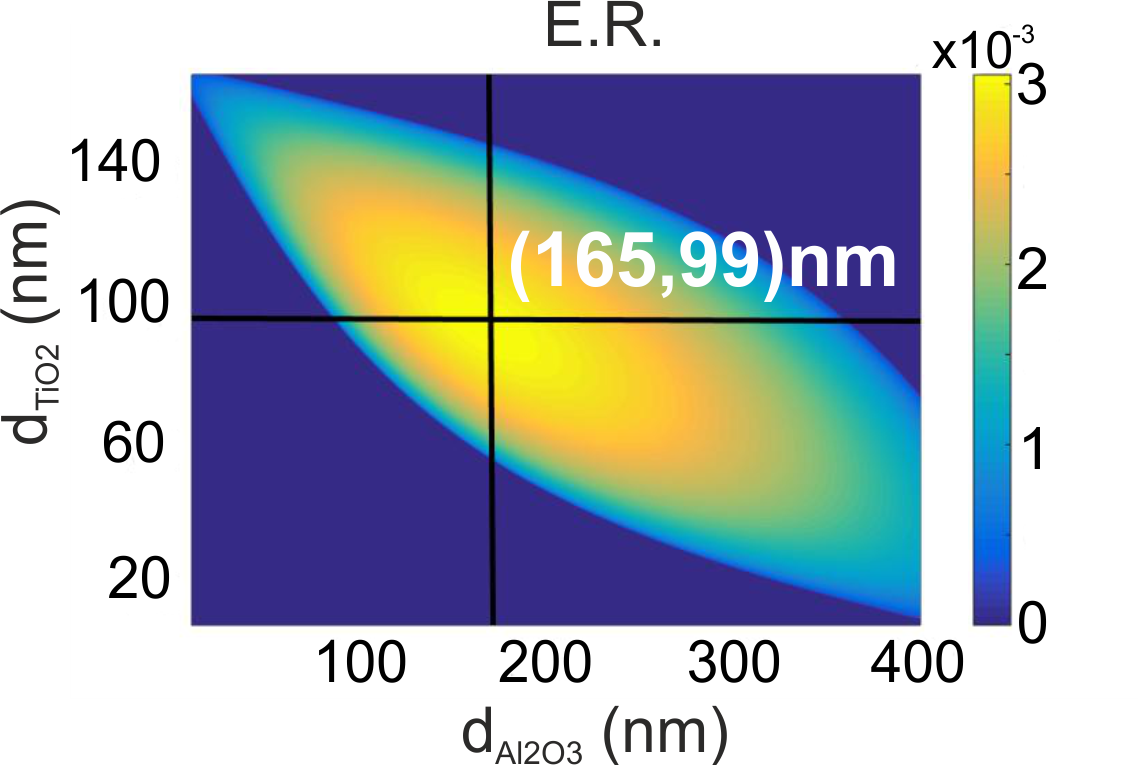}
	\caption{Extinction of the band gap as a function of layer thicknesses in 1DPhC.}
	\label{fig:Fig_5_3_layersr_opt}
\end{figure}

As we described in chapter 2, the 1DPhC can be considered as a stack of N layers (numbered j = 1, 2...N), with refractive indices $n_j$ and thicknesses $d_j$. Incident and external media should be considered as semi-infinite with refractive indices $n\rm{_{inc}}$ and $n\rm{_{ext}}$ respectively. In our calculations we choose Al$_2$O$_3$ (n$_{\rm{Al_2O_3}}$=1.67) as the low refractive index material and TiO$_2$ (n$_{\rm{TiO_2}}$=2.39) as the high refractive index material of the multilayer structure. The 1DPhC is schematically shown in Fig. \ref{fig:Fig_5_4_1DPhC}. For a given set of materials and a particular wavelength, we determine the thicknesses at which the maximum band gap extinction per length occurs and for which the PhC structure has minimum overall thickness. The extinction ratio ($E.R.$) is expressed in Eq. \ref{eq:refname11} (see chapter 2.1.2).

In Eq. \ref{eq:refname11} (see chapter 2.1.2) we consider $T$ as the transmission coefficient for one period of the 1DPhC. The desired values of thicknesses $d_1=d{\rm{_{Al_2O_3}}}=d{\rm{_{Al_2O_3max}}}$ and $d_2=d{\rm{_{TiO_2}}}=d{\rm{_{TiO_2max}}}$ are the thicknesses at which $E.R.$ reaches its maximum value. For Al$_2$O$_3$/TiO$_2$ stacks at $\lambda=632.8$~nm with $n\rm{_{Al_2O_3}}=1.67$ and $n\rm{_{TiO_2}}=2.39$ these values are 165 nm and 99 nm respectively (see Fig. \ref{fig:Fig_5_3_layersr_opt}). The $632.8$~nm wavelength was chosen as a wavelength from the visible part of spectrum in order to simplify the alignment of the setup and the experimental measurements. 

\begin{figure}[!t]
	\centering
	\includegraphics[width=4.2in]{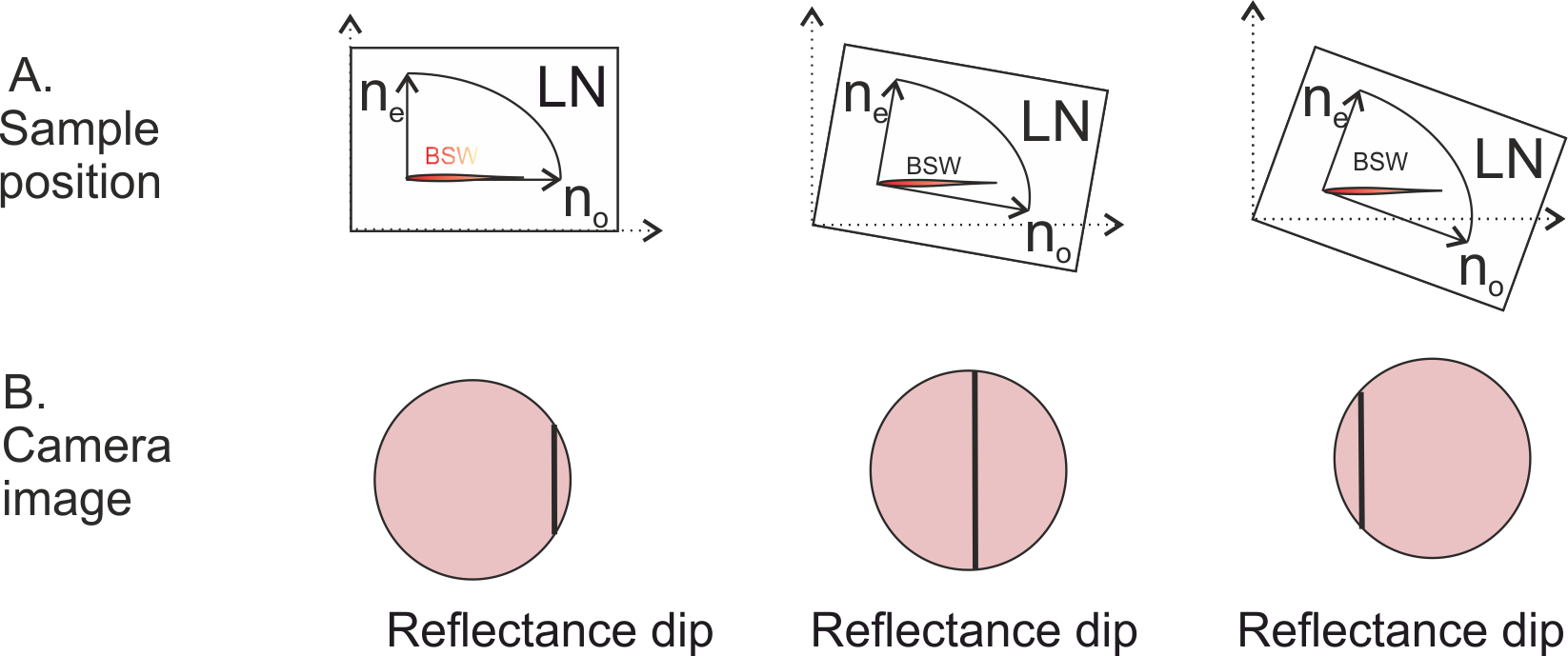}
	\caption{(a) Dispersion curve for extraordinary refractive index of LN; (b) Dispersion curve for ordinary refractive index of LN; (c) Calculated reflectance for ordinary and extraordinary refractive indices for TE polarized light at the TFLN/SiO$_2$ interface; (d) Field profile in the 1DPhC.}
	\label{fig:Fig_5_5_disp_curves}
\end{figure}

The best coupling of the incident wave into the surface wave depends on the total extinction in the layers (see Chapter 2.1.3). In our case, considering a SF11 glass prism as an incident medium and SiO$_2$ as external medium, the optimal coupling (reflectance equals 0) occurs for 3 pairs of Al$_2$O$_3$/TiO$_2$ layers.

The optimization in the 1DPhC was made for TE polarized incident light. Dispersion curves for ordinary and extraordinary refractive indices of TFLN are shown in Fig. \ref{fig:Fig_5_5_disp_curves}(a) and \ref{fig:Fig_5_5_disp_curves}(b) for different refractive indices of LiNbO$_3$, leading to the two extreme positions, in wavelength, of the BSW. Reflection dips are observed at $n_{\rm{eff(o)}}$=1.579 and $n_{\rm{eff(e)}}$=1.52 corresponding to ordinary and extraordinary light propagation (see Fig. \ref{fig:Fig_5_5_disp_curves}(c)). Reflectance dips at $n_{\rm{eff}}\geq1.457$ represent BSWs, dips for $n_{\rm{eff}}\leq1.457$ are related to guided modes in the SiO$_2$ substrate layer.

The thin film LN ($TFLN\approx700~nm$) was bonded to a silicon oxide layer ($SiO_2\approx2~\mu m$) which was attached to the bulk LN ($LN\approx500~\mu m$). The 1DPhC was deposited on the thin film LN (see Fig. \ref{fig:Fig_5_4_1DPhC}). That requires to consider a BSW propagation at the TFLN/SiO$_2$ interface, where SiO$_2$ is the external media.  In order to check whether the SiO$_2$ can be considered as semi-infinite we need to calculate the penetration length of the evanescent tail of the BSW \cite{konopsky:07}, which is

\begin{equation}
l=\frac{\lambda}{4 \pi \sqrt{n_{\rm{eff}}^2-{n_{\rm{eff}}^2}_{\rm{TIR}}}},
\label{eq:refname2}
\end{equation}

where ${n_{\rm{eff}}}_{\rm{TIR}}=n_{\rm{ext}}=1.457$. The 1DPhC is made to sustain a BSW at $\lambda=632.8$ nm. 
Therefore $l_e=115.3$ nm and $l_o=82.7$ nm for ordinary and extraordinary refractive indices of TFLN. These values are significantly smaller than the given thickness of SiO$_2$ (2$\mu$m). As a consequence, we can take our approximation into account. Field profile within the 1DPhC is shown at Fig. \ref{fig:Fig_5_5_disp_curves}(d). Here we can see that the field is being enhanced by the Al$_2$O$_3$/SiO$_2$ multilayers and that it is confined within the TFLN. We also observe the evanescent decay at the TFLN/SiO$_2$ interface. The evanescent tale lies inside the SiO$_2$ layer. 

From the results of dispersion curve shift, and therefore from the reflectance dip shift, we can theoretically predict anisotropy of 1DPhC introduced by the TFLN. 

\section{Experimental.}

\subsection{Multilayer manufacturing. Setup.}

The multilayer stack was fabricated by atomic layer deposition (ALD) by Dr. Markus H\"{a}yrinen in the Institute of Photonics, University of Eastern Finland. 

As it was mentioned in Chapter \ref{ALD}, ALD is a cycling deposition process based on a modified chemical vapor deposition technique \cite{Puurunen:14,Hayrinen:15}. It allows a conformal and homogeneous coating over large surfaces with a surface roughness below 0.2 nm. For this particular structure, the two deposited materials are TiO$_2$ and Al$_2$O$_3$, for which the precursors are TiCl$_4$ and water, and TMA and water respectively. The process temperature is set to 120$^\circ$C to ensure amorphous material layers and thus reduced propagation losses.

\begin{figure}[!t]
	\centering
		\includegraphics[width=4.8in]{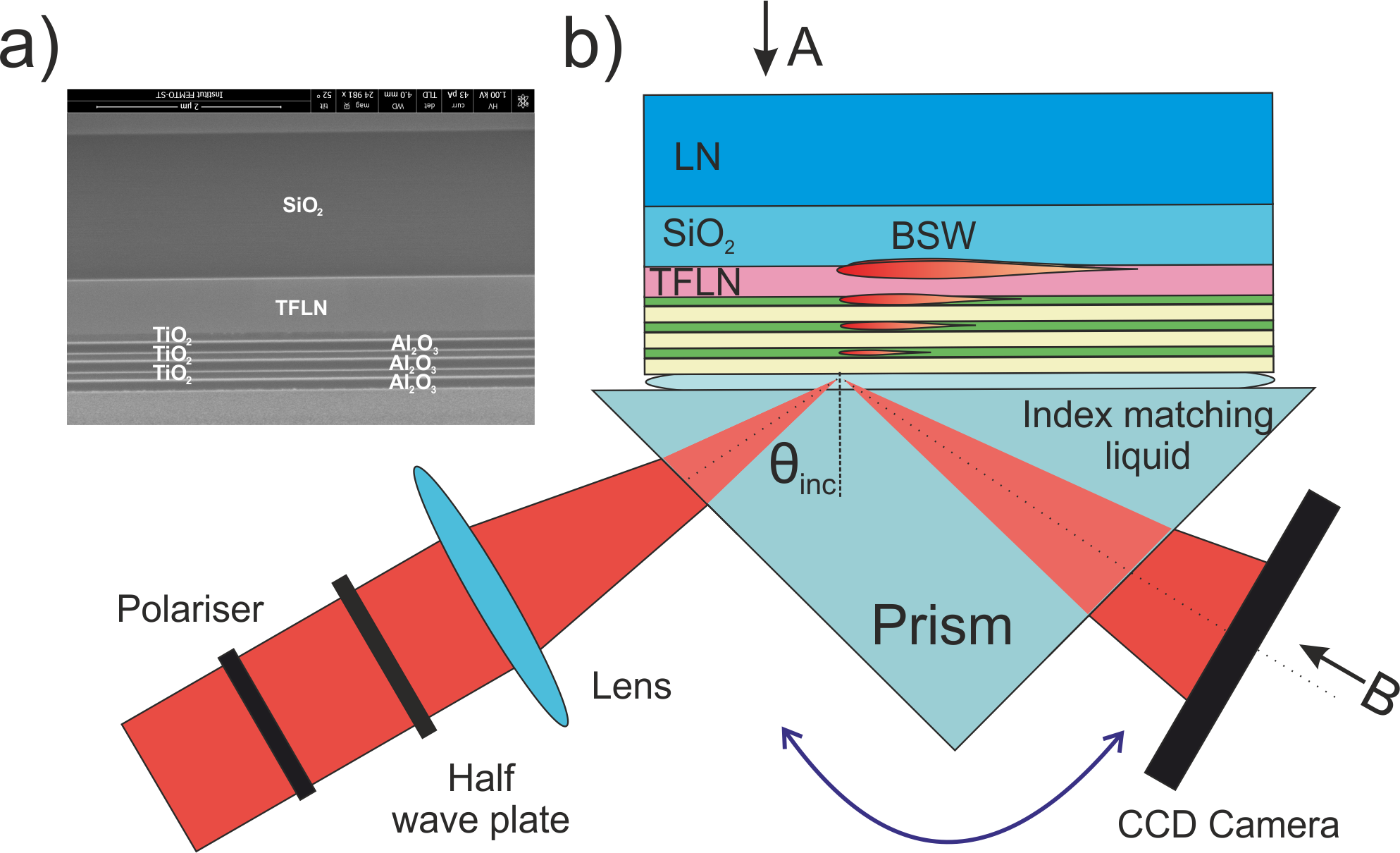}
		\caption{(a) FIB-SEM image of 1DPhC; (b) Experimental set up.}
	\label{fig:Fig_5_6_setup}
\end{figure}

\begin{figure}[!b]
	\centering
	\includegraphics[width=\linewidth]{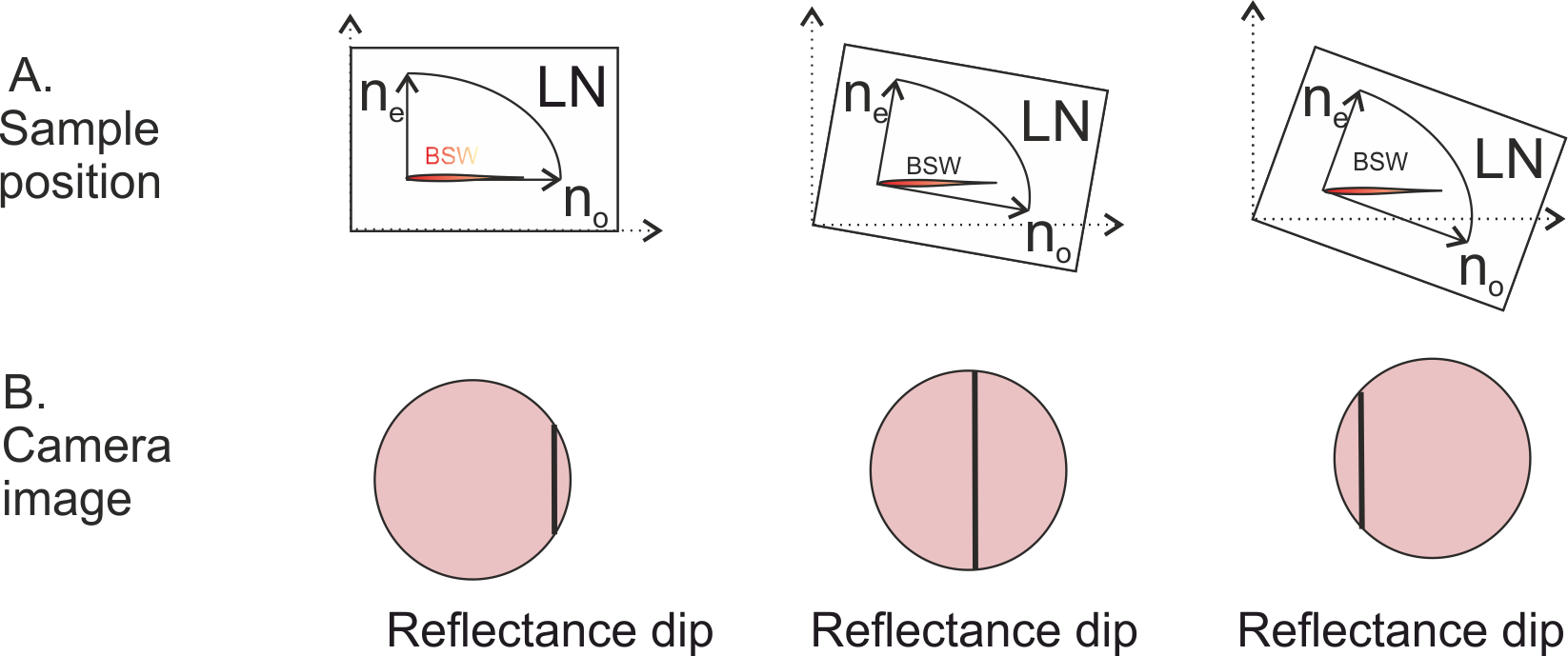}
	\caption{Sketch of the reflectance dip due to the sample orientation.}
	\label{fig:Fig_5_7_concept}
\end{figure}

In order to verify the quality and the thickness of the deposited layers and to minimize the damage of the sample, FIB-SEM measurements were done [Fig. \ref{fig:Fig_5_6_setup}(a)] The measurements were performed by Rolad Salut in FEMTO-ST. FIB was used in order to open up a small area for sample characterization. The experimental values slightly differ by 20 nm from the optimized value determined theoretically for Al$_2$O$_3$ and TiO$_2$, and by 50 nm for TFLN. This leads to some difference between theoretically predicted and experimental values of incident angle for BSW excitation.

To achieve experimental values for $\theta_{\rm{inc}}$, Kretschmann configuration was used for light coupling [Fig. \ref{fig:Fig_5_6_setup}(b)]. TE polarized light from a He-Ne laser (Melles Griot 25-LHP-925) was focused on the sample ($\lambda=632.8$~nm). A SF11 prism was fixed on a rotational stage, which allowed us to change the incidence angle until the BSW excitation. Reflected light was detected by a CCD camera (mV BlueFox 120GU). Refractive index matching oil was used to provide good contact and smooth optical transition at the interface between the prism and the 1DPhC.
Here it is important to mention that ALD technique provides perfect homogeneous layers of TiO$_2$ and Al$_2$O$_3$ in our case, though the required thicknesses of layers were close to the limit of of maximum possible. Therefore the manufactured 1DPhC had frequent delamination problems. Multilayers were pilled from the surface quite easily. It requires a very gentle handling in order to use the same sample for optical characterization tests. For example, sample cleaning from refractive index matching oil between the measurements only gentle rinsing in acetone, water and ethanol was possible. Even though after several times of cleaning the layers got delaminated and the sample had to be changed for a new identical one.

\begin{figure}[!t]
	\centering
	\includegraphics[width=5.2in]{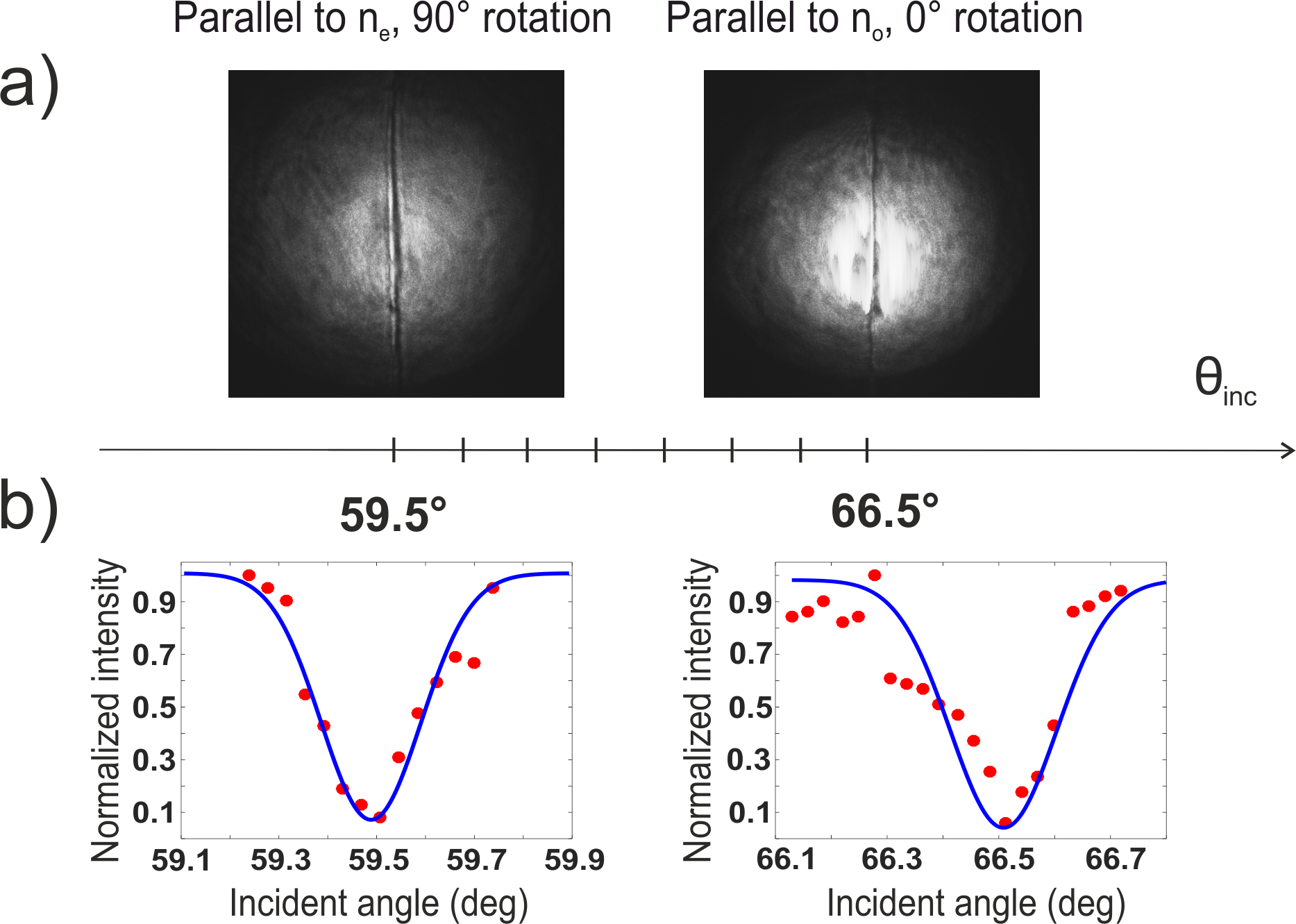}
	\caption{(a) Camera images of the reflectance dip as a function of sample orientation; (b) Measured intensity profile of the reflected light.}
	\label{fig:Fig_5_8_dips}
\end{figure}

The theoretically expected reflectance dip shift due to anisotropy of 1DPhC can be observed from the following tests: once the characteristic dip is detected, the sample can be rotated (on the surface of the prism, as explained in Fig. \ref{fig:Fig_5_7_concept}). Indeed, this behavior can be predicted from Fig. \ref{fig:Fig_5_5_disp_curves}(c). We know that $n_{{\rm{eff}}_{BSW}}$ is incident angle dependent. Therefore, at the fixed wavelength, by changing the refractive index of LN we will shift $n_{{\rm{eff}}_{BSW}}$ and as a consequence the incident angle of BSW excitation.

\subsection{Tunable BSW due to anisotropy of 1DPhC.}

Experimentally the BSW related absorption lines were detected at the following angles: 59.5$^\circ$ - for $n_e$; and 66.5$^\circ$ - for $n_o$, what makes a 7$^\circ$ difference [Fig. \ref{fig:Fig_5_8_dips}(a)]. For these angles intensity profile of reflected light was measured (see Fig. \ref{fig:Fig_5_8_dips}(b)). The reflected light at different positions of the dip was collected through the pinhole by a power meter (Thorlabs S130C sensor, PM100D power meter). The intensity was measured and normalized. Gaussian fit between measurement data is shown as a blue line at Fig. \ref{fig:Fig_5_8_dips}(b). As it was theoretically predicted we observe narrow lines of intensity dumping (up to 80\%) in the area of BSWs.

\begin{figure}[!t]
	\centering
	\includegraphics[width=5.2in]{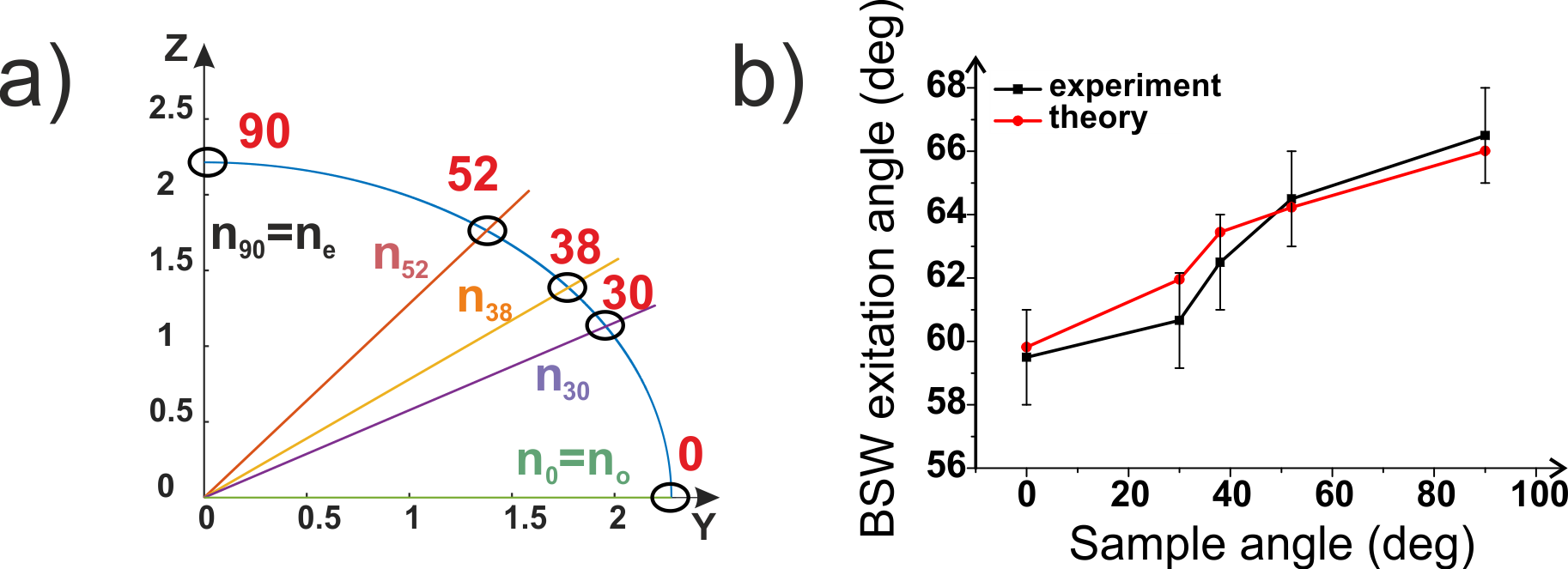}
	\caption{(a) Refractive index of LiNbO$_3$ for different sample positions; (b) Experimental results and theoretical dependance of the BSW excitation angle}
	\label{fig:Fig_5_9_rotation_prove}
\end{figure}

Additional measurements at the positions of the sample rotated for 30$^\circ$, 38$^\circ$ and 52$^\circ$ were done. In these cases, values of LN refractive index can be obtained graphically as the distance from the origin to the intersection of the line under the given angle with the index ellipsoid of LN (see Fig. \ref{fig:Fig_5_9_rotation_prove}(a)). BSW excitation angles for 0$^\circ$, 30$^\circ$, 38$^\circ$, 52$^\circ$ and 90$^\circ$ were calculated analytically, taking into account experimental values of the layers thicknesses. Experimentally obtained angles are in a good agreement with the theory (see Fig. \ref{fig:Fig_5_9_rotation_prove}(b)). The estimated error values of determined $\theta_{\rm{inc}}$ were $\pm 1.5^\circ$. Difference between theoretical and experimental values may be due to the difference of requested and experimentally achieved thickness of the layers as well as due to some inaccuracies of refractive indices values.

\section{Conclusion.}

In this part of the work, we have demonstrated a potential of BSWs on an anisotropic platform. LN based photonic crystal, which is able to sustain BSWs, was designed and fabricated. Characterization of such optical properties as ordinary and extraordinary refractive index of LiNbO$_3$ thin film was done. The structure of the 1DPHC was optimized for Al$_2$O$_3$ and SiO$_2$ multilayers at $\lambda=632.8$ nm by the impedance approach. The maximum extinction is achieved for 3 pairs of multilayers of thicknesses of TiO$_2$ of 99 nm and Al$_2$O$_3$ of 165 nm. For the 1DPHC of the following structure: SF11 glass / Al$_2$O$_3$ $\approx150$ nm / (TiO$_2$ $\approx$ 95 nm)$\times3$ / LiNbO$_3$ $\approx$ 700 nm / SiO$_2$ $\approx$  2 $\mu$m / LN $\approx$ 500 $\mu$m the far field measurements were done. The designed 1DPhC allows to obtain BSW at the TFLN/SiO$_2$ interface for TE polarized light. The experimentally achieved angles of excitation of Bloch surface waves along ordinary and extraordinary axis of the crystal are in an agreement with theoretical predictions and a 7$^\circ$ angle variation was observed.

In addition, the use of the nonlinear properties of LiNbO$_3$ opens up the possibility of creating BSW based active tunable devices if dispersion engineering is done on the LN as a top layer of 1DPhC.

\chapter{LiNbO$_3$ as a top layer of 1DPHC.}

Lithium niobabe is a nonlinear, birefringent material with electro- and thermo-optical properties, which is a widely used in photonics. However, the implementation of LiNbO$_3$ in real devices is not a trivial task due to difficulties in manufacturing and handling thin film LN. In this chapter, we investigate an optical device, where the Bloch surface wave propagates on the TFLN/air interface to access their properties. In order to sustain the BSW, one-dimensional photonic crystal is necessary to be fabricated under the condition of sub-wavelength thickness of layers. Therefore, the 1DPhC with LiNbO$_3$ as a part of the multilayer should be done on the base of the thin film LN. We consider two material platforms to realize such a device, bulk LN and commercial thin film LN. 

\vspace*{0.2cm}
\minitoc

\section{Motivation for 1DPhC with LN as a top layer.}

Lithium niobate is widely used for integrated optics and photonics \cite{Toney:15}. In order to improve the performance of integrated optical devices, several research groups have developed different structures such as ridge waveguides, photonic crystal waveguides, resonators, disks and periodically poled lithium niobate (PPLN) structures \cite{Sohler:08}. As a high refractive index material, LN as a top layer is used for enhanced light confinement for many devices \cite{Guarino:07} and the thin film LiNbO$_3$ will be used to improve the confinement even more.

Here we propose two different novel architectures that can generate Bloch surface waves at TFLN layers. As we have already demonstrated, light coupling can be easily achieved by a grating coupler \cite{Kovalevich:17} or by using the Kretschmann configuration \cite{konopsky:07,Dubey:j16}. Both coupling methods are simpler than the fiber-to fiber coupling technique that is employed to horizontally excite a guided mode on a TFLN \cite{Gerthoffer:14}. 

In this work, we propose a 1DPhC with a thin film of LiNbO$_3$ as the top layer of the multilayer structure. The bonding into the 1DPhC structure brings anisotropy into the whole crystal allowing the tunability of the BSW devices \cite{kovalevich:16}. In previously studied 1DPhC with TFLN the BSW is excited at the LiNbO$_3$/SiO$_2$ interface, where the SiO$_2$ is the external media. In that geometry, the BSW propagates on the surface of the LiNbO$_3$ thin film, which is embedded within the multilayer. This geometry limits the use of such a device and cannot, for instance, be applied for sensing applications, where the evanescent tail of the BSW needs to be in the contact with the external environment to be sensed. 

Direct electric field or temperature application cannot introduce a shift of LiNbO$_3$ refractive index significant enough to shift the dispersion curve of 1DPhC directly. However, the refractive index change up to $\Delta n=0.3$ is routinely produced by additional nano-structuring of LiNbO$_3$ \cite{Roussey:06}. Photonics crystals and Bragg gratings are some of the optical functions that can enhance the electro- and thermo- optical properties of LiNbO$_3$ based devices. To do so in the BSW platform the access to LiNbO$_3$ from the top of 1DPhC is required. 

Additionally nonlinear properties of BSWs is an attractive topic for studies. Recently phase-matched third-harmonic generation via doubly resonant optical surface modes \cite{Konopsky:2016} was achieved on the base of 1DPhC coated with a 15-nm GaAs film. Unfortunately GaAs is not transparent in the visible range of wavelengths what is not suitable for many optical applications. The use of LiNbO$_3$ would be a perfect solution for future studies of BSW induced nonlinearities on the TFLN surface. Once again to unlock the nonlinear properties of BSWs on LiNbO$_3$ it is necessary to have the interface between LiNbO$_3$ and air \cite{BEZPALY2017166}.

As we have mentioned before the 1DPhC that sustain surface waves requires sub-wavelength thickness layers. In the case of lithium niobate this requirement is challenging, especially because it needs to maintain its crystallinity in order to use the nonlinear properties of the material. Nowadays technologies such as sputtering, evaporation or epitaxial growth of LiNbO$_3$ only allow amorphous or polycrystalline films \cite{Joshkin:03,Wernberg:93,Makram:16}.

In this part of the work the multilayer fabrication is made by  PECVD in EPFL optics and photonics technology laboratory. This technology have been tested in our experiments and allows to overcome the delamination issues, which were observed in the case of previously studied 1DPhC prepared by Atomic Layer Deposition (ALD). Thus, in this work SiO$_2$ and Si$_3$N$_4$ are used for the multilayer instead of Al$_2$O$_3$ and TiO$_2$. 

In this chapter we present a BSW based device, which is able to sustain surface waves at the LiNbO$_3$/air interface. Two different geometries have been studied, fabricated and optically characterized. The first one is based on the LiNbO$_3$ membrane and the second one is held by a stable glass platform.

  \begin{figure}[!t]
	\centering
	\includegraphics[width=4in]{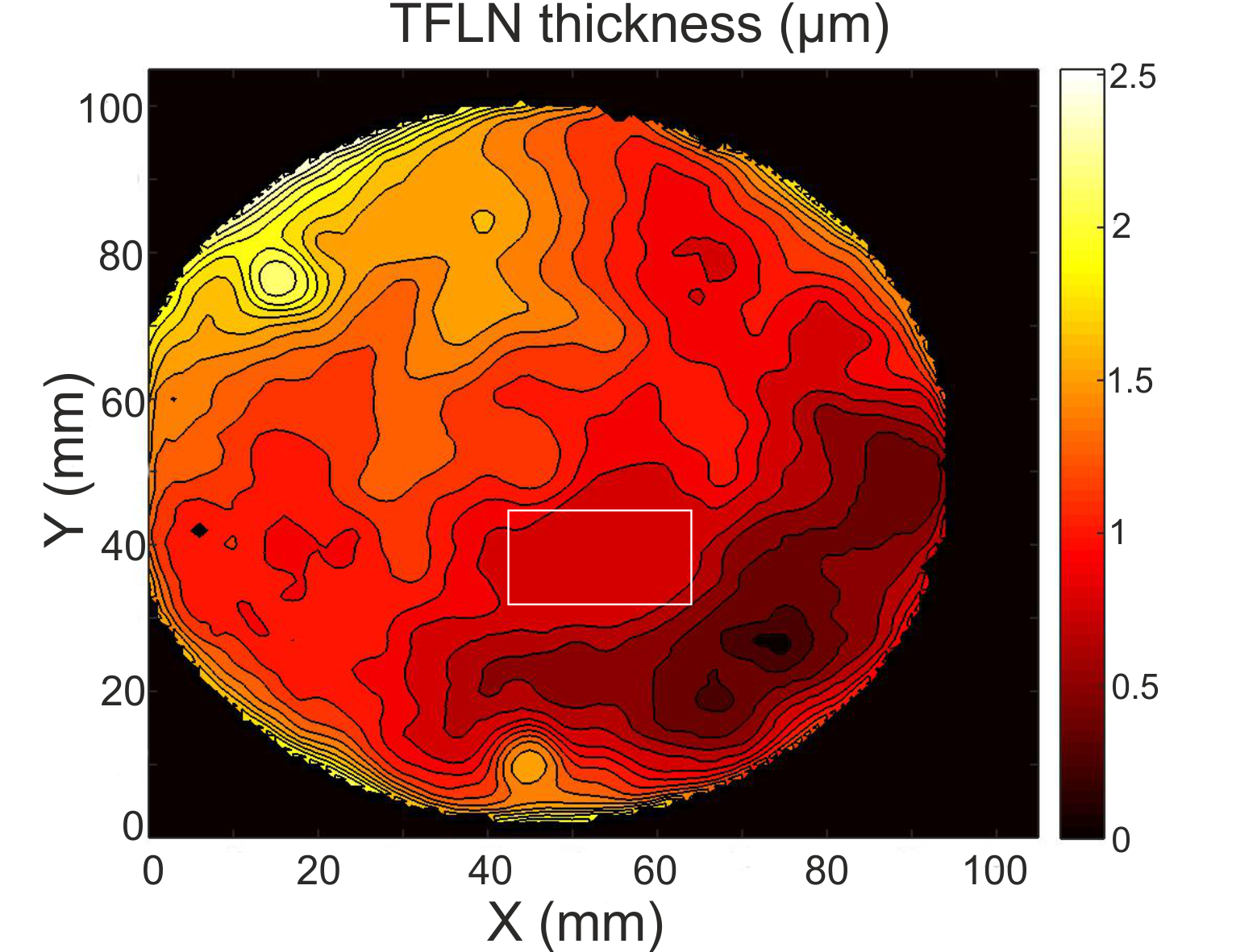}
	\caption{The thickness profile of the TFLN.}
	\label{fig:Fig_6_1_polishedLN}
\end{figure}

\section{1DPhC on membrane.}

The 1DPhC with on membrane configuration has been fabricated using a single-crystal TFLN bonded to an Cr and Au layer on a Si substrate \cite{bassignot2012acoustic} which is thereafter bonded to the 1DPhC. For this geometry the multilayer used was composed of Si/Au/Cr/Au/TFLN. The layers thickness  were 500 $\mu$m for Si, 20 nm for both layers of Cr, 400 nm for gold layer and about 2 $\mu$m for LiNbO$_3$ thin film respectively. In this case the TFLN was manufactured by polishing of bulk LiNbO$_3$ bonded to Si wafer by gold and chromium \cite{bassignot2012acoustic}. Bulk LN polishing and characterization of achieved thin film was performed at FEMTO-ST by Florent Bassignot and Ludovic Gauthier-Manuel. The thickness profile of the TFLN is shown in Fig. \ref{fig:Fig_6_1_polishedLN}. As it can be seen from Fig. \ref{fig:Fig_6_1_polishedLN} the TFLN thickness achieved by bulk LN polishing is not homogeneous. It has some areas, where LN was totally removed. For sample preparation we dice out the part, where the thickness of TFLN is the same. For the manufacturing of the 1DPhC, a sample with LN thickness of 1.1 $\mu$m was used.
 
\begin{figure}[!b]
	\centering
		\includegraphics[width=5.5in]{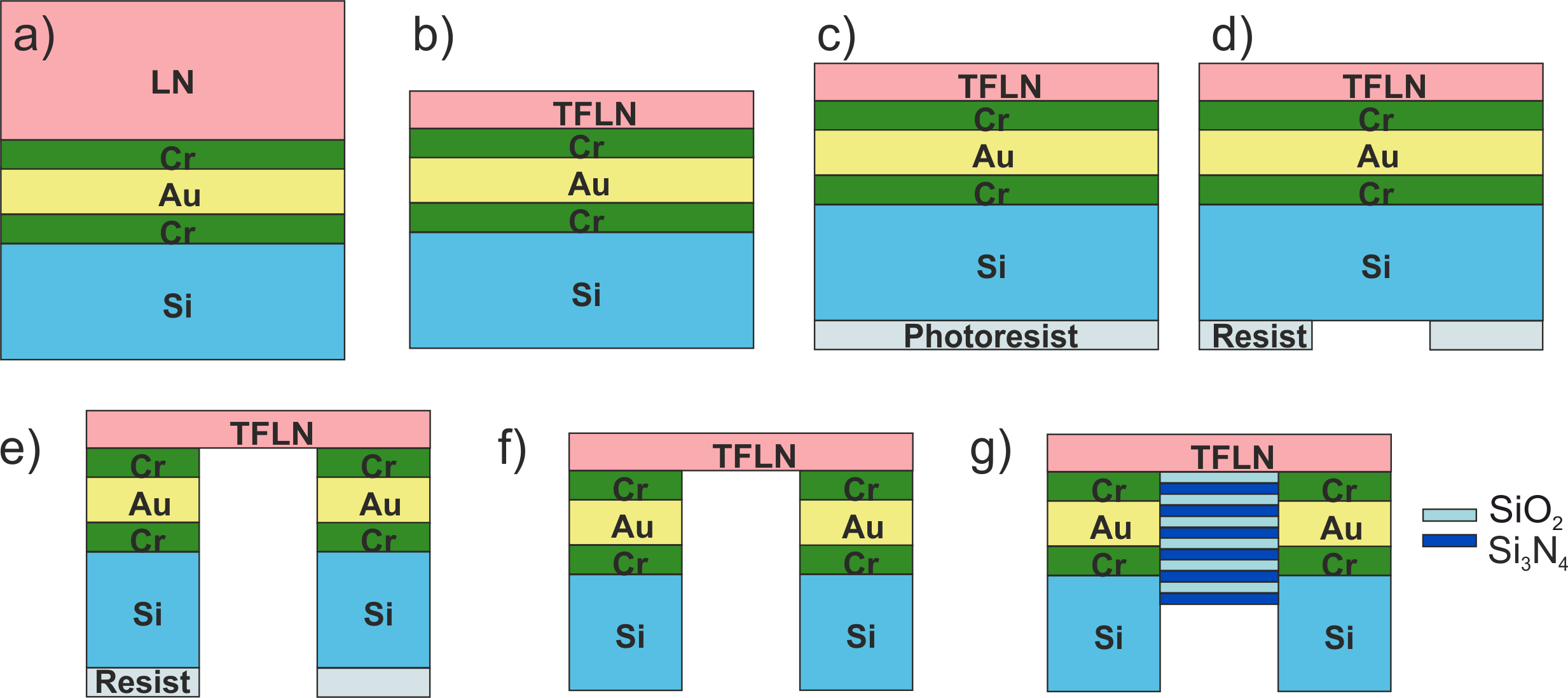}
		\caption{Schematic of the membrane based 1DPhC fabrication process: (a) Bonding of bulk LiNbO$_3$ to Si with Cr and Au, (b) LiNbO$_3$ polishing, (c) Photoresist deposition, (d) UV lithography of the photoresist, (e) DRIE etching of Si and wet etching of Cr and Au, (f) Photoresist removal, (g) Multilayer deposition.}
	\label{fig:Fig_6_2_MLonMembrane_steps}
\end{figure}

The 1DPhC was manufactured according to the steps shown in Fig. \ref{fig:Fig_6_2_MLonMembrane_steps}. After the thinning of the LiNbO$_3$, a positive photoresist AZ 9260 was used for UV lithography [Fig. \ref{fig:Fig_6_2_MLonMembrane_steps}(c,d)]. 

500 $\mu$m of Si layer were removed by deep reactive ion etching (DRIE) \cite{Labelle2004} on SPTS equipment tooled by dual plasma source (Rapier) [Fig. \ref{fig:Fig_6_2_MLonMembrane_steps}(e)]. Concerning the plasma etching process a Bosch process with three sequences (Teflon Deposition, Teflon removal in the trench bottom and silicon etching) was used \cite{Laermer:96}. These 3 sequences were repeated in order to have an anisotropic wall's profile (close to 90$^{\circ}$), the Teflon deposition was performed by C$_4$F$_8$ gas and the etching sequence is achieved by using SF$_6$ gas. The etch rate was 11 $\mu$m/min and the selectivity was 1:180 which is the ratio between the AZ 9260 photoresist etch rate to the Silicon etching rate. 

Chromium and gold were removed by a standard chemical wet etching process in chromium and gold etching solutions [Fig. \ref{fig:Fig_6_2_MLonMembrane_steps}(e)].

\begin{figure}[!t]
	\centering
		\includegraphics[width=\linewidth]{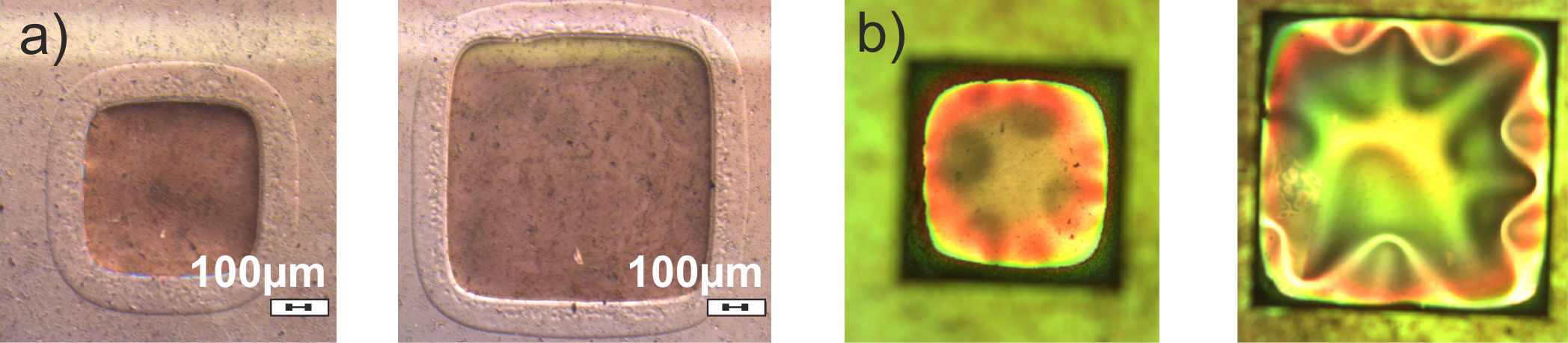}
		\caption{(a) The microscope images of the membranes. (b) The microscope images of the membranes after multilayer deposition.}
	\label{fig:Fig_6_3_LNmembr_MLmembr}
\end{figure}

Concerning the multilayer fabrication, six periods of Si$_3$N$_4$ and SiO$_2$ were alternately deposited on top of the TFLN membrane suspended in air by using plasma-enhanced chemical vapor deposition (PECVD, PlasmaLab 80 Plus by Oxford) [Fig. \ref{fig:Fig_6_2_MLonMembrane_steps}(g)]. The precursor gases for the deposition are silane (SiH$_4$), ammonia (NH$_3$) and nitrous oxide (N$_2$O), and the process temperature is 300$^{\circ}$C.

\begin{figure}[!b]
	\centering
	\includegraphics[width=\linewidth]{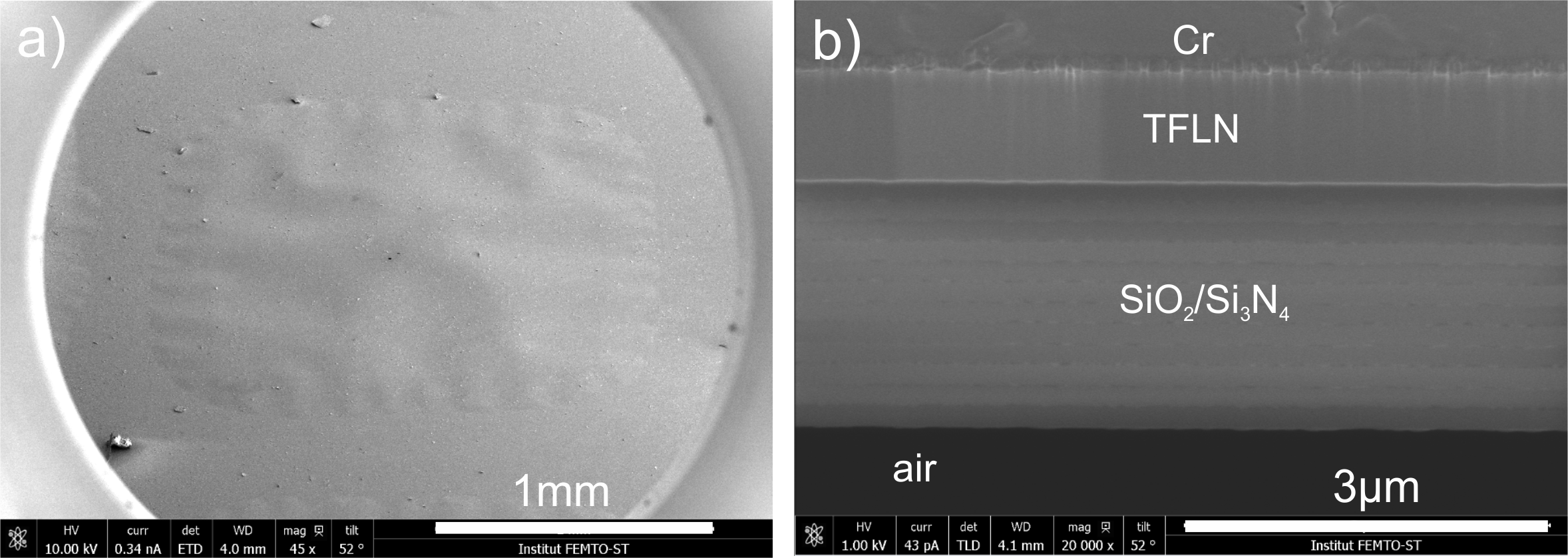}
	\caption{(a) FIB-SEM image of the membrane (with a whole multilayer). (b) FIB-SEM image of the 1DPhC (suspended membrane).}
	\label{fig:Fig_6_4_FIB_MLmembr}
\end{figure}

Membranes of two different areas were manufactured: $1\times1$mm$^2$ and $1.5\times1.5$mm$^2$. The microscope images of the membranes are shown in Fig. \ref{fig:Fig_6_3_LNmembr_MLmembr}(a). After the deposition of multilayers membranes undergo some stretching which is clearly seen from Fig. \ref{fig:Fig_6_3_LNmembr_MLmembr}(b). We suppose that these deformations are due to redistribution of the stress within the TFLN. Firstly, new forces appear in the suspended membrane after Si and SiO$_2$ removal. Then additional tensions appear after multilayer deposition. Important to mention, that the thickness of deposited layers changing towards the border of the membrane (can be seen from the change of color at Fig. \ref{fig:Fig_6_3_LNmembr_MLmembr}(b)). Thus, we may see that the membrane starches in different parts. The deformations are stronger for the bigger membrane. The curvature of the membrane leads to the change of the incident angle for the BSW excitation inducing uncertainties in the BSW propagation.   

FIB-SEM images [Fig. \ref{fig:Fig_6_4_FIB_MLmembr}] show the thickness of the layers. In the central part of one of the membranes the small area is milled by FIB. This allows to see the cross-section of the multilayer with SEM. The thicknesses of silicon dioxide were about 215 nm and of silicon nitride about 200 nm. The top layer of Cr seen in Fig. \ref{fig:Fig_6_4_FIB_MLmembr}(b), was deposited in order to avoid charge effects during the measurements and was removed by wet etching process.

\section{1DPhC on glass support.}

\begin{figure}[!b]
	\centering
	\includegraphics[width=5.5in]{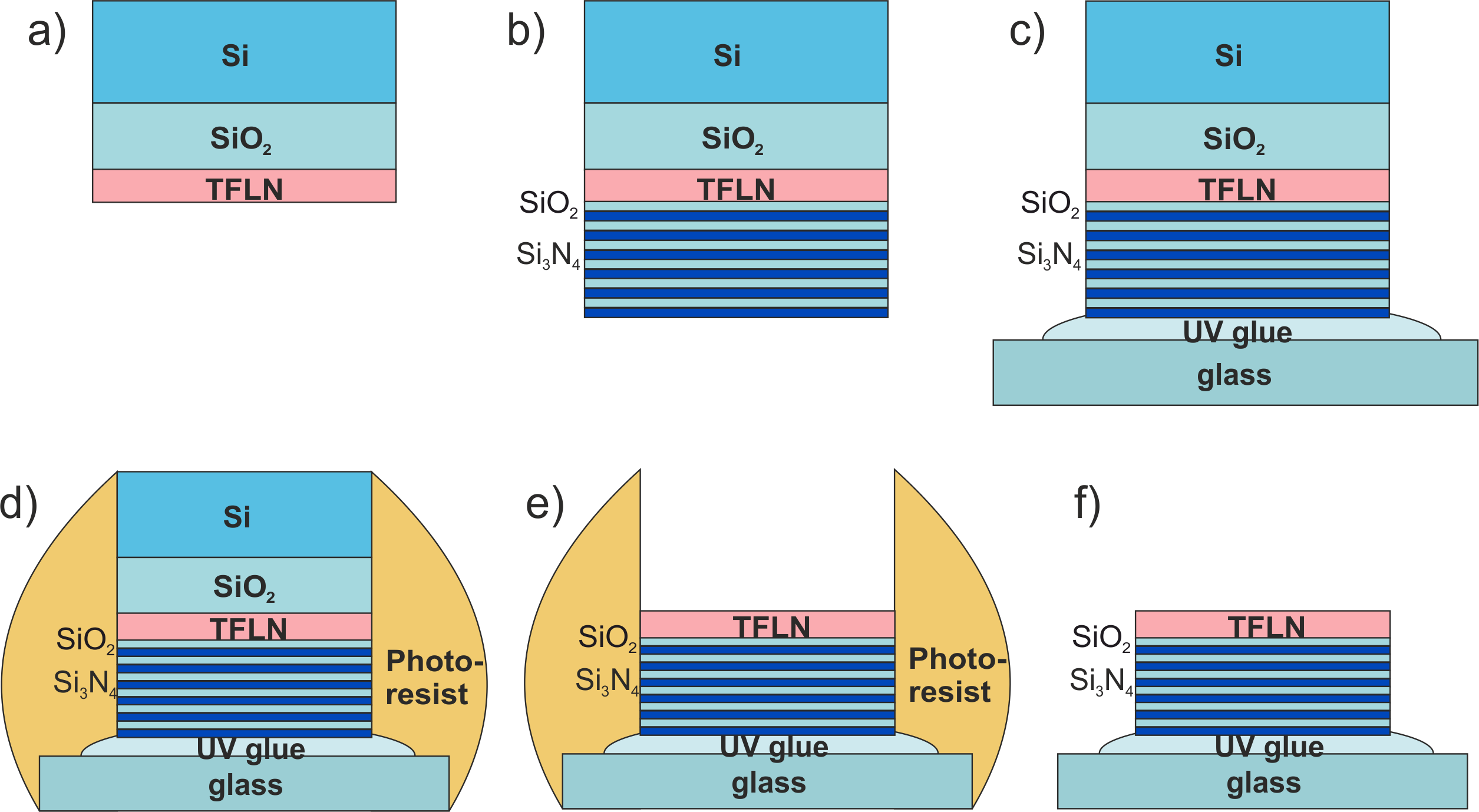}
	\caption{Schematic of the on-glass 1DPhC fabrication process: (a) Obtaining of TFLN with smart cut technology, (b) Multilayer deposition, (c) UV glue bonding to the glass substrate, (d) Protection of the sample with photoresist, (e) DRIE etching of Si and RIE etching of SiO$_2$, (f) Photoresist removal.}
	\label{fig:Fig_6_5_MLglass_steps}
\end{figure}

For the on-glass 1DPhC a single-crystal TFLN bonded to an SiO$_2$ layer on Si substrate \cite{Han:15} has been used. 
  
The manufacturing steps for the on-glass 1DPhC are shown in the Fig. \ref{fig:Fig_6_5_MLglass_steps}. In this case a single-crystal X-cut TFLN of 400 nm bonded to a SiO$_2$ layer on Si substrate has been used [Fig. \ref{fig:Fig_6_5_MLglass_steps}(a)]. The TFLN waver was prepared by a smart-cut technology at NANOLN company using He+ ion implantation technique \cite{Han:15}. As in the previous sample, the multilayer stack of alternating silicon dioxide and silicon nitride were deposited on TFLN by PECVD [Fig. \ref{fig:Fig_6_5_MLglass_steps}(b)]. The whole structure was bonded to the glass holder (500 $\mu$m) by UV glue [Fig. \ref{fig:Fig_6_5_MLglass_steps}(c)]. The UV glue (VITRALIT 6127) is chosen in such a way that its refractive index is close to the refractive index of glass. The top Si layer of Si was thinned down by polishing from 400 $\mu$m to 20 $\mu$m.

The whole stack was protected by S1813 photoresist [Fig. \ref{fig:Fig_6_5_MLglass_steps}(d)], and the 20 $\mu$m of Si were dry etched by DRIE [Fig. \ref{fig:Fig_6_5_MLglass_steps}(e)]. For this etching, the Bosch process was also employed. A 2 $\mu$m thickness layer of SiO$_2$ was etched by RIE with a mixture of fluorine gases (CHF$_3$ - 10 sccm and C$_2$F$_6$ - 5 sccm). After Si and SiO$_2$ etching, the remaining photoresist was chemically removed by using an acetone bath, followed by an ethanol bath and finally by DI water. After inspection on the optical microscope, some photoresist residuals were still present. O$_2$ plasma (power 600 Watts, O$_2$ flux = 90 sccm) during 15 minutes was used to remove them [Fig. \ref{fig:Fig_6_5_MLglass_steps}(f)].

\section{Samples characterization.}

After the manufacturing we obtained 2 different lithium niobate BSW based devices. The multilayer of the membrane based crystal was as following: air/6 pairs of Si$_3$N$_4$(200nm) and SiO$_2$(215nm)/TFLN(1.1$\mu$m)/air. 

\begin{figure}[!b]
	\centering
		\includegraphics[width=\linewidth]{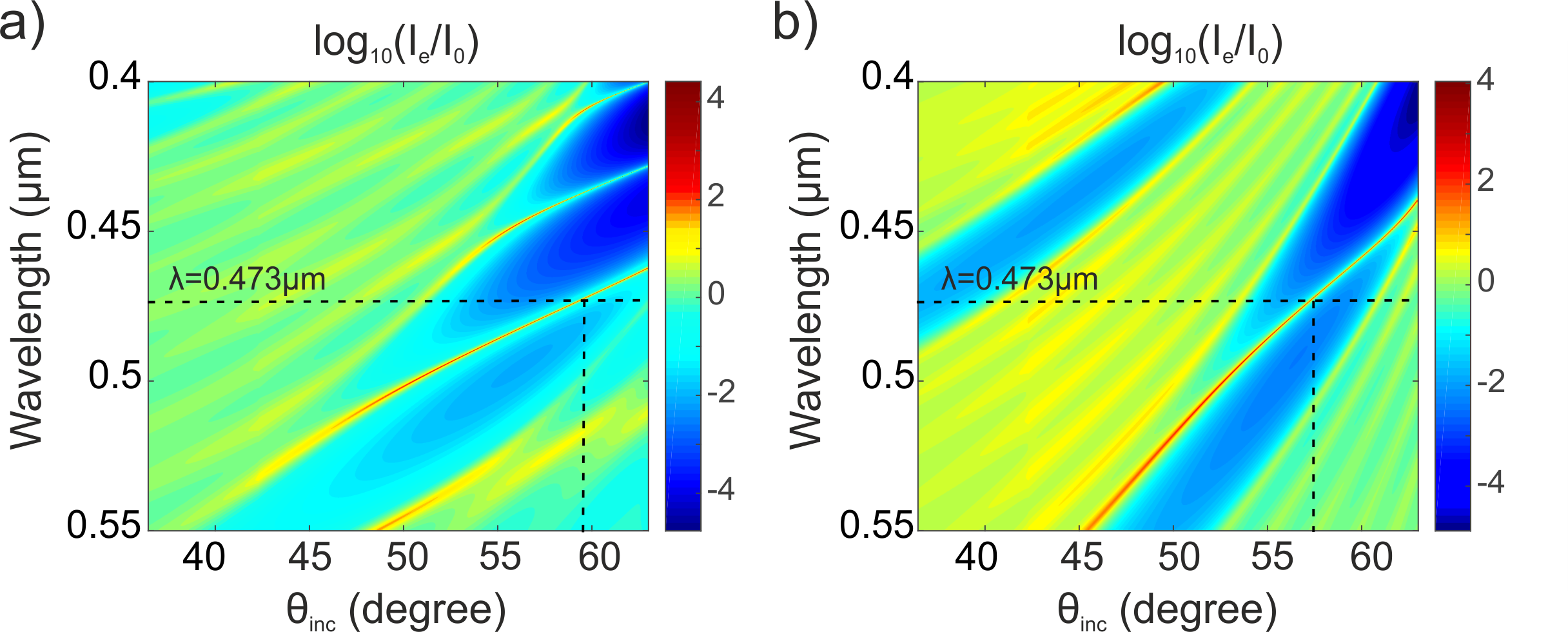}
		\caption{(a) Dispersion curves for the on-membrane 1DPhC. (b) Dispersion curves for the on-glass 1DPhC.}
	\label{fig:Fig_6_7_disp_curves}
\end{figure}

The multilayer of the glass supported crystal was as following: glass/UV glue/6 pairs of Si$_3$N$_4$(220 nm) and SiO$_2$(490 nm)/TFLN(386 nm)/air. In the green part of the spectrum [Fig. \ref{fig:Fig_6_7_disp_curves}] one can observe a band gap corresponding to the multilayer. Dispersion curves are shown in Fig. \ref{fig:Fig_6_7_disp_curves} for both samples. At the wavelength of 473 nm refractive indices of multilayer compounds are as n$_{\rm{glass}}$=1.52, n$_{\rm{Si_3N_4}}$=1.96, n$_{\rm{SiO_2}}$=1.47 and n$_{\rm{TFLN}}$=2.35. In order to obtain the dispersion curves a standard impedance approach was used \cite{konopsky:2010}. In the case of the membrane sample, light can be coupled into the Bloch surface wave at the incident angle of 59$^\circ$ and at the 56$^\circ$ for glass supported sample, as it can be seen from Fig. \ref{fig:Fig_6_7_disp_curves}.

\begin{figure}[!t]
	\centering
		\includegraphics[width=4.8in]{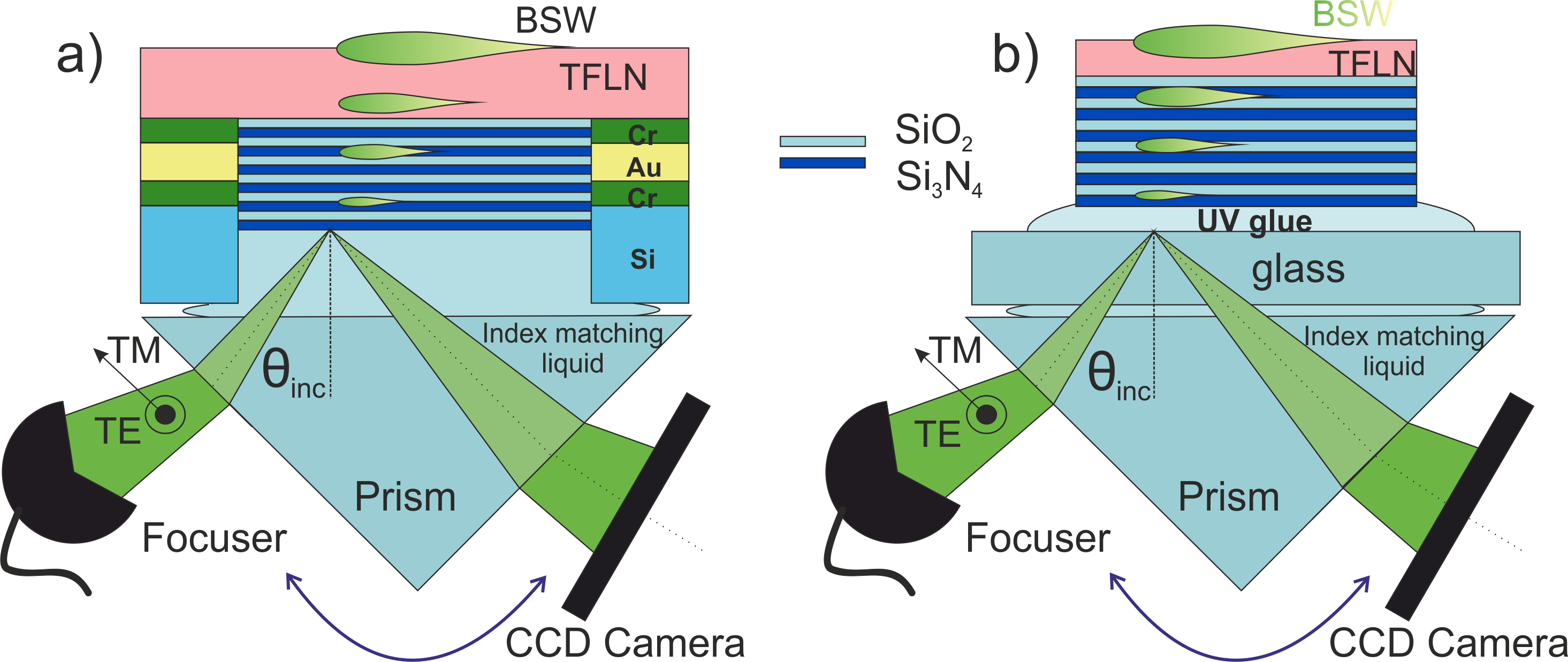}
		\caption{(a) Experimental setup for the on-membrane 1DPhC. (b) Experimental setup for the on-glass 1DPhC.}
	\label{fig:Fig_6_8_setups}
\end{figure}

In order to achieve experimental values for $\theta_{\rm{inc}}$ and to confirm them with the theoretical predictions, the Kretschmann configuration was used for light coupling (Fig. \ref{fig:Fig_6_8_setups}). TE polarized light from a diode laser (Spectra-Physics Excelsior) was focused on the sample ($\lambda=473$nm). The 473 nm wavelength was selected as a wavelength from the visible part of spectrum. Therefore it is easier to observe the BSW. A BK7 prism was fixed on a rotational stage, which allowed us to change the incidence angle until the obtention of BSW excitation. Refractive index matching oil was used for both samples as a connecting media between the sample and the prism. Reflected light was detected with a CCD camera (mV BlueFox 120GU). The BSW absorption lines were detected at 61$^\circ$ for the membrane and 55$^\circ$ for the glass supported sample. We can therefore observe that the experimentally achieved angles are slightly different from the theoretical ones. This shift may be due to the difference of requested and experimentally achieved thickness of the multilayers as well as due to some inaccuracies of refractive indices values and angle measurement.

\begin{figure}[!b]
	\centering
	\includegraphics[width=\linewidth]{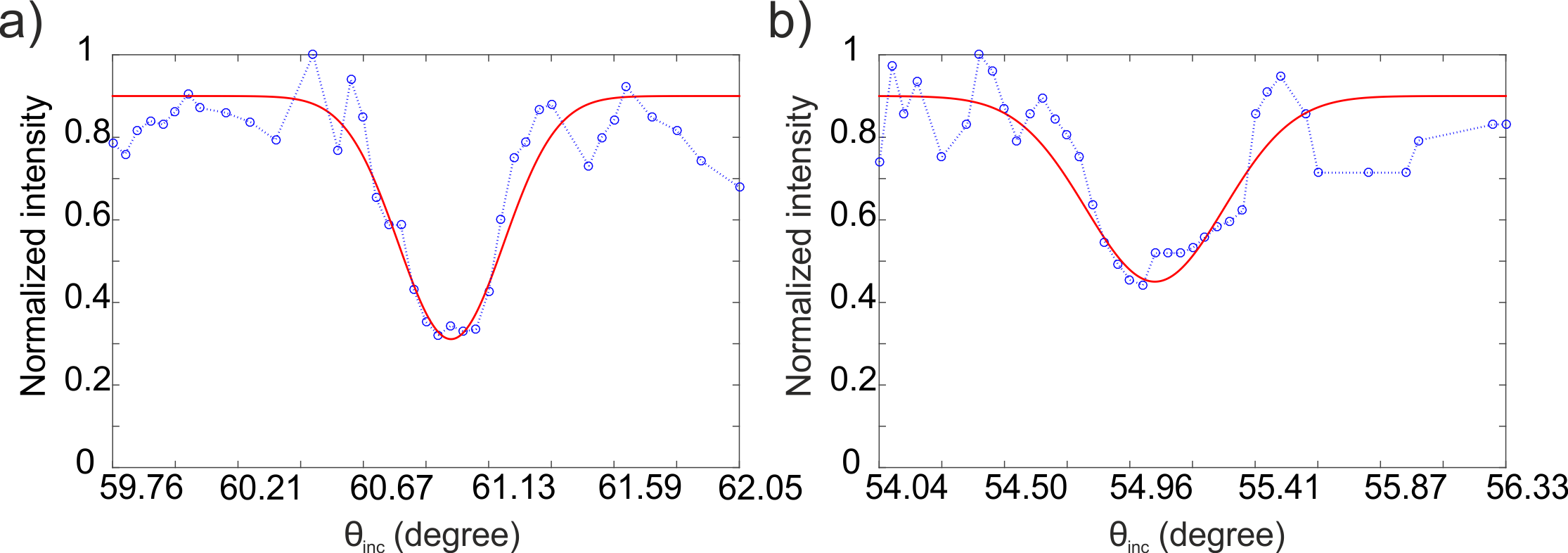}
	\caption{(a) Camera image intensity profile of the BSW related reflectance dip for the membrane based sample. (b) Camera image intensity profile of the BSW related reflectance dip for the on-glass sample. In blue - cross-sections of the reflected light images. In red - Gausian fit for experimental data.}
	\label{fig:Fig_6_9_refl_dips}
\end{figure} 

\begin{figure}[!t]
	\centering
	\includegraphics[width=5.1in]{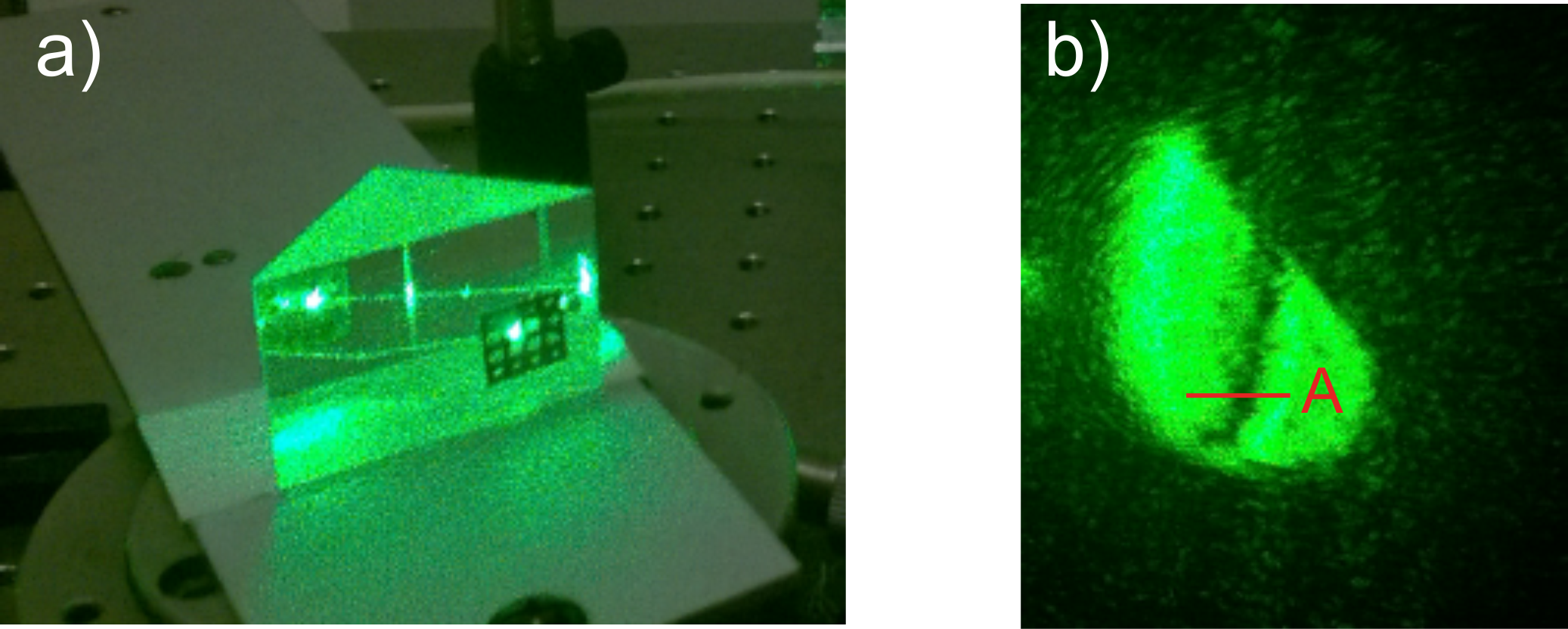}
	\caption{(a) BSW excited on the sample with matrix of 16 membranes in Kretschmann configuration; (b) Image of reflected light at the BSW coupling angle collected by camera for the membrane. A - selected line for the image cross-section.}
	\label{fig:dipsANDmembraneMatrix}
\end{figure}

\begin{figure}[!b]
	\centering
	\includegraphics[width=2.9in]{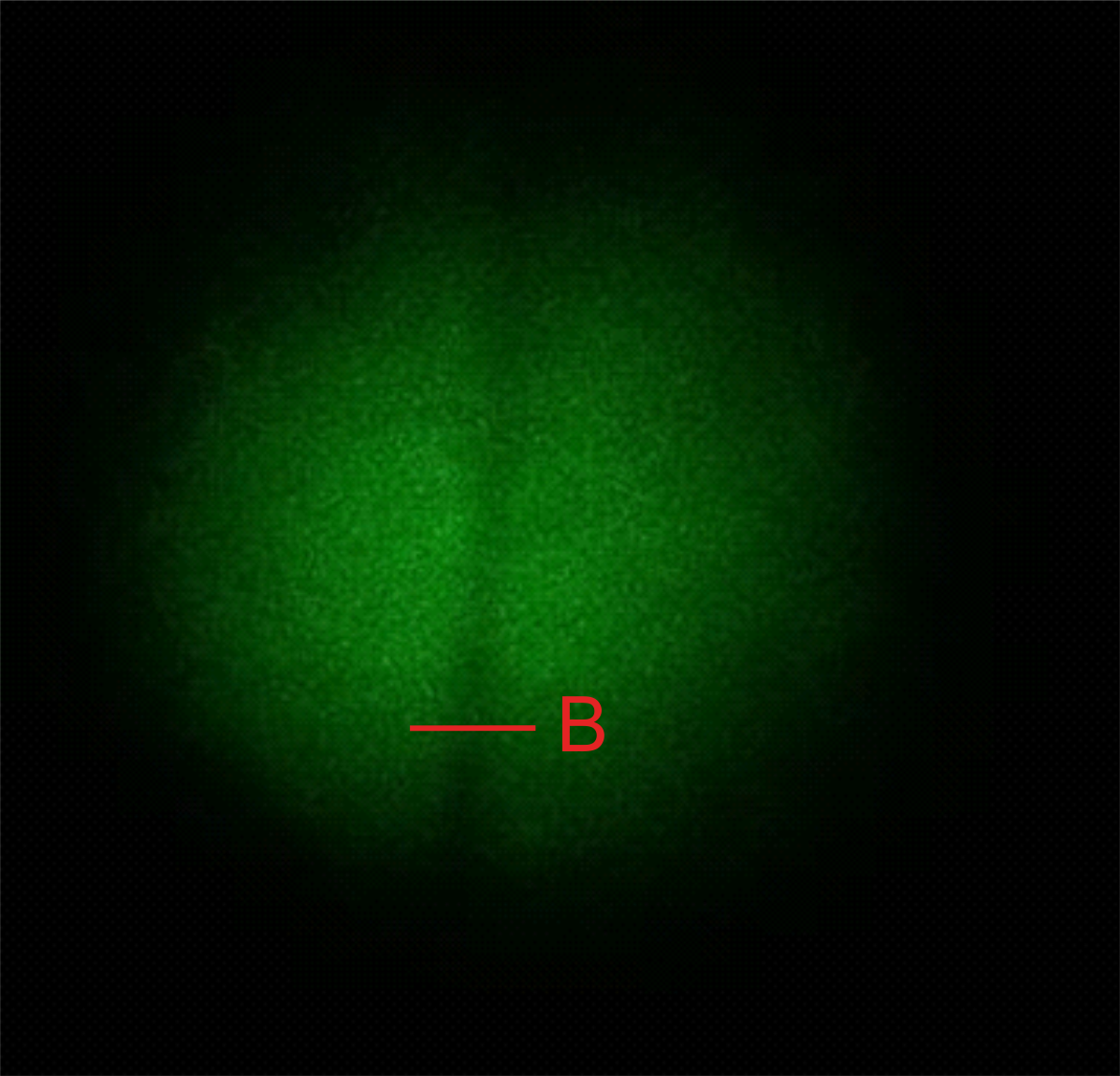}
	\caption{Image of reflected light at the BSW coupling angle collected by camera for the sample on glass support. B - selected line for the image cross-section.}
	\label{fig:ReflDipOnGlass}
\end{figure}

The images of the reflected light at the angle of BSW excitation were collected. In Fig. \ref{fig:Fig_6_9_refl_dips} the normalized intensity along the image profile at the dip area is shown for both samples. In blue lines we can see the image profiles of the reflected light for the on-membrane sample along selection A (see Fig. \ref{fig:dipsANDmembraneMatrix}(b)) and for the on-glass-support sample along selection B (see Fig. \ref{fig:ReflDipOnGlass}). Gaussian fit between average measurements along the image profile from Fig. \ref{fig:dipsANDmembraneMatrix}(b) and Fig. \ref{fig:ReflDipOnGlass} gives the red line in the Fig. \ref{fig:Fig_6_9_refl_dips}. In the case of the membrane the dip is narrower and the reflectance dumps for 60$\%$. For the glass supported sample the dip is 1.3 times wider and only 40$\%$ of reflectance dumping is observed. The bigger dumping of reflectance for the membrane sample indicates that in this case we have better coupling of light into the BSW than for the on-glass 1DPhC. However, in the case of the membrane significant aberrations due to the curvature of the membrane occur.

We can therefore conclude that the multilayer on the glass support is more stable and provides a bigger area for light manipulation that use the BSW (in our case $1\times1$cm$^2$ area). This advantage can be used for BSW based on-chip integrated optics \cite{dubey:16}. However, additional manipulations (such as polishing, plasma etching of Si and SiO$_2$ as well as photoresist removal) introduce some deformations into the layers structure. This leads to decrease of the coupling efficiency and propagation length of BSW, what can be observed from the broadening of the BSW related absorption dip. 

In the case of the membrane configuration, we have a limited access to the multilayer for the BSW coupling on the Kretschmann configuration (in our case a maximum $1.5\times1.5$mm$^2$ area is available). This limitation can be overcame for example by using a grating coupler \cite{Kovalevich:17}. In addition, the whole membrane with the multilayer can be transferred and bonded into a fiber tip. This fibered configuration can be used for BSW based lab-on-fiber systems with TFLN as a part of it. For the membrane multilayer a stronger BSW coupling was observed (as it can be seen from Fig. \ref{fig:Fig_6_9_refl_dips}(a)), though there are still additional perturbations for the light propagation which are due to the membrane curvature.

The images of the reflected light at the angle of BSW excitation are shown at the Fig. \ref{fig:dipsANDmembraneMatrix}(b) and Fig. \ref{fig:ReflDipOnGlass} for the 1DPhC on the membrane and on glass support respectively. For the SW excitation on membrane the sample of 1$\times$1 cm$^2$ was used. The mask with rows of 0.5$\times$0.5 mm$^2$ (top and bottom rows) and of 1$\times$1 mm$^2$ (two central rows) with 4 membranes in each was prepared for the lithography steps [Fig. \ref{fig:Fig_6_2_MLonMembrane_steps}(d-f)]. Thus we obtained a sample with 16 (4$\times$4) openings in Si and SiO$_2$ with suspended membranes of LiNbO$_3$. The multilayer was deposited from the openings side. The image of the final sample with membranes illuminated by light at 473 nm wavelength in Kretschmann configuration is shown at the Fig. \ref{fig:dipsANDmembraneMatrix}(a). Due to the curvature of the membrane it is not trivial to find a flat place where we can observe a clear reflectance dip and the walls of Si work as an obstacles (see Fig. \ref{fig:Fig_6_8_setups}). Therefore we do not obtain a symmetric circle on the camera image at Fig. \ref{fig:dipsANDmembraneMatrix} and can see a lot of aberrations. Meanwhile the reflectance dip for on glass substrate sample is clear and can be excited at any part of the sample area (1.5$\times$1.5cm$^2$). Due to the easiness in handling and manipulation of the on-glass-support platform we chose this configuration for our further investigations.

\section{Top surface improvement of 1DPhC bonded by UV glue.}

During our first experiments with BSW on TFLN we have successfully excited the surface wave in the visible part of spectrum at the TFLN/air interface on the sample with a multilayer bonded by glue to the glass platform. Though the detected reflectance dumping of only 40$\%$ and a broad reflectance dip represent high losses which occur during the BSW excitation. After thorough investigation we discovered that these losses appear due to the top surface imperfections.

\begin{figure}[!b]
	\centering
	\includegraphics[width=5in]{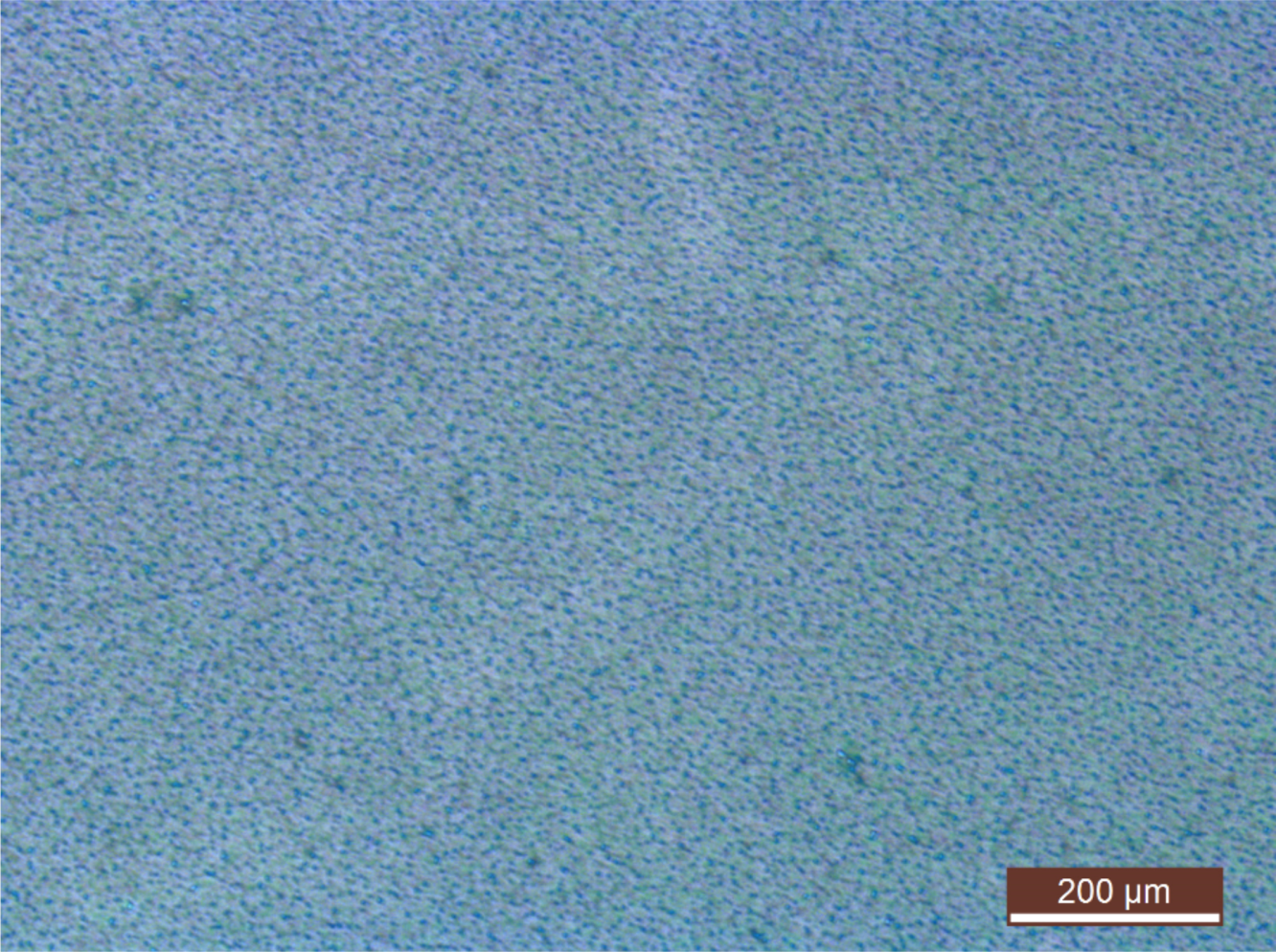}
	\caption{Microscope image of the TFLN surface of the sample on glass support after Si and SiO$_2$ removal with DRIE and RIE.}
	\label{fig:Sample_for_green_not_improved}
\end{figure}

\begin{figure}[!ht]
	\centering
	\includegraphics[width=4.2in]{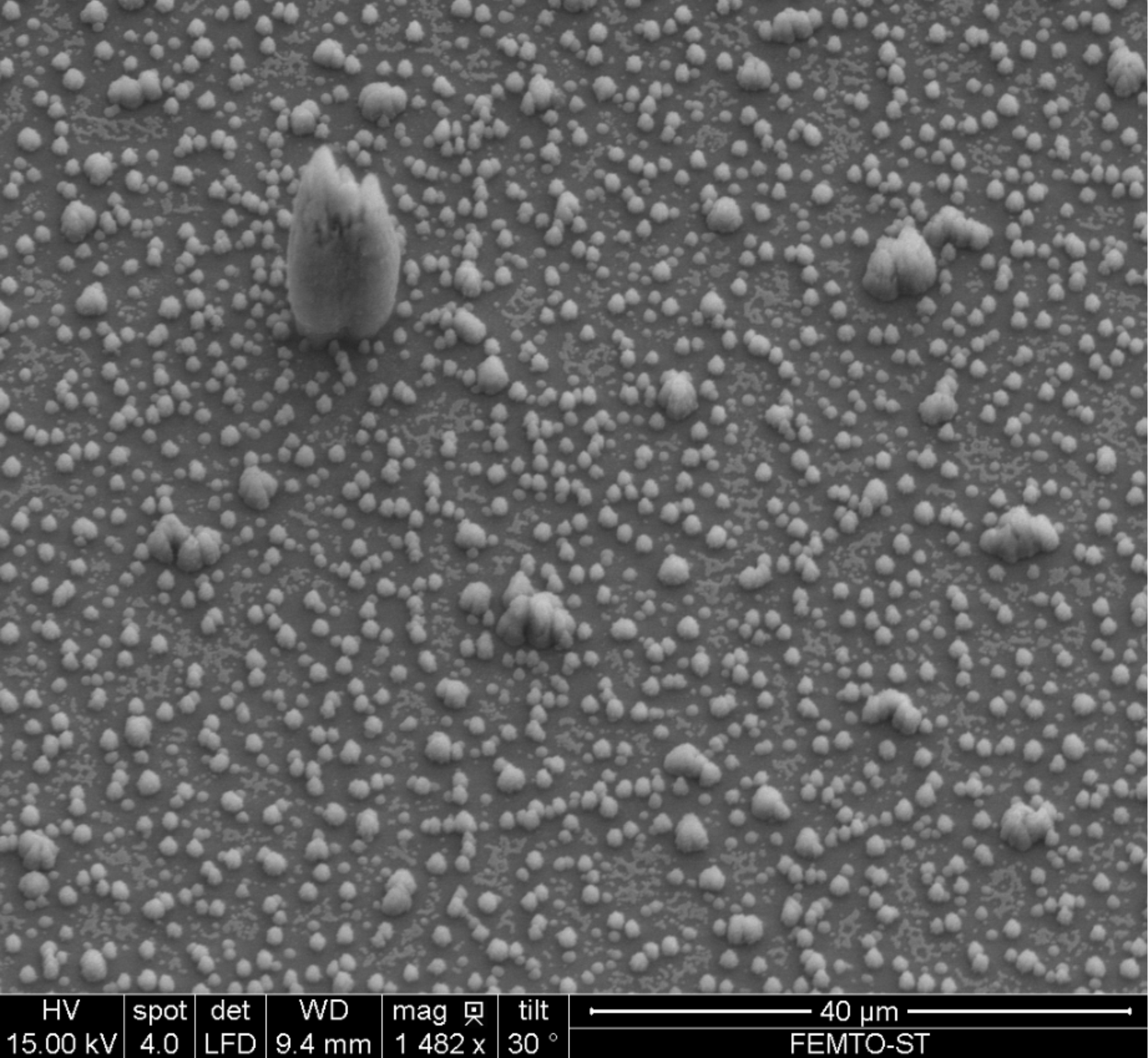}
	\caption{SEM image of the TFLN surface of the sample on glass support after Si and SiO$_2$ removal with DRIE and RIE.}
	\label{fig:SEM_of_unprocessed_sample}
\end{figure}

The microscope image of the sample surface prepared by the recipe described in chapter 6.3 is shown at the Fig. \ref{fig:Sample_for_green_not_improved}. Here even with a simple microscope characterization we may observe various inhomogeneities. After a closer look with SEM microscope [Fig. \ref{fig:SEM_of_unprocessed_sample}] the pillars of different hight ( up to 6 $\mu m$) can be observed. The thickness of SiO$_2$ layer of the TFLN waver from NANOLN is 2 $\mu m$. Here we can conclude that these inhomogeneities are actually pillars of Si and SiO$_2$ which remained after DRIE and RIE process.

\begin{figure}[!ht]
	\centering
	\includegraphics[width=4.2in]{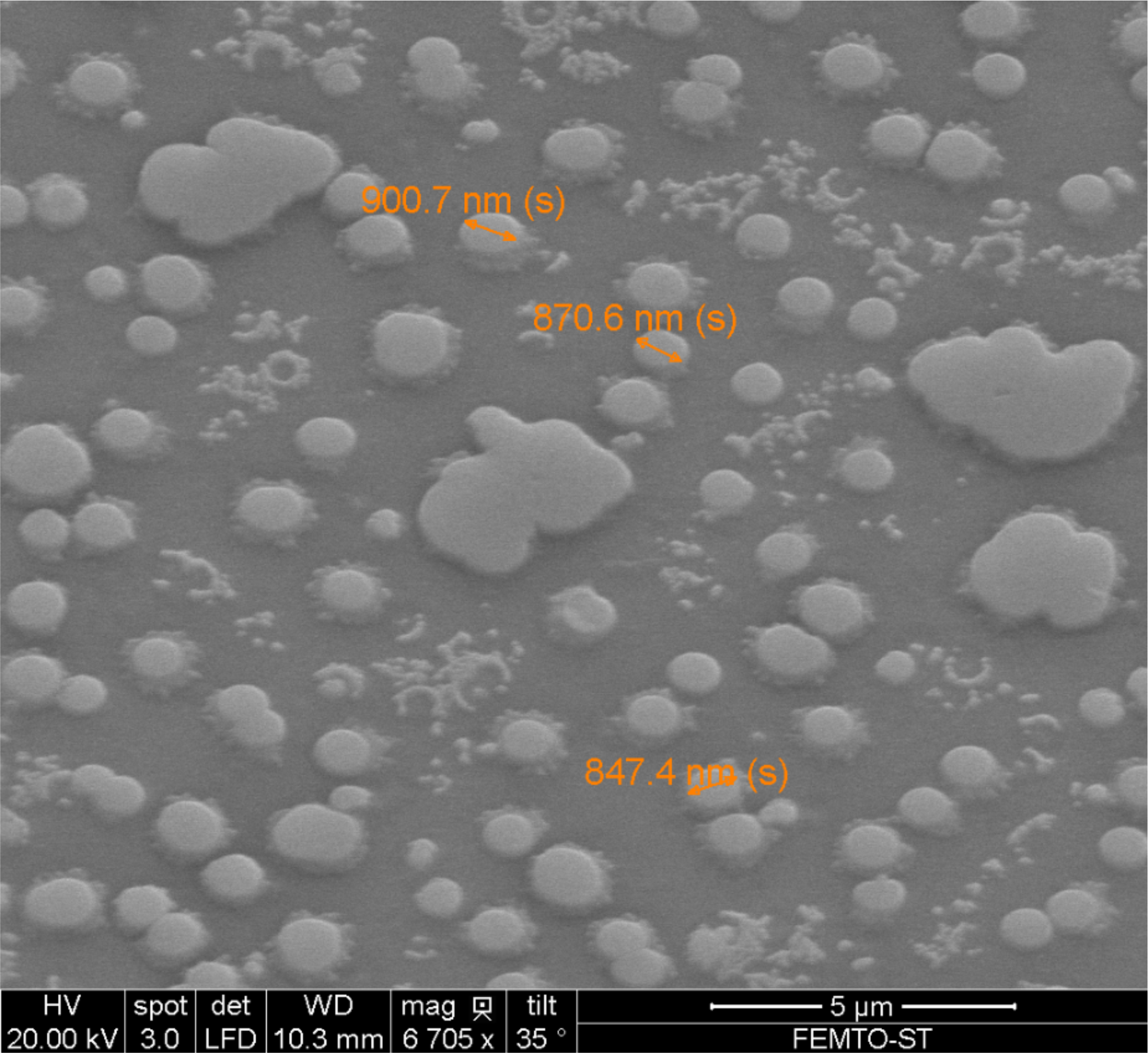}
	\caption{SEM image of the TFLN surface of the sample on glass support after Si and SiO$_2$ removal with DRIE, RIE and additional 57 min in BHF.}
	\label{fig:SEM_of_unprocessed_sample_and_57min_BHF}
\end{figure}

As a logical solution to improve the sample surface at this step may be additional wet etching of Si (top part of pillars) in KOH with following wet etching in HF of SiO$_2$. The KOH etching appeared to be too aggressive for the UV glue. With or without additional heating (which is usually used for standard KOH wet etching), the UV glue dissolves faster than Si, what leads to immediate multilayer delamination. After 3 hours in KOH with the 55$^{\circ}$C temperature the sample was totally destroyed. 

As a next step wet etching in buffered HF solution (BHF) (17$\%$) was tested. This is a fast process and it goes without additional heating required. Usually in order to etch completely 2 $\mu m$ of SiO$_2$ the sample should be left in BHF for 30 min. However, even after 57 min in BHF the bases of the pillars were not removed. Meanwhile the top part of the SiO$_2$ layer was successfully etched [Fig. \ref{fig:SEM_of_unprocessed_sample_and_57min_BHF}]. 

In addition, independently from how long the sample stays in the BHF solution, after the DRIE and RIE processes (maximum time tested is 1 hour 30 min) the bases of pillars were not removed. From here we conclude that apparently some high temperature regions have appeared during the plasma etching. Possibly, at the contact area between the support wafer of DRIE machine and the sample. This could have slightly changed the properties of SiO$_2$. Therefore, for future samples preparation additional temperature stabilization steps were added to DRIE process.

\begin{figure}[!b]
	\centering
	\includegraphics[width=5.3in]{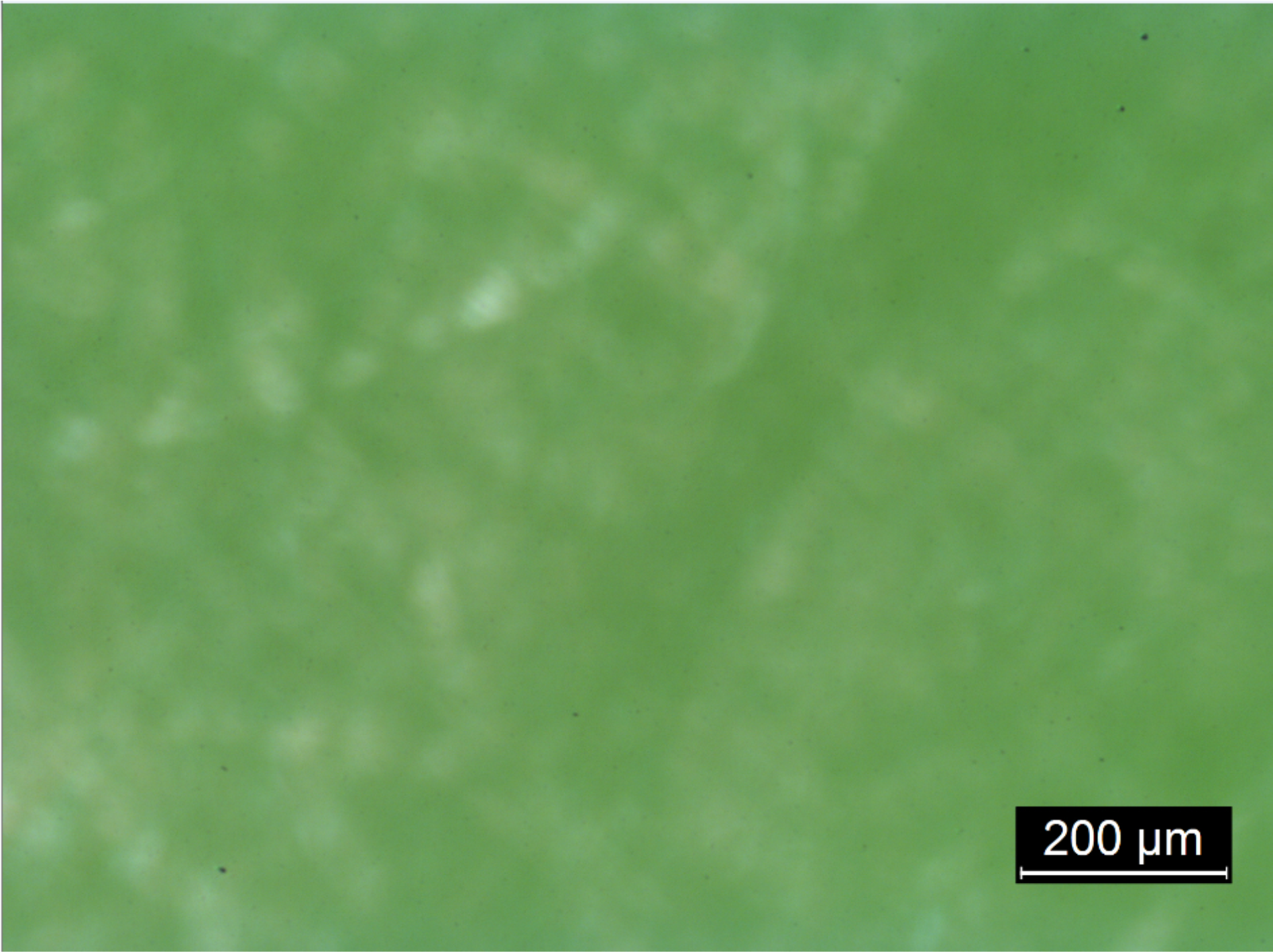}
	\caption{Microscope image of the TFLN surface of the sample on glass support after Si and SiO$_2$ removal with DRIE and RIE.}
	\label{fig:Microscope_improved_surface}
\end{figure}

\begin{figure}[!b]
	\centering
	\includegraphics[width=4.5in]{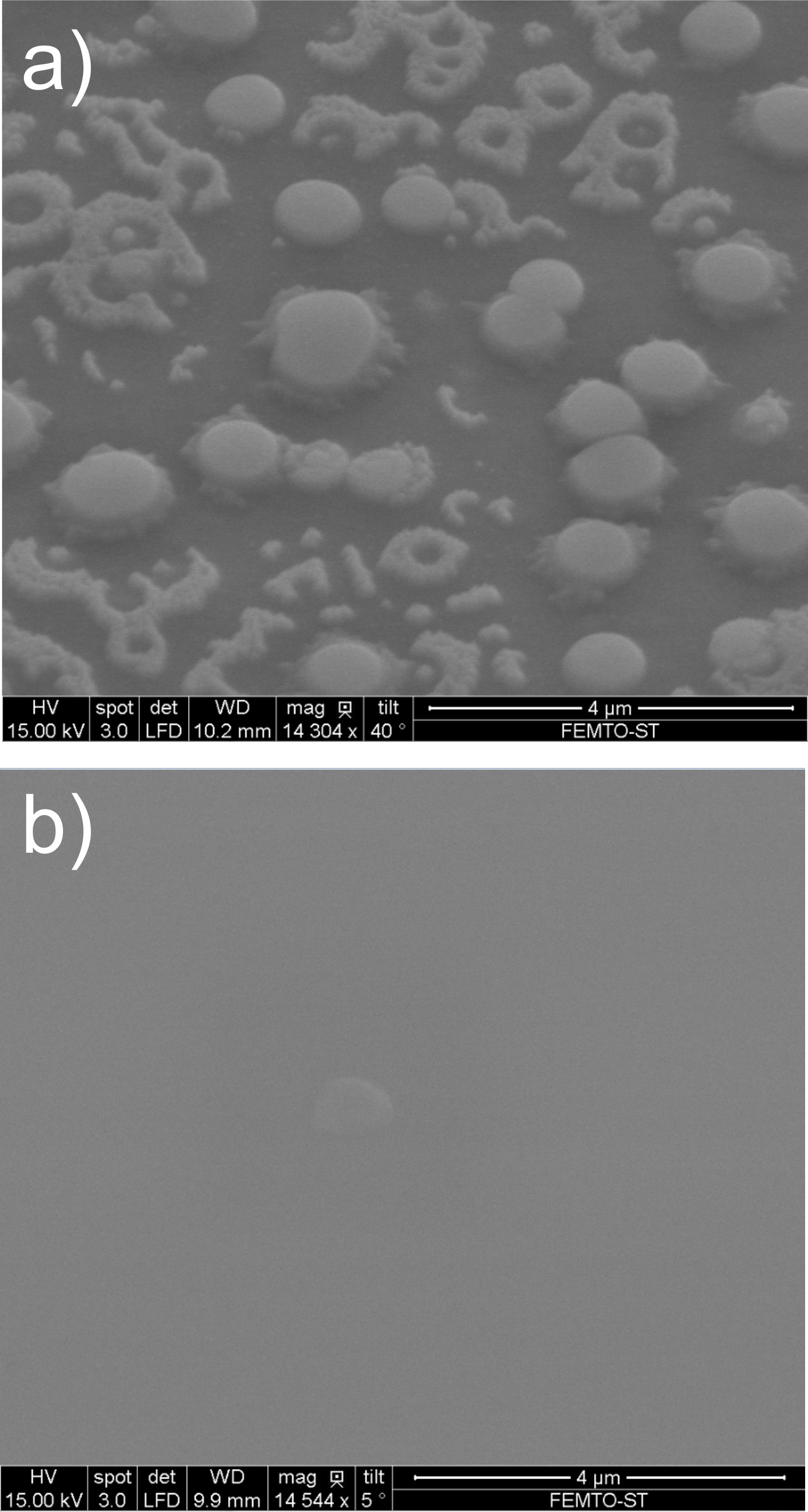}
	\caption{SEM images of 1DPhC with TFLN top surface before (a) and after (b) implementation of additional cleaning and temperature stabilization protocols.}
	\label{fig:Befor_and_after_extra_cleaning}
\end{figure}

Additionally, instead of RIE the 20 min wet etching of SiO$_2$ in HF solution (40$\%$) was used.

From Fig. \ref{fig:SEM_of_unprocessed_sample_and_57min_BHF} we see that the distance between the SiO$_2$ areas as well as their size varies up to 1-2 $\mu m$. For BSW excitation at 1550 nm wavelength (which is very commonly used in integrated optics) the size of the features is comparable with the wavelength of the surface wave. In the 1DPhCs designed to work at 1550 nm wavelength with same top surface roughness the BSW was not excited. Meanwhile for smaller wavelength (473 nm) the BSW related reflectance dip, which indicates the BSW excitation was clearly observed.  Such a wavelength-dependent attenuation could be attributed to a "band-gap" like effect in the random features at the surface, or the presence of Anderson localizations \cite{bla} at 1.55 microns (the random features become subwavelength) which would strongly dissipate the BSW due to their radiating dipole nature. It seems that the ratio between the island size and separation distance and the wavelength could be a central parameter in defining the degree of perturbation of the BSW. This could give precious information on the robustness of the BSW to an external random perturbation and could unveil two different coupling regimes of a BSW and a random structure. This merits further considerations in future works.

%Random media, even random features, at a surface could show some band gaps. From Fig. \ref{fig:SEM_of_unprocessed_sample_and_57min_BHF} we may see that the distance between the SiO$_2$ areas as well as their size varies up to 1-2 $\mu m$. For BSW excitation at 1550 nm wavelength (which is very commonly used in integrated optics) the size of the features is comparable with the wavelength of the surface wave. In the 1DPhCs designed to work at 1550 nm wavelength with same top surface roughness the BSW was not excited. Meanwhile for smaller wavelength (473 nm) the BSW related reflectance dip, which indicates the BSW excitation was clearly observed.  We expect that a band gap of the random features
%at the surface is responsible for the wavelength-dependent attenuation. 

In any case, if the 1DPhC with TFLN would be applied in biosensing, when the wavelengths from the visible range are commonly used, the surface must be clear from any contamination for homogeneous antibody and analyte distribution. If the 1DPhC is made for integrated optics applications the near infrared wavelengths are commonly used, therefore, once again the TFLN surface have to be improved.

The micro-pillars on the TFLN surface appear due to micro-masking process \cite{Shearn:10}, which happens during plasma etching because of some chamber or sample surface contamination. The cleaning process for the single LN membrane was developed in FEMTO-ST nano-optics group. Though the presence of the UV glue requires additional process development. The glue easily dissolves in aggressive chemicals, but the sample still have to be cleaned before DRIE process. Therefore we add and adjust extra cleaning steps for samples preparation.

These steps are as following:
1) standard piranha clean before the multilayer deposition (between steps (a) and (b) Fig. \ref{fig:Fig_6_5_MLglass_steps}); 
2) 2 min in acetone, 2 min in ethanol, water rinsing and piranha clean for the samples with multilayer and for the glass support (between steps (b) and (c) Fig. \ref{fig:Fig_6_5_MLglass_steps});
3) 2 min in acetone, 2 min in ethanol, water rinsing and piranha clean for the support wafer for DRIE (between steps (d) and (e) Fig. \ref{fig:Fig_6_5_MLglass_steps}).
 
Also the temperature stabilization before Si etching in DRIE machine should be increased. 

The microscope image of the sample surface prepared with the improved recipe is shown at the Fig. \ref{fig:Microscope_improved_surface}. Here we may see that the inhomogeneities are almost gone in comparison with Fig. \ref{fig:Sample_for_green_not_improved}, the sample is clean and transparent. Also the SEM image taken with the same tilt and magnification shows a drastic improvement [Fig. \ref{fig:Befor_and_after_extra_cleaning}].

These final steps of the TFLN surface improvement now indeed make possible all the further nano-structuring of the top part of 1DPhC. Also the device is ready to be used in any domains where the BSW can be applied.

\section{Membranes transportation.}

As a side part of the study we have discovered different ways to manipulate and transport membranes. 

\begin{figure}[!b]
	\centering
	\includegraphics[width=5in]{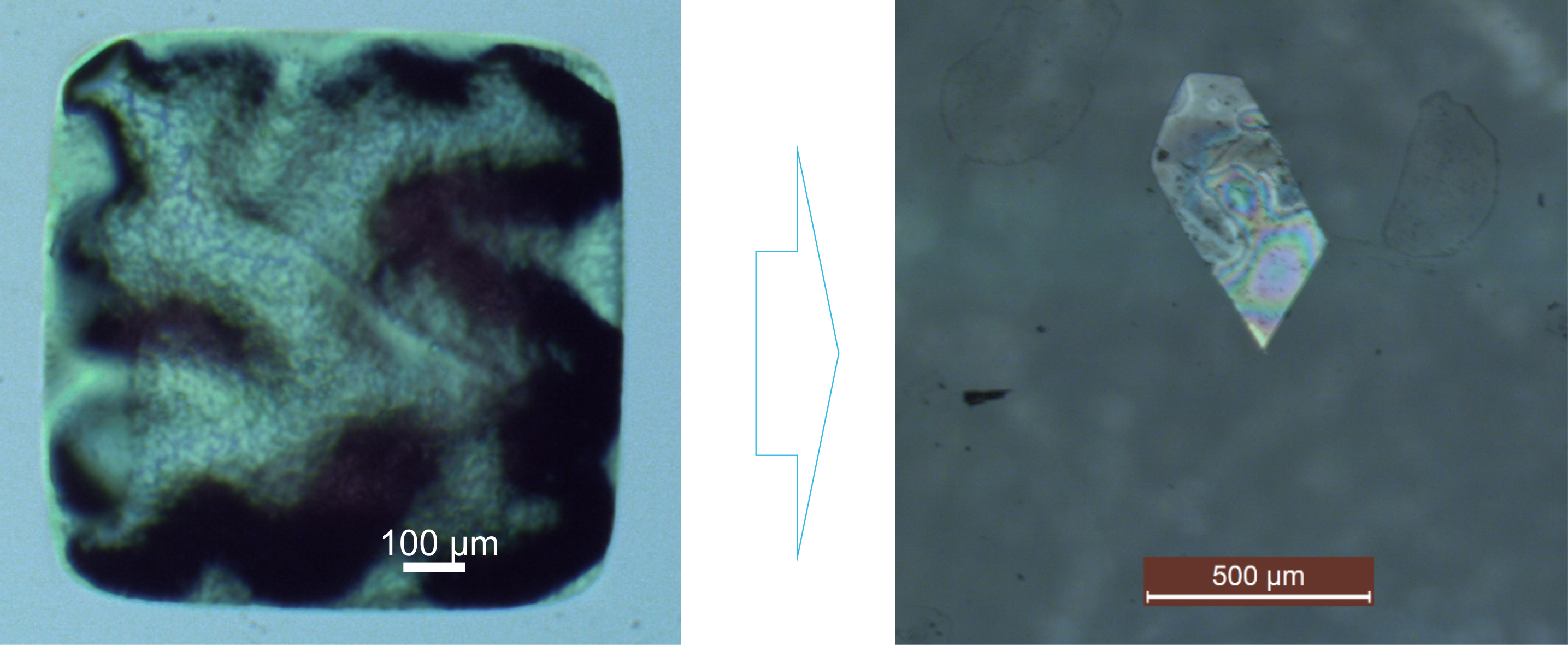}
	\caption{(a) Single TFLN membrane suspended in air; (b) TFLN membrane transported to the microscope cover glass.}
	\label{fig:LNmembrSeparated}
\end{figure}

For example the 500 nm TFLN membrane of 1.5$\times$1.5 mm$^2$ size can be cut out with the needle and transported without any damage by an optical fiber to any other surface (microscope cover glass in our case) thanks to the static forces which appear when we touch the membrane with a fiber tip. The example of the transported TFLN membrane of 0.5$\times$0.25 mm$^2$ size is shown at the Fig. \ref{fig:LNmembrSeparated}.

During TFLN top surface improvement tests we have discovered that the samples with the UV glue as a part of the multilayer are very sensitive to aggressive chemicals and temperature. Once the sample is ready additional piranha clean should be avoided, if possible, as well as long (more than 2 min) rinsing in acetone. If 1.5$\times$1.6 cm$^2$ sample stays in acetone for a time longer than 3 hours (sometimes an additional heating at 30/35 $^\circ$C may be required) the bonding UV glue layer dissolves. This gives us an opportunity to obtain a single membrane made only out of the multilayer. Such a membrane is robust and can be transported by fiber tip or simply by tweezers. 

\begin{figure}[!t]
	\centering
	\includegraphics[width=5in]{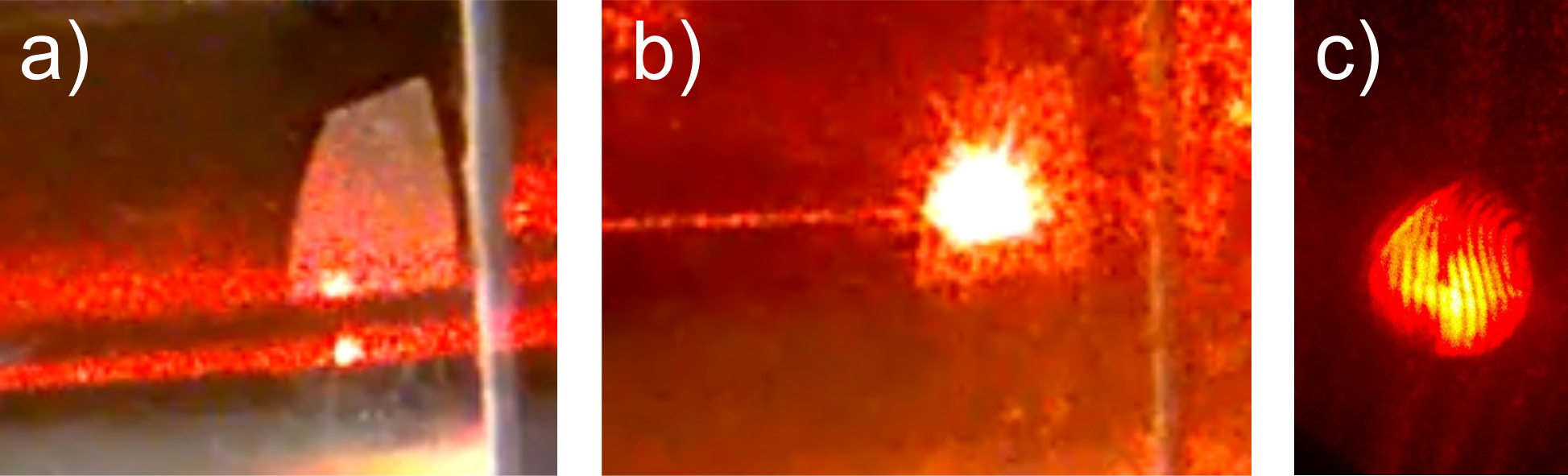}
	\caption{(a) Multilayer with TFLN on prism; (b) BSW excited on the single membrane; (c) BSW related reflectance dip.}
	\label{fig:LMmembrONprizm}
\end{figure}

Depending on the thickness of TFLN a single 1.5$\times$1.5 mm$^2$ LiNbO$_3$ membrane almost reaches the size limit, when it can be manipulated without being damaged during the manufacturing steps described in Fig \ref{fig:Fig_6_2_MLonMembrane_steps}(c-f). Therefore if we deposit the multilayer on TFLN suspended membrane the size of the structure to be transported is also about 1.5$\times$1.5 mm$^2$ and it is also quite curved as it was discussed earlier.

If the membrane is prepared out of the bonded to glass multilayer we may reach much bigger sizes. Figure \ref{fig:LMmembrONprizm} shows the example of the 2.5$\times$3 mm$^2$ size membrane made of 6 pairs of Si$_3$N$_4$(220 nm) and SiO$_2$(490 nm) and TFLN(386 nm) and transported to the BK7 prism. A small drop of refractive index matching liquid was applied on the prism surface with an optical fiber to provide a good contact between the prism and the membrane. For this structure the BSW was excited by the red light ($\lambda=632.8~nm$). At the Fig. \ref{fig:LMmembrONprizm}(b) we may see a BSW related light enhancement on the membrane and the BSW related reflectance dip [Fig. \ref{fig:LMmembrONprizm}(c)].

% \begin{figure}[!t]
% 	\centering
% 		\includegraphics[width=4in]{LMmembrONprizm.png}
% 		\caption{SEM + 57 min BHF}
% 	\label{fig:LMmembrONprizm}
% \end{figure}

 \section{Conclusion.}

In this part of the work, we have shown theoretically and experimentally the excitation of BSWs at a TFLN/air interface, what introduces all the potentialities of the LiNbO$_3$ functionalities through  BSW light guiding.

We have demonstrated BSWs on the top of thin film LiNbO$_3$ in two different configurations. LN based photonic crystals, which are able to sustain BSWs were designed and fabricated at the base of thin film LiNbO$_3$ membrane and on a glass support. In order to compare two different 1DPhCs far field measurements were done. The designed 1DPhCs allow to obtain BSWs at the TFLN/air interface for TE polarized light at 473 nm wavelength. Such designs of the 1DPhCs together with the use of the nonlinear properties of LiNbO$_3$ opens up the possibility of creating BSW based active tunable devices.

In addition we have achieved a drastic improvement of the 1DPhC with TFLN top surface and described different ways to manipulate the TFLN membranes (with and without multilayer). The BSW at the TFLN/air interface for TE polarized light at 632.8 nm wavelength was excited on a single multilayer membrane in Kretschmann configuration.

\chapter{Perspectives for BSW on photonic crystals with LiNbO$_3$.}

In this chapter we will describe a concept and several major preparation steps, which were made for electro-optical tuning of BSW on TFLN. We design and fabricate the 1DPhC with TFLN on top, which is able to support the BSW at 1550 nm wavelength and develop special electrodes deposition technique which does not damage the UV glue in the sample. 

\vspace*{0.2cm}
\minitoc

\section{Concept.}

Photonic crystals and Bragg gratings are some of the optical functions that can enhance the electro- and thermo-optical properties of LiNbO$_3$ based devices. The refractive index change up to $\Delta n=0.3$ can be obtained by additional nano-structuring of LiNbO$_3$ \cite{Roussey:06}. With the improved surface of the TFLN on the top of 1DPhC it is now possible to pattern the refractive index change of the BSW in any convenient way, just by following the laws of 2D "flatland" optics \cite{yu:13,wu:14,dubey:16}.

The Pockels effect is a commonly used phenomena for the refractive index change in crystals that lack inversion symmetry, such as LiNbO$_3$ \cite{Jelinkova:04}. This effect in a bulk LiNbO$_3$ can be drastically enhanced by photonic crystals. For example, it was used to create an electro-optically tunable photonic crystal linear cavity on a 200 nm lithium niobate ridge waveguide. The photonic crystal, of area 4$\times$0.8 $\mu m^2$, has been engineered to work in a slow light configuration so that the electro-optic effect is 20 times stronger than in bulk material \cite{Lu:12}. Also this effect was applied to demonstrate how slow group velocities that are easily attainable at the band edge of photonic crystals can drastically enhance the electro-optical effect on tunable photonic crystal components. This property opened up the possibility of micro-sized nonlinear devices with low power requirement. The enhancement of nonlinear effects have been used to fabricate a 13$\times$13$\mu m^2$ sized lithium niobate photonic crystal intensity modulator that shows an enhanced electro-optic effect 312 times bigger than the one predicted by the classical Pockels effect for an equivalent device in bulk material \cite{Roussey:06}.

To implement similar techniques for BSW tunability additional notion about electrodes deposition on the multilayer sample is required. Therefore we develop a recipe for electrodes manufacturing on the top of multilayer with TFLN, bonded to glass support by UV glue.

As a perspective, we expect that the implementation of electro-optical tunability of BSW on 1DPhC with TFLN may be realized according to the following steps. Firstly we should excite the BSW at 1DPhC surface. Then, the BSW should be coupled into the waveguide with 2DPhC. The 2DPhC should provide the band gap in such a wavelength range, that $\lambda_{BSW}$ would be at the edge of the bang gap. The waveguide should be positioned between two electrodes. Thus, the applied voltage will generate the Pockels effect. As a consequence, we may expect the band gap shift in the 2DPhC, what would lead to some modulations in transmission of BSW through the 2DPhC.

Several major steps on the way to the concept realization would be described below.

 \section{1DPhC for 1550nm with TFLN on top.}

\begin{figure}[!b]
	\centering
	\includegraphics[width=3.8in]{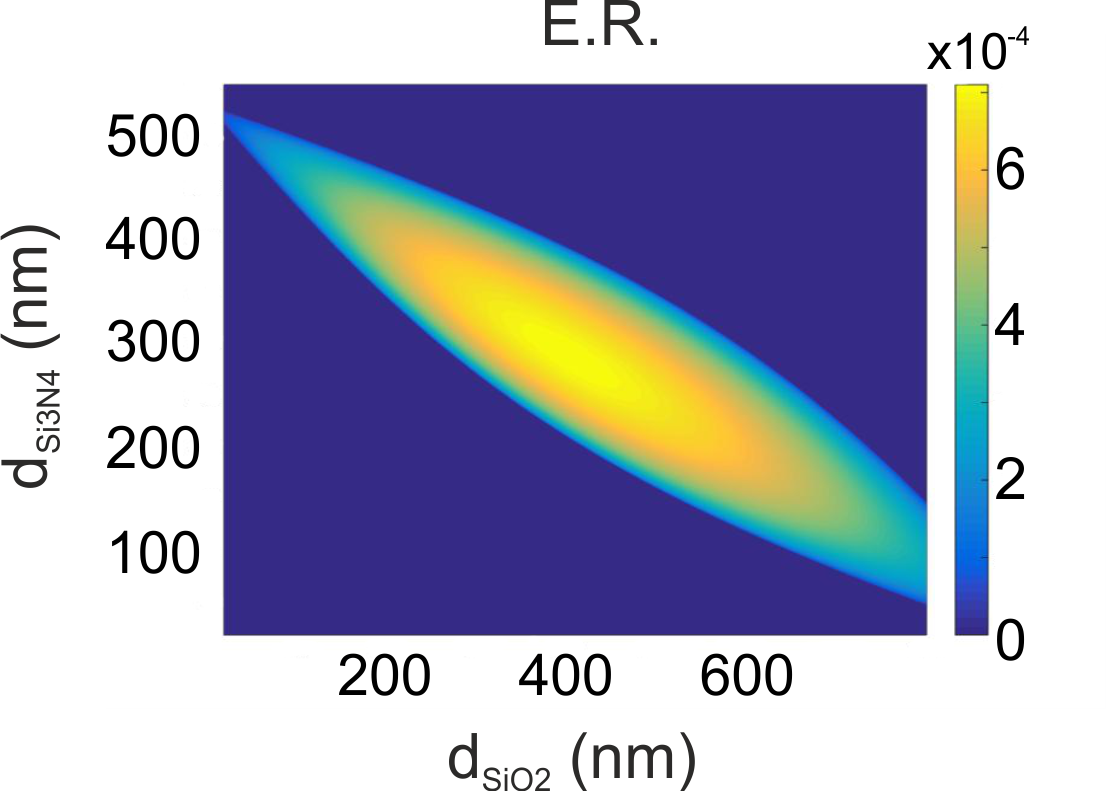}
	\caption{Extinction of the band gap as a function of layer thicknesses in 1DPhC.}
	\label{fig:Fif_7_1_find_d1_d2}
\end{figure} 

First of all we design the 1DPhC, which is optimized to work at the 1550 wavelength. This wavelength was chosen as a commonly used one for integrated optics applications \cite{Sun:07,Boltasseva:05}. It is also a common wavelength for which other BSW "flat-land" devises such as curved waveguides \cite{wu:14}, ring resonators \cite{dubey:16}, 2D lenses \cite{yu:14}, etc. were developed. 

For this part of study we once again chose Si$_3$N$_4$, SiO$_2$ and 450nm X-cut TFLN as materials for multilayer. TFLN bonded by 2 $\mu m$ of SiO$_2$ to Si wafer is provided by NANOLN company. Si$_3$N$_4$ and SiO$_2$ were chosen as materials which proved stable during our previous experiments.

As it is described in chapter 2.1.2 we determine the thicknesses of the Si$_3$N$_4$ and SiO$_2$ by solving the Eq. \ref{eq:refname11}. With $T$ - the transmission coefficient for one period of silicon oxide and silicon nitride, the desired values of thicknesses $d_1=d{\rm{_{Si_3N_4}}}=d{\rm{_{Si_3N_4max}}}$ and $d_2=d{\rm{_{SiO_2}}}=d{\rm{_{SiO_2max}}}$ are the thicknesses at which $E.R.$ reaches its maximum value. Graphical solution for $E.R.$ for Si$_3$N$_4$/SiO$_2$ stacks at $\lambda=1550$~nm with $n\rm{_{Si_3N_4}}=1.79$ and $n\rm{_{SiO_2}}=1.44$ is shown at the Fig. \ref{fig:Fif_7_1_find_d1_d2}.

\begin{figure}[!t]
	\centering
	\includegraphics[width=3.6in]{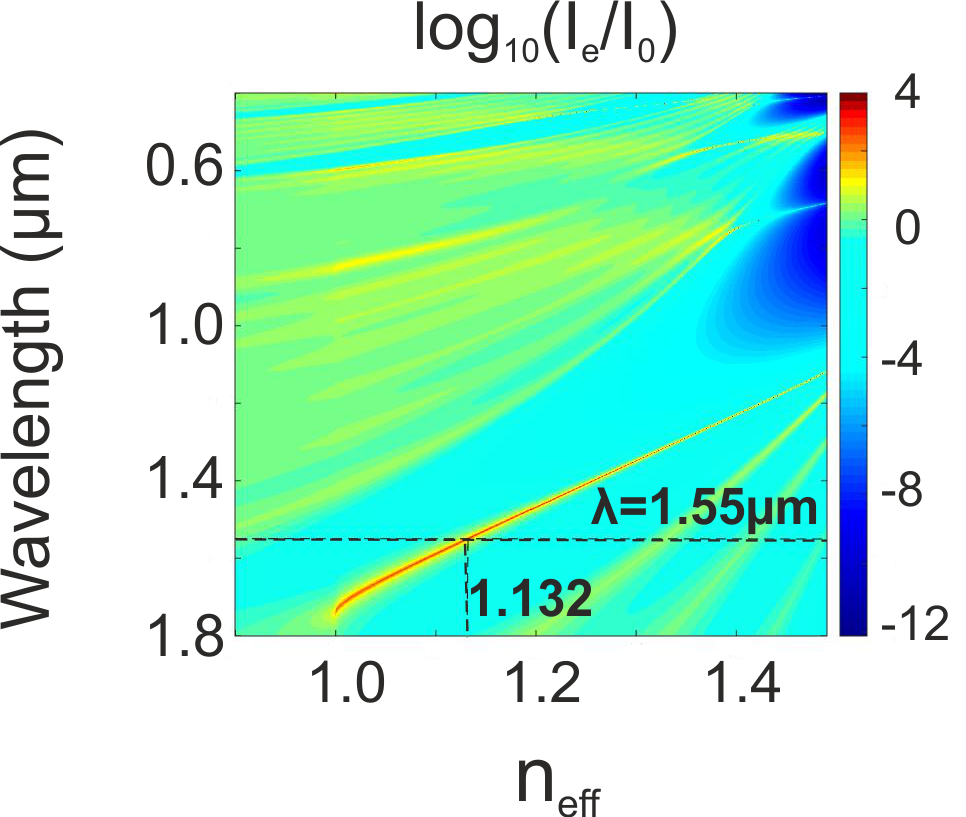}
	\caption{Dispersion curve for designed 1DPhC.}
	\label{fig:Fif_7_2_disp_curve_no}
\end{figure} 

\begin{figure}[!b]
	\centering
	\includegraphics[width=\linewidth]{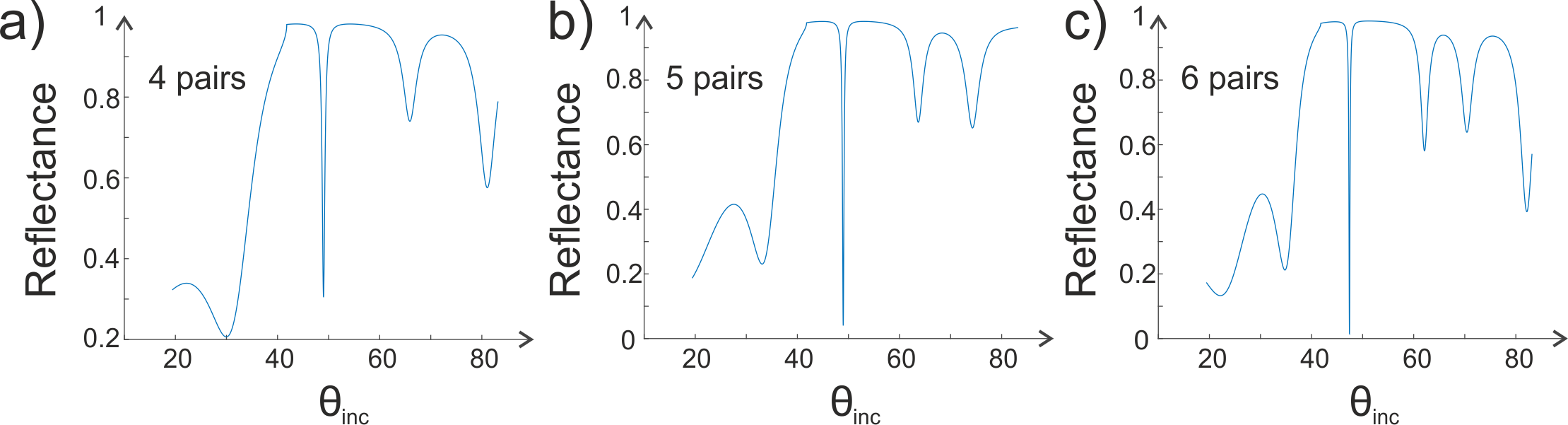}
	\caption{Calculated reflectance in 1DPhC for (a) 4 pairs, (b) 5 pairs and (c) 6 pairs of Si$_3$N$_4$ and SiO$_2$.}
	\label{fig:Fif_7_3_refl_no}
\end{figure}

For our final multilayer design we want to chose the configuration which can utilize the largest electro-optic coefficient, what would bring a higher change of the refractive index. For LiNbO$_3$ it is the coefficient $r_{33}=30.8$ pm/V at the wavelength $\lambda=1550$ nm. In this case for the X-cut TFLN the electrodes should be deposited along Y axis of the LiNbO$_3$. Thus the BSW also should propagate along the Y direction. 

\begin{figure}[!b]
	\centering
	\includegraphics[width=\linewidth]{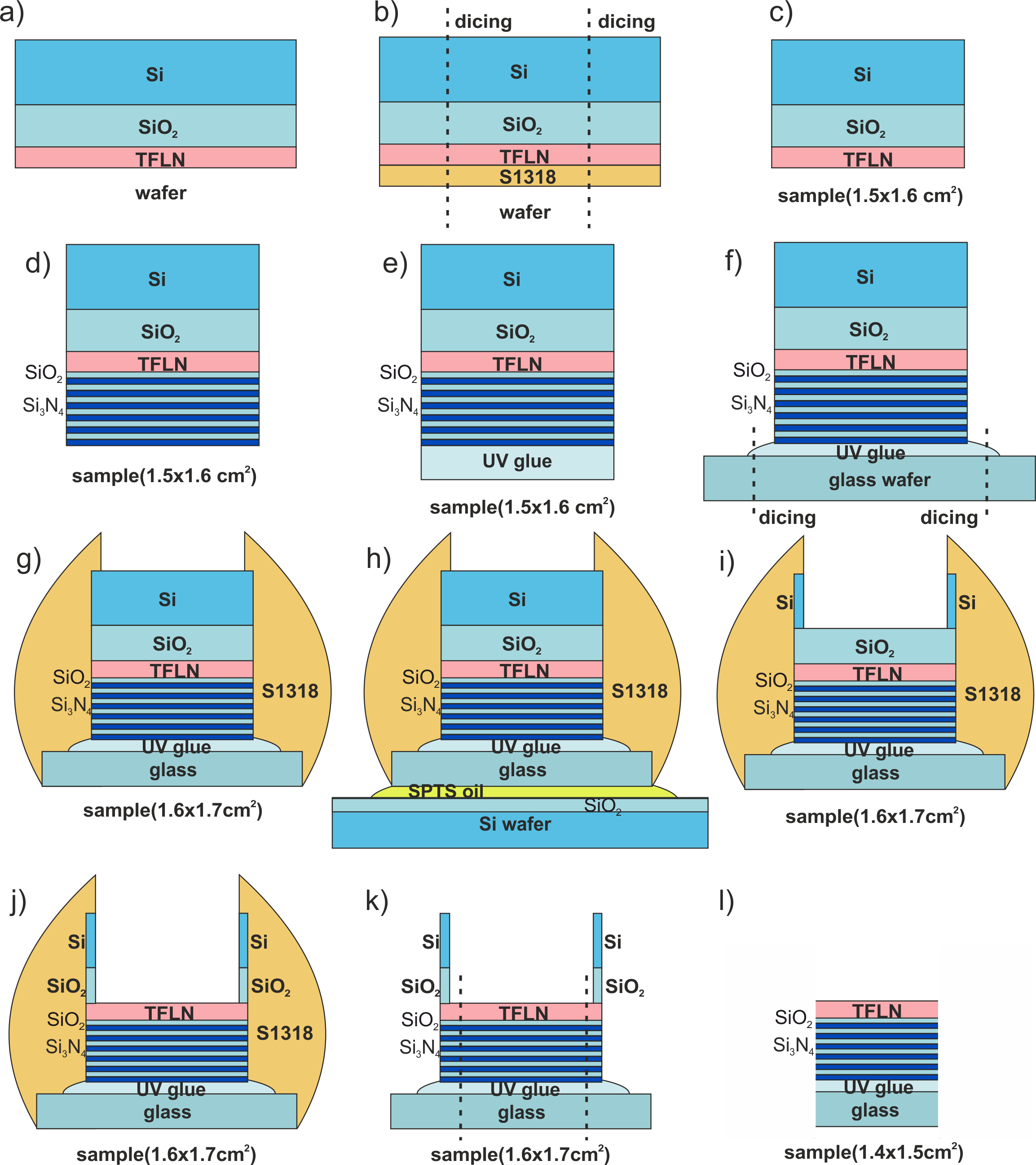}
	\caption{(a) Original TFLN wafer; (b) TFLN protection with S1318 for dicing; (c) Sample prepared for multilayer deposition; (d)  Multilayer deposition; (e) UV glue coating; (f) UV glue bonding to glass support wafer, dicing of the bonded sample out of the wafer; (g) Covering of the walls of the sample with S1318; (h) Sample positioning on the Si support wafer with SiO$_2$ top layer for DRIE process; (i) Si plasma etching; (j) SiO$_2$ wet etching in HF; (k) S1318 removal in acetone and dicing away of the Si/SiO$_2$ walls; (l) Final sample.}
	\label{prosecc}
\end{figure}

The dispersion curve for 1DPhC with this propagation conditions is shown at the Fig. \ref{fig:Fif_7_1_find_d1_d2}.  The curve is obtained for the multilayer on a BK7 glass support made out of six pairs of alternating Si$_3$N$_4$ (250 nm thick) and SiO$_2$ (450 nm thick) layers with TFLN (450 nm thick) on top. Air is considered to be an external media. Here we also take into account the fact that on the 3 inch wafer the thickness of TFLN is not totally homogeneous and it varies within 50 nm depending on the position on the wafer. with above-mentioned thicknesses of Si$_3$N$_4$ and SiO$_2$ the BSW would be still excited (for $d{\rm{_{TFLN}}}$ in the range from 440 nm to 490 nm), but at a slightly different angle with a slightly different efficiency. The dispersion curve would shift but still remain within the band gap.

The coupling of the incident wave into surface wave depends on the total extinction in the layers and can be verified by observing reflection dumping (see Chapter 2.1.3). In this part of the study we experimentally test samples with different number of Si$_3$N$_4$ and SiO$_2$ pairs. We check samples for four, five and six pairs. In our case, for TE polarized incident light reflection dips are observed at the 49$^\circ$ incident angle for all the samples [Fig. \ref{fig:Fif_7_3_refl_no}]. Refractive indices used for reflectance calculations for BK7 glass, Si$_3$N$_4$, SiO$_2$ and LiNbO$_3$ at 1550 nm wavelength are $n_{\rm{0}}$=1.5007, $n_{\rm{Si_3N_4}}$=1.790+0.001i, $n_{\rm{SiO_2}}$=1.44+0.001i and $n_{\rm{LiNbO_3}}$=2.199+0.001i respectively. Data for refractive indices of all the multilayer materials but lithium niobate was taken from refractive index database, $n_{\rm{LiNbO_3}}$ for the TFLN was obtained from ellipsometry measurements (see Chapter 5.1). Imaginary part of ($k=0.00i$) was introduced to refractive indices of the multilayer materials as losses at the boundary.  

Figure \ref{fig:Fif_7_3_refl_no} shows reflectance dips which appear due to the BSW coupling at the 49$^\circ$ in the following 1DPhC: $BK7/(Si_3N_4-250~nm/SiO_2-450~nm) \times N/ TFLN-450~nm/air$, where $N$ is number of pairs in multilayer and equals to 4, 5 and 6. Calculated reflectance values are R=0.30, R=0.040 and R=0.043 for samples with 4, 5 and 6 pairs of silicon oxide and silicon nitride respectively. Calculated field enhancement at the TFLN/air interface is T=63, T=86 and T=85 for samples with 4, 5 and 6 pairs. Here we can conclude that for all the samples we would be able to excite the BSW under the same experimental conditions but with a slightly lower coupling efficiency for the 4 pairs of Si$_3$N$_4$ and SiO$_2$.

For the samples' preparation we used a manufacturing process with additional cleaning steps described in a previous chapter. 

These final steps of multilayer fabrication are as following:

\vspace{-0.1cm}
\begin{enumerate}[{1)}]
\item We take a 3 inch 450 nm X-cut TFLN wafer from NANOLN company. TFLN is bonded to $\approx$ 400 $\mu m$ of Si by $\approx$ 2 $\mu m$ of SiO$_2$ [Fig. \ref{prosecc}(a)].
\vspace{-0.1cm}
\item We protect the TFLN surface with S1318 photo-resist on the spin coater. Spin-coating conditions: speed - 4000 turns/min; acceleration - 4000 turns/(min$\cdot$s); time - 30 s; baking on a hot plate at 120$^\circ$ during 2 min [Fig. \ref{prosecc}(b)].
\vspace{-0.1cm}
\item Then we dice a wafer into 1.5$\times$1.6 cm$^2$ pieces. The bigger side of the sample is along the $Y$ axis. Along this axis we will deposit electrodes later [Fig. \ref{prosecc}(b)].
\vspace{-0.1cm}
\item We remove the S1318 photo-resist from the samples [Fig. \ref{prosecc}(c)].
\vspace{-0.1cm}
\item Samples should be cleaned in acetone (2 min); rinsed in water; in ethanol (2 min); rinsed in water; in piranha solution (2 min); rinsed in water; dried with a nitrogen gun.
\vspace{-0.1cm}
\item Then we deposit SiO$_2$ and Si$_3$N$_4$ layers with PECVD. The deposition is done in EPFL by Dr. Myun-Sik Kim [Fig. \ref{prosecc}(d)].
\vspace{-0.1cm}
\item The UV glue VITRALIT 6127 applied on the multilayer surface with a spin coating. Spin-coating conditions: speed - 3000 turns/min; acceleration - 3000 turns/(min$\cdot$s); time - 30 s. Then the sample is flipped over and glued to any standard glass wafer, which was separately cleaned before the bonding. The UV glue is exposed with a pressure on the sample [Fig. \ref{prosecc}(e,f)].

The glass wafer cleaning is as following: cleaned in acetone (2 min); rinsed in water; in ethanol (2 min); rinsed in water; in piranha solution (2 min); rinsed in water; dried with a nitrogen gun.
\vspace{-0.1cm}
\item We dice samples of the 1.6$\times$1.7 cm$^2$ size from the glass waver [Fig. \ref{prosecc}(f)].
\vspace{-0.1cm}
\item Even though the UV glue is sensitive to aggressive chemicals, a thorough cleaning is necessary at this step: cleaning in acetone (2 min); rinsing in water; in ethanol (2 min); rinsing in water; cleaning in piranha solution (2 min); rinsing in water; drying with a nitrogen gun.
\vspace{-0.1cm}
\item We apply the S1318 photo-resist with a brush on the side walls of the sample. The photo-resist should be baked in the stove during 4 hours at 60$^\circ$C temperature. Then it should rest during 2-3 hours [Fig. \ref{prosecc}(g)].
\vspace{-0.1cm}
\item After the sample cooled down it is ready for DRIE etching of Si. We prepare a support Si waver with 2 $\mu m$ SiO$_2$ on the top. The support wafer should be cleaned: cleaning in acetone (2 min); rinsing in water; in ethanol (2 min); rinsing in water; cleaning in piranha solution (2 min); rinsing in water; drying with a nitrogen gun. The sample can be bonded to the support wafer by oil [Fig. \ref{prosecc}(h)]. 

\begin{figure}[!t]
	\centering
	\includegraphics[width=3.5in]{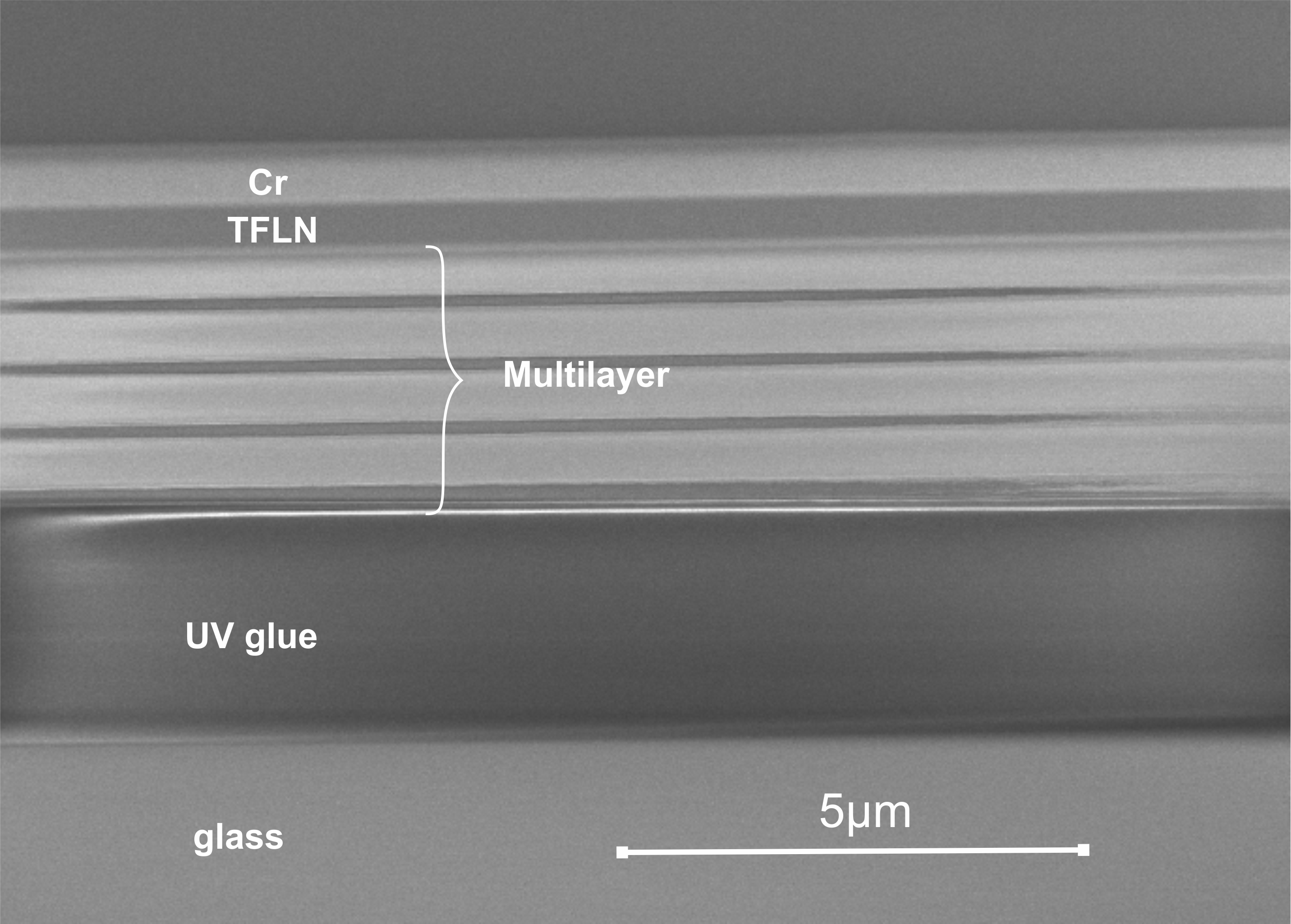}
	\caption{FIB-SEM image of 1DPhC with TFLN top layer on the glass support.}
	\label{fig:UVglass1550}
\end{figure}

Note: Very small amount of oil should be put in the center of the support wafer. Once the sample is glued to the waver minimum of the oil should remain uncovered by the sample, otherwise it would lead to contamination and micro-masking effect. However, the oil have to be everywhere under the sample, thus it would provide a good thermal conductivity between the sample and the support waver.

Taking all above listed into account top layer of Si can be removed by DRIE plasma etching.

After DRIE the sample can be detached from the support waver by Remover 1165, followed by acetone, ethanol and water rinsing [Fig. \ref{prosecc}(i)].
\vspace{-0.1cm}
\item Finally the 2 $\mu m$ of SiO$_2$ can be removed in HF (40\% solution) during $\approx$ 20 min. Then the sample should be thoroughly rinsed in water [Fig. \ref{prosecc}(j)].  

Note: As an improvement step it would be interesting do develop a dry plasma etching process of SiO$_2$ with LiNbO$_3$ as a stop layer in order to avoid HF as a dangerous chemical solution. 
\vspace{-0.1cm}
\item If any S1318 still left it can be removed in acetone [Fig. \ref{prosecc}(k)].
\vspace{-0.1cm}
\item Remained walls of Si and SiO$_2$ can be diced away [Fig. \ref{prosecc}(l)].
\end{enumerate}

The FIB-SEM image of the sample with 4 pairs of Si$_3$N$_4$ and SiO$_2$ layers and with a TFLN on the top of 1DPhC is shown at the Fig. \ref{fig:UVglass1550}. The whole stack is bonded to the glass support by UV glue. Thick layer of Cr was deposited on the top of 1DPhC in order to avoid charge effect during the FIB milling of the small opening on the sample surface. The Cr layer was removed by wet etching in the etch-Cr solution after FIB-SEM characterization.

\begin{figure}[!t]
	\centering
		\includegraphics[width=1.9in]{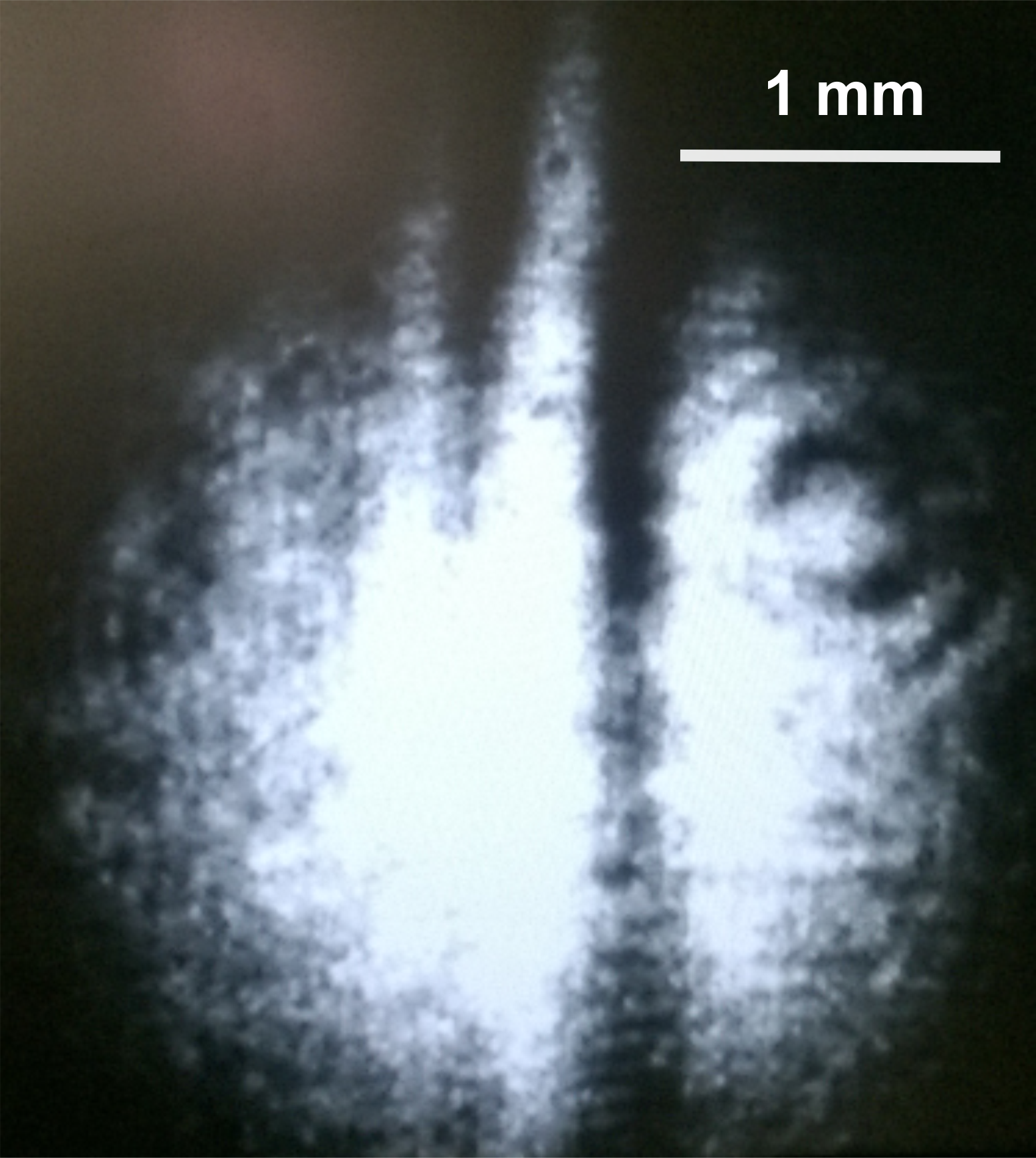}
		\caption{Image of reflected light at the BSW coupling angle collected by IR camera (the image is almost the same for the samples with 4, 5 and 6 pairs of Si$_3$N$_4$ and SiO$_2$).}
	\label{fig:Fig_7_reflectance_line.png}
\end{figure}

For all the samples with 4, 5 and 6 pairs of Si$_3$N$_4$ and SiO$_2$ BSW excitation tests on Kretschmann configuration were performed. The setup is the same as in the case for the BSW at 473 nm wavelength (see Fig. \ref{fig:Fig_6_8_setups}(b)) but the laser was changed to the one emitting light at 1550 nm. Clear dark line at the camera image of the reflected light represents the BSW coupling (see Fig. \ref{fig:Fig_7_reflectance_line.png}).

\section{BSW propagation on the top of TFLN.}

In this part of work we will provide some preliminary results about the BSW propagation on the surface of manufactured samples. On the top of 1DPhC we have milled various waveguides in order to get the estimation about BSW wave-guiding properties at 1550 nm.

Thus for preliminary tests we prepared set of 200 $\mu m$ long waveguides along both $Y$ and $Z$ crystalline directions of LiNbO$_3$: three narrow (1.7 $\mu m$) waveguides along $Z$ direction (WG. 1-3) and three narrow (1.7 $\mu m$) waveguides along $Y$ direction (WG. 4-6), one wide (12.1 $\mu m$) waveguide along $Z$ direction (WG. 9) and two wide (12.1 $\mu m$) waveguides along $Y$ direction with (WG. 8) and without (WG. 7) test 2DPhC in the middle of the waveguide.

All geometrical parameters of waveguides and 2DPhC were chosen arbitrary for preliminary tests on BSW propagation.

In this part of work we excite the BSWs at the wavelength around 1550 nm. Therefore, in order to detect the BSW propagation in the waveguide from the far field we mill a grating decouplers in the end of the waveguides (3 periods of grating). The period of grating is chosen in such a way that it also may work as a BSW coupler for the 1550 nm wavelength at the normal incidence, when the BSW propagates along Y direction ($\Lambda=1.55/1.132=1.37\mu m$ see Chapter 3.3.1). Width of the groves is equal to the half of the period, depth of the droves is about 530 nm. Therefore we manufactured some waveguides with 10 periods of grating in the end for potential BSW grating coupling tests.

The waveguides themselves were isolated from the surrounding 1DPhC surface by $2-3~\mu m$ wide grooves.

\begin{figure}[!t]
	\centering
	\includegraphics[width=4.6in]{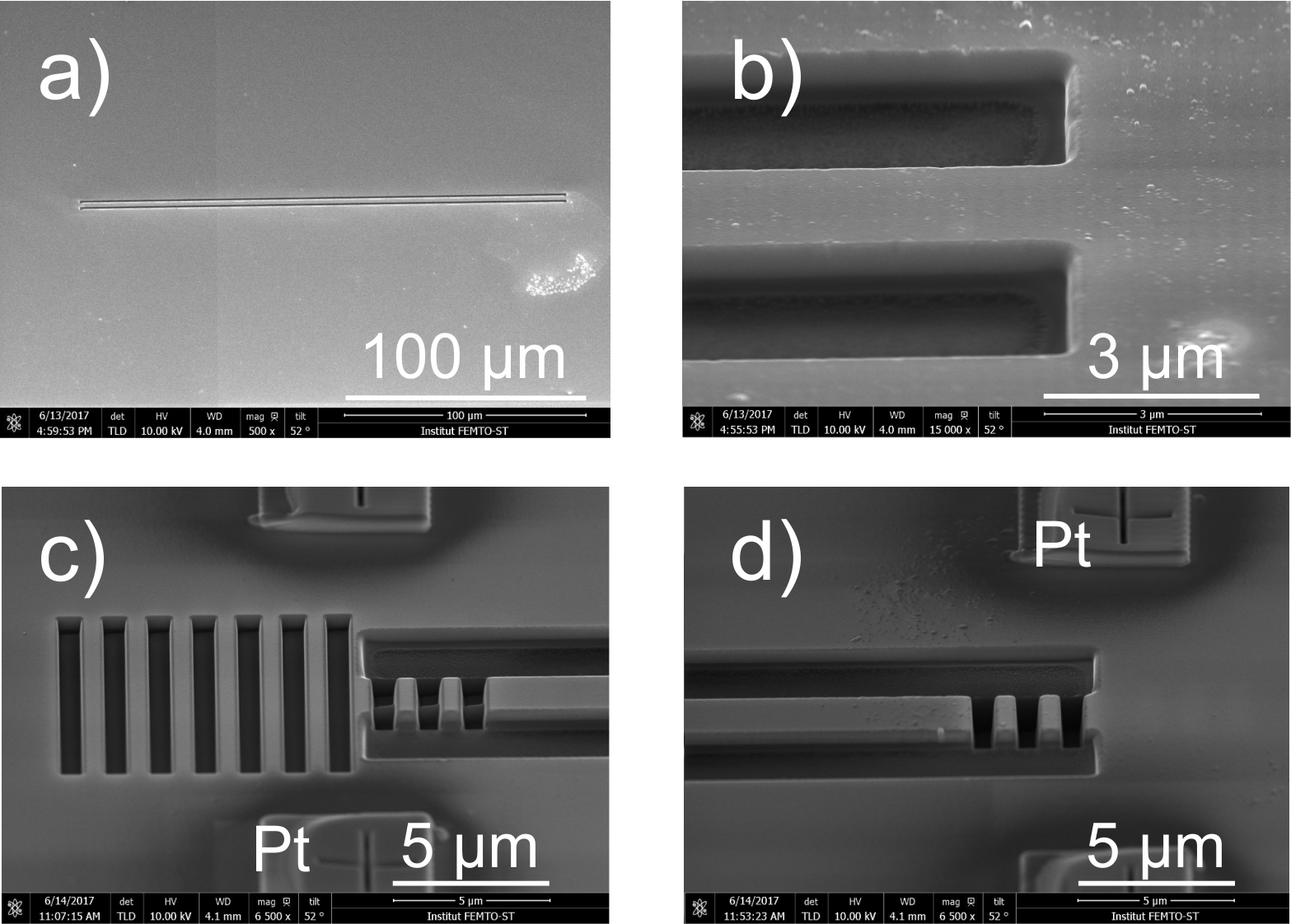}
	\caption{(a) FIB-SEM image of the narrow waveguide. Variants of the narrow waveguide endings: (b) free end, (c) BSW grating coupler / decoupler made out of 10 periods (7 long and 3 short) and (d) grating decoupler (3 periods of short grooves).}
	\label{fig:Fig_7_6_smal_wg_fib}
\end{figure}  

\begin{figure}[!b]
	\centering
	\includegraphics[width=3in]{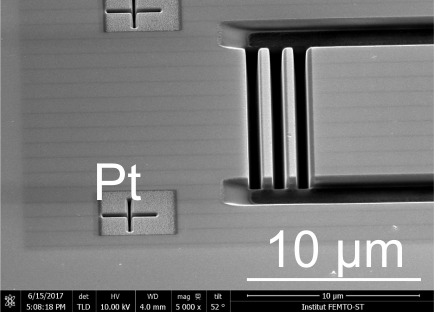}
	\caption{FIB-SEM image of the multi-mode waveguide with a grating decoupler (3 periods of short grooves).}
	\label{fig:Fig_7_6_big_wg_fib}
\end{figure}  

All the nano-structuring of the top surface of 1DPhC with TFLN was done by FIB milling at FEMTO-ST by Gwenn Ulliac. As usually an additional 15 nm thin layer of Al was deposited on the sample made out of nine layers: 4 pairs of silicon nitride and silicon oxide and TFLN.
Al layer was removed by wet etching after the nano-structuring was complete. 

Figure \ref{fig:Fig_7_6_smal_wg_fib} shows the $200 \mu m$ long waveguide of  $1.7 \mu m$ width. The waveguide is created by milling of two  $200 \mu m$ long grooves with $2~\mu m$ width and $530~nm$ depth. The endings of the waveguides have several different configurations: a free end [Fig. \ref{fig:Fig_7_6_smal_wg_fib}(b)] - both ends in WG 1 and WG 4; three period grating decoupler [Fig. \ref{fig:Fig_7_6_smal_wg_fib}(c)] - in WG 2, WG 3 and WG 5; ten period grating decoupler / coupler [Fig. \ref{fig:Fig_7_6_smal_wg_fib}(d)] - in WG 6.

On the Fig. \ref{fig:Fig_7_6_big_wg_fib} the $12.1~\mu m$ wide waveguide is shown. The waveguide is created by milling of two  $200 \mu m$ long grooves with $2~\mu m$ width and $530~nm$ depth. In the end of the WG we make a decoupler with 3 periods of grating - WG 7 and WG 9.

 \begin{figure}[!t]
 	\centering
 		\includegraphics[width=4.6in]{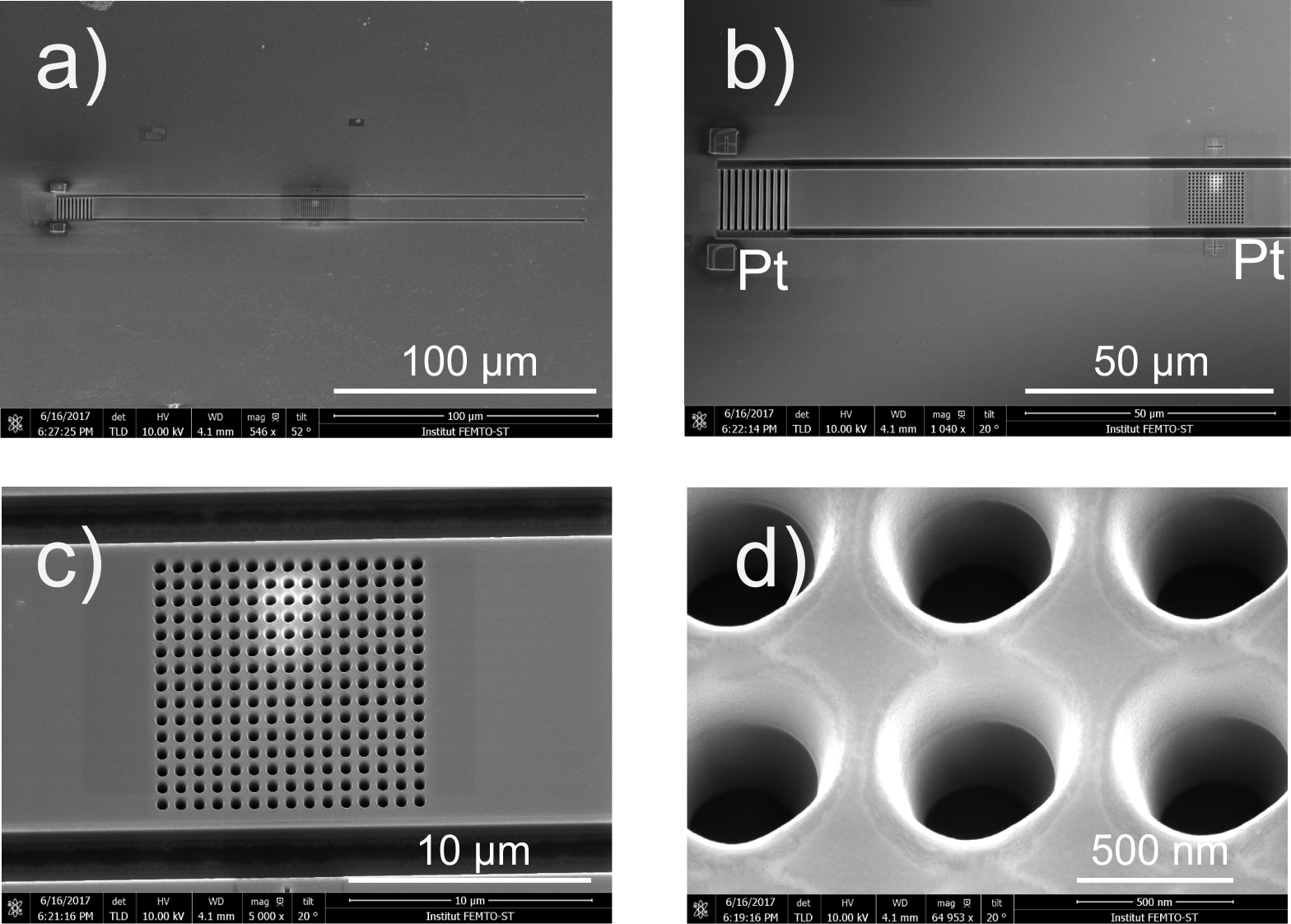}
 		\caption{(a) FIB-SEM image of the multi-mode waveguide with 2DPhC; (b) Pt alignment and shift control marks for the grating and 2DPhC milling; (c) 15$\times$15 rows 2DPhC; (d) Conicity of the walls introduced by FIB milling during 2DPhC manufacturing.}
 	\label{fig:Fig_7_6_big_wg_fib_2DPhC}
 \end{figure} 

Finally we mill one waveguide (WG 8) along $Y$ axis of LiNbO$_3$ with a 2DPhC in the middle [Fig. \ref{fig:Fig_7_6_big_wg_fib}]. The waveguide is created by milling of two  $200 \mu m$ long grooves with $2~\mu m$ width and $530~nm$ depth. The ending of the waveguide has 10 periods of the grating for BSW decoupling [Fig. \ref{fig:Fig_7_6_big_wg_fib} (a,b)]. 2DPhC with 15 rows of holes with periodicity $a$=740 nm  and filling ratio $r/a$=0.3 ($r$ is a hole radius and equals to 222 nm) is milled in the center of the WG 8. Depth of the holes is $\approx 530~ mn$Figure \ref{fig:Fig_7_6_big_wg_fib_2DPhC}(d) shows a top view of the 2DPhC. The 2DPhC was fabricated in order to check potential problems, which may appear during fabrication. Indeed, here we can see that the top part of the holes is a bit digger and has a conical shape, what may bring some mismatch between theoretical predictions and experimental results in the 2DPhC behavior.

It is also important to mention that additional small squares of Pt [Fig. \ref{fig:Fig_7_6_smal_wg_fib}(c,d), Fig. \ref{fig:Fig_7_6_big_wg_fib} and Fig. \ref{fig:Fig_7_6_big_wg_fib_2DPhC}(a,b)] were deposited in-situ next to the gratings and next to the 2DPhC for alignment purposes and also as a control layer according to which the shift adjustment is done (this prevents from the shift of focused ion beam during the milling).

\begin{figure}[!t]
 	\centering
 		\includegraphics[width=\linewidth]{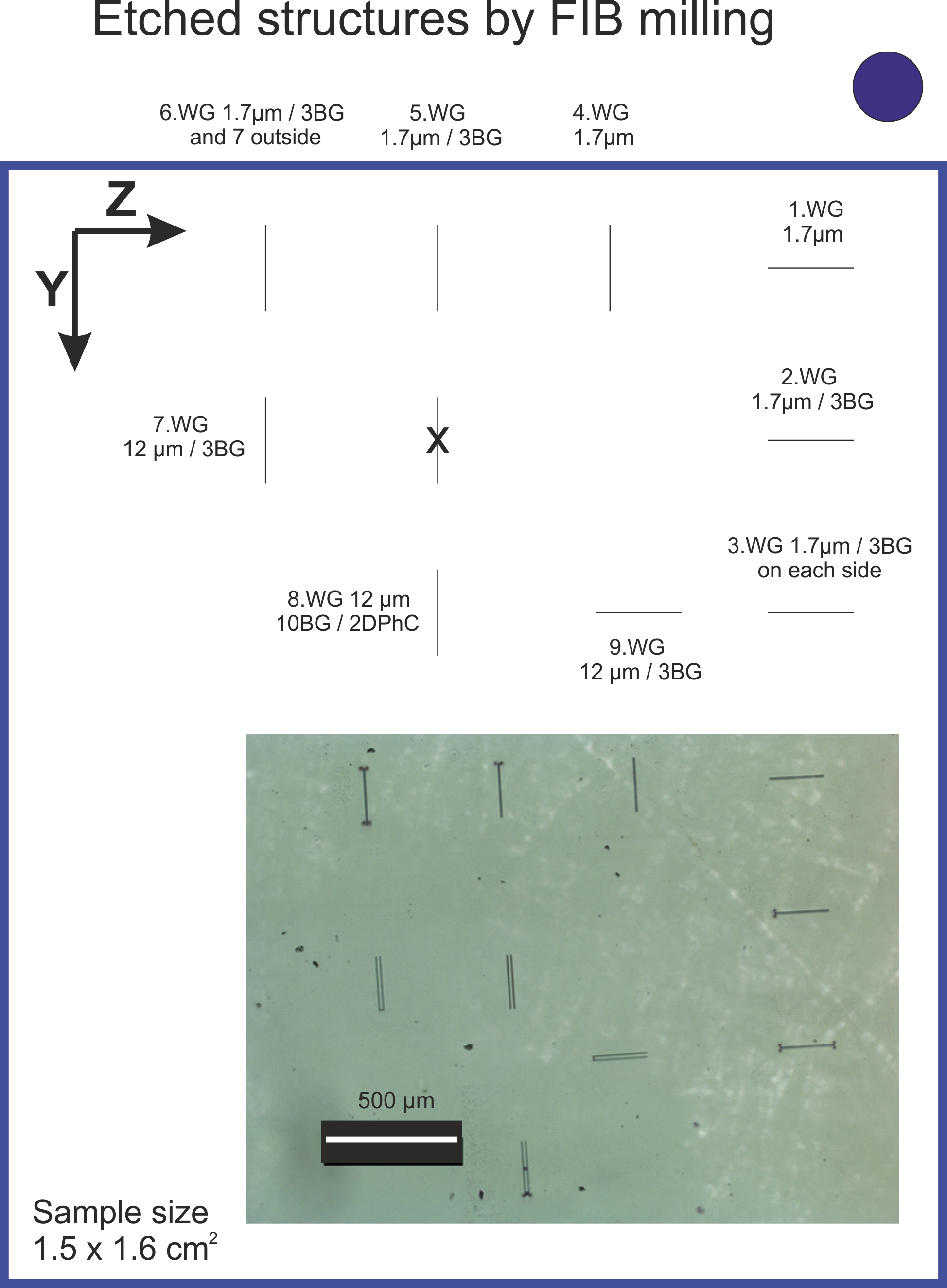}
		\caption{Sketch of the waveguides positions on the sample surface and the microscope image of fabricated structures.}
 	\label{fig:Fig_7_5_wg_pozitions_on_LN1}
 \end{figure}

All the waveguides were milled on the sample as it is shown at the Fig. \ref{fig:Fig_7_5_wg_pozitions_on_LN1}. The distance between any of the waveguides is 0.5 mm in $Y$ and $Z$ directions. The blue bot in the right top corner is an indicator of the corner where the WGs are milled. 

\section{Experimental.}

\begin{figure}[!t]
	\centering
	\includegraphics[width=3in]{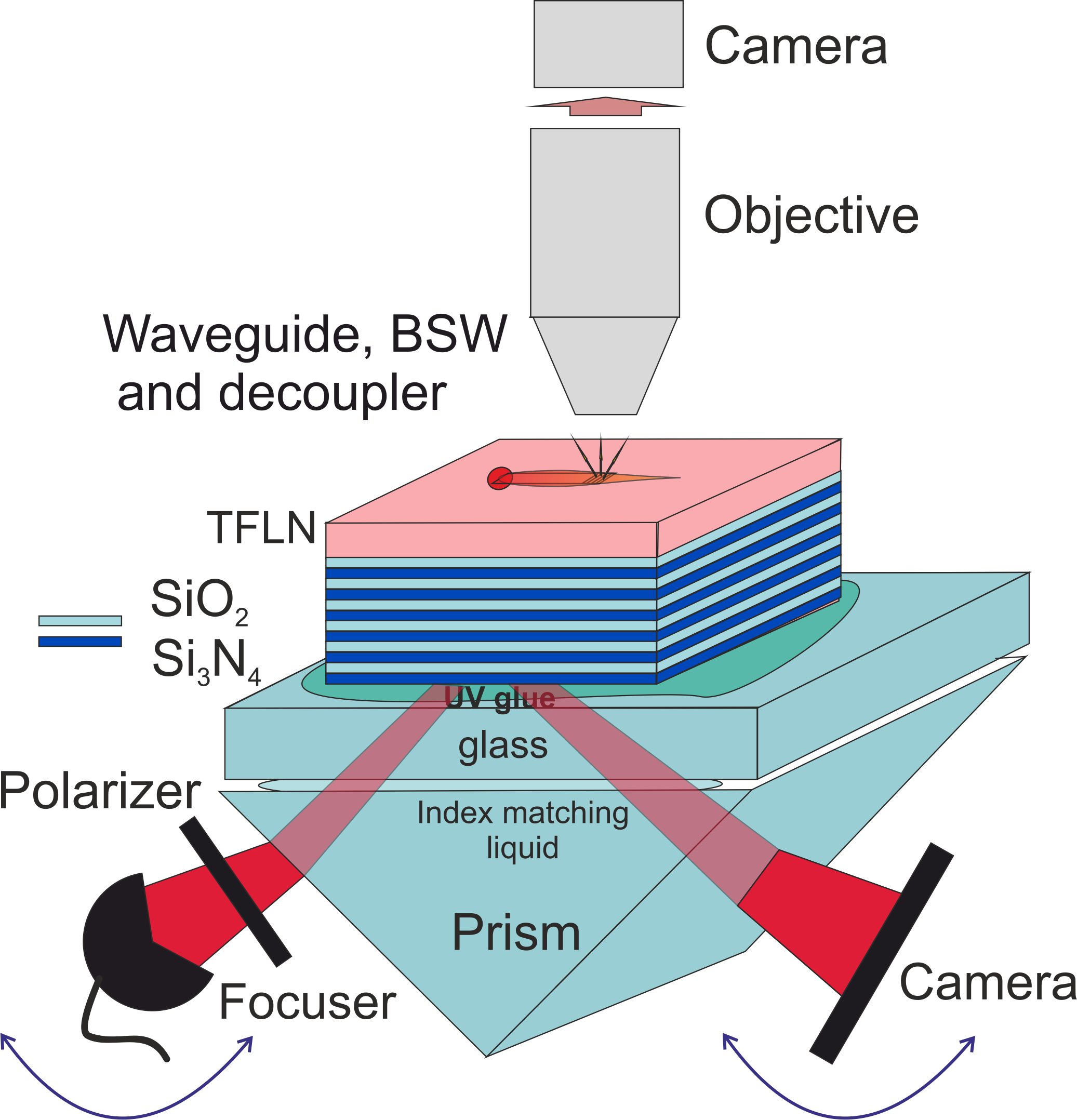}
	\caption{Experimental setup for the wave-guiding characterization.}
	\label{fig:Fig_7_7_setup1.png}
\end{figure}

  \begin{figure}[!b]
	\centering
	\includegraphics[width=3in]{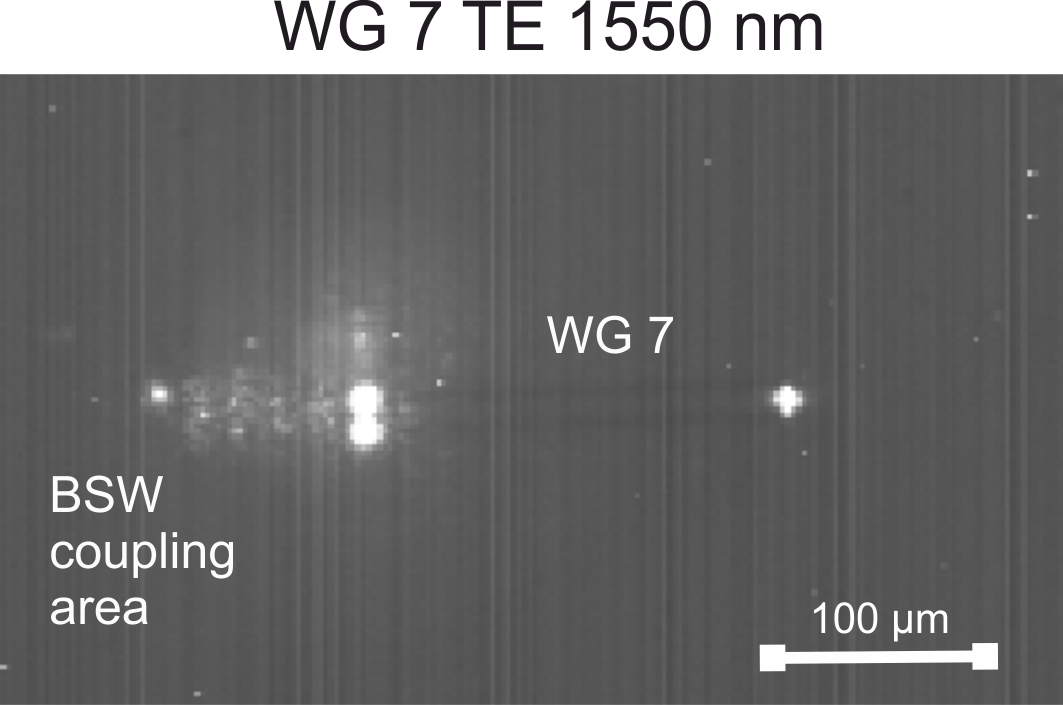}
	\caption{IR camera image collected by x20 objective for the BSW propagation on WG 7 at 1550 nm.}
	\label{fig:Fig_7_big_wg_BSW}
\end{figure}  

\begin{figure}[!b]
	\centering
	\includegraphics[width=\linewidth]{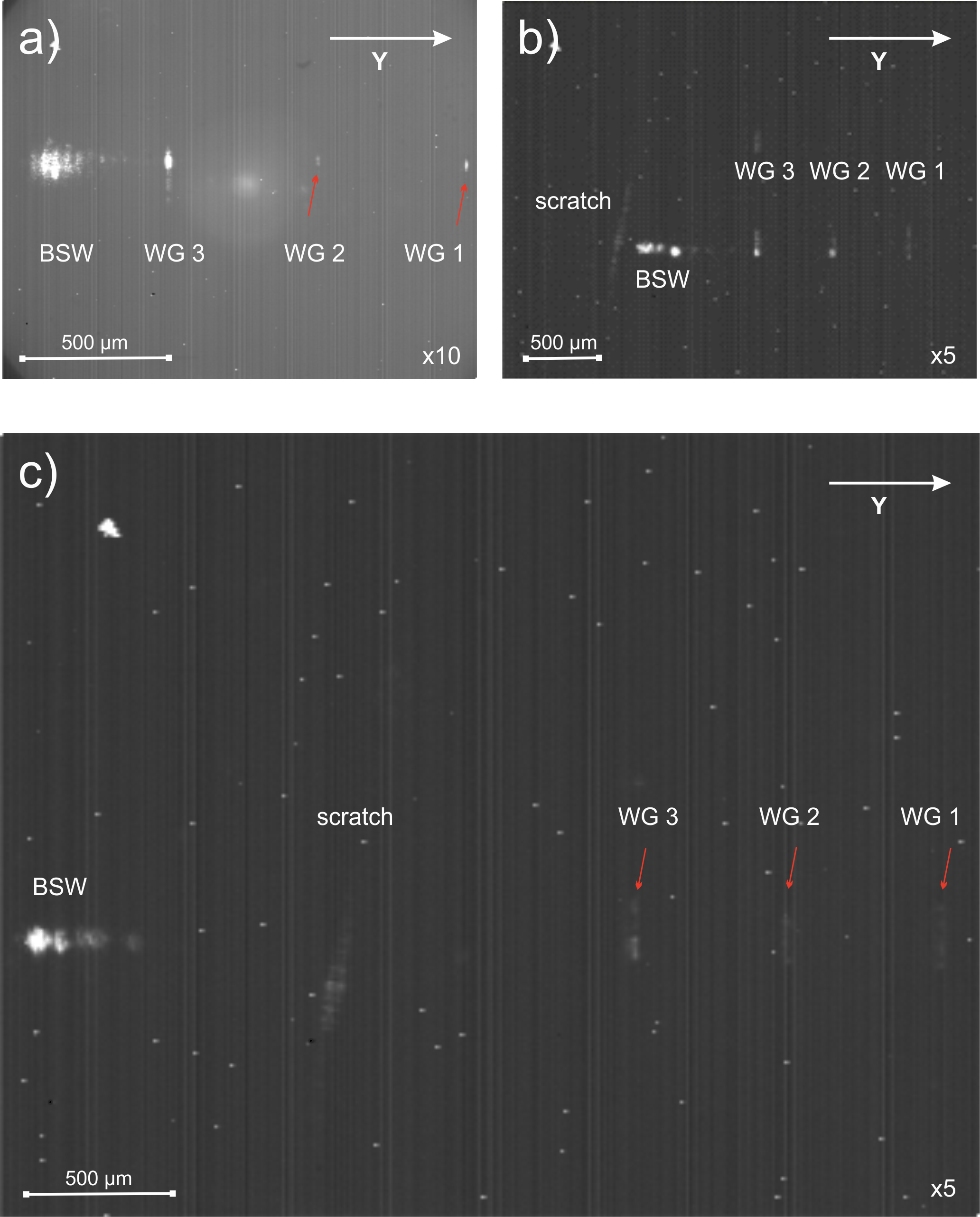}
	\caption{(a) Camera image collected by x10 objective indicating the BSW decoupling on WG 1, WG 2 and WG 3; (b) Camera image collected by x5 objective indicating the BSW decoupling on WG 1, WG 2 and WG 3; (c) Camera image collected by x5 objective indicating the BSW decoupling on WG 1, WG 2, WG 3 and the scratch indicating maximum observed BSW propagation distance from the far field.}
	\label{fig:Fig_7_12_propagarion_length}
\end{figure} 

For the experimental characterization of the waveguides the Kretschmann configuration was used see Fig. \ref{fig:Fig_7_7_setup1.png}. Light from tunable laser (1480-1570nm) was focused on the sample through the prism (10-30 $\mu m$ diameter on the surface of 1DPhC). Polarization of the incident beam was controlled by a Glan polarizer. Reflected light was collected by IR camera (camIR 1550, applied scintillation) in order to verify the presence of the BSW (when the reflectance dip is observed). The BSW excited at the TFLN surface propagates along the waveguide and decouples on the grooves in the end of the waveguide. Decoupled light is collected by the long working distance objective (magnification 5,10,20 and 50) and analyzed by another IR camera (Xenix XEVA-2232). The experiments were held in collaboration with the optics and photonics technology lab in EPFL.

We focus our study on the BSW propagation along $Y$ axis.

Clear wave-guiding for the 12 $\mu m$ wide WG 7 is observed [Fig. \ref{fig:Fig_7_big_wg_BSW}]. The BSW was excited about 50 $\mu m$ away from the WG. When it reaches the grooves which define the WG, it scatters (2 bright spots in the beginning of the WG). Still some part of the BSW propagates along the WG and decouples on the grooves in the end of the WG (bright spot on the left part of Fig. \ref{fig:Fig_7_big_wg_BSW}).

Additionally during wave-guiding tests we estimated the propagation length of the BSW in the designed 1DPhC with 4 pairs of alternating silicon nitride and silicon oxide and with 450 nm TFLN on the top.

In our case if we launch the BSW along the $Y$ axis in such a way that WG 1, WG 2 and WG 3 would work as an obstacles (see Fig. \ref{fig:Fig_7_12_propagarion_length}) we can see that the BSW partially decouples on each of of them and keeps propagating. Thus knowing that the distance between the waveguides, which is 0.5mm, we may estimate the propagation length of the  surface wave.

Firstly we obtained the Fig. \ref{fig:Fig_7_12_propagarion_length}(a) from which we may conclude that that our BSW excited at 1550nm wavelength can propagate over the distance more then 1.5 mm. 

As a next step we change the objective from magnification 10 to magnification 5 and move the BSW further away from the waveguides [Fig. \ref{fig:Fig_7_12_propagarion_length}(b)].
Finally we reach a point when the BSW crosses the scratch on the surface of 1DPhC. Th surface wave partially decouples on the scratch and keep propagating till the WG 1. Finally we reach the place when the field of view of the objective is not enough for further measurements [Fig. \ref{fig:Fig_7_12_propagarion_length}(c)]. From this tests we can conclude that our 1DPhC supports the BSW with the propagation length more than 3 mm. This is a very impressive result for the surface wave which is obtained by a simple microscopy.

Also here we experimentally demonstrate the possibility of BSW to propagate through the obstacles, such as scratches, grooves and other inhomogeneities of the surface. This feature may be found interesting in integrated optics applications in terms of robustness of propagation. 

Thus we can sum up the obtained results in this section. We have definitely obtained the propagation of BSW in our multi-mode waveguides. Additionally, manufactured 1DPhC sustains a BSW with an impressive propagation length.

\section{Electrodes deposition.}

As it was mentioned before, in order to exploit electro-optical properties of 1DPhC with LN on top it is necessary to develop a special electrodes deposition recipe.

Steps of the electrodes deposition on the top of 1DPhC are as following:

\vspace{-0.1cm}
\begin{enumerate}[{1)}]
\item The sample surface should be absolutely flat in order to have a good contact between mask and the sample during lithography steps. Generally the edges or the samples prepared as it previously described (Chapter 7.1) have remained 400 $\mu m$ Si walls. The edges of the sample are being coved by photoresist and not etched by plasma during DRIE step. Therefore we dice out central part of the sample, where there is only TFLN on top. 
Before dicing we protect the TFLN surface with S1813 photoresist. Spin-coating conditions: speed - 4000 turns/min; acceleration - 4000 turns/(min$\cdot$s); time - 30 s; baking on a hot plate at 120$^\circ$ during 2 min.
\vspace{-0.1cm}
\item Then we clean the sample and the mask. The mask is cleaned for 5min in acetone, then rinsed in water; cleaned in ethanol for 5 min, thoroughly rinsed in water and cleaned in piranha solution during 2 min. It should be rinsed in water and  dried with a nitrogen gun.

The sample meanwhile can mot be cleaned with piranha. The cleaning process consists of 5 min in acetone; rinsing in water; 5 min in ethanol; rinsing in water; drying with an nitrogen gun. 
\vspace{-0.1cm}
\item Once the sample is cleaned we deposit 40 nm of Cr and 160 nm of Au on the top of 1DPhC by sputtering.
\vspace{-0.1cm}
\item Then we use a positive photoresist for lithography steps.

Spin-coating of Ti prime conditions: speed - 3000 turns/min; acceleration - 3000 turns/(min$\cdot$s); time - 20 s; baking on a hot plate at 120$^\circ$ during 2 min.

Spin-coating of S1813 conditions: speed - 4000 turns/min; acceleration - 4000 turns/(min$\cdot$s); time - 30 s; baking on a hot plate at 120$^\circ$ during 2 min. deposit electrodes later.
\vspace{-0.1cm}
\item Then we expose the resist through the electrodes mask. Dose = 55 mJ$\cdot$cm$^{-2}$. Resist is developed by MP-26-A developer during 40 s. Then we rinse the sample in water for 5 min and proceed to the final baking of resist on a hot plate at 120$^\circ$ during 2 min. 
\vspace{-0.1cm}
\item Then we wet etch Au during 1 min and Cr during 15 s. 
\vspace{-0.1cm}
\item Remained resist is removed during the final cleaning steps: acetone (5 min), ethanol (5 min), water rinsing, drying with the nitrogen gun.
\end{enumerate}

% \begin{figure}[!ht]
% 	\centering
% 		\includegraphics[width=4in]{Fig_7_electrodes_resipe.png}
% 		\caption{(a) Refractive index of LiNbO$_3$ for different sample positions; (b) Experimental results and theoretical dependance of the BSW excitation angle}
% 	\label{fig:Fig_7_electrodes_resipe}
% \end{figure}  

\begin{figure}[!t]
	\centering
		\includegraphics[width=3.8in]{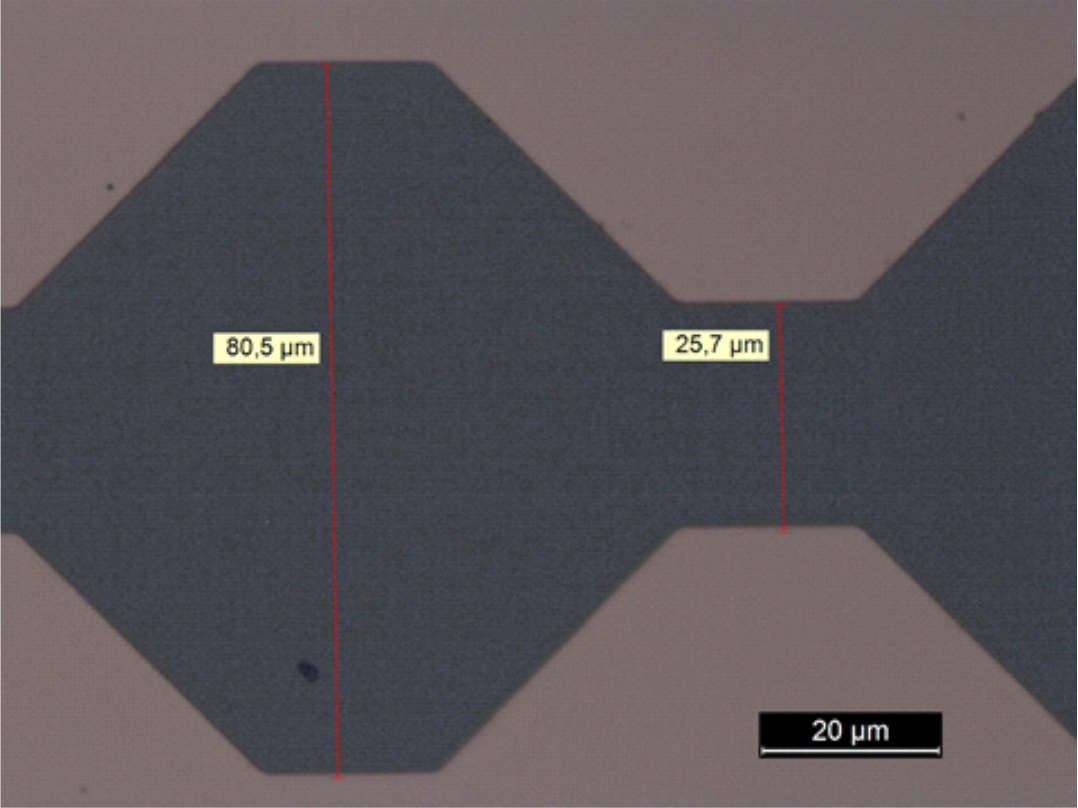}
		\caption{Narrowing part of the mask for electrodes deposition.}
	\label{fig:Fig_7_electrodes_mask_and_dummy_sample}
\end{figure}

The mask for the electrodes deposition contained 4 lines of 1 mm wide electrodes the distance between the electrodes on the mask is 80 $\mu m$. At several positions we create a narrowing down to 25 $\mu m$ between 2 electrodes (see Fig. \ref{fig:Fig_7_electrodes_mask_and_dummy_sample}). This narrow area is a place for the 2D PhC. Following the recipe of electrodes deposition the real distance between the electrodes is slightly bigger: 33$\mu m$ for the narrow area and 88 $\mu m$ for the wide area. 

\section{Perspectives.}

Additionally to fabrication technology development, we performed some preliminary deice characterization tests on BSW excited at 1550 nm on the TFLN/air interface. The BSW was coupled into 12 $\mu m$ wide waveguide, propagated through test 2DPhC crystal and decoupled at the grating in the end of the waveguide. The waveguide was milled by FIB between two Cr/Au electrodes.

\begin{figure}[!b]
	\centering
	\includegraphics[width=5.2in]{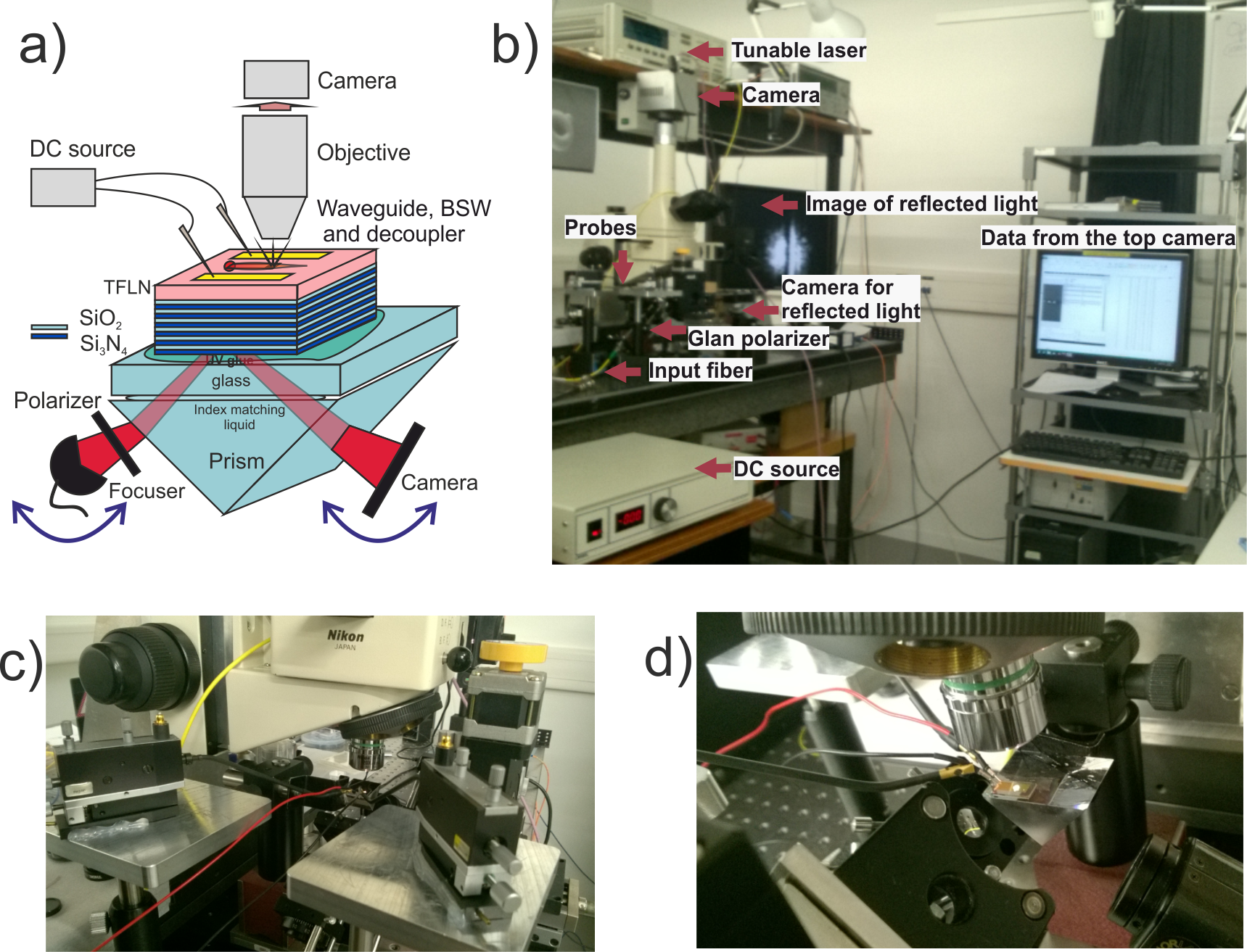}
	\caption{(a) Schema of experimental setup for sample characterization; (b) Optical table with the setup; (c) and (d) detailed image of the probes contact with the sample.}
	\label{fig:Fig_7_electrodes_setuo}
\end{figure}

The setup for the electro-optical tests on BSW is shown at the Fig. \ref{fig:Fig_7_electrodes_setuo}. Figure \ref{fig:Fig_7_electrodes_setuo}(a) illustrates the sketch of the setup with added DC source and probes. The experiments were held in collaboration with the optics and photonics technology lab in EPFL. 

Light from a tunable laser (1480-1570nm) was focused on the sample through the prism (15 $\mu m$ diameter on the surface of 1DPhC). Polarization of the incident beam was controlled by a Glan polarizer. Reflected light was collected by IR camera (camIR 1550, applied scintillation) in order to verify the presence of the BSW (when the reflectance dip is observed). The BSW excited at the TFLN surface propagates along the waveguide and decouples on the grooves in the end of the waveguide. Decoupled light is collected by the long working distance objective (magnification 20) and analyzed by another IR camera (Xenix XEVA-2232). Electric probes were fixed to the arm of micro-manipulator and connected to the DC source, which provided potential on electrodes for Pockels effect generation in LiNbO$_3$.

Figure \ref{fig:Fig_7_electrodes_setuo}(b) shows a general look of the setup, Fig. \ref{fig:Fig_7_electrodes_setuo}(c) and (d) shows a closer look on the probes integrated into the setup.

\begin{figure}[!t]
	\centering
		\includegraphics[width=5.8in]{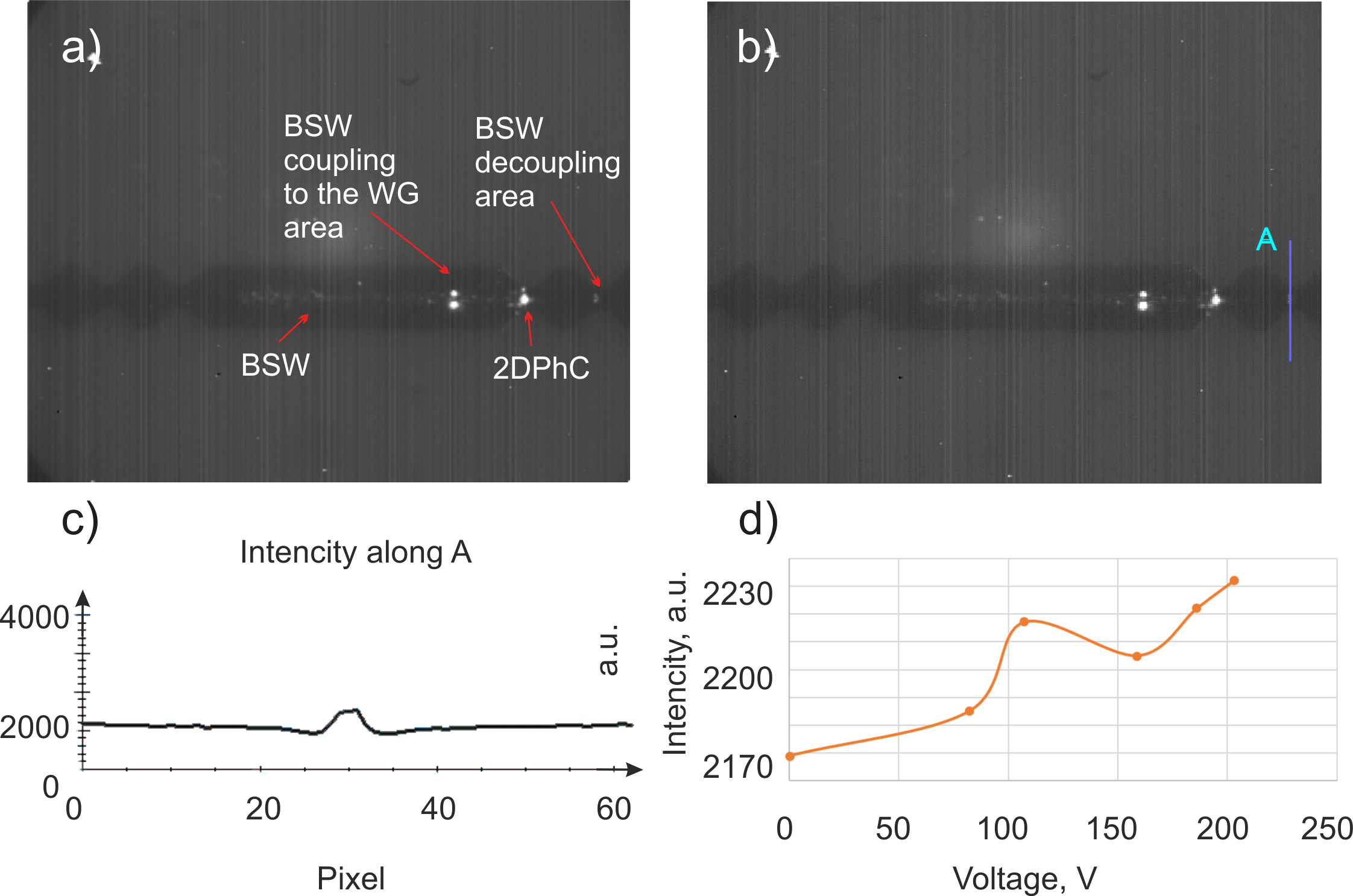}
		\caption{(a) Camera image of the BSW propagating through the waveguide with 2DPhC in the middle between the electrodes; (b) Cross-section line at the decoupler (line A); (c) Intensity profile along the line A; (d) Plot of the intensity of decoupled light with respect to applied voltage.}
	\label{fig:Fig_7_electrodes_with2DPhC_tuning}
\end{figure}

In order to verify whether it is indeed possible to modulate the amount of BSW light transmitted through the 2DPhC by external field application, we first of all excite the BSW on the TFLN area between 2 electrodes. We excite the BSW at 1550 nm some distance away from the waveguide. It propagates freely for few hundreds of nanometers and then couples into the waveguide. We can observe that some light decouples at the beginning of the WG on the edges of the grooves, which define the WG (2 bright dots at the Fig. \ref{fig:Fig_7_electrodes_with2DPhC_tuning}(a) and (b)). The BSW confined within the WG then skaters on the 2DPhC (single bright dot at the Fig. \ref{fig:Fig_7_electrodes_with2DPhC_tuning}(a) and (b)). Still some light is being transmitted through the 2DPhC and reaches the decoupler in the end of the waveguide. We measure the intensity of the light scattered from the decoupler. We set a cross-section line along which this intensity can be measured (see Fig. \ref{fig:Fig_7_electrodes_with2DPhC_tuning}(b)). As we can see from the Fig. \ref{fig:Fig_7_electrodes_with2DPhC_tuning}(c) a clear intensity pick at the decoupling area can be observed.

Then we have increased the voltage. For 5 different voltage values we make 5 measurements. The average between them becomes an intensity value which we use to plot the intensity on voltage dependence curve (see Fig. \ref{fig:Fig_7_electrodes_with2DPhC_tuning}(d)). From this curve the modulation of 3$\%$ is observed. 

\begin{figure}[!ht]
	\centering
		\includegraphics[width=2.5in]{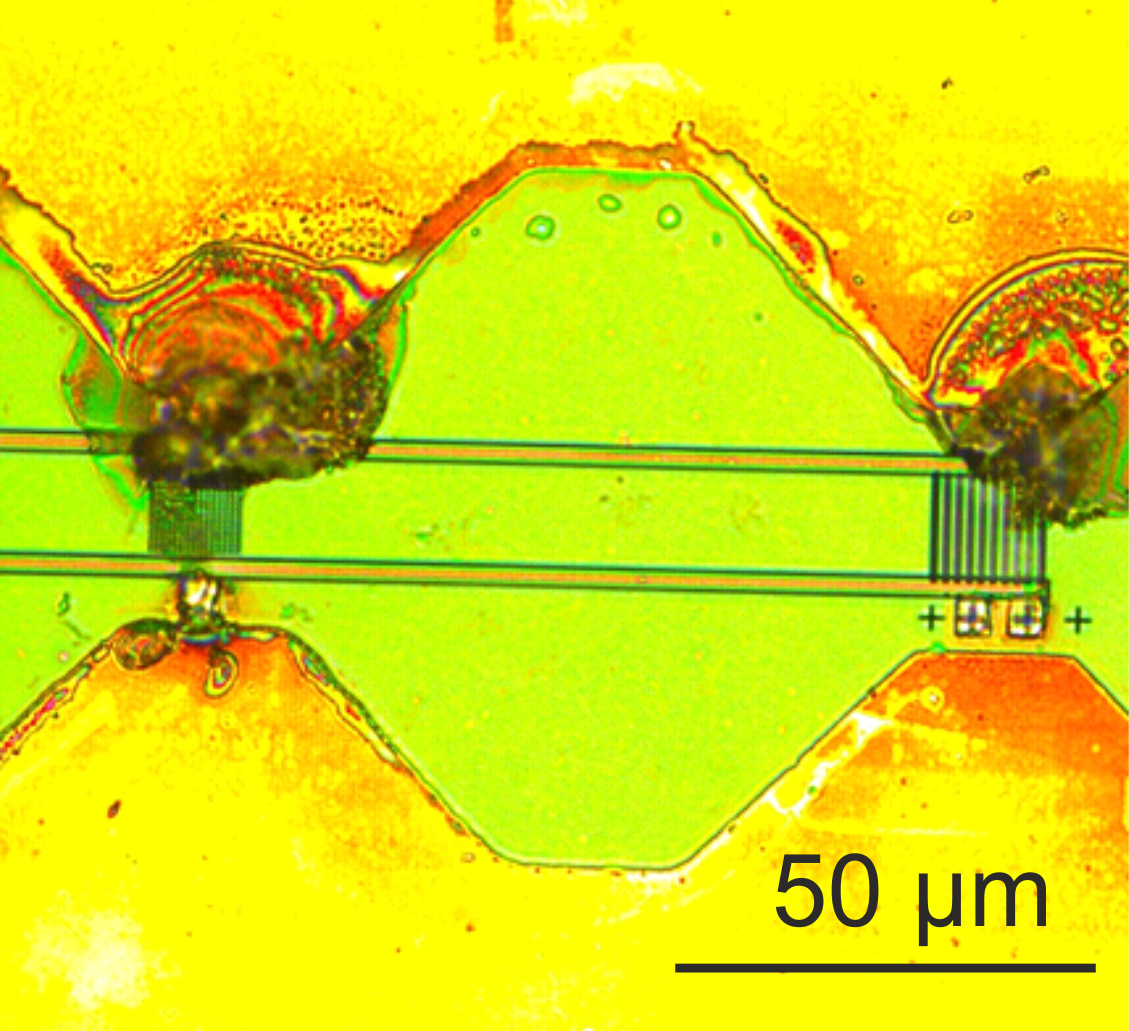}
		\caption{Microscope image of the sample with 2DPhC after voltage application.}
	\label{fig:Fig_7_electrodes_with2DPhC_burned}
\end{figure}  

Unfortunately we were not able to reproduce achieved modulation values. The IR camera image does not provide enough resolution for clear surface observation and when we checked the WG after the measurements under the microscope (see Fig. \ref{fig:Fig_7_electrodes_with2DPhC_burned}), we saw that some parts of the electrodes got burned (even though we did not go over the allowed voltage value (250 V) verified during the preliminary tests). 

Additional tests on electro-optical tunability of BSWs are one of the objectives for our future work. Also, in perspective, nonlinear properties of BSWs may be explored (especially now when it's known how to operate with LN as a part of the multilayer).

In this work we consider several different oxides for 1DPhC fabrication, though many other materials can be used. For example electro-optical polymers or VO$_2$ (unique material, which performs a transformation from metal to dielectric and vice-versa by application of temperature or current \cite{Cueff:15}). 

\begin{figure}[!b]
	\centering
	\includegraphics[width=5in]{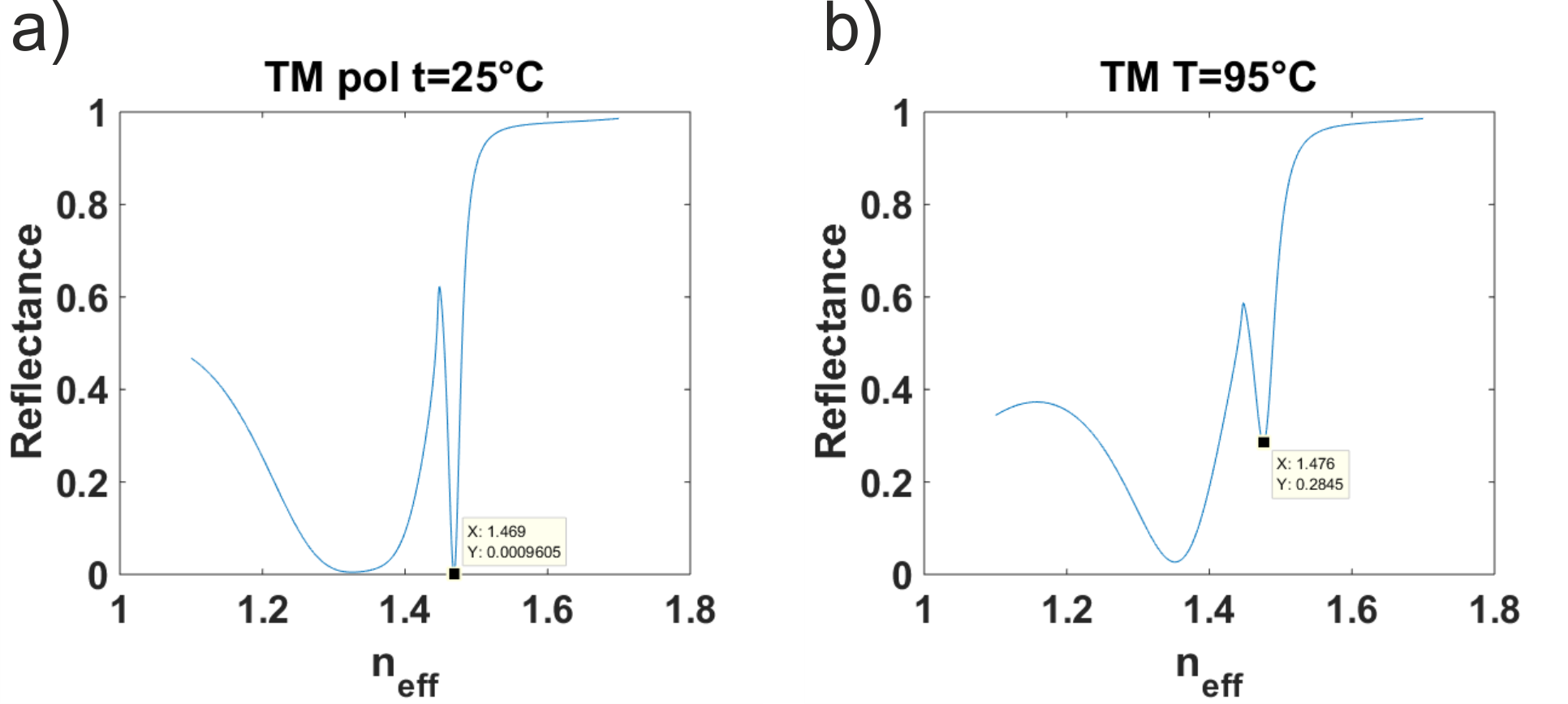}
	\caption{(a) Reflectance for the multilayer with VO$_2$ at 25$^\circ$C temperature; (b) Reflectance for the multilayer with VO$_2$ at 95$^\circ$C temperature.}
	\label{concludeVO2}
\end{figure}

As preliminary tests we modeled a 1DPHC made out of 2 pairs of silicon oxide (500 nm) and amorphous silicon (400 nm) with VO$_2$ (50 nm) on top. SF11 prism was considered as an incident media and SiO$_2$ was considered as an external media. Such a multilayer stack performs a surface at 1314 nm wavelength for TM polarization. With a temperature switch from 25$^\circ$C to 95$^\circ$C we may observe the modulation in reflectance of 20 $\%$ [Fig. \ref{concludeVO2}].

These results are just a quick estimation of possible modulation in 1DPhCs with VO$_2$ as a part of the multilayer and can be a starting point for further exploration of this material for BSW applications.

\section{Conclusion.}

In this chapter we have introduced the concept of electro-optical tuning of the BSW, which propagates on the top of 1DPhC with TFLN on top. 

On the way to the concept realization we have designed and manufactured a 1DPhC which contains the single-crystalline X-cut LiNbO$_3$. The 1DPhC supports a BSW at 1550 nm with a remarkably long propagation length (over 3 mm). Also we indicated a capability of the BSW to propagate through the obstacles on the surface, such as scratches or grooves.

We investigate the BSW propagation properties through 12 $\mu m$ wide waveguides and develop an electrodes deposition technology for the 1DPhC bonded to the glass substrate by UV glue. 

All these manufacturing and characterization steps are important for further investigation of BSW based active devices. Additionally, we demonstrated first results on BSW modulation by external electric field and described perspectives for future work.

\chapter*{Conclusion.}

In this thesis, we mainly focused on developing various tunable BSW based devices. We have proposed several ways to achieve the tunability and designed various 1DPhCs which are able to sustain the BSW at different wavelengths and in a different external media.

1DPhCs were designed and manufactured with BSW excited at the interface between a multilayer and air, water and silicon oxide. We generated BSWs in visible and near infra-red parts of spectrum at 473 nm, 633 nm , 808 nm and 1550 nm wavelengths.

Various materials prepared with different fabrication techniques were used for multilayer stacks. We studied samples, which contain silicon nitride and silicon oxide, aluminum oxide and titanium oxide as well as crystalline X-cut lithium niobate. SiO$_2$ and Si$_3$N$_4$ were deposited by plasma enhanced chemical vapor deposition, Al$_2$O$_3$ and TiO$_2$ were prepared by atomic layer deposition, LiNbO$_3$ thin films were fabricated with a smart-cut technology or by polishing bulk LN waver down to micrometer thickness.

We have demonstrated, designed and fabricated a grating coupler on a BSW sustaining platform for integrated optics applications. A special cross-grating configuration of the BSW launcher allows us to demonstrate directionality of the BSW propagation depending on polarization of the incident light. We were capable to achieve a complete switching from one polarization (P1) to another (P2) and therefore a complete switching from one direction of propagation to the orthogonal one.  

We also studied BSWs excited on the grating coupler which is applied for particles manipulation. We have demonstrated the latex beads interaction with BSW in water and shown the feasibility to use a cross-grating coupler for lab-on-a-chip trapping devices.

Also we have investigated BSWs on an anisotropic platform. LN based photonic crystal, which is able to sustain BSWs was designed and fabricated. We have proved that anisotropy of one of the layers in 1DPhC leads to anisotropy of the whole structure. Thus we have achieved the tunadility of the BSW coupling angle depending on the orientation of 1DPhC. For the first time we have studied this effect on the multilayer made out of Al$_2$O$_3$ and TiO$_2$ with TFLN, when the BSW was propagating on the TFLN/SiO$_2$ interface.

As a next step, we have shown theoretically and experimentally the excitation of BSWs at a TFLN/air interface, what introduces all the potentialities of the LiNbO$_3$ functionalities through BSW light guiding.
LN based 1DPhCs, which are able to sustain BSWs were designed and fabricated at the base of thin film LiNbO$_3$ membrane and on a glass support.

We have introduced the concept and demonstrated the feasibility of electro-optical tuning of the BSW, which propagates on the top of 1DPhC with TFLN on top. We have designed and manufactured a 1DPhC which contains the single-crystalline X-cut LiNbO$_3$. The 1DPhC supports a BSW at 1550 nm with a remarkably long propagation length (over 3 mm) and which is able to propagate through the obstacles on the surface. For the fabricated multilayer a special electrodes deposition technique was developed. 

Additionally, we have performed preliminary sample characterization tests, which have indeed shown a change in propagation of BSW depending on applied voltage. BSW propagated through the 12 $\mu m$ wide and 200 $\mu m$ long waveguide with 2D PhC in the middle and decoupled on the grating in the end of the waveguide. All the nano-structuring on the TFLN surface of the multilayer was made by FIB milling.
 
Finally, we have described the objectives for a future work. We propose to improve BSW based optical trapping and electro-optical modulation. Also a nonlinear properties of 1DPhC with TFLN is an interesting objective for future studies. For tunable devices we suggest to explore new materials as a part of the multilayer. For example, VO$_2$ can be a very interesting candidate for this study.

%Vous pouvez définir votre propre environnement pour décrire un théorème, un lem, etc.
%Ce type d'environnement doit être déclaré dans le préambule de votre document avec la
%macro \texttt{{\textbackslash}declareupmtheorem} (voir l'exemple dans le préambule de
%ce squelette).

%\begin{mytheorem}[Theorem of Everything]
%This is the theorem of Evereything.
%\end{mytheorem}

%\`A la fin de votre document, vous pourrez alors ajouter un chapitre listant les théorèmes présents dans votre document: \texttt{{\textbackslash}listofmytheorems}

%% Citation from the general bibliography
%\cite{key}

%% Citation from the PERSO bibliography
%\citePERSO{key}

%%--------------------
%% Start the end of the thesis
\backmatter

%%--------------------
%% Bibliography

%% PERSONAL BIBLIOGRAPHY (use 'multibib')

%% Change the style of the PERSONAL bibliography

%\bibliographystylePERSO{phdthesisapa}

%% Add the chapter with the PERSONAL bibliogaphy.
%% The name of the BibTeX file may be the same as
%% the one for the general bibliography.
%\bibliographyPERSO{biblio.bib}

%% Below, include a chapter for the GENERAL bibliography.
%% It is assumed that the standard BibTeX tool/approach
%% is used.

%% GENERAL BIBLIOGRAPHY

%% To cite one of your PERSONAL papers with the style
%% of the PERSONAL bibliography: \cite{key}

%% To force to show one of your PERSONAL papers into
%% the PERSONAL bibliography, even if not cited in the
%% text: \nocite{key}

%% The following line set the style of
%% the GENERAL bibliogaphy.
%% The "phdthesisapa" is a "apalike" style with the following
%% differences:
%% a) The titles are output with the color of the institution.
%% b) The name of the PhD thesis' author is underlined.
%\bibliographystyle{phdthesisapa}
%% The following line may be used in place of the previous
%% line if you prefer "numeric" citations.
%\bibliographystyle{phdthesisnum}

%% Link the GENERAL bibliogaphy to a BibTeX file.
%\bibliography{biblio.bib}

%%--------------------
%% List of figures and tables

%% Include a chapter with a list of all the figures.
%% In French typograhic standard, this list must be at
%% the end of the document.
\listoffigures

%% Include a chapter with a list of all the tables.
%% In French typograhic standard, this list must be at
%% the end of the document.
\listoftables
%% Include a chapter with a list of all the tables.
%% In French typograhic standard, this list must be at
%% the end of the document.

\chapter*{publications}

Journal Publications

\begin{itemize}
\item \textbf{T. Kovalevich}, A. Ndao, M. Suarez, S. Tumenas, Z. Balevicius, A. Ramanavicius...  T. Grosjean, M.P. Bernal, “Tunable Bloch surface waves in anisotropic photonic crystals based on lithium niobate thin films,” Optics Letters, \textbf{41}(23), 5616-5619 (2016).
\item \textbf{T. Kovalevich}, P. Boyer, M. Suarez, R. Salut, M.S. Kim, H.P. Herzig, M.P. Bernal, T. Grosjean, “Polarization controlled directional propagation of Bloch surface wave,” Optics Express, \textbf{25}(5), 5710-5715 (2017).
\item \textbf{T. Kovalevich}, M.S. Kim, D. Belharet, L. Robert, H.P. Herzig, T. Grosjean, M.P. Bernal, “Experimental evidence of Bloch surface waves on photonic crystals with thin film LiNbO3 as a top layer,” Photonics Research, \textbf{5}, 649-653 (2017).
\item M. Wang, H. Zhang, \textbf{T. Kovalevitch}, R. Salut, M.-S. Kim, M. Suarez, M.-P. Bernal, H.P. Herzig, H. Lu, T. Grosjean “Magnetic spin-orbit interaction directs Bloch surface waves,” (Submitted).
\end{itemize}	

International conferences

\begin{itemize}
\item \textbf{T. Kovalevich}, A. Ndao, M. Suarez, M.-S. Kim, H.-P. Herzig, M. Roussey, T. Grosjean, M.-P. Bernal, “Demonstration of Bloch Surface Waves in crystals with LiNbO3 thin films.” Poster presented at EOSAM 2016, Trends in Resonant Nanophotonics. European Optical Society Annual Meeting, Berlin, Germany, 26-30 September, 2016.
\item \textbf{T. Kovalevich}, P. Boyer, M.-P. Bernal, M.-S. Kim, H.P. Herzig, T. Gosjean, “Polarization controlled directional excitation of Bloch Surface Waves.” Oral presentation for OP16, Nanoscience and Engineering. Optics and Photonics 2016, San Diego, California, United States, 28 August - 1 September, 2016. \textit{SPIE officer travel grant winner}.
\item \textbf{T. Kovalevich}, , D. Belharet, L. Robert, T. Grosjean, M.-S. Kim, H.P. Herzig, M.-P. Bernal, “Photonic crystals for Bloch surface wave propagation based on lithium niobate thin films” Oral presentation for SPIE Photonics West OPTO 2018, San Francisco, California, United States, 27 January - 1 February, 2018, (Accepted).
\item M. Wang, H. Zhang, \textbf{T. Kovalevich}, R. Salut, M.-S. Kim, M.-A. Suarez, M.-P. Bernal, H.P. Herzig, T. Grosjean, “Tunable unidirectional coupling of Bloch surface waves controlled by the magnetic field of light” Poster presentation for SPIE Photonics West OPTO 2018, San Francisco, California, United States, 27 January - 1 February, 2018, (Accepted).
\end{itemize}

\vspace{2in}

Other presentations

\begin{itemize}
\item “Lithium niobate Bloch surface waves” – Oral presentation, BSW workshop, Neuchâtel, Switzerland, 18 - 19 May, 2017.
\item “Tunable Bloch surface waves devises” – Oral presentation, Laser Seminar, ETH Zurich, Switzerland, 03 April, 2017
\item “Demonstration of Bloch Surface Waves in crystals with LiNbO$_3$ thin films” – Poster, Atelier SMYLE, Neuchatel, Switzerland, 22-23 September, 2016.
\item “Bloch Surface Waves in photonic crystals with thin film of lithium niobate” – Poster, Dijon, France, 2015
\end{itemize}

%%--------------------
%% Include a list of definitions
%\listofdefinitions

%%--------------------
%% Appendixes
%\appendix
%\part{Annexes}

%\chapter{Premier chapitre des annexes}

%\chapter{Second chapitre des annexes}

\end{document}